\documentclass{aa}  

\usepackage{graphicx}
\begin{document}

   \title{Effects of environmental gas compression on the multiphase ISM and star formation% \titlerunning
}
\titlerunning{Environmental gas compression}
   \subtitle{The Virgo spiral galaxies NGC~4501 and NGC~4567/68}

   \author{F. Nehlig\inst{1},  B. Vollmer\inst{1}, \and J.~Braine\inst{2}}      
   \institute{CDS, Observatoire astronomique de Strasbourg, UMR 7550, 11, rue de l'universit\'e, 67000 Strasbourg, France \and
     Univ. Bordeaux, Laboratoire d'Astrophysique de Bordeaux, UMR 5804, 33270 Floirac, France}

%\\ \email{francois.nehlig@astro.unistra.fr}

   \date{Received ; accepted }

% \abstract{}{}{}{}{} 
% 5 {} token are mandatory
 \abstract{The cluster environment can affect galaxy evolution in different ways: via ram pressure stripping or by gravitational perturbations caused by galactic encounters. Both kinds of interactions can lead to the compression of the interstellar medium  (ISM) and its associated magnetic fields, causing an increase in the gas surface density and the appearance of asymmetric ridges of polarized radio continuum emission. New IRAM 30m HERA CO(2-1) data of NGC~4501, a Virgo spiral galaxy currently experiencing ram pressure stripping, and NGC~4567/68, an interacting pair of galaxies in the Virgo cluster, are presented. 
We find an increase in the molecular fraction where the ISM is compressed. The gas is close to self-gravitation in compressed regions. This leads to an increase in gas pressure and a decrease in the ratio between the molecular fraction and total ISM pressure. The overall Kennicutt Schmidt relation based on a pixel-by-pixel analysis at $\sim$1.5 kpc resolution is not significantly modified by compression. However, we detected continuous regions of low molecular star formation efficiencies in the compressed parts of the galactic gas disks. The data suggest that a relation between the molecular star formation efficiency $SFE_{\rm H_{2}}=SFR/M({\rm H_2})$ and  gas self-gravitation ($R_{\rm mol}/P_{\rm tot}$ and Toomre $Q$ parameter) exists. Both systems show spatial variations in the star formation efficiency with respect to the molecular gas that can be related to environmental compression of the ISM.
An analytical model was used to investigate the dependence of $SFE_{\rm H_{2}}$ on self-gravitation. The model correctly reproduces the correlations between $R_{\rm mol}/P_{\rm tot}$, $SFE_{\rm H_{2}}$, and $Q$ if different global turbulent velocity dispersions are assumed for the three galaxies. We found that variations in the $N_{\rm H_2}/I_{\rm CO}$ conversion factor can mask most of the correlation between $SFE_{\rm H_{2}}$ and the Toomre $Q$ parameter.     
Dynamical simulations were used to compare the effects of ram pressure and tidal ISM compression. These models give direct access to the volume density.
We conclude that a gravitationally induced ISM compression has the same consequences as ram pressure compression: (i) an increasing gas surface density, (ii) an increasing molecular fraction, and (iii) a decreasing $R_{\rm mol}/P_{\rm tot}$ in the compressed region due to the presence of nearly self-gravitating gas. The response of $SFE_{\rm H_{2}}$ to compression is more complex. While in the violent ISM-ISM collisions (e.g., Taffy galaxies and NGC~4438) the interaction makes star formation drop by an order of magnitude, we only detect an $SFE_{\rm H_{2}}$ variation of $\sim 50$\,\%  in the compressed regions of the three galaxies. We suggest that the decrease in star formation depends on the ratio between the compression timescale and the turbulent dissipation timescale.
In NGC~4501 and NGC~4567/68 the compression timescale is comparable to the turbulent dissipation timescale and only leads to  minor changes in the molecular star formation efficiency.
}


   \maketitle
%
%________________________________________________________________

\section{Introduction}

Environment plays an important role in galaxy evolution, affecting the gas content and star formation of cluster galaxies. Wide statistical studies of galaxy evolution in group and cluster environments (Peng et al. 2010) have pointed out that a dense environment quenches the star formation rate (SFR) on long timescales (several Gyr). Nevertheless, on a shorter timescale (hundreds of Myr) the interstellar medium (ISM) can be compressed by the interactions and thus the SFR can be enhanced during this relatively short timescale.

Two types of interactions that can modify the SFR in dense environments can be distinguished. The first type includes  gravitational interactions such as flybys, galaxy-galaxy encounters, or galaxy harassment (Moore et al. 1996,1999). This class of interactions affects both the stellar and the gaseous contents of galaxies. Tidal interactions can be compressing, and in this way can enhance both the gas and the stellar surface densities. The second type includes hydrodynamical interactions between the tenuous and hot ($n \sim 10^{-4}$~cm$^{-3}$, $T \sim 10^{7}$~K; Sarazin 1986) intra-cluster gas and ISM of galaxies, known as ram pressure striping (Gunn \& Gott 1972). This interaction is most efficient when the galaxy enters the central part of a cluster where its velocity relative to the cluster mean is highest and the intra-cluster medium is dense (Vollmer et al. 2001). 

The nearby Virgo cluster is an ideal laboratory to study the effect of a dense environment on galaxies and thus to explore the physical process of star formation. The distance of $17$~Mpc allows us to reach a resolution of about $1$~kpc with radio telescopes.
The spiral population in the Virgo cluster shows disturbed and truncated H{\sc i} disks (Cayatte et al. 1990, 1994; Chung et al. 2009) caused by tidal interactions and ram pressure stripping. Ram pressure stripping naturally leads to truncated gas disks. During a ram pressure stripping event the ISM can be compressed (e.g., Vollmer 2009),  especially when ram pressure increases and the vector of the galaxy's motion is parallel to the galactic disk, i.e., the ram pressure wind is edge-on.

Otmianowska-Mazur \& Vollmer et al. (2003) have shown that the polarized radio continuum emission increases significantly whenever the ISM is compressed or sheared. In most cases shear motions can be  detected in the H{\sc i} velocity field (e.g., NGC~4535, Chung et al. 2009). Therefore, a ridge of enhanced polarized emission located at the outer edge of the gas disk in the absence of a sheared velocity field is a clear sign of gas compression (Vollmer et al. 2007, 2012, 2013). We  identified such ridges in the Virgo spiral galaxies NGC~4501 and NGC~4568. Whereas NGC~4501 is affected by ram pressure (Vollmer et al. 2008), NGC~4568 is tidally interacting with its close neighbor, NGC~4567.

\begin{figure}
   \centering
\includegraphics[width=9cm]{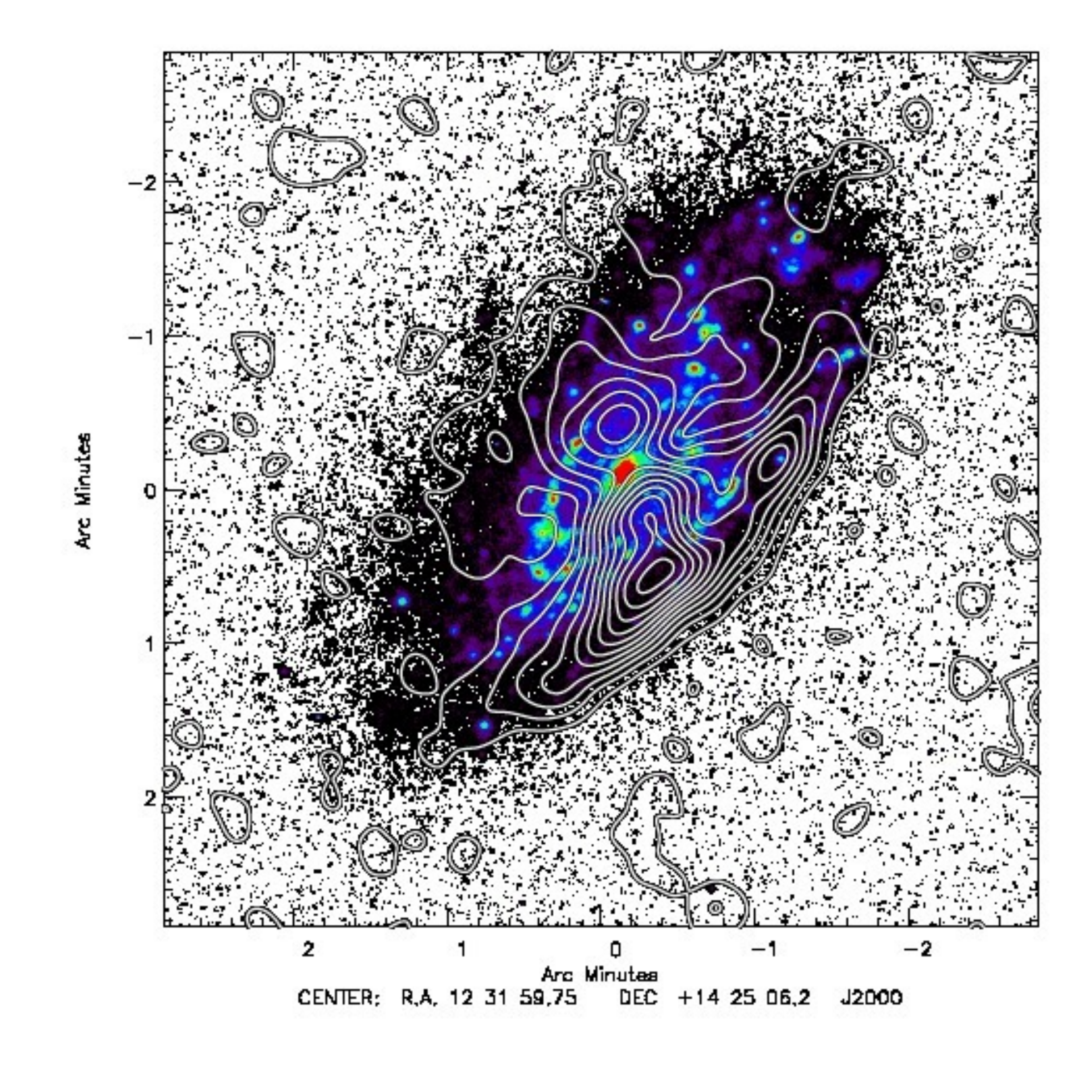}
  \caption{NGC~4501: 6~cm polarized emission Vollmer et al. 2003) on H$_{\alpha}$ emission (from GOLDMINE Gavazzi et al. 2003).}
\label{4501Ha+pol}%
\end{figure}

Vollmer et al. (2008) has found an enhanced polarized emission in the southeastern region of NGC~4501 (see Fig.~\ref{4501Ha+pol}) due to a nearly edge-on intracluster medium (ICM)-ISM ram pressure stripping event. The compressed region is oriented toward M87 indicating that this galaxy is falling into the dense region of the cluster for the first time. Contrary to gravitational perturbations, ram pressure stripping does not perturb the stellar disk. The stellar disk of NGC~4501 shows symmetric spiral arms with obscured dust inter arm features (Fig.~\ref{4501G+H2}).

\begin{figure}
   \centering
\includegraphics[width=9cm]{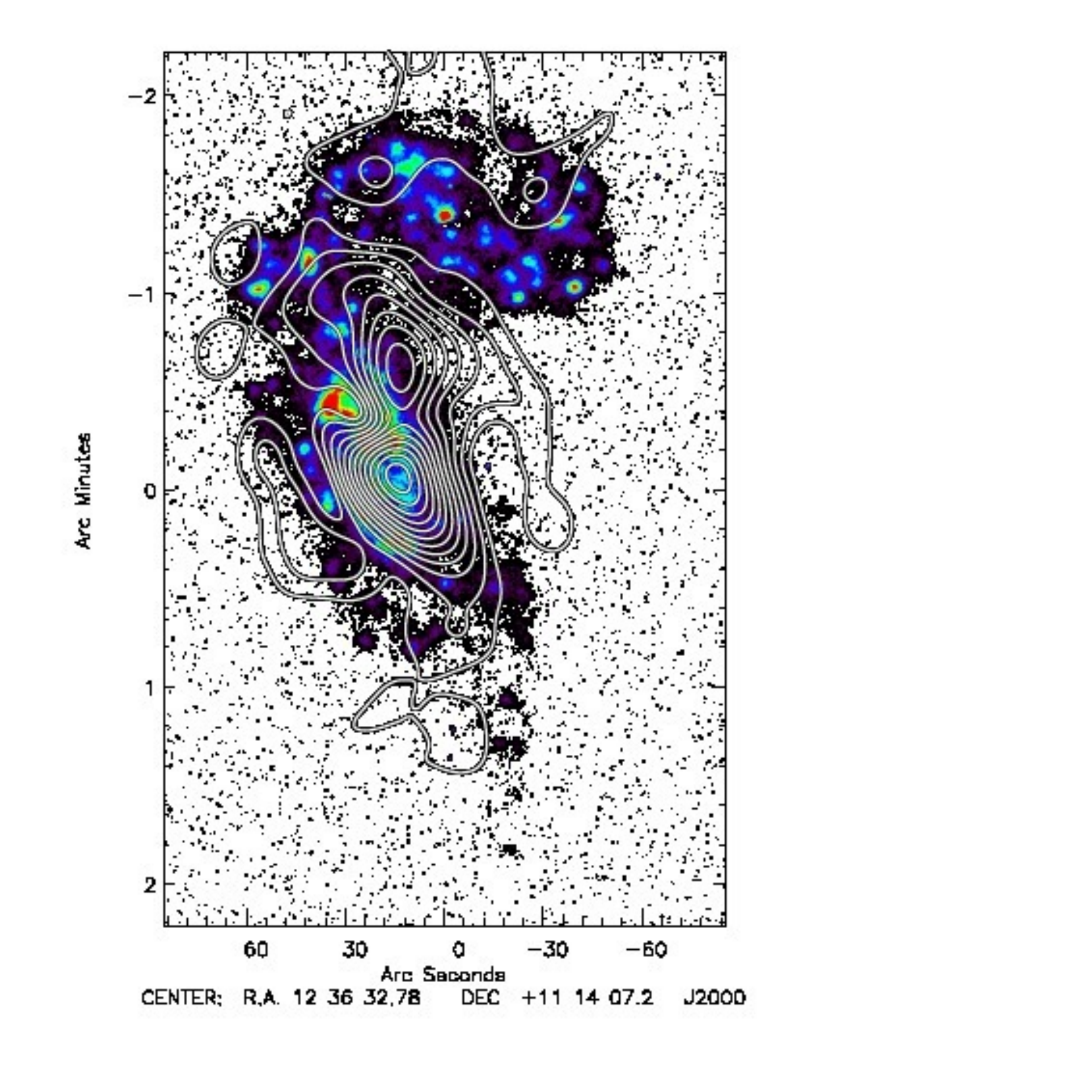}
  \caption{NGC~4567/68: 6~cm polarized emission (Vollmer et al. 2013) on H$_{\alpha}$ emission (from GOLDMINE Gavazzi et al. 2003).}
\label{456768Ha+pol}%
\end{figure}

NGC~4567/68 are two close Virgo spiral galaxies. The centers of NGC~4567 and NGC~4568 are separated in projection by $\sim6$ kpc and have comparable radial velocities ($v_r=2255$~km\,s$^{-1}$) suggesting that they are gravitationally bound. The absence of any obvious large-scale optical tidal features in either galaxy prompted Iono et al. (2005) to qualify this system as a young interacting pair.
The stellar component of NGC~4568 (Fig.~\ref{456768G+H2}) is only mildly disturbed and that of NGC~4567 has a faint external ring structure, indicating that the interaction is in an early stage. The interaction has not destroyed the inner spiral structures of NGC~4568 and NGC~4567, but does affect the outer stellar disk. 
In the northern interaction region the H{\sc i} velocities of NGC~4567 and NGC~4568 are comparable. A low surface brightness H{\sc i} tail extending to the north is detected in NGC~4568 (Chung et al. 2009).
While NGC~4567 has a fairly normal symmetric large scale magnetic field structure (Fig.~\ref{456768Ha+pol}), NGC~4568 shows an asymmetric 6cm polarized emission with its maximum in the south and northwest of the galaxy center (Vollmer et al. 2013). The tidal compression region with an enhanced surface brightness of H{\sc i} and 6cm polarized emission is located northwest of the galaxy center.

In this article we investigate the influence of large-scale ($\sim 1$~kpc) compression on ISM properties (density and molecular fraction) and its efficiency in forming stars
based on new IRAM 30m CO$(2-1)$ observations of NGC~4501 and NGC~4567/68 and existing GALEX, Spitzer, and Herschel data. The basic properties of the two systems are given in Tables~\ref{tab1} and \ref{tab2}.

This article is organized in the following way. In Section~\ref{Obs} the IRAM 30m CO$(2-1)$ observations of NGC~4501 and NGC~4567/68 and the data reduction process are presented. The observational results are given in Section~\ref{res} and discussed in Section \ref{dis}. In Section~\ref{analy} the analytical model of Vollmer \& Leroy (2011) is used to interpret our observations. In section \ref{model} the results of a dynamical model of NGC~4501 and NGC~4567/68 are compared to all existing observations. We give our conclusions in Section~\ref{conclu}. We use a distance of $D=17$~Mpc for the Virgo cluster.

\begin{table}
\caption{Integrated properties NGC~4501 and NGC~4567/68 }           % title of Table
\label{tab1}    
\centering      
\begin{tabular}{@{}l c c c c@{}}       
\hline\hline
Galaxy      & SFR                           & $SFE_{\rm H_{2}}$$^{\star}$    & M$_{\rm H_{2}}$    & M$_{\rm HI}$   \\     
            &   $($M$_{\odot}$yr$^{-1}$$)$  &$($Gyr$^{-1}$$)$                 & $($M$_{\odot}$$)$  &$($M$_{\odot}$$)$\\
\hline 
NGC~4501    &        1.94                   &         0.61                   &   $3.57\times10^9$ & $1.78\times10^9$\\  
NGC~4567    &        1.38                   &         0.72                   &   $5.97\times10^8$ & $2.90\times10^8$\\  
NGC~4568    &        2.34                   &         0.74                   &   $2.04\times10^9$ & $1.23\times10^9$\\

\hline                                   %inserts single line
\end{tabular}
\tablefoot{$^{\star}$ From the chi-square fit in the SFR-$\Sigma_{\rm H_{2}}$ relation. The central part of NGC~4501 was excluded from the fit. }
\end{table}

\begin{table*}
\caption{Basic properties of NGC~4501, NGC~4567/68, and NGC~4321 }           % title of Table
\label{tab2}    
\centering      
\begin{tabular}{@{}l r r r r r r r r@{}}      
\hline\hline                 
Galaxy   & Gal type $^{1}$ & Inc $^{2}$ & Pa $^{3}$  &     V$_{r}$ $^{4}$    &    R$_{25}$ $^{5}$ & B $^{6}$ & d$_{M87}$ $^{7}$ & $V_{rot}$ $^{8}$ \\    % table heading 
          &               &  (deg)      &    (deg)   &     (km s$^{-1}$)    &           (')    & (mag)      & (Mpc) &    (km s$^{-1}$)  \\   
\hline                                                     

NGC~4501  & SA(rs)b       & $62.9$      & $137.7$    &  $+2282$   &    $4.35 $    &     $ 10.19$ $\pm$ $0.13$&  $ 0.55$     &        $272.1$ $\pm $ $5.0$     \\
NGC~4567  & SA(rs)bc      & 39.4      & 90.0      &  $+2255$   &    1.37          & 12.11 $\pm$ 0.20         &   0.49       &    222.8 $\pm$ 6.9     \\
NGC~4568  & SA(rs)bc      & 67.5      & 32.3     &  $+2255$   &    2.13           &      11.69 $\pm$ 0.12     &   0.49      & 165.5 $\pm$ 7.8     \\NGC~4321  & SABb          & 23.4      & 0.0      &   $+1574$  &     $1.23$        &   $10.02\pm 0.13$        &   1.19      &   283.7 $\pm$ 4.8       \\
\hline                                   %inserts single line
\end{tabular}
\tablefoot{(1) Type of galaxy, (2) Inclination of the galaxy, (3) Position angle of the galaxy, (4) Radial velocity, (5) half light radius in the B band, (6) Total B-band magnitude, (7) projected distance to the central Virgo galaxy M87, (8) Maximum rotation velocity corrected for inclination}
\end{table*}

\section{Observations and data reduction \label{Obs}}

NGC~4501 and NGC~4567/68 CO(2-1) were observed in December 2012 at the IRAM 30m single-dish telescope at Pico Veleta, Spain. The CO(2-1) emission line was mapped with the HERA multi-beam heterodyne receiver (Schuster et al. 2004) in on-the-fly mode. The scans were realized along the minor and major axes of the galaxies with a scanning speed of $5"$s$^{-1}$. The scans were done with two derotator positions (+90 and -90 deg), thus the same position in the sky was scanned with two different pixels of HERA. The resulting maps were oversampled with  spectra taken at least every $3.8"$.

The WILMA backend autocorrelator with a spectral resolution of 2~MHz and a bandwidth of 1024~MHz was used. Four $2.6$~km\,s$^{-1}$ velocity channels were binned together yielding a 10.4~km\,s$^{-1}$ channel separation. We observed each system for about 20~h, to reach an average rms noise level of 8~mK and 7~mK in the 10.4~km\,s$^{-1}$ channel for NGC~4501 and NGC~4567/68, respectively. The calibration was done with MIRA, using the TIME mode calibration, which decreases the rms noise level compared to the default calibration mode AVER. Bad spectra were filtered by comparing the theoretical system temperature $\sigma_{theo}=T_{sys}/\sqrt{\Delta V t}$ ($\Delta V=10.4$~km s$^{-1}$ the channel separation and $t=0.5$~s the integration time) to the real noise level. Spectra with an rms noise in excess of $1.1\times \sigma_{theo}$ were rejected. A velocity window was selected for NGC~4501 ($1950$-$2700$~km\,s$^{-1}$) and for NGC~4567/68 ($2000$-$2550$~km\,s$^{-1}$). A constant baseline was adjusted outside the velocity window and subtracted from the spectrum. A data cube at a resolution of $12''$ was created using the CLASS GILDAS xy\_map routine with a pixel size of $3"$. 

In order to compute the total intensity map, the atomic H{\sc i} data from VIVA (Chung et al. 2009) was used. For each spectrum, we selected the velocity window where the H{\sc i} flux was higher than 4.5 times the rms of the H{\sc i} spectrum. For NGC~4501, no H{\sc i} emission was detected near the galactic center. In this case, the CO(2-1) window was computed by fitting a Gaussian to the CO spectra and selecting a spectral window from $\mu - 3.2\sigma$ to $\mu + 3.2\sigma$ , where $\sigma$ and $\mu$ are the standard deviation and peak position of the Gaussian. At the center of NGC~4501, the CO(2-1) emission is strong and a Gaussian shape fits  the CO(2-1) line well. After windowing the CO(2-1) line, a third-order baseline was subtracted from the CO spectra. First- and second-order baseline subtractions were tried on both NGC~4501 and NGC~4567/68 and gave a higher rms noise than the third-order baseline subtraction. The rms maps of NGC~4501 and 4567/68 did not show any systematic residuals (Fig.~\ref{RMS4501}).

For comparison with a symmetric spiral galaxy, we used the face-on Virgo spiral galaxy NGC~4321 since both CO(2-1) and VLA H{\sc i} data were available. The CO(2-1) data cube is taken from the HERACLES data (Leroy et al. 2009) with a spatial resolution of $13"$ and a typical rms of $20-40$~mK in the $10.4$~km\,s$^{-1}$ channel. The H{\sc i} data are from the VIVA survey (Chung et al. 2009) with a beam size of $30''$.

\section{Results}
\label{res}

\subsection{Molecular gas}

Following Leroy et al. (2008) the molecular gas surface density $\Sigma_{{\rm H}_{2}}$ was derived using a galactic CO-to-H$_2$ conversion factor ${\rm X}_{{\rm co}}=2 \times 10^{20} \rm cm^{-2}$ or $\alpha_{\rm co}=4.4$ M$_{\odot}$(K km/s pc$^2$)$^{-1}$ and a fixed line ratio $I_{\rm CO}(2-1)/I_{\rm CO}(1-0)=0.8$ (see equation (\ref{eqq1})). A factor of 1.36 was included to take the presence of helium into account, 
\begin{equation}
 \Sigma_{\rm H_{2}} =  \frac{4.4}{0.8}  \times I_{\rm CO}(2-1) \times \cos(i)\ {\rm M}_{\odot}{\rm pc}^{-2}, \label{eqq1}
\end{equation}
where $I_{\rm CO}$(2-1) is the integrated intensity of the CO(2-1) in K~km\,s$^{-1}$ and $i$ the inclination angle of the galaxy. The typical uncertainties in the molecular gas surface density are about $\sim1$~${\rm M}_{\odot}{\rm pc}^{-2}$, which corresponds to the $3\sigma$ level per $10$~km s$^{-1}$ channel. The implications of using a ${\rm X}_{\rm co}$ factor is discussed in Section~\ref{XCO}. 

\begin{figure}
   \centering
\includegraphics[width=9cm]{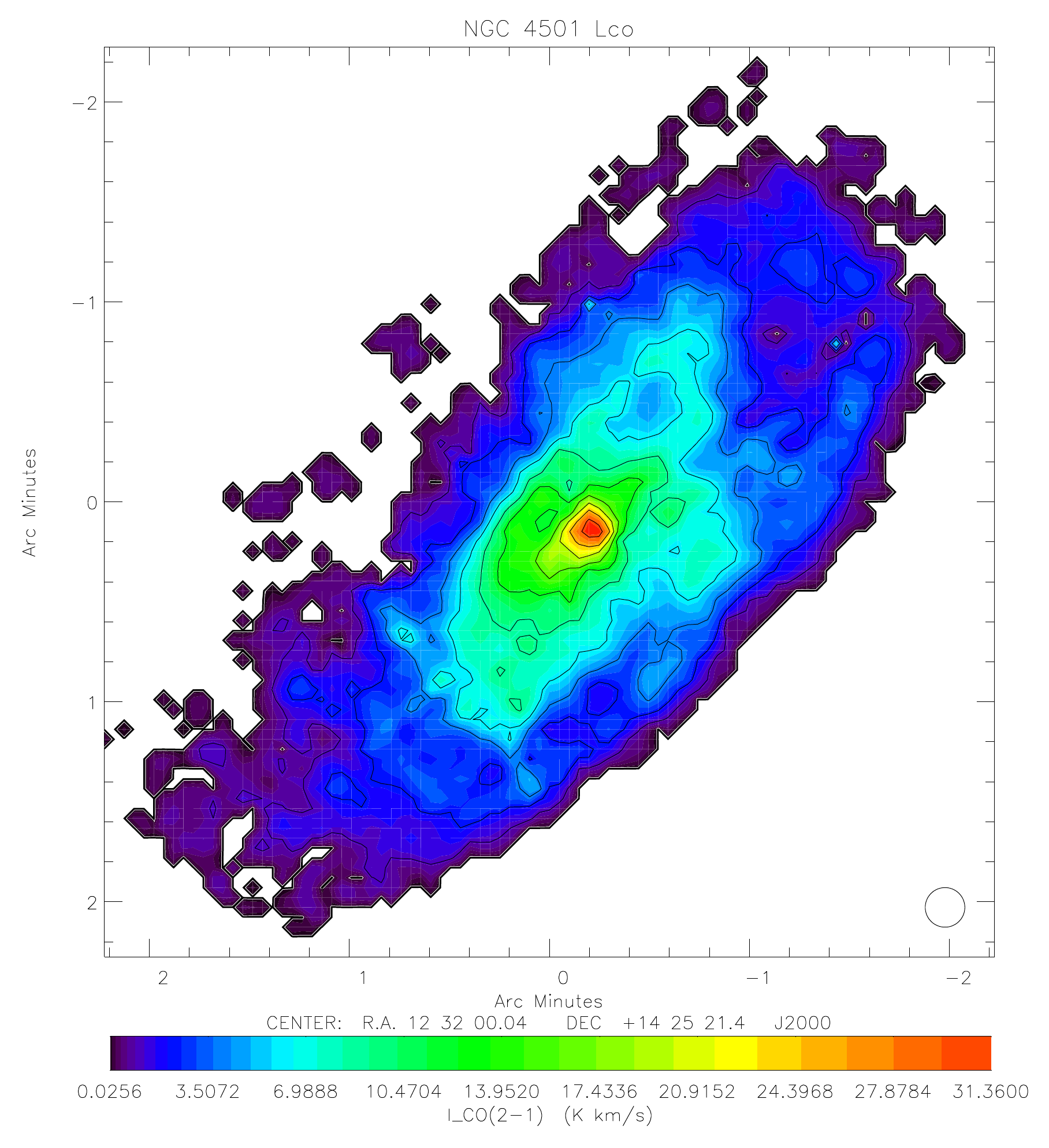}
   \caption{ NGC~4501: CO(2-1) emission map. Contour levels are 1.0, 2.25, 4.0, 6.25, 9.0, 12.25, 16.0, 20.25, 25.0, and 30.25~K~km\,s$^{-1}$. The white contour corresponds to $0.1$~K~km\,s$^{-1}$.}
\label{CO4501}%
\end{figure}

\begin{figure}
   \centering
\includegraphics[width=9cm]{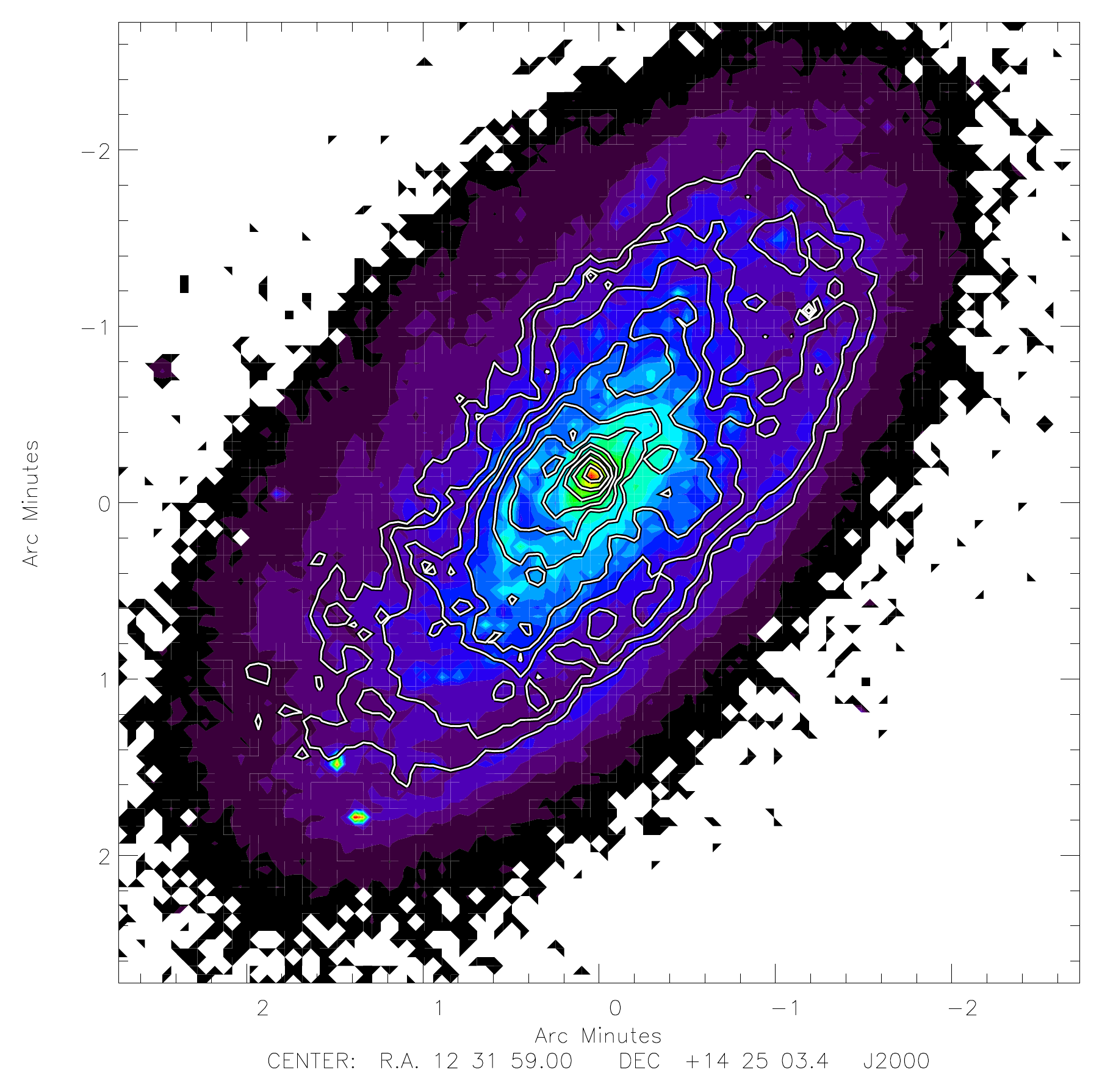}
   \caption{NGC~4501: CO(2-1) emission on SDSS g'-band image. Contour levels are 1.0, 2.25, 4.0, 6.25, 9.0, 12.25, 16.0, 20.25, 25.0, and 30.25)~K~km\,s$^{-1}$.}
\label{4501G+H2}%
\end{figure}

For NGC~4501 the CO(2-1) emission extends over a radius of $\sim 2.5'=11$~kpc (Fig. \ref{CO4501}). This is over half the optical radius R$_{25}=4.35'$. In Fig.~\ref{4501G+H2} the molecular gas is presented on the g'-band emission from SDSS. 
The emission map shows a well-defined southwestern edge, compared to the more diffuse northeastern side of the galactic disk.
The inner disk of NGC~4501 has surface densities in excess of 30~M$_{\odot}$pc$^{-2}$ (light blue contours in Fig.~\ref{CO4501}). In this region two spiral arm structures are present, the first one extending to the south and curved toward the southwestern edge, the second extending to the north and curved to the northeast. At the compressed western side a region of high surface density is found west of the galaxy center with surface density of $40$-$60$~M$_{\odot}$pc$^{-2}$.
The southeastern quadrant of the disk has an overall lower surface density than the northwestern quadrant. 

\begin{figure}
   \centering
\includegraphics[width=9cm]{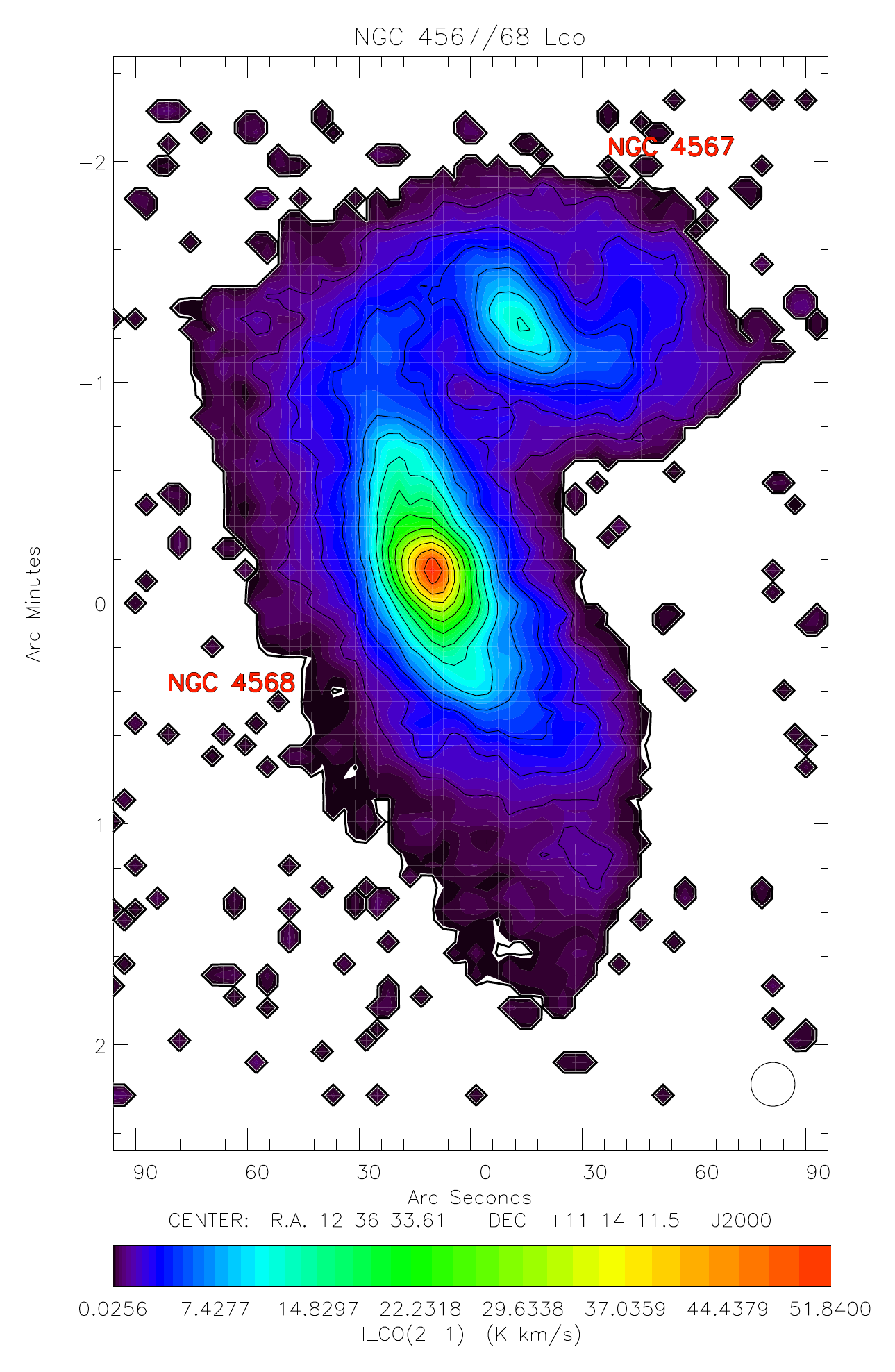}
   \caption{NGC~4567/68: CO(2-1) emission map. Contours are 1.0, 2.25, 4.0, 6.25, 9.0, 12.25, 16.0, 20.25, 25.0, 30.25, 36.0, 42.25, 49.0, 56.25, and 64.0)~K~km\,s$^{-1}$. The white contour corresponds to $0.1$~K~km\,s$^{-1}$.}
\label{CO4567/68}%
\end{figure}

The CO(2-1) emission distribution of NGC~4567 (Fig.~\ref{CO4567/68}) shows two spiral arms with high molecular gas surface densities. The first   extends to the southwest and is curved to the north; the second is shorter, extends to the north, and bends toward NGC~4568.

\begin{figure}
   \centering
\includegraphics[width=9cm]{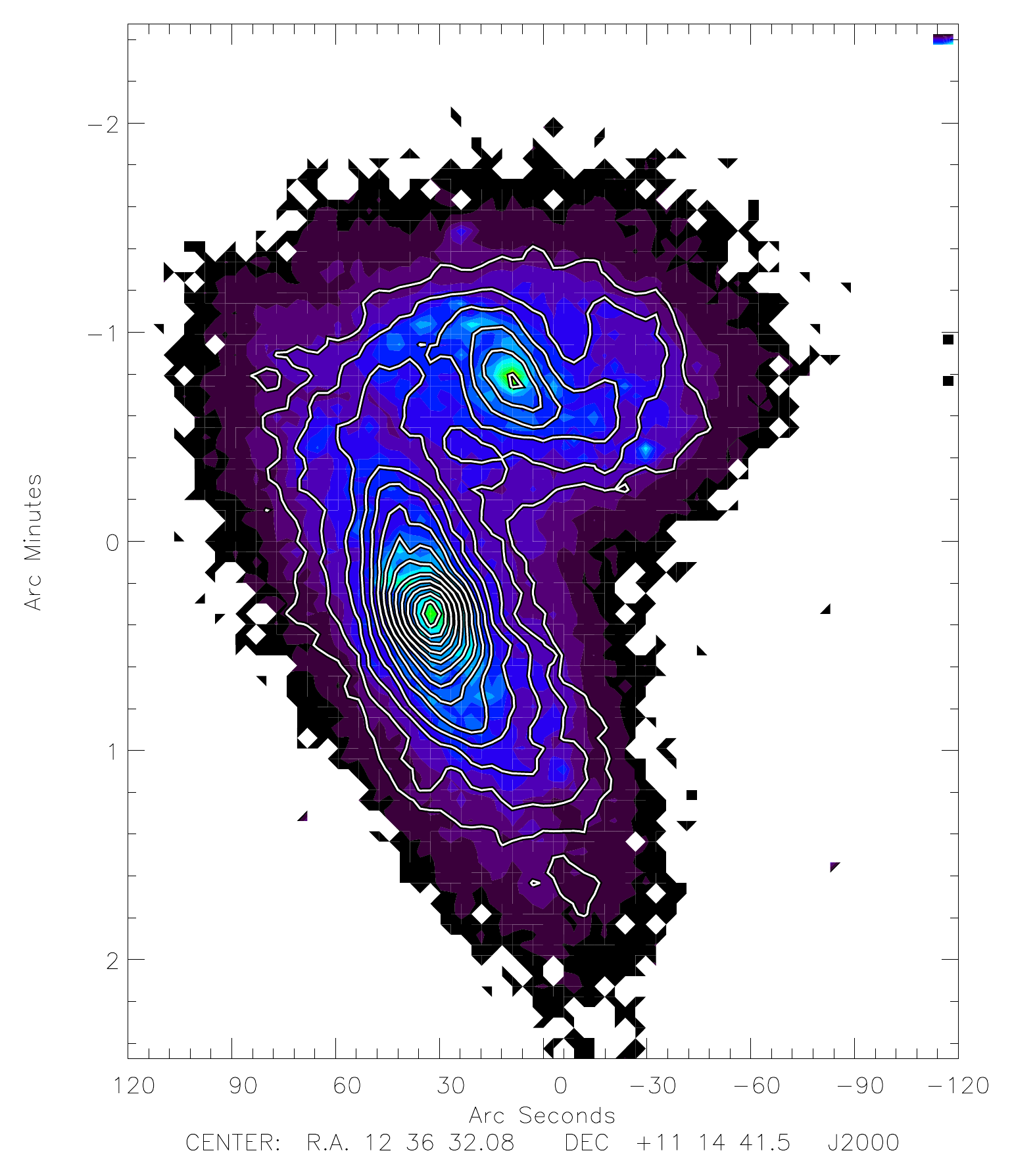}
   \caption{NGC~4567/68: CO(2-1) emission  on SDSS g'-band image. Contour levels are 1.0, 2.25, 4.0, 6.25, 9.0, 12.25, 16.0, 20.25, 25.0, 30.25, 36.0, 42.25, 49.0, 56.25, and 64.0~K~km\,s$^{-1}$.}
\label{456768G+H2}%
\end{figure}

The southern part of the CO(2-1) emission of NGC~4568 is more extended than the northern part. The northern part of the disk has higher molecular gas surface densities than the more extended southern part.
A peculiar molecular gas arm is observed in NGC~4568. It extends to the south and curves to the west. The northern arm first extends to the northeast and then bends to the west, toward NGC~4567.
The latter part of the northern arm might well be  associated with the eastern spiral arm of NGC~4567.
The molecular gas map of NGC~4567/68 is presented over the g' SDSS band emission in Fig.~\ref{456768G+H2}.

\subsection{Total gas}

Total gas surface density maps $\Sigma_{\rm g}$ were computed by combining the atomic gas surface density $\Sigma_{\rm H_{\rm I}}$ and the molecular gas surface density $\Sigma_{\rm g} = \Sigma_{\rm H_{\rm I}} +\Sigma_{\rm H_{2}}$. The atomic gas surface density $\Sigma_{\rm H_{\rm I}}$ was computed for these two galactic systems using VLA ${\rm H}_{\rm I}$ data from Chung et al. 2008. Following Leroy et al. (2008),
\begin{equation}
\Sigma_{{\rm H}_{\rm I}} = 0.020 \times I_{21 {\rm cm}} \times \cos(i)  \ {\rm M}_{\odot}{\rm pc}^{-2} \ ,  
\label{eqq2}
\end{equation}
where ${\rm I}_{21\rm cm}$ is the integrated intensity of the 21cm line in K~km\,s$^{-1}$ and $i$ the inclination angle of the galaxy. Eq.~(\ref{eqq2}) includes a factor of 1.36 to reflect the presence of helium.
The typical uncertainties in the atomic gas surface density obtained here are $\sim 0.5 {\rm M}_{\odot}{\rm pc}^{-2}$, which corresponds to the $3\sigma$ level per $10$~km s$^{-1}$ channel. 
Before summing, the CO observations of NGC~4501 and NGC~4567/68 were convolved to the H{\sc i} resolution ($17"$ for NGC~4501 and NGC~4567/68, $30"$ for NGC~4321).  

\begin{figure}
   \centering
\includegraphics[width=9cm]{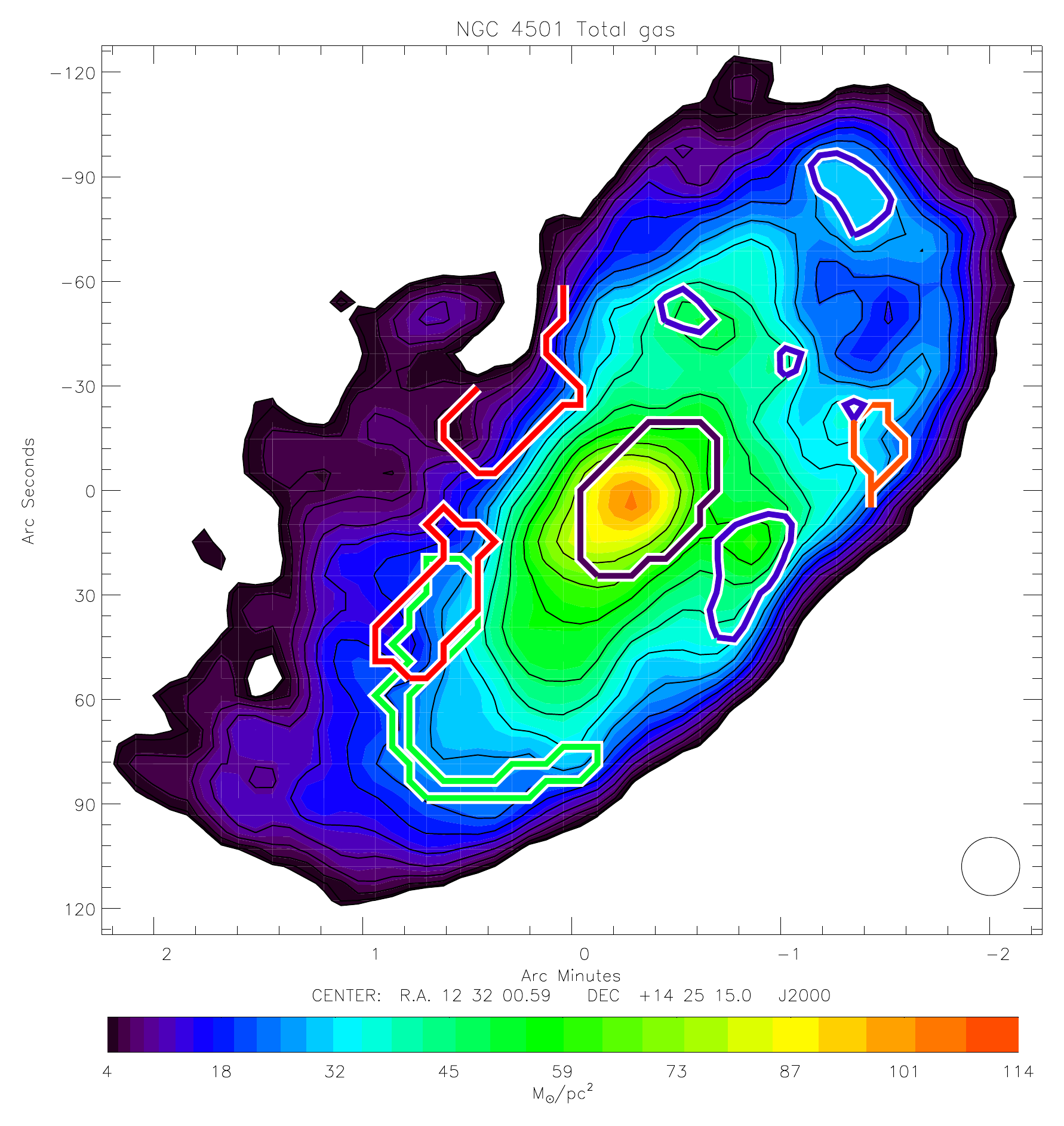}
   \caption{NGC~4501 total gas surface density $\Sigma_{\rm g}=\Sigma_{{\rm H}_{\rm I}}+\Sigma_{{\rm H}_{2}}$. The thin black contour levels are 4.0, 4.36, 5.44, 7.24, 9.76, 13.0, 16.96, 21.64, 27.04, 33.16, 40.0, 47.56, 55.84, 64.84, and 74.56~M$_{\odot}$pc$^{-2}$. The resolution is $17"$. Blue contour: high atomic gas surface density regions. Green contour: southern ``U'' shaped region with low $SFE_{\rm H_{2}}$. Orange contour: northern low $SFE_{\rm H_{2}}$ region. Red contour: high $R_{\rm mol}
/P_{\rm tot}$ region. Purple contour: central region defined from the stellar surface density excluded from the fit. The beam size is shown in the lower right corner.}
\label{totgas++}%
\end{figure}

The total gas surface density maps of NGC~4501 (Fig.~\ref{totgas++}) shows three low surface density spirals arms on the eastern side of the disk, curved to the northwest. Moreover, the southwestern side of NGC~4501 has a well-defined edge as seen in $\Sigma_{{\rm H}_{2}}$. The northern part of the NGC~4501 outer disk shows higher gas surface density then the southern part.

\begin{figure}
   \centering
\includegraphics[width=9cm]{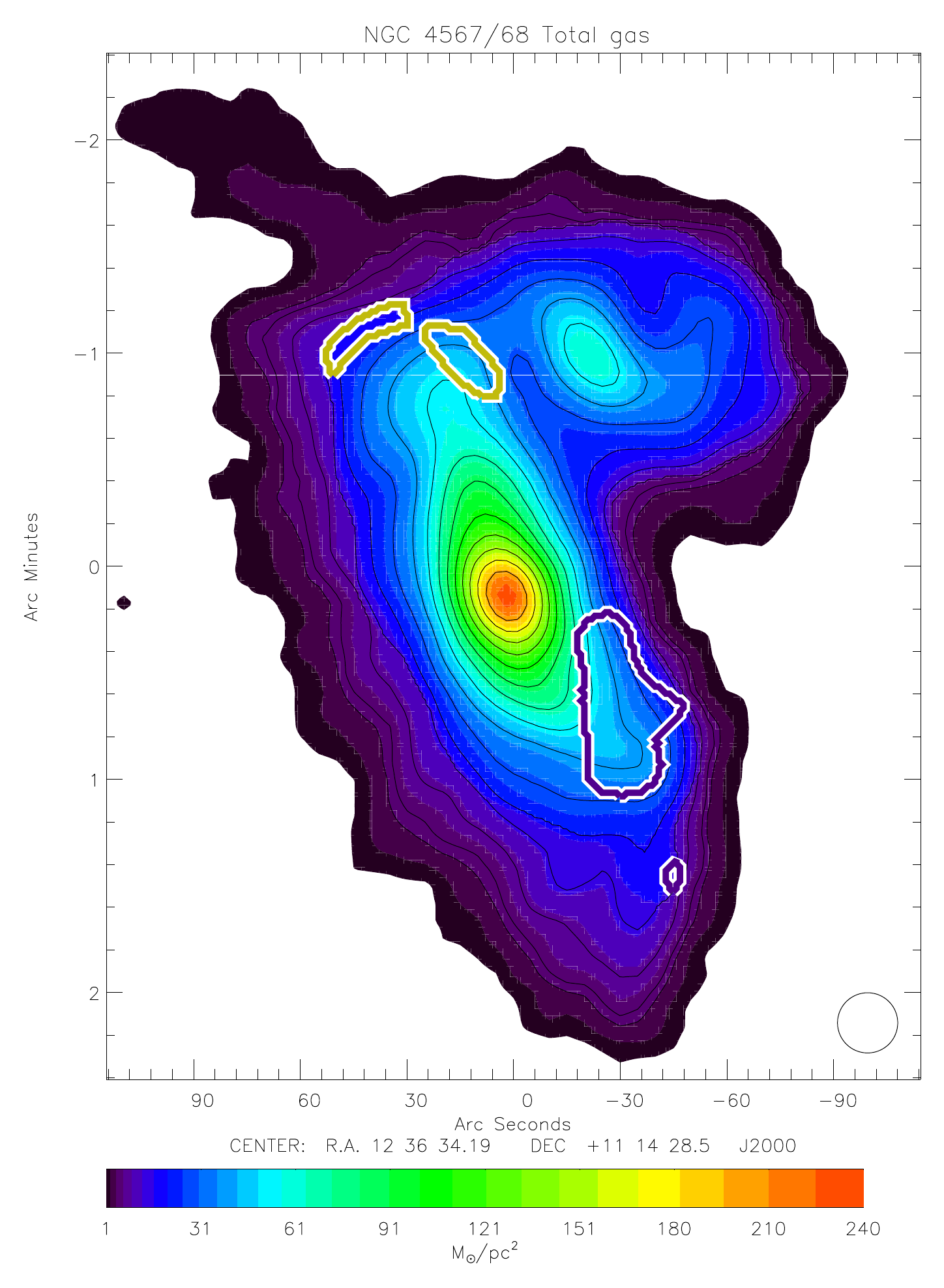}
   \caption{NGC~4567/68 total gas surface density map. The thin black contours are 4, 5, 8, 13, 20, 29, 40, 53, 68, 85, 104, 125, 148, 173, and 200~M$_{\odot}$pc$^{-2}$. The resolution is $17''$. Yellow contour: northern low $SFE_{\rm H_{2}}$ region. Purple contour: southern low $SFE_{\rm H_{2}}$ region. The beam size is shown in the lower right corner.}
\label{totgas4567}%
\end{figure}

NGC~4567/68 total gas surface density is dominated by the molecular gas surface density inside R$ \sim 0.5 \times R_{25}$. Therefore, the total gas surface density distribution of NGC~4567 and NGC~4568 show the same spiral arms as in molecular gas (Fig.~\ref{totgas4567}).

The total gas surface density of NGC~4568 is more extended to the south; it has a  high surface density southern arm curved to the west and a southern low surface density component curved to the east, which causes an east-west asymmetry in the south of NGC~4568. 

The northern low surface density ($\sim 2$ M$_{\odot}{\rm pc}^{-2}$) tail of NGC~4567/68 does not have a detected CO counterpart.

\subsection{Molecular fraction}

\begin{figure}
   \centering
\includegraphics[width=9cm]{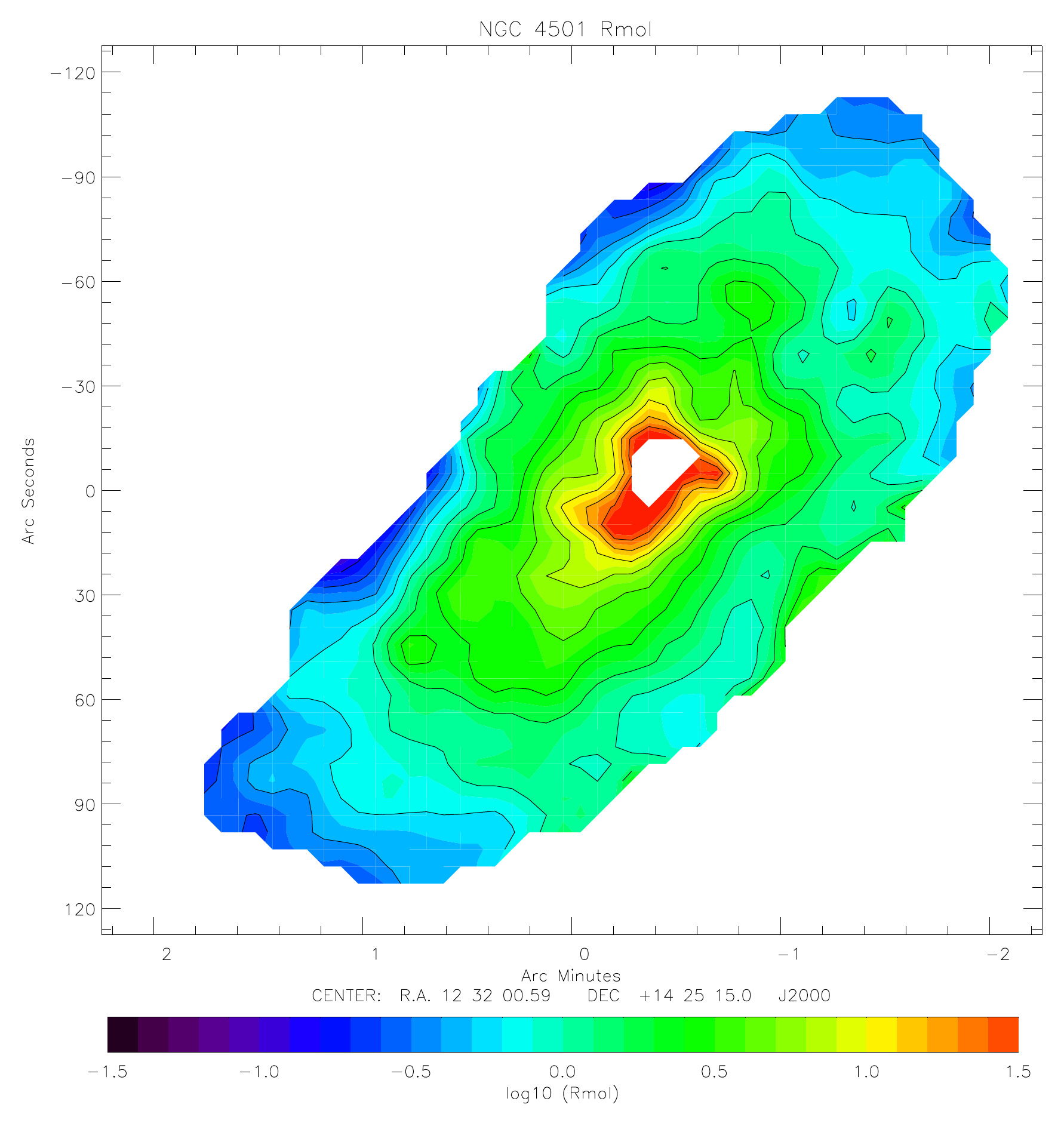}
   \caption{NGC~4501 molecular fraction $\rm R_{\rm mol}=\Sigma_{\rm H_{2}}/\Sigma_{\rm H_{\rm I}}$. The resolution is $17"$. Contour levels are 0.04, 0.06, 0.1, 0.16, 0.25, 0.4, 0.63, 1, 1.58, 2.51, 3.98, 6.31, 10, 15.84, and 25.11.}
\label{rmol4501}%
\end{figure}

\begin{figure}
   \centering
\includegraphics[width=9cm]{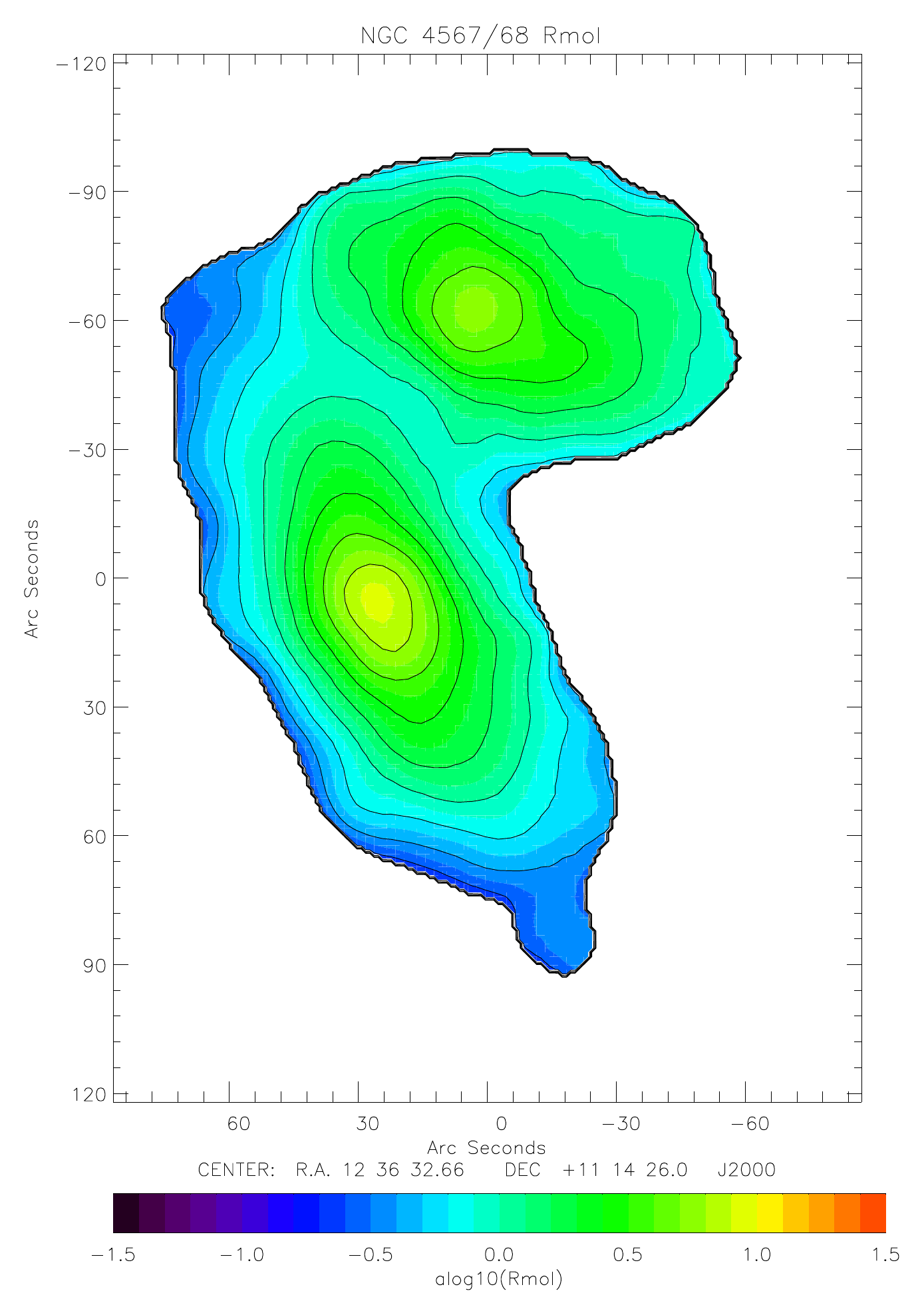}
   \caption{NGC~4567/68 molecular fraction $\rm R_{\rm mol}=\Sigma_{\rm H_{2}}/\Sigma_{\rm H_{\rm I}}$. The resolution is $17"$. Contour levels are 0.04, 0.06, 0.1, 0.16, 0.25, 0.4, 0.63, 1, 1.58, 2.51, 3.98, 6.31, 10, 15.84, and 25.11.}
\label{rmol4567}%
\end{figure}

The molecular fraction $\rm R_{\rm mol}=\Sigma_{\rm H_{2}}/\Sigma_{\rm H_{\rm I}}$ was derived using the convolved $17"$ resolution data sets.
This ratio is presented for NGC~4501 and NGC~4567/68 in Fig.~\ref{rmol4501} and Fig.~\ref{rmol4567}.

For NGC~4501 the molecular fraction increases toward the center where no atomic gas is present. The molecular fraction is significantly higher in the western ridge of compressed gas ($\rm R_{\rm mol} \sim 1$) than  at similar radii in the northeast, east, and southwest of the galaxy ($R_{\rm mol} \sim 0.1$).

In NGC~4567/68 the molecular fraction increases toward the center of the disk, where the molecular gas surface density is about 10 times that of the atomic gas surface density. Between the two galaxies the molecular fraction is high (about one). However, in this region the molecular fraction is sensitive to projection effects. In NGC~4567 the molecular fraction is higher in the spiral arms. The southern outer gas disk has a very low molecular fraction ($R_{\rm mol} < 0.1$).

\subsection{Interstellar gas pressure P$_{\rm ext}$}

Elmegreen (1989) derived the total mid-plane gas pressure $P_{\rm tot}$ from the numerical solution of hydrostatic equilibrium for a gas plus star disk,
\begin{equation}
P_{\rm tot}= \frac{\pi}{2} G \Sigma_{\rm g} (\Sigma_{\rm g} + \Sigma_{\star} \frac{\sigma_{\rm g}}{\sigma_{\star,{\rm z}}})=\rho_{\rm g} {\rm v}_{\rm turb}^2, \label{eqq5}
\end{equation}
where $\rho_{\rm g}$ is the gas density, ${\rm v}_{\rm turb}$ the gas velocity dispersion, $G$ the gravitational constant, $\Sigma_{\star}$ the stellar surface density, $\sigma_{\rm g}$  the gas velocity dispersion, and $\sigma_{\star,{\rm z}}$  the vertical stellar velocity dispersion. This expression is accurate to within 10\% (Elmegreen 1989).

The $3.6$~$\mu {\rm m}$ emission, observed with the IRAC instrument on the {\em Spitzer} spatial telescope, was used to compute the stellar surface density of these two galactic systems. 
Following Leroy et al. (2008),
%The conversion, used here, from  ${\rm I}_{3.6\mu {\rm m}}$ to $\Sigma_{\star}$ was described in detail in Leroy et al. 2008 (appendix C) and summarised here in equation (\ref{eqq3}) (Leroy et al. 2008 equation C1).
\begin{equation}
 \Sigma_{\star}= \gamma^{\rm K}_{\star} \langle{  \frac{\rm I_{\rm K}}{\rm I_{3.6\mu {\rm m}}}  \rangle} {\rm I}_{3.6\mu {\rm m}}\cos(i) = 280\, {\rm I}_{3.6\mu {\rm m}} \cos(i) \ {\rm M}_{\odot}{\rm pc}^{-2}, \label{eqq3} 
\end{equation}
where $\gamma^{\rm K}_{\star}=0.5$ M$_{\odot}/{\rm L}_{\odot,{\rm K}}$ is the K-band mass-to-light ratio, I$_{\rm K}$ is the K-band intensity in MJy~ster$^{-1}$, and I$_{3.6\mu {\rm m}}$ is the $3.6$ $\mu$m intensity in MJy~ster$^{-1}$.
A fixed ratio I$_{\rm K}/{\rm I}_{3.6\mu {\rm m}}=1.818$ and a solar K-band magnitude of 3.28~mag (Binney \& Merrifield 1998) were assumed.     
The vertical stellar velocity dispersion can be derived from the stellar scale length ${\rm l}_{\star}$ (following Leroy et al. (2008) and reference therein):
\begin{equation}
\sigma_{\star,z}=\sqrt{\frac{2 \pi G l_{\star}}{7.3}}\Sigma_{\star}^{\frac{1}{2}}. \label{eqq6}
\end{equation}
Eq. (\ref{eqq6}) assumes (1) a constant exponential stellar scale height h$_{\rm star}$ within the disk, (2) l$_{\star}/{\rm h}_{\star}=7.3 \pm 2.2$ (Kregel et al. 2002), and (3) an isothermal disk in the vertical direction (see van der Kruit 1988, van der Kruit \& Searle 1981).
We derived the stellar scale length of NGC~4501, NGC~4567/68, and NGC~4321 from radial profiles of their stellar surface density. The inner bulge component was excluded for the fit. We found l$_{\star}=3.3$ kpc for NGC~4501, l$_{\star}=1.6$ kpc for NGC~4567, l$_{\star}=2.1$ kpc for NGC~4568, and l$_{\star}=4.9$ for NGC~4321.

\begin{figure}
   \centering
\includegraphics[width=9cm]{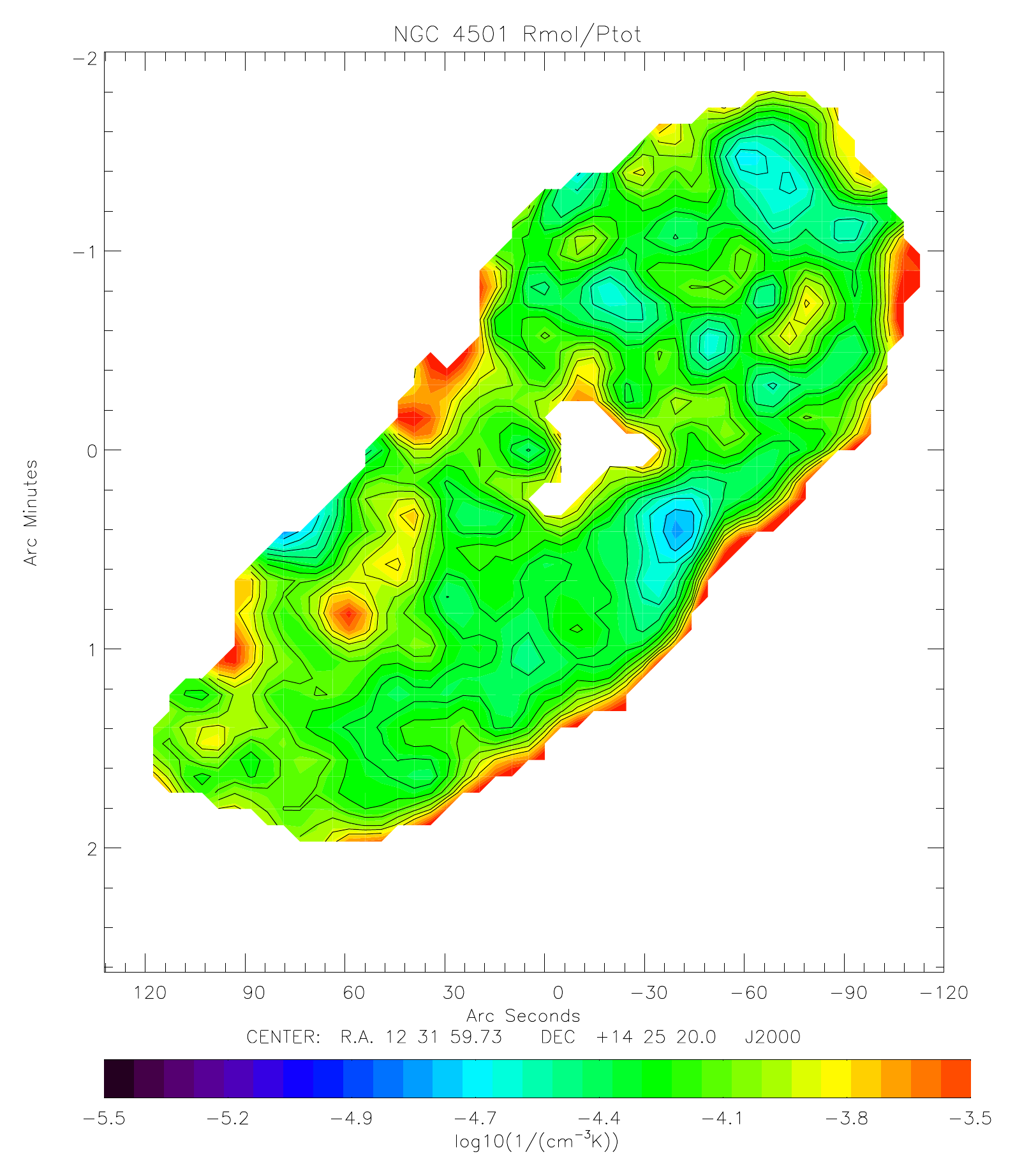}
\includegraphics[width=9cm]{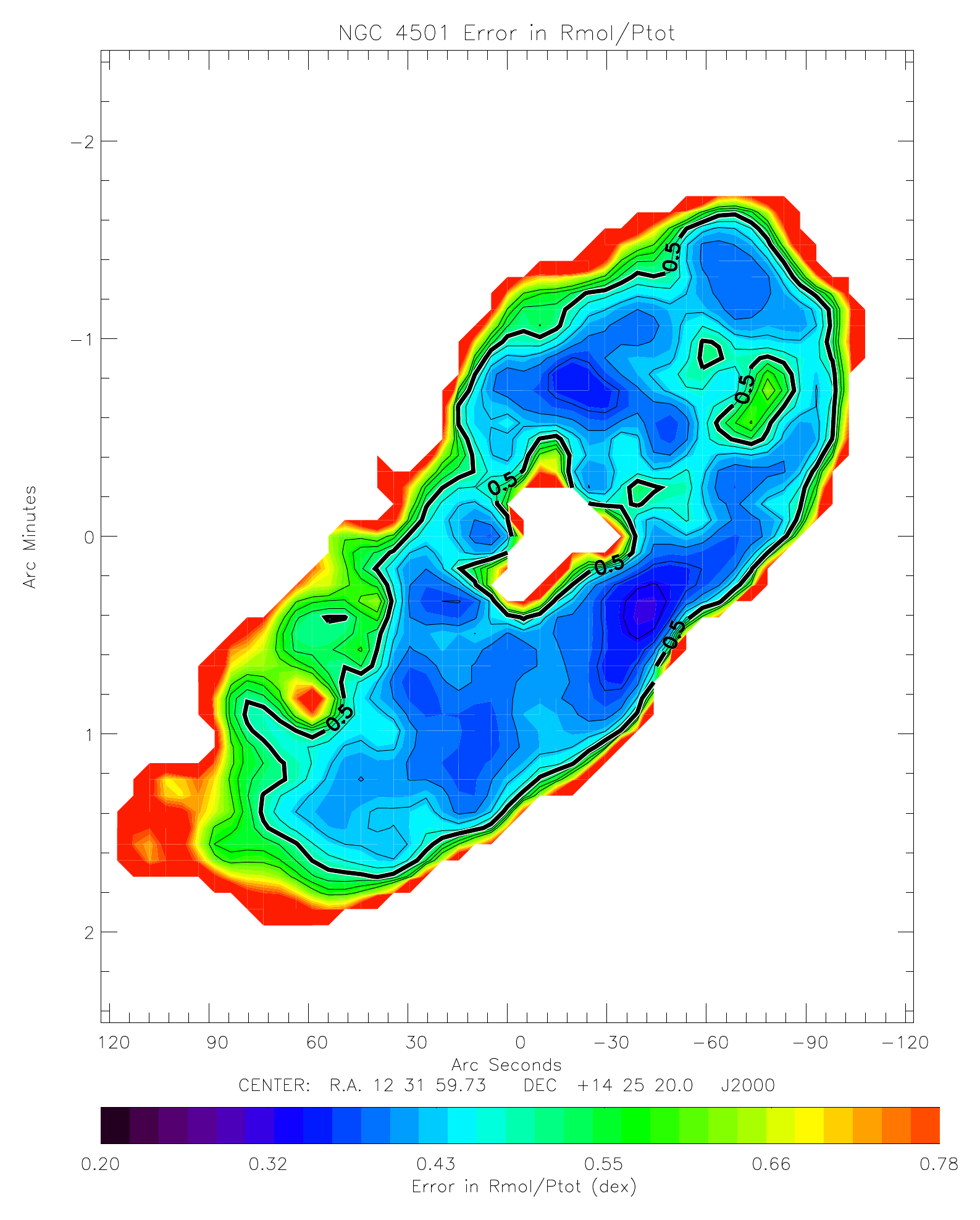}
\caption{NGC~4501. {\em Upper panel}: Molecular fraction divided by the ISM pressure $R_{\rm mol}/P_{\rm tot}$. The resolution is $17"$. The ISM pressure and the molecular fraction are deprojected. Contour levels are (2, 2.5, 3.2, 4, 5, 6.3, 7.9, 10, 12.6, 15.8) $\times 10^{-5}$ cm$^3$ K$^{-1}$. {\em Lower panel}: Error on $R_{\rm mol}/P_{\rm tot}$ (in dex). The thick black contour corresponds to 0.5 dex. Contour levels are 0.3, 0.33, 0.37, 0.4, 0.43, 0.47, 0.5, 0.53, 0.57, and 0.6 dex.}
\label{FFIGRmolP1}%
\end{figure}

\begin{figure}
   \centering
\includegraphics[width=9cm]{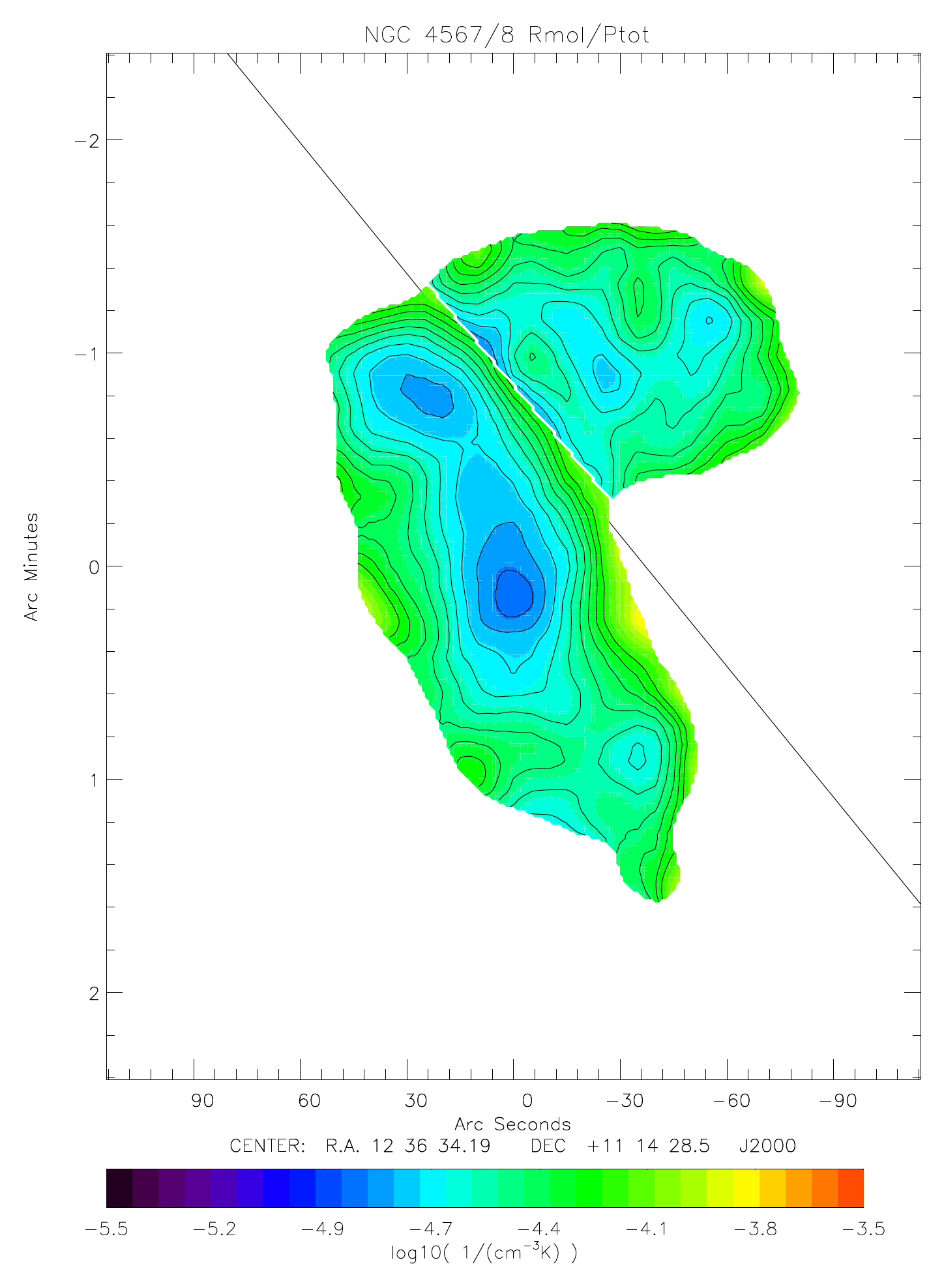}
\includegraphics[width=9cm]{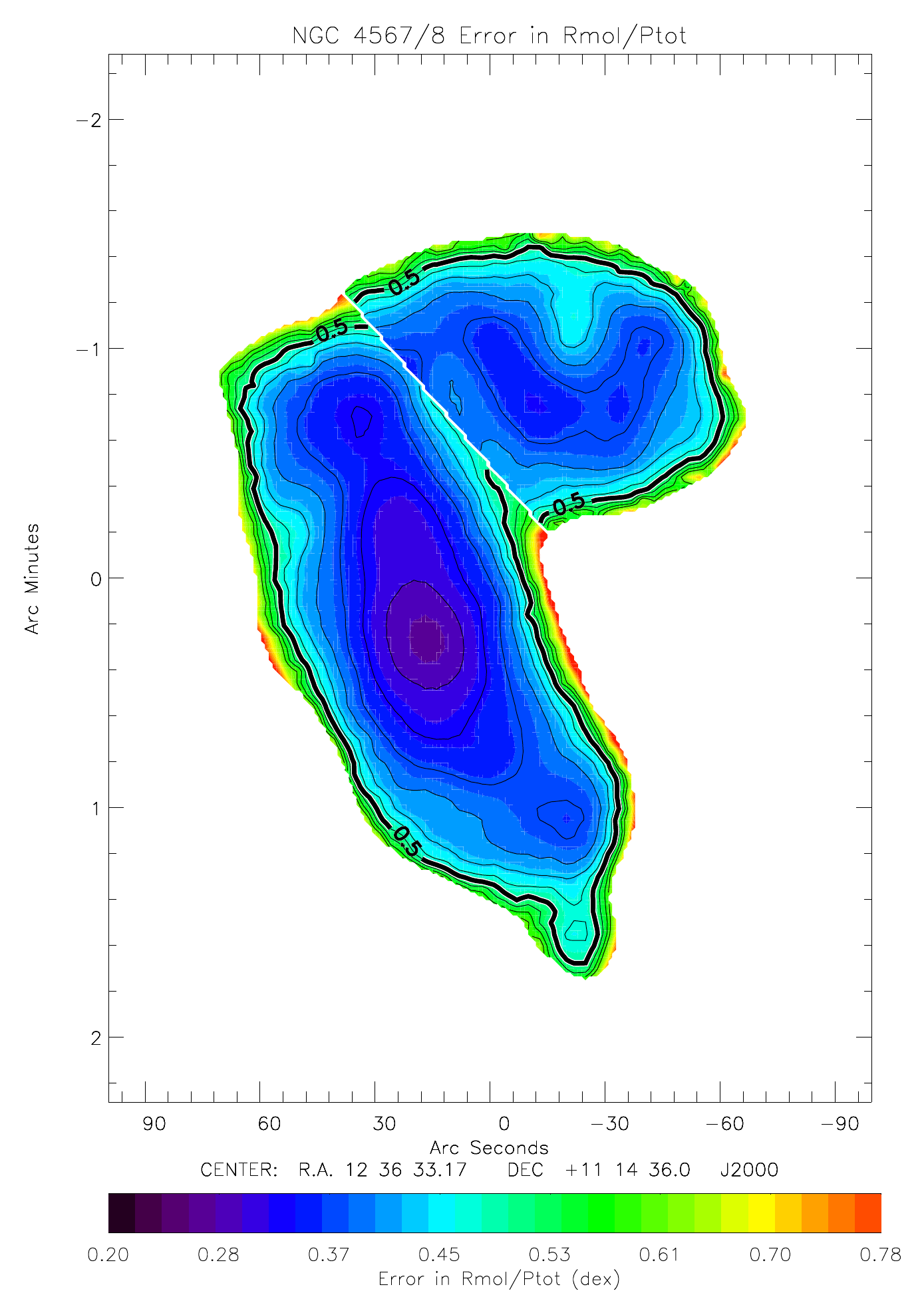}
   \caption{NGC~4567/68. {\em Upper panel}: Molecular fraction divided by the ISM pressure $R_{\rm mol}/P_{\rm tot}$. The resolution is $17"$. The ISM pressure and the molecular fraction are deprojected. Contour levels are (1.2, 1.4, 1.7, 2.1, 2.5, 3, 3.6, 4.4, 5.2, 6.3) $\times 10^{-5}$ cm$^3$ K$^{-1}$. {\em Lower panel}: Error on $R_{\rm mol}/P_{\rm tot}$ (in dex). The thick black contour corresponds to 0.5 dex. Contour levels are 0.3, 0.33, 0.37, 0.4, 0.43, 0.47, 0.5, 0.53, 0.57, and  0.6 dex.}
\label{FFIGRmolP2}%
\end{figure}

\begin{figure}
   \centering
\includegraphics[width=9cm]{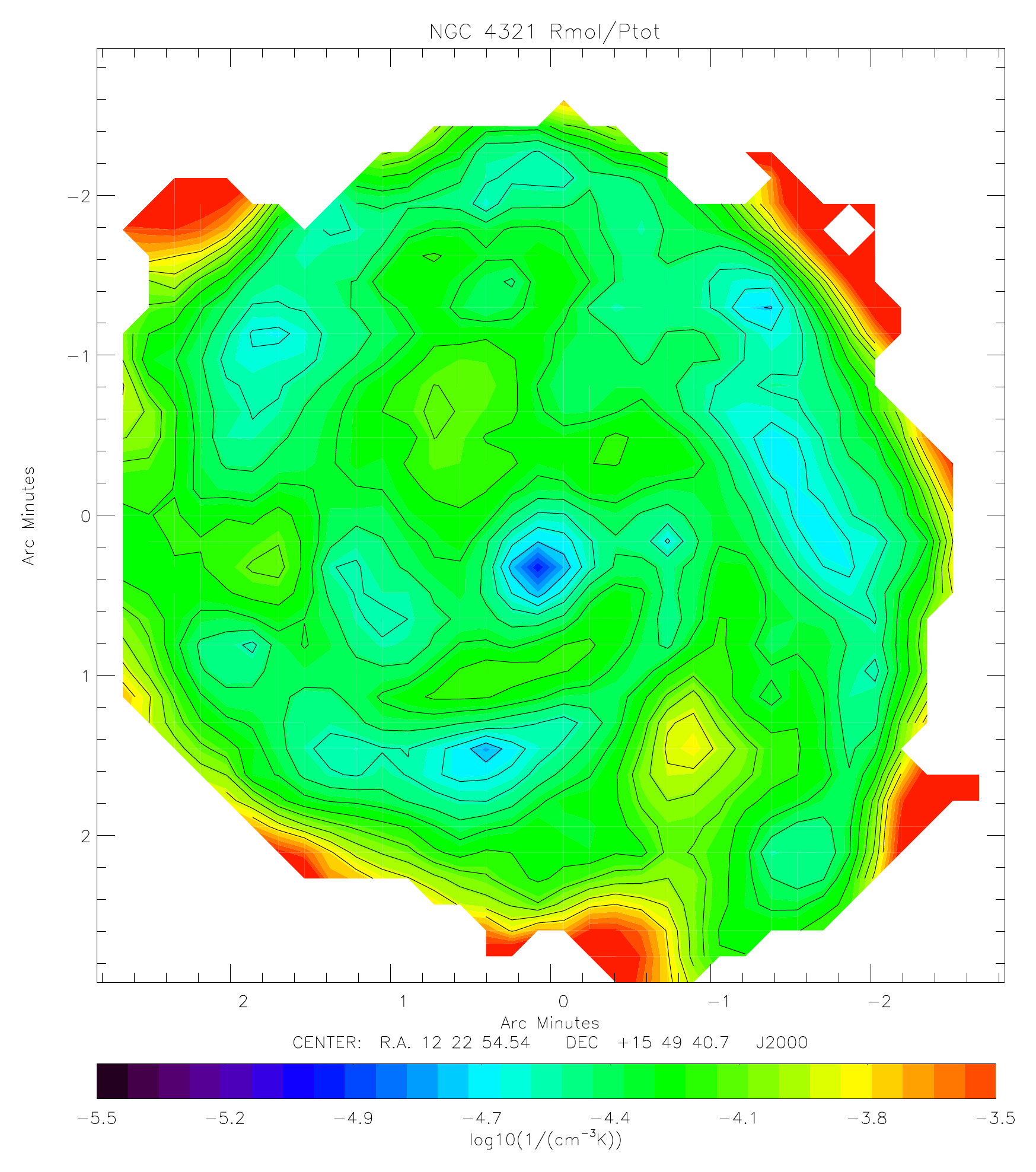}
\includegraphics[width=9cm]{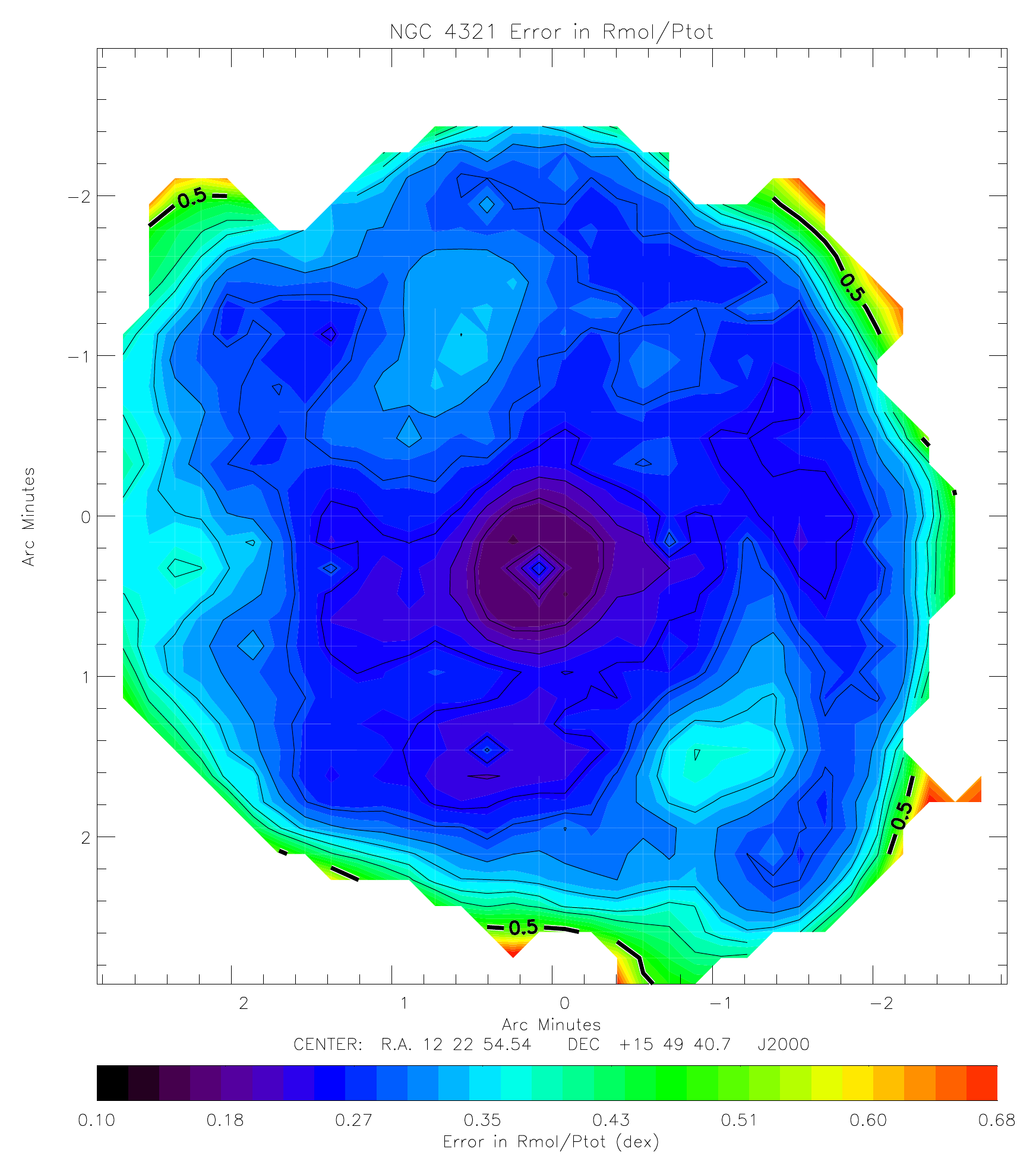}
   \caption{NGC~4321. {\em Upper panel}: Molecular fraction divided by the ISM pressure $R_{\rm mol}/P_{\rm tot}$. The resolution is $17"$. The ISM pressure and the molecular fraction are deprojected. Contour levels are (1.2, 1.4, 1.7, 2.1, 2.5, 3, 3.6, 4.4, 5.2, 6.3)$\times 10^{-5}$ cm$^3$ K$^{-1}$. {\em Lower panel}: Error on $R_{\rm mol}/P_{\rm tot}$ (in dex). The thick black contour corresponds to 0.5 dex. Contour levels are 0.1, 0.13, 0.17, 0.2, 0.23, 0.27, 0.3, 0.33, 0.36, and 0.4 dex.}
\label{FFIGRmolP3}%
\end{figure}

For all four galaxies, the ISM pressure $P_{\rm tot}$ was computed with deprojected data.
Since both, the ISM pressure (Eq.~\ref{eqq5}) and the molecular fraction mainly depend on the gas density (see, e.g., Leroy et al. 2008), 
we expect a close correlation between these parameters. Indeed, Blitz \& Rosolowsky (2006) found $R_{\rm mol} \propto P_{\rm tot}^{0.92 \pm 0.07}$.
The molecular fraction divided by the ISM pressure ($R_{\rm mol}/P_{\rm tot}$) maps are presented in the upper panels of Fig.~\ref{FFIGRmolP1}, Fig.~\ref{FFIGRmolP2}, and Fig.~\ref{FFIGRmolP3} for NGC~4501, NGC~4567/68, and NGC~4321, respectively.
Error maps of $R_{\rm mol}/P_{\rm tot}$ are presented in the lower panels of Figs.~\ref{FFIGRmolP1}, \ref{FFIGRmolP2}, and \ref{FFIGRmolP3}. The thick black contours indicate errors greater than 0.5 dex (a factor of $\sim$ 3).   

For NGC~4501, $R_{\rm mol}/P_{\rm tot}$ is nearly constant all over the disk. Nevertheless, some regions deviate from the average value; $R_{\rm mol}/P_{\rm tot}$ is significantly lower in the west of the galaxy center (blue region) and in three distinct northern regions where the atomic gas surface density is also high. Moreover, in the southeastern part of NGC~4501 we detect a region where $R_{\rm mol}/P_{\rm tot}$ is high (although with a high associated error). 
In NGC~4568 (Fig.~\ref{FFIGRmolP2}) a clear north-south asymmetry is detected, $R_{\rm mol}/P_{\rm tot}$ being lower in the northern part of the disk. In NGC~4567 the spiral arms have a lower $R_{\rm mol}/P_{\rm tot}$ than the inter-arm regions.
In NGC~4321 (Fig.~\ref{FFIGRmolP3}) the spirals arms show a $\sim$ 0.2 dex lower $R_{\rm mol}/P_{\rm tot}$ than inter-arm regions as in NGC~4567. 

\begin{figure}
   \centering
\includegraphics[width=9cm]{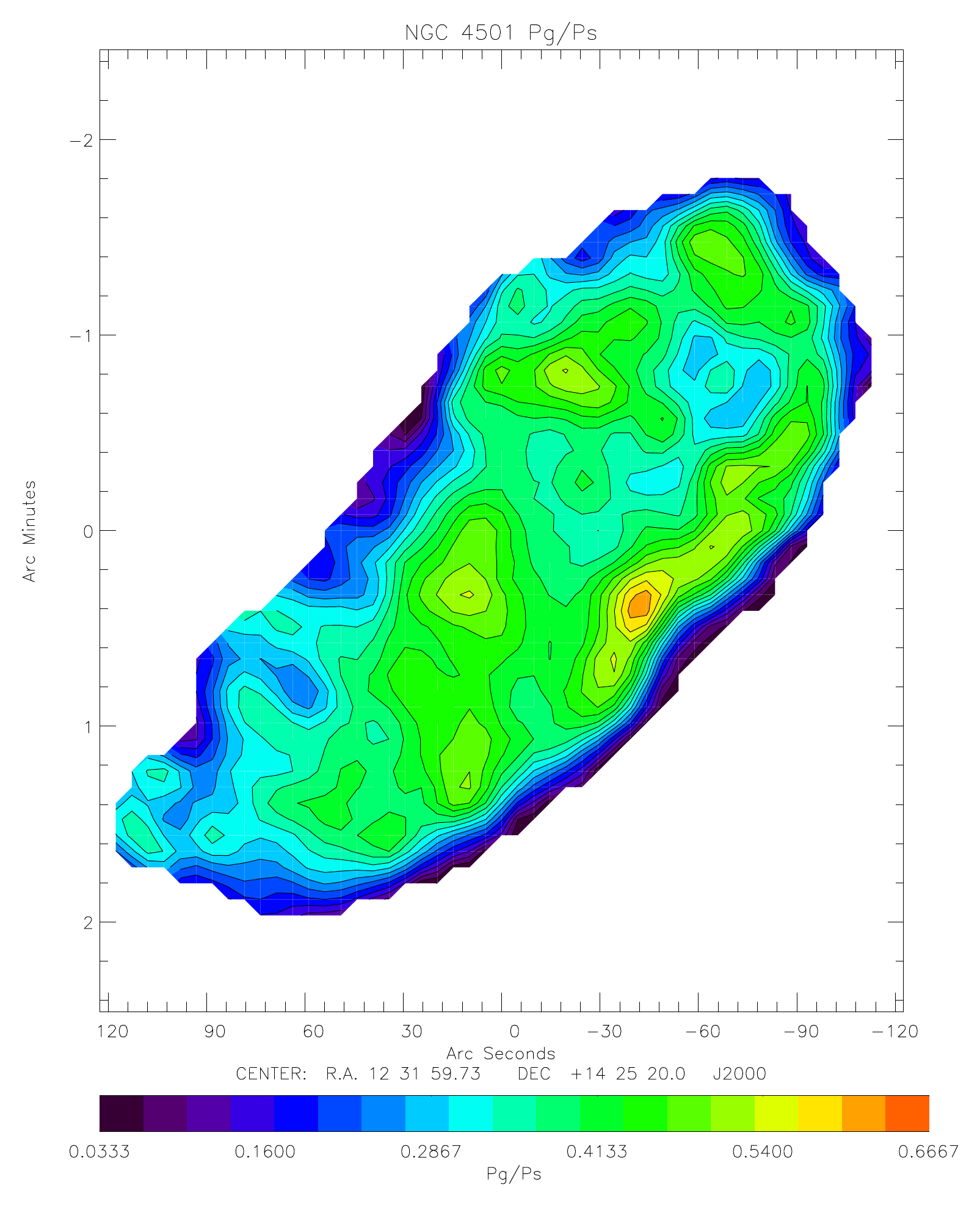}
   \caption{NGC~4501: ISM pressure due to the gravitational potential of the gas divided by the ISM pressure due to the stellar potential (P$_{\rm g}$/P$_{\rm s}$). The resolution is $17"$. Contour levels are (0.33, 0.67, 1, 1.33, 1.67, 2, 2.33, 2.67, 3, 3.33, 3.67, 4, 4.33, 4.67, 5, 5.33, 5.67, 6, 6.33, 6.67)$\times10^{-1}$.}
\label{Pgps01}%
\end{figure}

\begin{figure}
   \centering
\includegraphics[width=9cm]{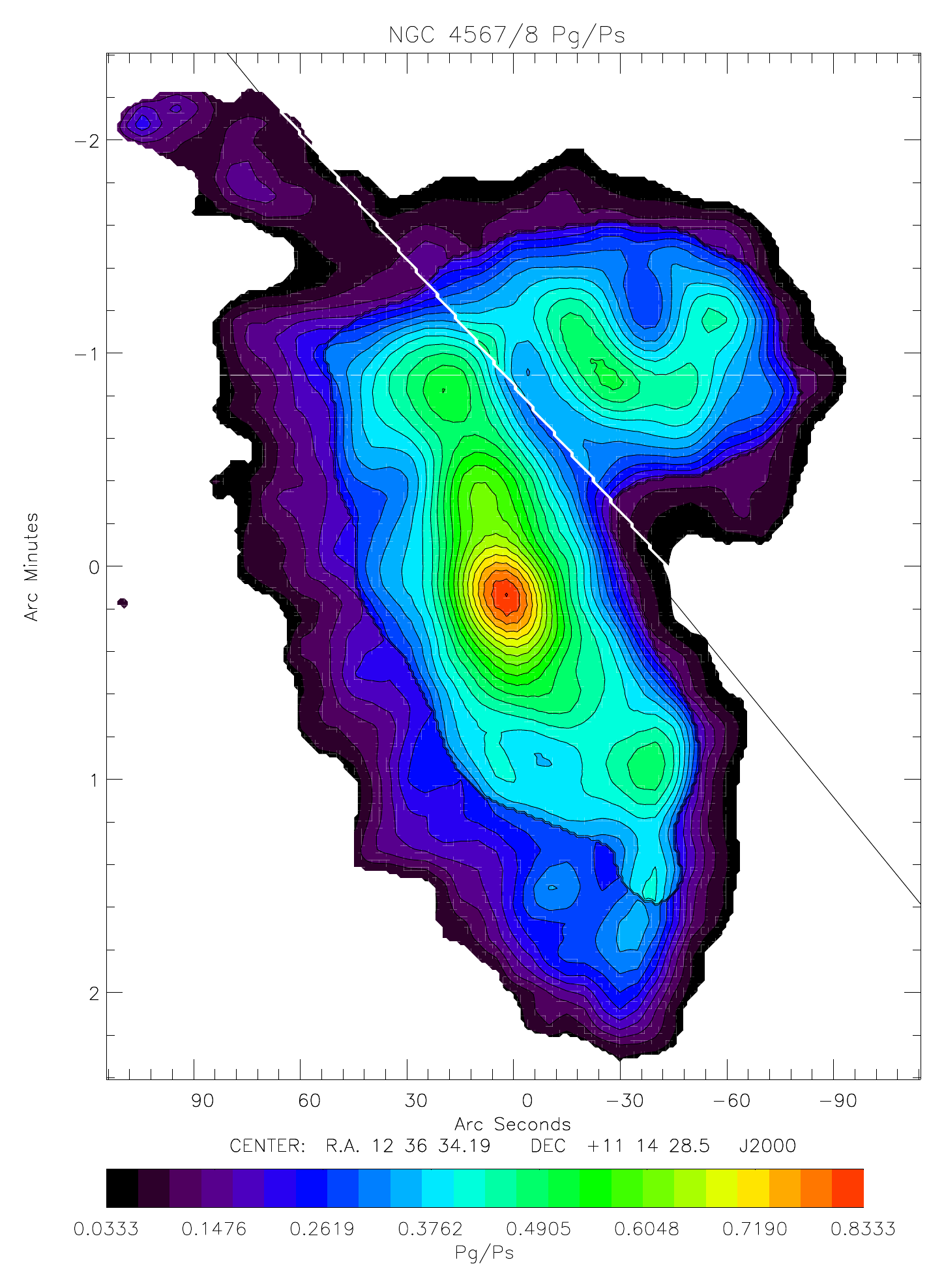}
   \caption{NGC~4567/68: ISM pressure due to the gravitational potential of the gas divided by the ISM pressure due to the stellar potential (P$_{\rm g}$/P$_{\rm s}$). The resolution is $17"$. Contour levels are (0.33, 0.67, 1, 1.33, 1.67, 2, 2.33, 2.67, 3, 3.33, 3.67, 4, 4.33, 4.67, 5, 5.33, 5.67, 6, 6.33, 6.67, 7, 7.33, 7.67, 8, 8.33)$\times10^{-1}$.}
\label{pgps68}%
\end{figure}

To investigate the role of self-gravitating gas, we determined the ratio of gas pressure due to the gravitational potential of the gas $P_{\rm g}=(\pi/2)G\Sigma_{\rm g}^{2}$ to that due to the stellar potential $P_{\rm s}=(\pi/2)G\Sigma_{\rm g}(\sigma_{\rm g}/\sigma_{\star,{\rm z}})\Sigma_{\star}$. 
The ratio $P_{\rm g}/P_{\rm s} > 1$ indicates that the gravitational potential of the gas disk dominates that of the stellar disk. For $P_{\rm g}/P_{\rm s} \gg 1$ the ISM is self-gravitating. The pressure ratio maps $P_{\rm g}/P_{\rm s}$ of NGC~4501 and NGC~4567/68 are presented in Figs.~\ref{Pgps01} and \ref{pgps68}.

In NGC~4501 the ISM pressure is mostly dominated by the stellar potential. However, west of the galaxy center, within the ridge of compressed gas, the gravitational potential of the gas becomes more important than the stellar potential. Moreover, in the eastern region of the disk the pressure due to the stellar component dominates the ISM pressure budget. Contrary to the inner regions of NGC~4501, the central part of NGC~4568 has a high pressure ratio $P_{\rm g}/P_{\rm s} > 0.7$, i.e., the gas is almost self-gravitating.
The spiral arms of NGC~4567 and NGC~4568 also show a higher $P_{\rm g}/P_{\rm s} \sim 0.6$ than the inter-arm or more external regions where $P_{\rm g}/P_{\rm s} \sim 0.3$.

%%%%%%%%%%%%%%%%%%%%%%%%%%%%%%%%%%%%%%%%%%%%%%%%%%

\subsection{SFR and $SFE_{\rm H_{2}}$}

For both galaxy systems, the star formation rate surface density ($\dot{\Sigma_{\star}}$) was computed with the {\em Galaxy Evolution Explorer} (GALEX, Martin et al 2005) far-ultraviolet (FUV) emission and the 24~$\mu$m emission from {\em Spitzer} (Werner et al. 2004) following Leroy et al. (2008),
\begin{equation}
\dot{\Sigma_{\star}}(FUV+24 \mu m)=(8.1\times 10^{-2}I_{FUV}+3.2\times 10^{-3}I_{24})\times \cos(i), \label{eqqfuv}
\end{equation}
where both intensities I$_{\rm FUV}$ and I$_{24}$ are given in MJy ster$^{-1}$ and $\dot{\Sigma_{\star}}$(FUV+24 $\mu$m) is given in M$_{\odot}$ kpc$^{-2}$ yr$^{-1}$.
While the FUV flux traces unobscured star formation from O, B, and A stars, the 24 $\mu$m emission is related to the UV heated dust. Thus, adding the 24$\mu$m to the FUV emission fills the gap of shielded O and B stars. The FUV heated dust emission can also be tracked using the total infrared (TIR) emission (see Hao et al. 2011, Galametz et al. 2013, and Boquien et al. 2013).

We tried three different ways of computing the TIR emission and concluded that  24~$\mu {\rm m}$ traces the TIR within  0.09 dex. This motivated our choice to use  24~$\mu {\rm m}$ to trace the obscured star formation (see discussion in Appendix~\ref{TTIIRR}). We also made sure that the SFR computed with the H$\alpha$ does not deviate significantly from the SFR computed with the FUV (see Appendix~\ref{HA} for the detailed study).
Finally, we chose the FUV+24 $\mu$m for an easier comparison with other studies of nearby galaxies (e.g., Leroy et al. 2008, Bigiel et al. 2008). In the following $\dot{\Sigma_{\star}}$  refers to $\dot{\Sigma_{\star}}(FUV+24 \mu m)$.

\begin{figure}
   \centering
\includegraphics[width=9cm]{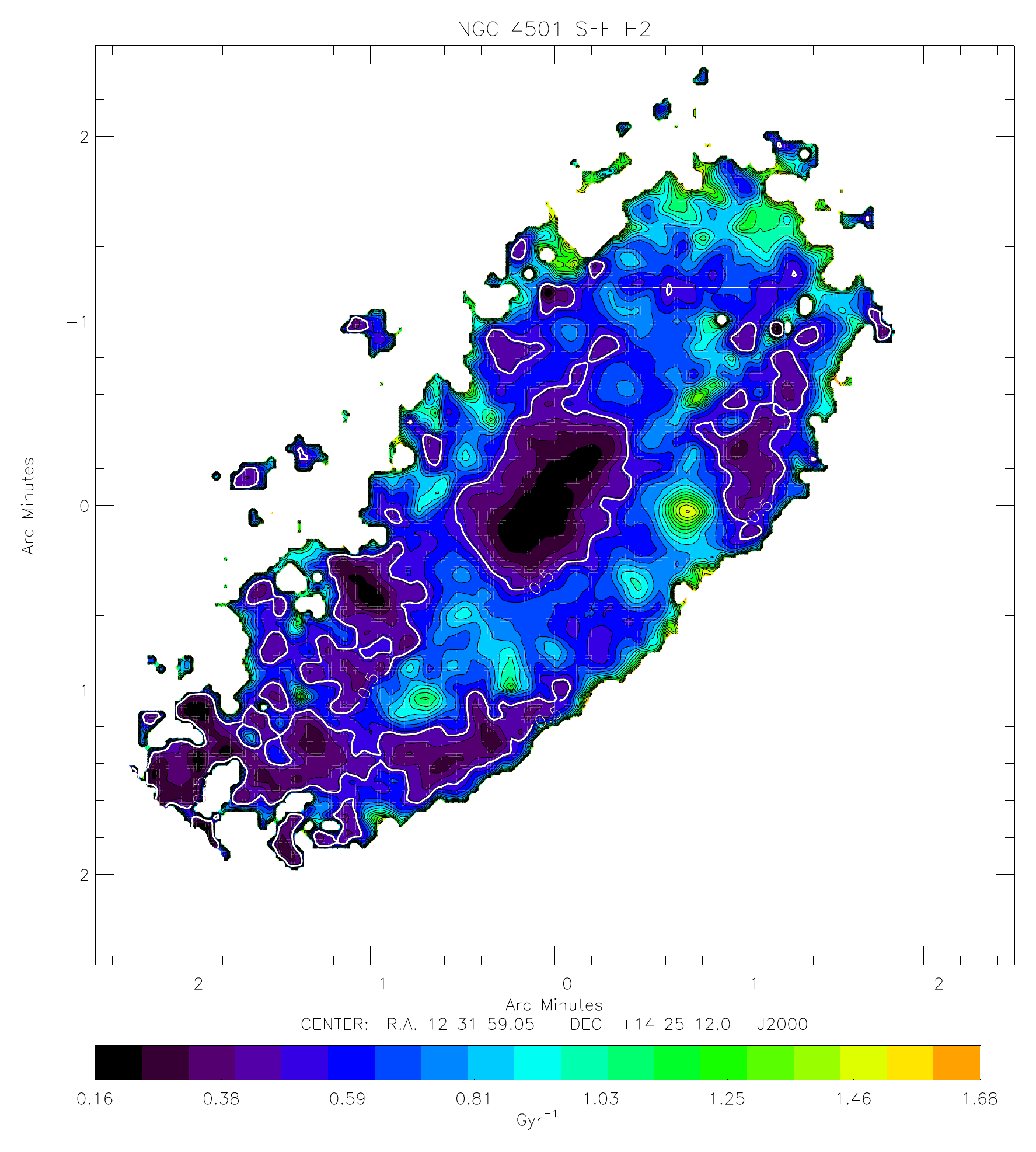}
\includegraphics[width=9cm]{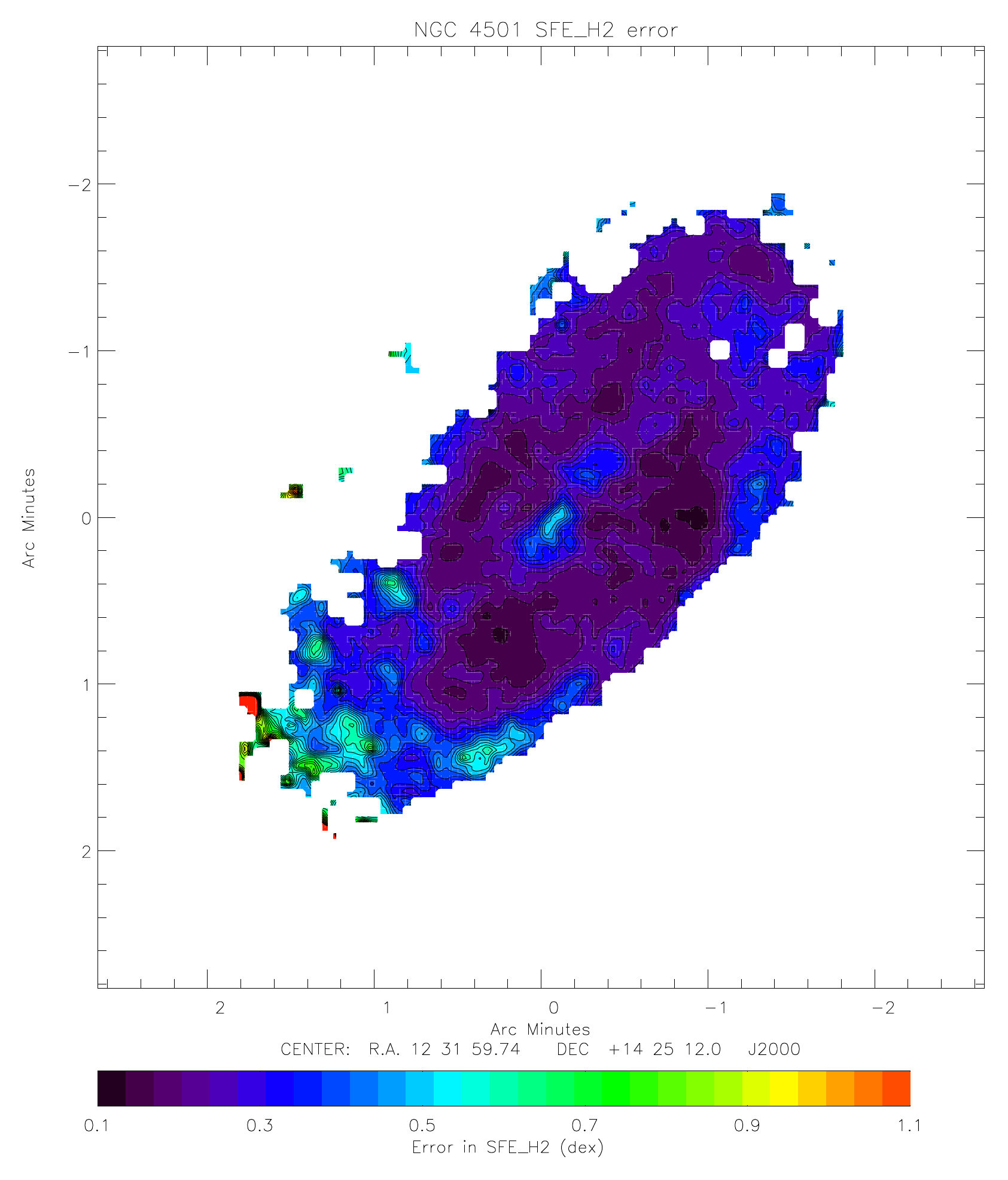}
   \caption{NGC~4501. {\em Upper panel}: Star formation efficiency with respect to the molecular gas (SFE$_{{\rm H}_{2}}={\rm SFR}/\Sigma_{{\rm H}_{2}}$). Contour levels are (6, 12, 18, 24, 29, 36, 42, 48, 53, 59, 65, 72, 78, 84, 90, 96, 102, 107, 113, 119)$\times 10^{-2}$ Gyr$^{-1}$. The white contour corresponds to $SFE_{\rm H_{2}}$=0.5~Gyr$^{-1}$ or a molecular gas depletion time of 2~Gyr. {\em Lower panel}:  Error of the star formation efficiency with respect to the molecular gas (dex).}
\label{SFEh24501}%
\end{figure}

\begin{figure}
   \centering
\includegraphics[width=9cm]{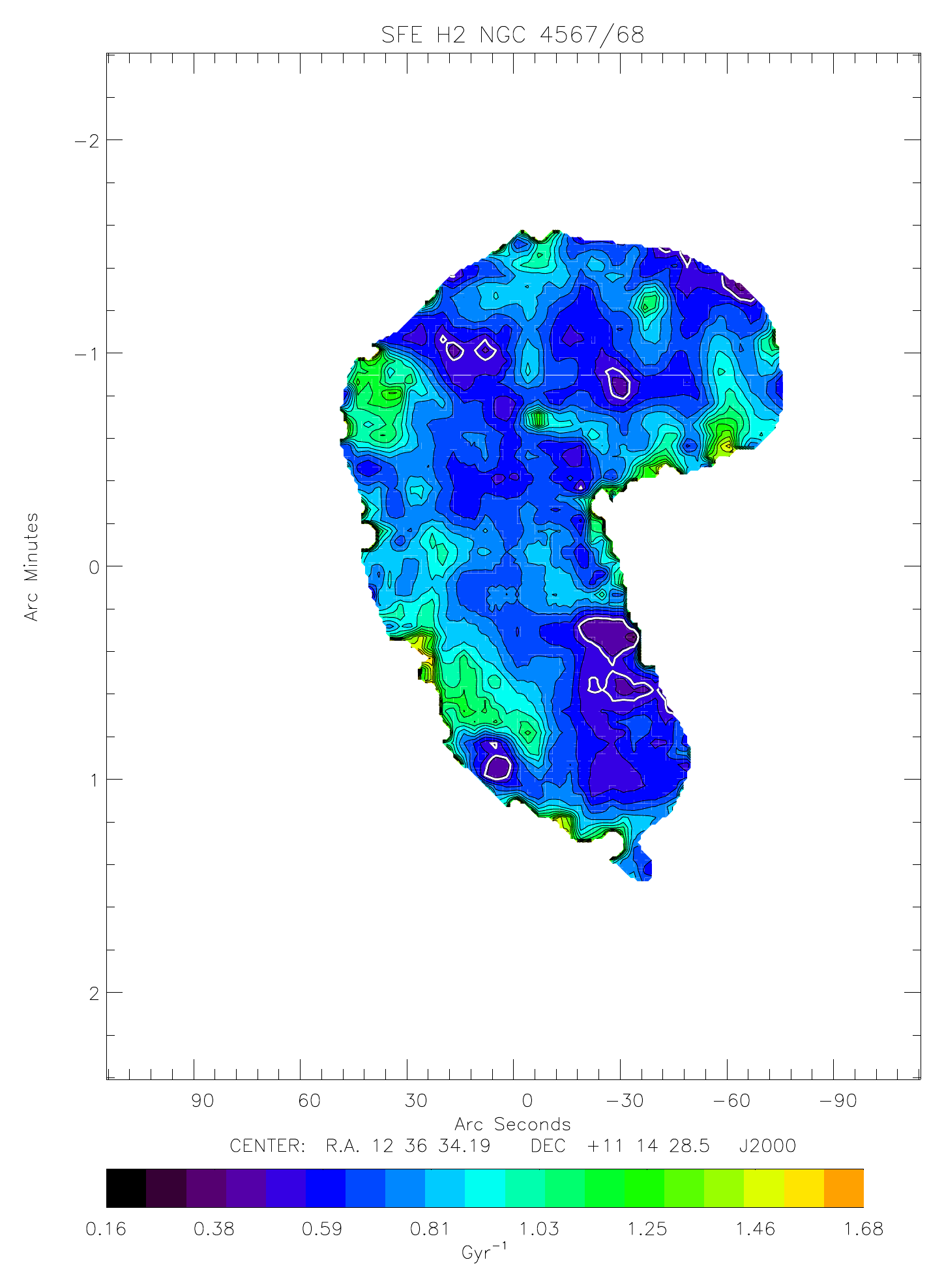}
\includegraphics[width=9cm]{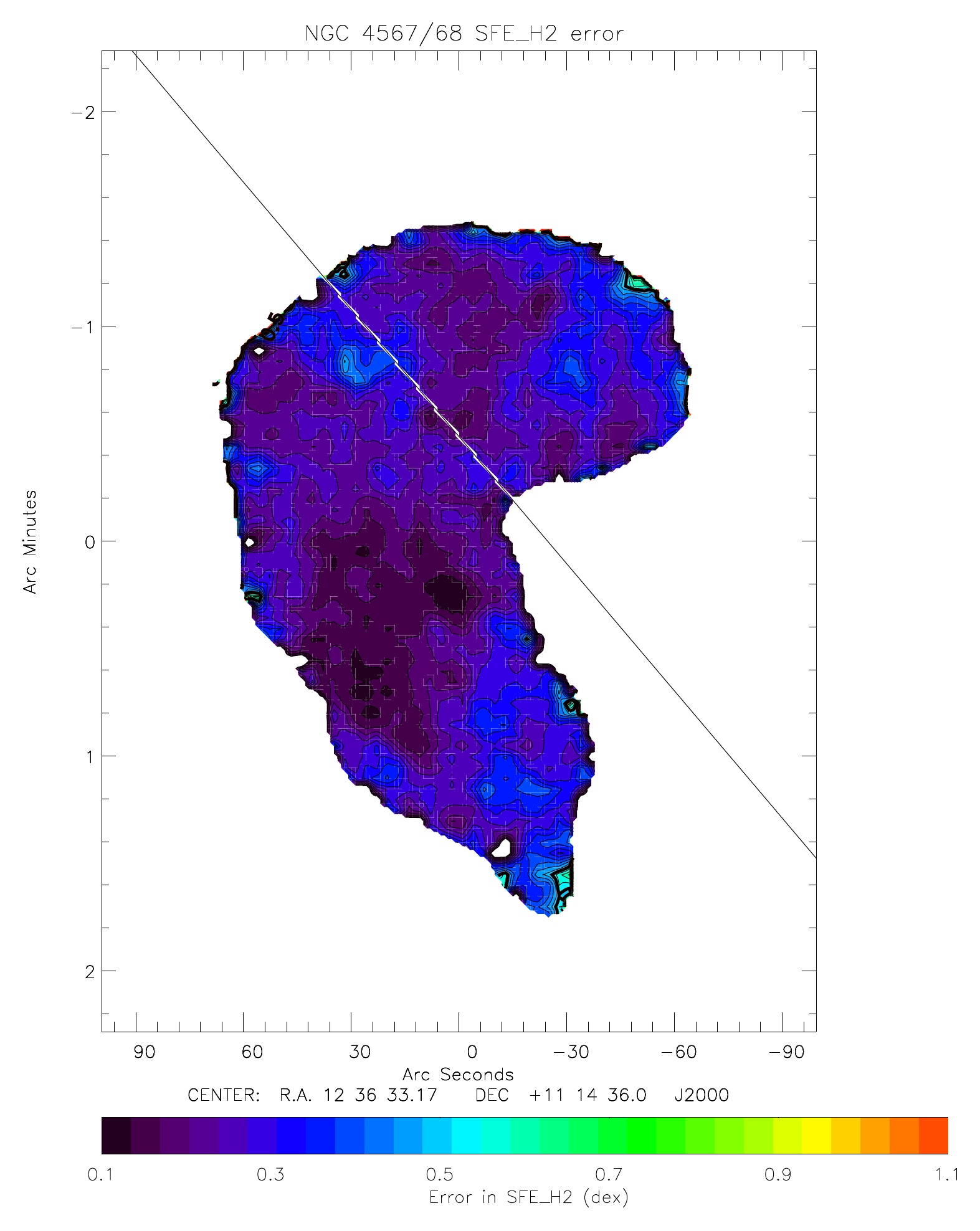}
   \caption{NGC~4567/6. {\em Upper panel}: NGC~4567/68 star formation efficiency with respect to the molecular gas (SFE$_{{\rm H}_{2}}={\rm SFR}/\Sigma_{{\rm H}_{2}}$. Contour levels are (6, 12, 18, 24, 29, 36, 42, 48, 53, 59, 65, 72, 78, 84, 90, 96, 102, 107, 113, 119)$\times 10^{-2}$ Gyr$^{-1}$. The white contour corresponds to $SFE_{\rm H_{2}}=0.5$~Gyr$^{-1}$ or a molecular gas depletion time of 2~Gyr. {\em Lower panel}:  Error of the star formation efficiency with respect to the molecular gas (dex).}
\label{SFEh24567}%
\end{figure}

For the two galactic systems, the star formation efficiency maps with respect to the molecular gas $SFE_{{\rm H}_{2}}={\dot{\Sigma_{\star}}}/\Sigma_{{\rm H}_{2}}$ are presented in the upper panels of Figs.~\ref{SFEh24501} and \ref{SFEh24567}. The corresponding errors maps of $SFE_{\rm H_{2}}$ are presented in the lower panels for NGC~4501 and NGC~4567/68, respectively.

The $SFE_{\rm H_{2}}$ of NGC~4501 shows spatial variations within the disk from $\sim 0.2$ to $\sim 1.5$ Gyr$^{-1}$
with most of the disk being close to the standard value of $\sim 0.5$~Gyr$^{-1}$ ($H_2$ consumption time of 2~Gyr).
The high molecular $SFE_{\rm H_{2}}$ regions correspond to regions where the molecular surface density is high, except in the very center of NGC~4501 where the $SFE_{\rm H_{2}}$ is low and $\Sigma_{\rm H_{2}}$ is high, as found by Sandstrom et al. (2013) in other nearby star-forming spiral galaxies.  Most probably, the molecular gas surface density in the central region is overestimated by the assumed constant X$_{\rm co}$ factor (see Section~\ref{XCO} for details on X$_{\rm co}$ issues). At the northeastern edge, the $SFE_{\rm H_{2}}$ is high with a low $\Sigma_{\rm H_{2}}$. The southern edge of the disk shows a low $SFE_{\rm H_{2}}$. This U-shaped region corresponds to regions of low $\Sigma_{\rm H_{2}}$. Within the compressed western region, the region west of the galaxy center has an average $SFE_{\rm H_{2}}$, whereas the northwestern part shows a low  $SFE_{\rm H_{2}}$.

The $SFE_{\rm H_{2}}$ of NGC~4567/68 is on average higher than that of NGC~4501.
The $SFE_{\rm H_{2}}$ is lower in the northwestern tidally interacting part of the disk and in the southwest of NGC~4568. The $SFE_{\rm H_{2}}$ is higher in the eastern part of NGC~4568 than in the western part of the disk. In the center and in the spirals arms of NGC~4567, the $SFE_{\rm H_{2}}$ is lower than in the  inter-arm regions.  

The molecular star formation efficiency of NGC~4321 is almost constant over the entire galactic disk.

\section{Discussion \label{dis}}

\subsection{ X$_{\rm CO}$ conversion factor \label{XCO}}

It has been found that techniques for estimating the H$_2$ gas density according to CO emission are tricky as they rely mainly on three key theoretical  assumptions (Dickman et al. 1986): 
(1) all GMCs have the same mean surface density (overlapping clouds can lead to a smaller value of $I_{\rm CO}$, which means that H$_2$ mass can be underestimated);
(2) the mean brightness temperature of the CO is well defined and does not vary with cloud size (CO emission is assumed to be optically thick); and 
(3) GMCs have a known mass spectrum and follow a size-linewidth relation. Moreover, the conversion factor is both expected and observed to vary with metallicity (van Dishoeck \& Black 1988; Wolfire et al. 2010) leading to lower value at higher metallicity. Thus, in spiral galaxies the well-known metallicity gradient (Shaver et al. 1983, Pilkington et al. 2012 and references therein) may lead to an overestimation of $\Sigma_{\rm H_{2}}$ in the central regions where the metallicity is expected to be higher. Sandstrom et al. (2013) investigated the radial variations of the conversion factor in a sample of 26 nearby star-forming galaxies. These authors found a significantly lower conversion factor in the nuclear regions of their sample. However, they found only a weak dependency of the conversion factor on metallicity. The conversion factor is also observed to sharply increase at metallicities below one-half solar. Nevertheless, the variation of X$_{\rm CO}$ with metallicity is still poorly constrained, thus we prefer to use  a simpler, first-order description of X$_{\rm CO}$. Bolatto et al. (2013) recommend the use of X$_{\rm CO(1-0)}=2\times10^{20}$cm$^{-2} (\alpha_{CO(1-0)}=4.4{\rm M}_{\odot} $(K km/s pc$^2$)$^{-1}$) with an estimated $\pm30$\% uncertainty. 
Assuming a line ratio R$_{21}=0.8$, this gives a value of $\alpha_{\rm CO(2-1)}=5.5~{\rm M}_{\odot}$(K km/s pc$^2$)$^{-1}$ including He.

\subsubsection{ Star formation efficiency}

The Kennicutt-Schmidth relation (KS) (Schmidt 1959, 1963 and  Kennicutt 1989, 1998b) links the gas reservoir of galaxies to the SFR using integrated values of SFR and $\Sigma_{\rm g}=\Sigma_{\rm HI}+\Sigma_{\rm H_{2}}$. Kennicutt (1998b) found a non-linear relation (considering disk averages SFR and gas densities) $\Sigma_{\rm SFR}={\rm A} \times \Sigma_{\rm g}^{N}$ with $N=1.4\pm0.15$ from a sample of normal star-forming spirals and starburst galaxies. Many other works found $N$ in the range of $1.0$-$2.0$. Nevertheless, the use of $\Sigma_{\rm g}=\Sigma_{\rm H_{\rm I}}+\Sigma_{\rm H_2}$ in this relation is puzzling since stars form in molecular clouds and thus the information of the transition from H$_{\rm I}$ to H$_{2}$ is hidden when using $\Sigma_{\rm g}$. These studies consider average SFR and gas surface densities.

Other studies have found shallower power laws ($N\sim 1$) using only the molecular gas and SFR (Wong \& Blitz 2002; Kennicutt et al. 2007; Bigiel et al 2008; Leroy et al. 2008, 2013);
 Wong \& Blitz (2002) consider azimuthally averaged surface densities, while Kennicutt et al. (2007), Bigiel et al. (2008), and Leroy et al. (2008, 2013) have studied the KS law at kpc or sub-kpc resolution.  

 A weaker correlation, or even no correlation, between atomic gas and SFR is found by Bigiel et al. (2008) in their sample of 18 nearby galaxies. Moreover, they observe that the atomic gas surface density in spirals and dwarf galaxies saturates at $\sim 10$ M$_{\odot}$pc$^{2}$ so that the correlation between atomic gas and SFR  shows the saturation of the H{\sc i} line (Blitz \& Rosolowsky (2006), Krumholz et al. (2009)).

\begin{figure*}
   \centering
\includegraphics[width=15cm]{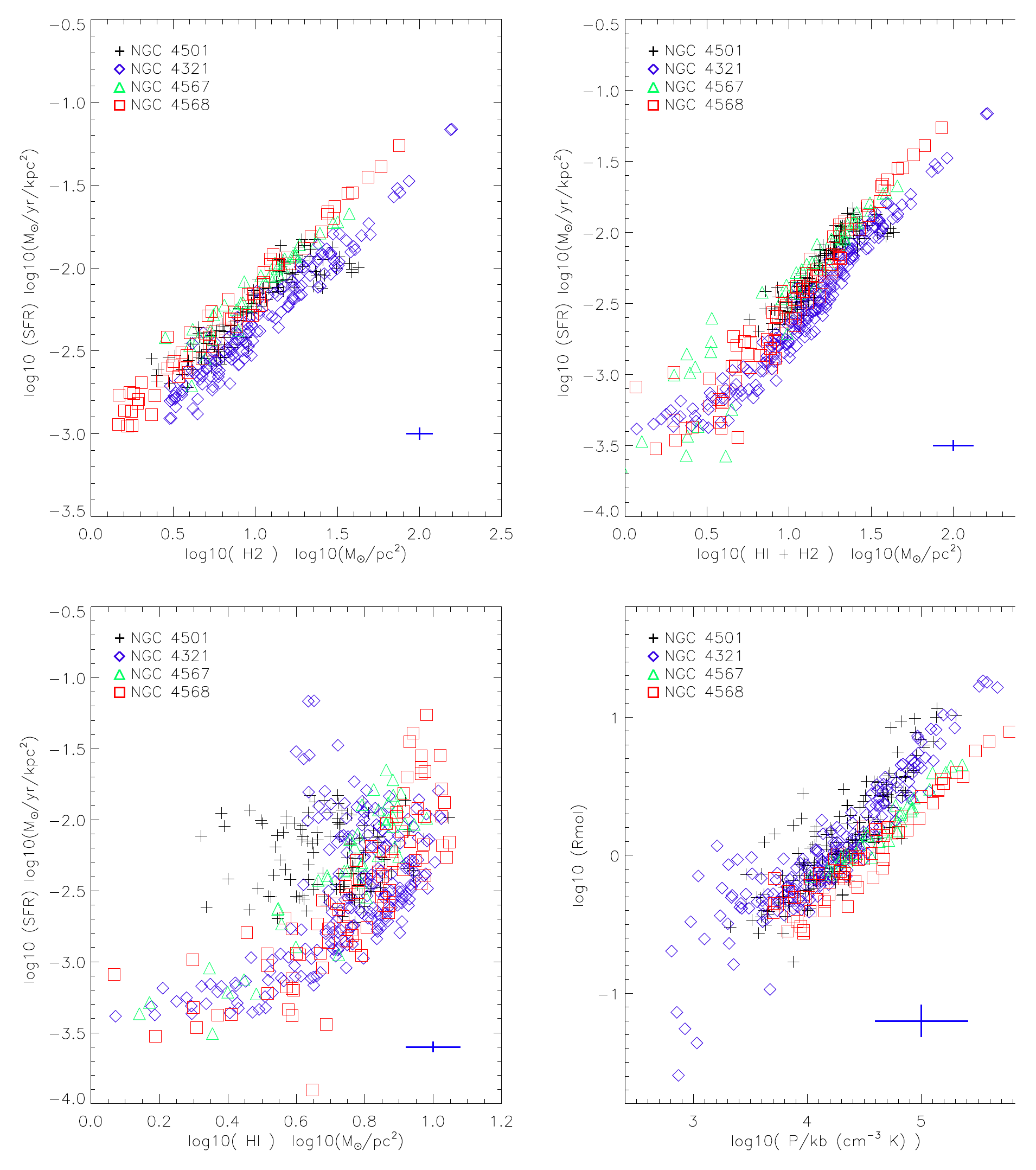}
   \caption{NGC~4501, NGC~4567/8, and NGC~4321. {\em Upper left panel}: SFR as a function of $\Sigma_{H_2}$. {\em Upper right panel}: SFR as a function of total gas surface density. {\em Lower left panel}: SFR as a function of atomic gas surface density. {\em Lower right}: Molecular fraction as a function of the ISM pressure $P_{\rm tot}$. Typical errors bars are given for each relation.}
\label{plot4gal1}%
\end{figure*}

We investigated the spatially resolved SFR-$\Sigma_{\rm g}$, SFR-$\Sigma_{\rm HI}$, and SFR-$\Sigma_{\rm H_{2}}$ relations for the four galaxies (Fig.~\ref{plot4gal1}). Each point corresponds to a resolution element of $17''$ or $\sim1.5$ kpc. The resulting plots for each galaxy are presented in Appendix~\ref{annex44} (Fig.~\ref{plot4501}, \ref{plot4567}, \ref{plot4568}, and \ref{plot4321} for NGC~4501, NGC~4567, NGC~4568, and NGC~4321, respectively).

%;;;;;
%fig 18-21 --> renvoie annexe fig 22
%;;;;;;

As expected, we do not see any clear correlation between SFR and $\Sigma_{\rm HI}$ for NGC~4501 and NGC~4321. Instead, for NGC~4567/68 and NGC~4321 two regimes can be seen: (1) where log$_{10}$($\Sigma_{\rm HI})<0.9$  (8 M$_{\odot}$ pc$^{-2}$) and (2) where log$_{10}$($\Sigma_{\rm HI})>0.9$. The first regime shows a shallower slope than the second. In the central part of NGC~4501 the H{\sc i} surface density decreases radially to the center where no H{\sc i} is detected. Thus, the correlation between H{\sc i} and SFR breaks down in the central part of NGC~4501. This behavior is also observed in the field spiral galaxy NGC~628 (Leroy et al. 2008);  the H{\sc i} surface density also strongly decreases toward the galaxy center where the gas is mostly molecular.
         
A power law was fitted to the SFR-$\Sigma_{\rm H_{2}}$ and the SFR-$\Sigma_{\rm g}$ relations:
\begin{eqnarray}
\log_{10}({\rm SFR}) & = & {\rm n_1} \times \log_{10}( \Sigma_{\rm H_2}) +\gamma_1  \label{fit1} \\
\log_{10}({\rm SFR}) & = & {\rm n_2} \times \log_{10}( \Sigma_{\rm g}  )   +\gamma_2  \label{fit2}.
\end{eqnarray}
We used an MCMC approach for a linear regression with errors on both X and Y data (IDL routine LINMIX\_ERR). This Bayesian approach gives the probability density distribution of both the slope and the offset of the fit. We also computed the Spearman rank correlation $\rho$ for the two relations (Table~\ref{tab2b}). The uncertainty in the Spearman rank correlation $\rho$ was estimated  via the bootstrap method and by redistributing the data within the uncertainties. The errors obtained with the two methods agree within a factor of three. To be conservative, we only present the maximum of these error estimates.

\begin{table*}
\caption{\bf Results of  fitting analysis I}           % title of Table
\label{tab2b}    
\centering      
\begin{tabular}{ c c c c c }       
\hline\hline                 

 Name       &             &  SFR-$\Sigma_{H_{2}}$       &   SFR-$\Sigma_{HI+H_{2}}$     &   $R_{\rm mol}$-$P_{\rm tot}$                       \\    % table heading 
                                  
\hline                                                     
NGC~4501     & slope      &  ${\rm n}_1=0.91\pm0.08  $  & ${\rm n}_2= 1.53\pm0.27 $ &  $\alpha=1.02\pm0.27$                                \\
             & offset       &  $\gamma_1=-3.13\pm0.09 $   & $\gamma_2= -4.06\pm0.32 $ &  $P_0=1.67\pm0.27 \times 10^{4}$ cm$^{-3}$ K  \\
             & rms (dex)  &  $0.065 $                   & $ 0.052 $                 &  $0.150$                                 \\
             & $\rho$      &   $0.930\pm0.014$     & $ 0.937\pm0.036$    &  $0.838\pm0.049$          \\                      
\hline                                                  
NGC~4567     & slope      &  ${\rm n}_1=0.85\pm0.09  $  & ${\rm n}_2= 1.11\pm0.15 $ &  $\alpha=0.70\pm0.40$                                \\
             & offset       &  $\gamma_1=-2.98\pm0.11 $   & $\gamma_2= -3.48\pm0.20 $ &  $P_0=2.50\pm1.20 \times 10^{4}$ cm$^{-3}$ K  \\
             & rms (dex)  &  $0.055 $                   & $ 0.083 $                 &  $0.048$                                 \\
             & $\rho$      &  $0.978\pm0.017$           & $ 0.981\pm0.024$          &  $0.962\pm0.103$          \\                      
\hline                                                   
NGC~4568     & slope      &  ${\rm n}_1=1.00\pm0.04  $  & ${\rm n}_2= 1.33\pm0.10 $ &  $\alpha=0.76\pm0.15$                                \\
             & offset       &  $\gamma_1=-3.14\pm0.05 $   & $\gamma_2= -3.82\pm0.14 $ &  $P_0=3.20\pm0.72 \times 10^{4}$ cm$^{-3}$ K  \\
             & rms (dex)  &  $0.061 $                   & $ 0.101 $                 &  $0.100$                                 \\
             & $\rho$      &  $0.977\pm0.009$           & $ 0.968\pm0.017 $         &  $0.912\pm0.054$          \\                      
\hline  
NGC~4321     & slope      &  ${\rm n}_1=0.98\pm0.06  $  & ${\rm n}_2= 1.30\pm0.07 $ &  $\alpha=0.86\pm0.09$                                \\
             & offset       &  $\gamma_1=-3.37\pm0.09 $   & $\gamma_2= -3.98\pm0.10 $ &  $P_0=1.60\pm0.24 \times 10^{4}$ cm$^{-3}$ K  \\
             & rms (dex)  &  $0.066 $                   & $ 0.071 $                 &  $0.246$                                 \\
             & $\rho$      &  $0.962\pm0.035$           & $ 0.980\pm0.031 $         &  $0.924\pm0.045$          \\                      
\hline  
\end{tabular}
%\tablefoot{  }
\end{table*}

For NGC~4501 the central area was excluded from the fit since the conversion factor is not reliable and no H{\sc i} is detected in the center.
For all three galaxies we found a power law index n$_1$ for the molecular KS relation ($\Sigma_{\rm H_{2}}$-SFR) in the range of $0.9$-$1$, as has been found in other non-interacting galaxies. For the $\Sigma_{\rm g}$-SFR relation a power law index n$_2$ of $1.1 -1.5$ has been found. Moreover, the rms of the molecular and total gas KS relations are less then $20$\,\%. 
Those two results suggest that gas compression, either from ram pressure or from gravitational interaction, do not significantly change the KS relation.  

For a direct comparison between galaxies, the SFR-$\Sigma_{H_{2}}$,  SFR-$\Sigma_{HI}$, SFR-$\Sigma_{\rm g}$, and $R_{\rm mol}$-$P_{\rm tot}$ relations are presented in Fig.~\ref{plot4gal1}. NGC~4321 shows a $\sim$0.2 dex lower molecular $SFE_{\rm H_{2}}$ and total $SFE_{\rm g}$. In the inner galactic disks (at high $P_{\rm tot}$) of NGC~4567 and entire gas disk of NGC~4568 the
molecular fraction is lower than that of the other galaxies. 

\subsubsection{ Molecular fraction}

Wong \& Blitz (2002) and  Blitz \& Rosolowsky (2006) (BR06) investigated the ISM pressure-molecular fraction relation with a pixel-by-pixel analysis . They fitted a relation between $\log(P_{\rm tot}/{\rm k_B})$ and $\log($$R_{\rm mol})$, where k$_{\rm B}$ is the Boltzmann constant. These authors find that the molecular fraction is correlated with the ISM pressure mostly in regions where $R_{\rm mol}>1$,
\begin{equation}
R_{\rm mol}= \left(\frac{P_{\rm tot}}{P_0}\right) ^{0.92\pm0.07},
\end{equation}
with $P_0/{\rm k_B}=3.5 \pm 0.6 \times 10^4$ cm$^{-3}$K the pressure of the interstellar medium, where $R_{\rm mol}=1$.
They found that $P_0$ shows an offset for H{\sc i}-poor systems, such as interacting galaxies. The index of the power law does not show any systematic behavior when comparing the interacting to the non-interacting population in their sample.  Blitz \& Rosolowsky (2006) computed the ISM pressure in the following way: $P_{\rm tot}^{BR06}=272.\Sigma_{\rm g} \Sigma_{\star}^{0.5}{\rm v}_{\rm disp}{\rm h}_{\star}^{-0.5} $, with a mean gas velocity dispersion ${\rm v}_{\rm disp}=8$ km\,s$^{-1}$. We added the $\Sigma_{\rm g}^2$ term to take the self-gravitation of the gas into account, and took ${\rm v}_{\rm disp}=10$ km\,s$^{-1}$. As did BR06, we used a constant conversion factor $\alpha_{\rm co}=4.4~$M$_\odot$ (K km\,s$^{-1}$ pc$^2$)$^{-1}$.

We investigated the spatially resolved molecular fraction-ISM pressure relation for NGC~4501, NGC~4567/68, and NGC~4321. The resulting relations are presented in Fig.~\ref{plot4501}, \ref{plot4567}, \ref{plot4568}, and \ref{plot4321}.
We fitted the following power law to these relations:
\begin{eqnarray}
R_{\rm mol}    & = & \left(\frac {\rm P_{\rm tot}}{\rm P_0}\right)^{\alpha}  \label{fit3}.
\end{eqnarray}
The results of the fit and Spearman correlation coefficient $\rho$ are given in Tab. \ref{tab2b}. 
For NGC~4501 the power law index  ($\alpha=1.02\pm0.27$) is higher than (but within the uncertainty) the values of NGC~4567 ($\alpha=0.70\pm0.40$), NGC~4568 ($\alpha=0.76\pm0.15$), and NGC~4321 ($\alpha=0.86\pm0.09$).
Nevertheless, the rms is higher and the Spearman coefficient lower for NGC~4501 ($\rho=0.85$) than that of the three other galaxies ($\rho=0.96$, $\rho=0.91$, $\rho=0.92$). 
Our derived power law indices are in agreement with $\alpha= 0.8-0.9$ found in other studies (Leroy et al. 2008; Wong \& Blitz 2002; BR06).

Leroy et al. (2008) find  a large scatter in both $\alpha$ and $P_0$ in their sample when looking at galaxies individually. Considering the errors, the $\alpha$ and $P_0$ values we find are in agreement with the values found by Wong \& Blitz (2002), BR06 and Leroy et al. (2008).

By calculating the ISM pressure with the BR06 recipe ($P=Ps$ and v$_{\rm turb}=8$ km/s), we find similar values for $P_{0}$ in NGC~4501 and a somewhat higher but still compatible value of $P_{0}$ for NGC~4321 (see Tab. \ref{tab2c}). Nevertheless, when computing the ISM pressure using $P_{\rm tot}=P_{\rm g}+P_{\rm s}$, we systematically find significantly higher values of $P_0$ compared to the calculation with $P_{\rm tot}=Ps$: $P_{0}=1.6 \times 10^4$~cm$^{-3}$K for NGC~4321 and NGC~4501, $P_{0}=2.5 \times 10^4$~cm$^{-3}$K for NGC~4567, and $P_{0}=3.2 \times 10^4$~cm$^{-3}$K for NGC~4568. We thus conclude that the gravitational potential of the gas is often significant compared to the stellar potential. In unperturbed gas-rich galaxies this typically occurs at $R > 0.5 \times R_{25}$. The values of $P_0$ of the tidally interacting galaxies are significantly higher than those of NGC~4321 and NGC~4501.    
\begin{table*}
\caption{Pressure where $\Sigma_{\rm H_{2}} = \Sigma_{\rm HI}$.}           % title of Table
\label{tab2c}    
\centering      
\begin{tabular}{l c c c}       
\hline\hline                 

 Galaxy  &  BR06                         &   $P_{\rm tot}$$=$P$_{\rm s}$  &     $P_{\rm tot}$ $=$ P$_{\rm s}$+P$_{\rm g}$ \\ 
         & P$_0$ ($10^{4}$ cm$^{-3}$ K) & P$_0$ ($10^{4}$ cm$^{-3}$ K)& P$_0$($10^{4}$ cm$^{-3}$ K) \\
\hline                                                    
NGC~4501 &  1.2              & $1.24\pm0.25$ & $1.67\pm0.27$ \\ 
\hline                                                   
NGC~4567 &  ...                         & $1.78\pm1.15$ & $2.50\pm1.20$  \\        
\hline                                                   
NGC~4568 &  ...                            & $2.20\pm0.70$ & $3.20\pm0.72$\\        
\hline                                                      
NGC~4321 & 0.7                 & $1.04\pm0.22$ & $1.60\pm0.24$  \\            
\hline   
\end{tabular}
\end{table*}

%%%%%%%%%

The molecular fraction can be linked to the stellar and gas surface density. A linear relation between $R_{\rm mol}$ and $\Sigma_{\star}$ has been found by Leroy et al. (2008).
These authors stated that ``the simplest explanation is that present-day star formation roughly follows past star
formation. A more aggressive interpretation is that the stellar potential well or feedback is critical to bring gas to high densities.''
Moreover, we investigated the dependence of the molecular fraction $R_{\rm mol}$ on the total gas surface density $\Sigma_{\rm g}$.
We show the $R_{\rm mol}$-$\Sigma_{\star}$ and $R_{\rm mol}$-$\Sigma_{g}$ relations in the upper panels of Fig.~\ref{plotall_2} for all four galaxies and in Appendix~\ref{annexe55} for the individual galaxies (Fig.~\ref{plot4501_2}, \ref{plot4567_2}, \ref{plot4568_2}, and \ref{plot4321_2} for NGC~4501, NGC~4567/68, and NGC~4321, respectively).

\begin{figure*}
   \centering
\includegraphics[width=7cm]{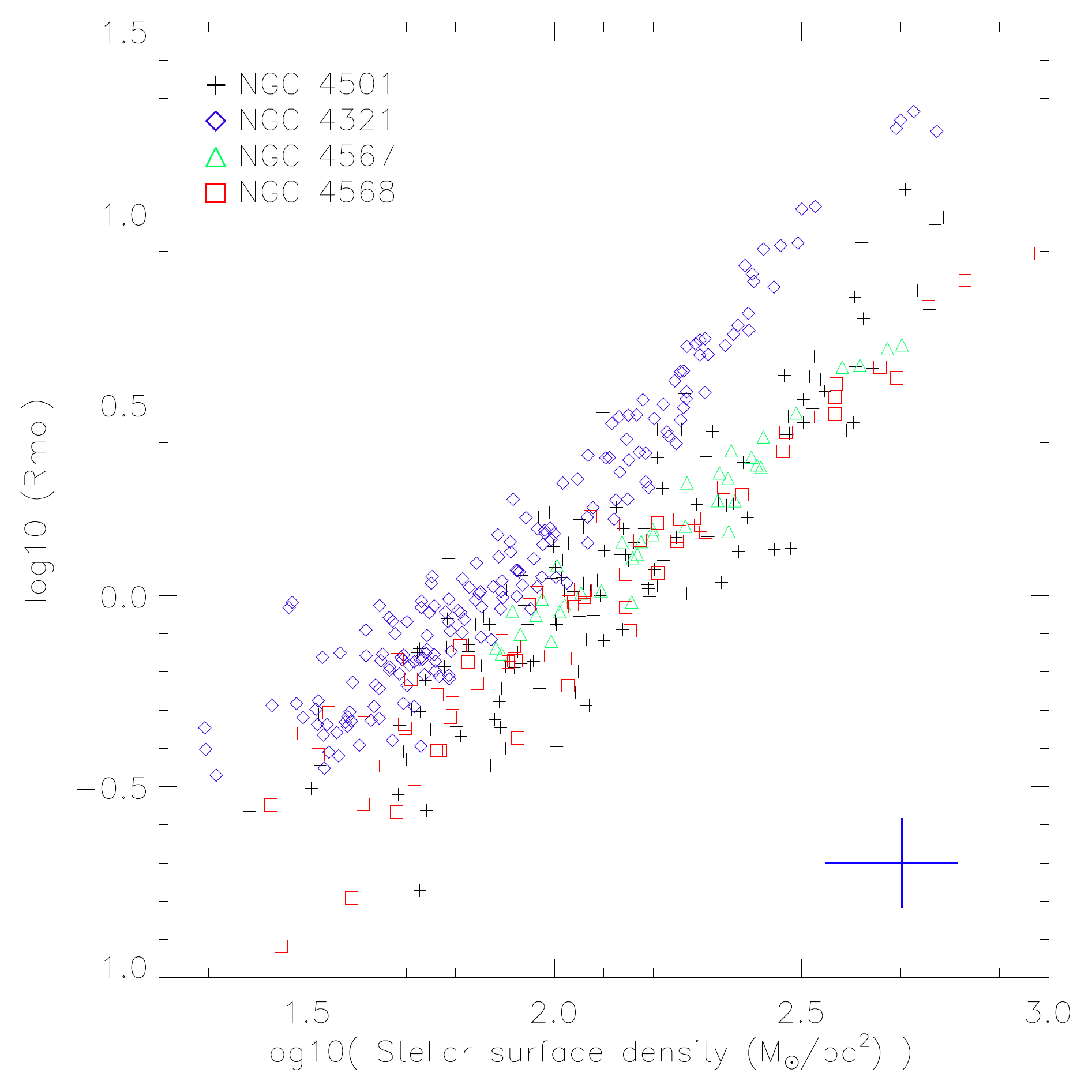}
\includegraphics[width=7cm]{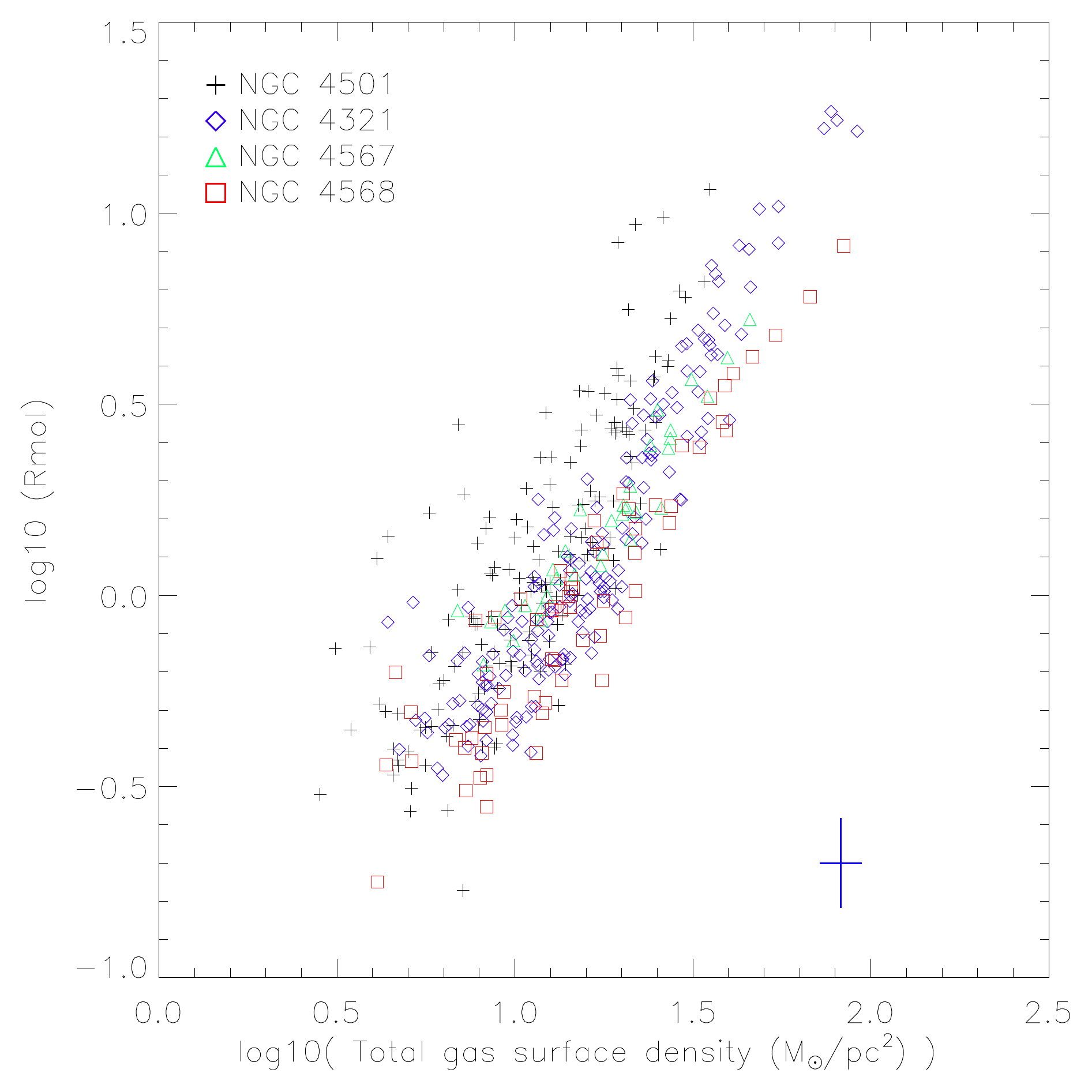}
\includegraphics[width=7cm]{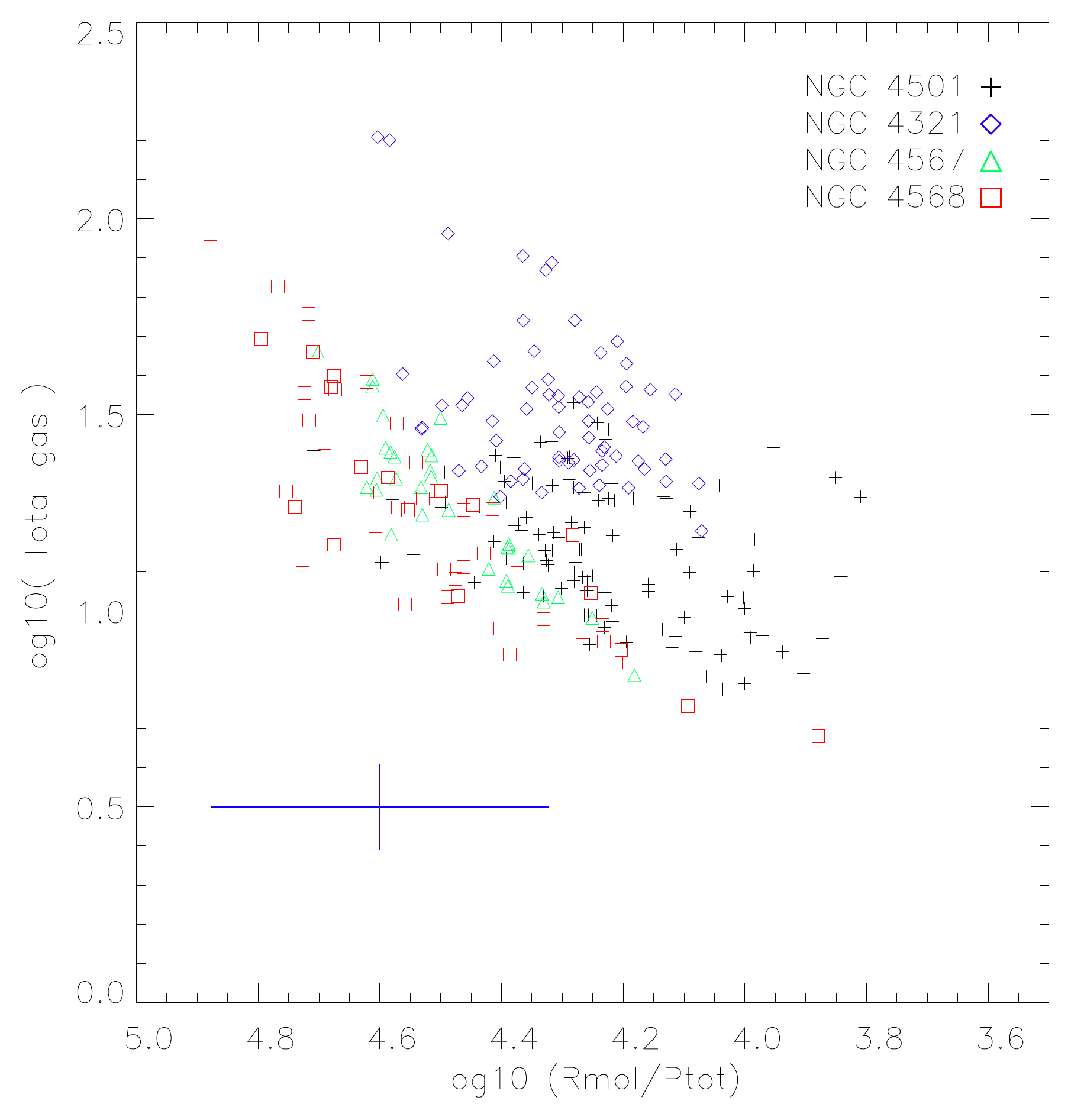}
\includegraphics[width=7cm]{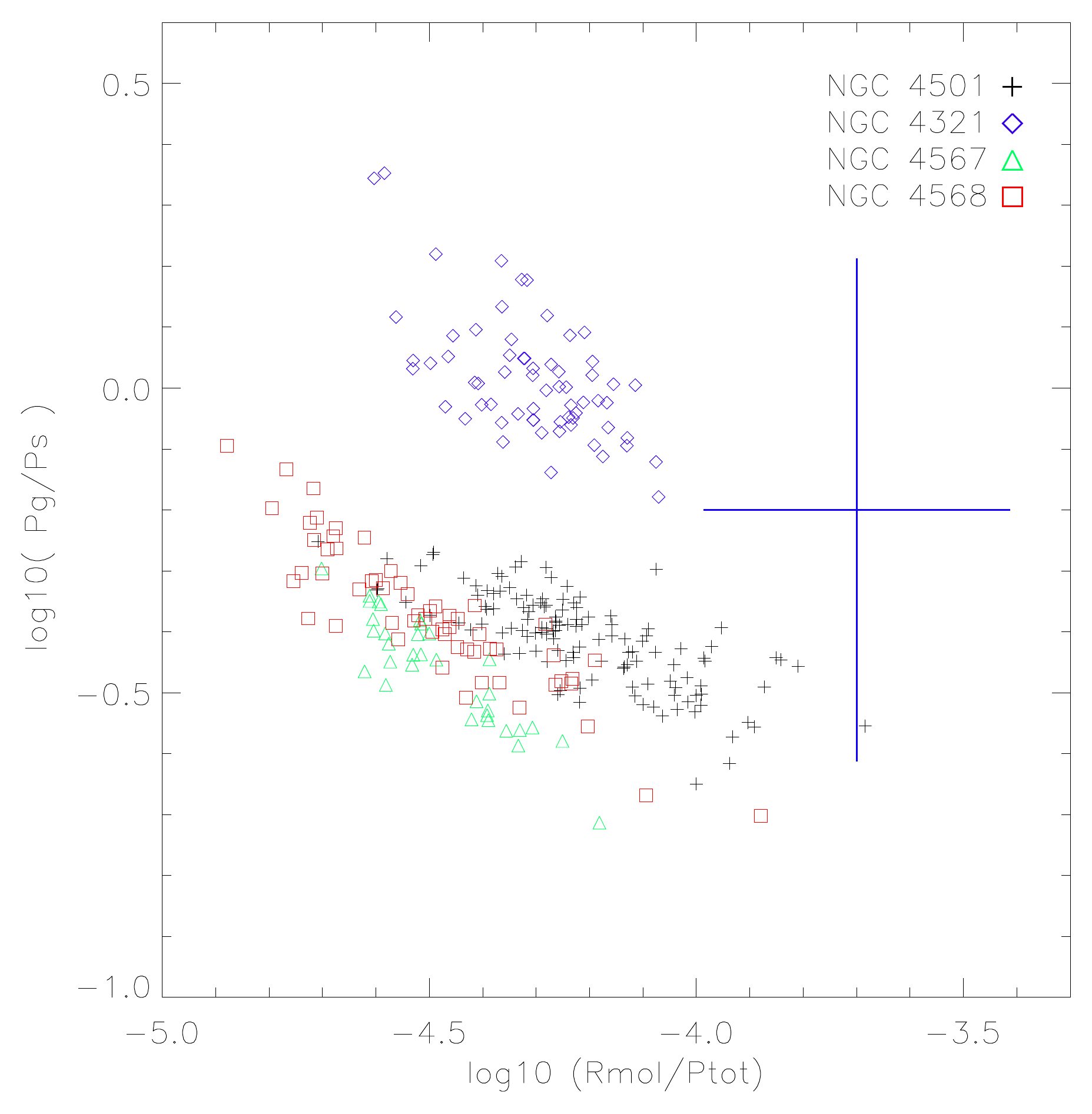}
   \caption{ {\em Upper left panel}: Molecular fraction as a function of stellar surface density. {\em Upper right panel}: Molecular fraction as a function of the total gas surface density.  {\em Lower left panel}: Total gas as a function of molecular fraction divided by ISM pressure. {\em Lower right}: Pressure due to the gravitational potential of the gas divided by the pressure due to the potential of the stellar disk ($P_{\rm g}/P_{\rm s}$) as a function of the molecular fraction divided by ISM pressure. Typical errors bars are given for each relation.}
\label{plotall_2}%
\end{figure*}

The following power laws were fitted to the data:
\begin{eqnarray}
\log_{10}({\rm R_{\rm mol}}) & = & {\rm n_3} \times \log_{10}( \Sigma_{\star}  )   +\gamma_3  \label{fit4} \\
\log_{10}({\rm R_{\rm mol}}) & = & {\rm n_4} \times \log_{10}( \Sigma_{\rm g}  )   +\gamma_4  \label{fit5}. 
\end{eqnarray}
The results of these fits and the corresponding Spearman coefficients are presented in Table~\ref{tab3}.

As has been found by Leroy et al. (2008), we observe an approximately linear correlation between the molecular fraction $R_{\rm mol}$ and the stellar surface density $\Sigma_{\star}$.
A slightly steeper slope of n$_3=1.24\pm0.26$ is found for NGC~4501 compared to n$_3=1.03\pm0.55$ and n$_3=1.00\pm0.22$ for NGC~4567 and NGC~4568, respectively. We find somewhat, but not significantly higher Spearman coefficients and a lower rms for the $R_{\rm mol}$-$\Sigma_{\star}$ relations compared to the $R_{\rm mol}$-$\Sigma_{g}$ relations (mean $<\rho>_{sample}=0.930$ for $R_{\rm mol}$-$\Sigma_{\star}$ compared to a mean $<\rho>_{sample}=0.895$ for $R_{\rm mol}$-$\Sigma_{g}$ ). Thus, as found by Leroy et al. (2008) the best prediction of $R_{\rm mol}$ is determined by using $\Sigma_{\star}$ or $P_{\rm tot}$.

\begin{table*}
\caption{Results of  fitting analysis II}           % title of Table
\label{tab3}    
\centering      
\begin{tabular}{ c c c c}       
\hline\hline                 
 Name       &             &  $R_{\rm mol}$-$\Sigma_{\star}$    &  $R_{\rm mol}$-$\Sigma_{\rm g}$  \\    % table heading                              
\hline                                                     
NGC~4501     & slope      &  ${\rm n}_3=1.24\pm0.26  $         & ${\rm n}_4= 1.26\pm0.14 $                 \\
             & offset       &  $\gamma_3=-2.51\pm0.55$           & $\gamma_4= -1.27\pm0.16 $               \\
             & rms (dex)  &  $0.12  $                          & $ 0.135 $                 \\
             & $\rho$      &  $0.864\pm0.043$                           & $ 0.799 \pm0.035$                \\

\hline          
NGC~4567     & slope      &  ${\rm n}_3=1.03\pm0.55  $         & ${\rm n}_4= 1.21\pm0.34 $                 \\
             & offset       &  $\gamma_3=-2.10\pm1.23$           & $\gamma_4= -1.30\pm0.44 $               \\
             & rms (dex)  &  $0.04  $                          & $ 0.051 $                 \\
             &  $\rho$     &  $0.969\pm0.138$                  & $ 0.959\pm0.060 $                \\
\hline          
NGC~4568     & slope      &  ${\rm n}_3=1.00\pm0.22  $         & ${\rm n}_4= 1.28\pm0.14 $                 \\
             & offset       &  $\gamma_3=-2.05\pm0.47$           & $\gamma_4= -1.53\pm0.19 $               \\
             & rms (dex)  &  $0.08  $                          & $ 0.086 $                 \\
             &$\rho$     &  $0.943\pm0.067$                           & $ 0.905\pm0.034 $                \\
\hline           
NGC~4321     & slope      &  ${\rm n}_3=1.24\pm0.18  $        & ${\rm n}_4=1.59\pm0.13 $                 \\
             & offset       &  $\gamma_3=-2.20\pm0.35$        & $\gamma_4= -1.85\pm0.18 $               \\
             & rms (dex)  &  $0.084$                          & $0.091$                 \\
             &$\rho$     &  $0.945\pm0.040$                           & $0.916\pm0.037$                \\
\hline                                    %inserts single line
\end{tabular}
%\tablefoot{  }
\end{table*}

\subsubsection{Correlations implying ratios}

Because a linear correlation exists between $R_{\rm mol}$ and $P_{\rm tot}$, the correlations between the ratio $R_{\rm mol}/P_{\rm tot}$ and observed quantities can be studied.
Using the Toomre $Q$ parameter defined as 
\begin{equation}
Q = \frac{\Omega v_{\rm turb}}{\pi G \Sigma_{\rm g}}\ ,\label{Qequa}
\end{equation}
where $\Omega=v(R)/R$ is the angular velocity and $v_{\rm turb}$ the velocity dispersion of the gas. Substituting $v_{\rm turb}=(Q \pi G \Sigma_{\rm g})/\Omega$ in Eq.~\ref{eqq5} leads to
\begin{equation}
\rho_{\rm g} = \frac{\Omega^2}{2 \pi G Q^2 }  + \frac{\Omega \Sigma_{\star}} {2 Q \sigma_{\star}} \ .
\label{eqqq}
\end{equation}
Thus, $\rho_{\rm g}$ increases for decreasing $Q$ or increasing gas self-gravitation. Since both $R_{\rm mol}$ and $P_{\rm tot}$ depend on $\rho_{\rm g}$ (see Eq.~\ref{eqq5}), we expect to find a correlation between  $R_{\rm mol}/P_{\rm tot}$ and $Q \propto \Omega/\Sigma_{\rm g}$ if a constant velocity dispersion of the gas is assumed: $R_{\rm mol}/P_{\rm tot} \propto Q^{1.4}$ (see Sect.~\ref{analy}).

\begin{figure}
   \centering
\includegraphics[width=9.5cm]{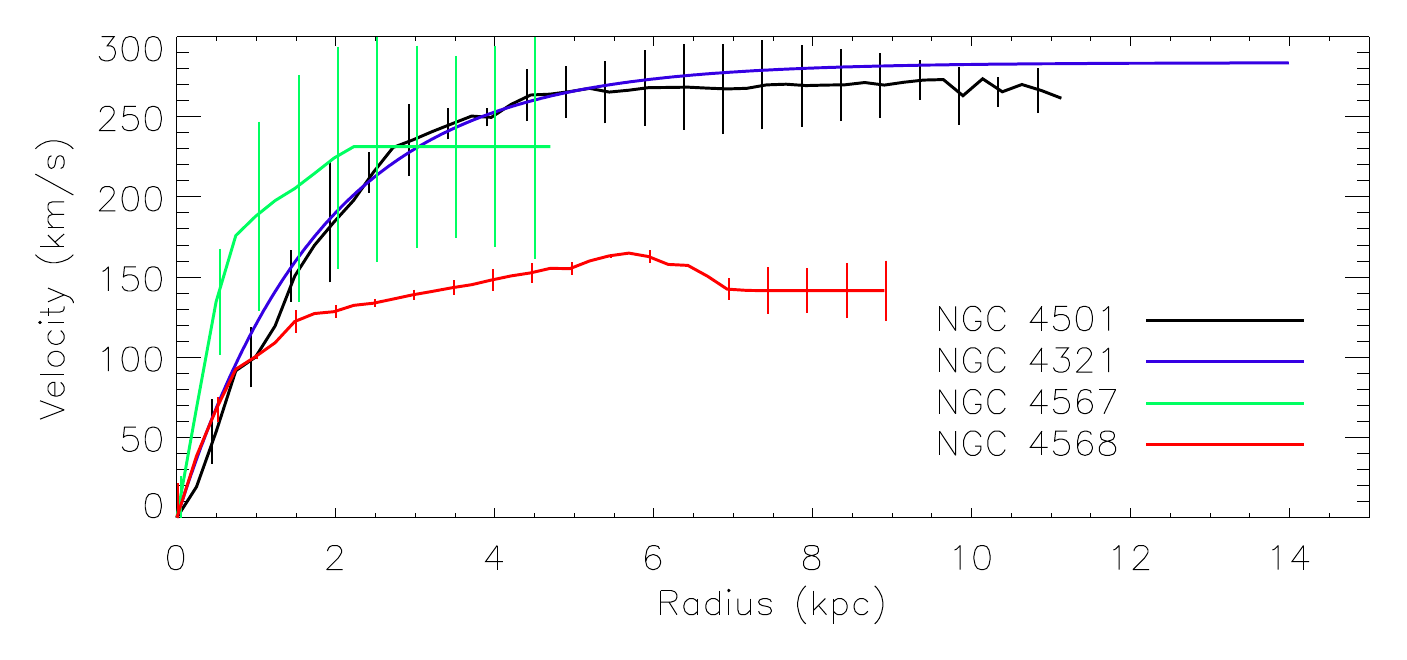}
   \caption{Rotation curves of the four galaxies from CO(2-1) data for NGC 4501 and NGC 4567/68 and from the two-parameter fit from Leroy et al. 2013 for NGC 4321.}
\label{rotcurv}%
\end{figure}

The rotation curves are derived from the IRAM CO(2-1) data along the major axis of the galaxies for NGC 4501 and NGC 4567/68. For NGC 4321 we used a two-parameter fit of the rotation curve from HERACLES and THINGS data $v_{\rm rot}=v_{\rm flat}\times(1-exp(-R_{\rm gal}/l_{\rm flat}))$ with $V_{\rm flat}=283.7$~km s$^{-1}$ and $l_{\rm flat}=1.8$~kpc (Leroy et al. 2013). These rotation curves are presented in Fig. \ref{rotcurv}.

We show $R_{\rm mol}/P_{\rm tot}$ as a function of $Q$ (assuming a turbulent velocity dispersion of $10$~km\,s$^{-1}$ for the gas) in the four galaxies (see Fig.~\ref{4gal255}). For the individual galaxies the plots are presented in Figs.~\ref{plot4501_3}, \ref{plot4567_3}, \ref{plot4568_3}, and \ref{plot4321_3} for NGC~4501, NGC~4567, NGC~4568, and NGC~4321, respectively.
We found a correlation in the four galaxies (Spearman coefficient $0.6<\rho<0.9$). The mean slope is $1.0$ and the typical rms of the correlations are less then 0.1 dex (a factor of 1.26). Thus, as expected, $R_{\rm mol}/P_{\rm tot}$ is low where the gas is more self-gravitating and $Q$ is low.

In a second step we looked at the relation between $R_{\rm mol}/P_{\rm tot}$ and $P_{\rm g}/P_{\rm s}$ to investigate the importance of self-gravitating gas (see Figs.~\ref{plot4501_2}, \ref{plot4567_2}, \ref{plot4568_2}, and \ref{plot4321_2}).
We find an anti-correlation (Spearman $\rho<-0.7 $ for NGC~4501 and NGC~4567/68) indicating that when the gas is at the limit of self-gravitation (high $P_{\rm g}/P_{\rm s}$), $R_{\rm mol}/P_{\rm tot}$ is low because the increase in total pressure is more important than the increase in the molecular fraction. For NGC~4321 the pressure due to gravitational potential of the gas  dominates everywhere ($P_{\rm g}/P_{\rm s} > 1$). The correlation is thus less well defined (Spearman $\rho= 0.53$).

The relation between the Toomre $Q$ parameter and the molecular $SFE_{\rm H_{2}}$ is shown in the lower panels of Fig.~\ref{plot4501_3}, \ref{plot4567_3}, \ref{plot4568_3}, and \ref{plot4321_3}. We find Spearman $\rho$ coefficients of less then 0.1 (see Tab. \ref{tab4}) indicating that these quantities are not correlated within the galaxies. 

For completeness we correlated  $R_{\rm mol}/P_{\rm tot}$ with $\Sigma_{\rm g}$. We find no correlation for NGC~4501 and NGC~4321 (Spearman $\rho\sim-0.3$). For NGC~4567 and NGC~4568 we find an anti-correlation (Spearman $\rho\sim-0.8$). However, $R_{\rm mol}/P_{\rm tot}$ is better correlated to the Toomre $Q$ parameter than to $\Sigma_{\rm g}$. 

Finally, we tested the correlation between the two ratios $SFE_{\rm H_{2}}$ and $R_{\rm mol}$/$P_{\rm tot}$. We find no clear sign of correlation when looking at galaxies individually (see left panel of Fig.~\ref{plot4501_3}, \ref{plot4567_3}, \ref{plot4568_3}, and \ref{plot4321_3} and Table~\ref{tab4} for Spearman $\rho$ coefficients).

We conclude that the ratio $R_{\rm mol}/P_{\rm tot}$ is approximately proportional to the Toomre $Q$ parameter. In regions of low $R_{\rm mol}/P_{\rm tot}$ the ISM is self-gravitating. 
The ratio $R_{\rm mol}/P_{\rm tot}$ can thus be used to detect regions where self-gravitation of the ISM is important.

\begin{table*}
\caption{Results of  fitting analysis III}           % title of Table
\label{tab4}    
\centering      
\begin{tabular}{l c c c c c c c c c  }       
\hline\hline                 

 NGC    &\multicolumn{4}{c}{Q-$R_{\rm mol}/P_{\rm tot}$}&\multicolumn{4}{c}{$P_{\rm g}/P_{\rm s}$-$R_{\rm mol}/P_{\rm tot}$ } \\ 
         &$\rho$ &  rms (dex)    & slope        & offset                        & $\rho$  &  rms (dex)  & slope        & offset                             \\
\hline                                                    
4501 &$0.629\pm0.080$&  0.090  & $0.84\pm0.36$& $3.80\pm1.56$              & $-0.759\pm0.089$& $0.050$   &$-0.43\pm0.93$&$-2.22\pm3.94$                \\ 
\hline                                                   
4567 &$0.638\pm0.170$&  0.084  & $1.05\pm0.71$& $7.20\pm3.20$            &  $-0.914\pm0.179$& $0.032$  & $-0.73\pm1.44$& $-3.71\pm6.45$                           \\        
\hline                                                    
4568 &$0.758\pm0.102$&  0.093  & $1.32\pm0.59$& $6.15\pm2.71$            &  $-0.850\pm0.129$&   $0.044$& $-0.60\pm0.75$ & $-3.07\pm3.45$                         \\        \hline                                                      
4321 &$0.904\pm0.055$& 0.036   & $0.90\pm0.62$& $4.25\pm2.37$            &  $-0.530\pm0.128$&   $0.069$&  $-0.82\pm1.43$ & $-3.52\pm6.16$                            \\            
\hline   

                              %inserts single line
\end{tabular}
\end{table*}

\begin{table*}
\caption{Results of  fitting analysis VI}           % title of Table
\label{tab44}    
\centering      
\begin{tabular}{l c c c}       
\hline\hline    
NGC   &$\Sigma_{\rm g}$-$R_{\rm mol}/P_{\rm tot}$  &  $SFE_{\rm H_{2}}$-$R_{\rm mol}/P_{\rm tot}$   & $SFE_{\rm H_{2}}$-Q \\  

       &          $\rho$               &     $\rho$                        &   $\rho$                      \\
\hline                                                    
4501   &   $-0.315\pm0.092$                    &     $-0.339\pm0.089$            & $+0.007\pm0.094$                      \\
\hline 
4567   &   $-0.881\pm0.164$                    &     $+0.558\pm0.180$             &  $+0.077\pm0.184$  \\
\hline                                                    
4568   &   $-0.818\pm0.102$                    &     $+0.016\pm0.132$                & $-0.076\pm0.125$  \\
\hline  
4321   &   $-0.272\pm0.124$                    &     $-0.548\pm0.123$            &$+0.036\pm0.081$  \\
   \hline                              %inserts single line
\end{tabular}
\end{table*}

\subsection{Continuous regions deviating from the correlations}

Continuous regions within the galaxies which deviate from the established correlations can potentially give insight into the physics of the ISM and the associated star formation.
In NGC~4501, three spatial regions were selected with the following criteria (see Fig.~\ref{totgas++}):
\begin{itemize} 
\item 
regions with high atomic gas surface densities where $\Sigma_{\rm H_{\rm I}} >$ 8 $ {\rm M}_\odot {\rm pc}^{-2}$ compared to the mean H{\sc i} surface density of 5 ${\rm M}_\odot {\rm pc}^{-2}$ (blue squares in Fig.~\ref{plot4501} and blue contours in Fig.~\ref{totgas++}). These regions are located in the western and northwestern part of the disk and show (i) a higher $SFE_{\rm H_{2}}$ ($\sim$ 1 Gyr$^{-1}$) than the mean molecular $SFE_{\rm H_{2}}$ of 0.61 Gyr$^{-1}$ and (ii) a lower ratio of molecular fraction to ISM pressure  (higher P$_0$). The molecular fraction in this region is not lower than the mean at this radius. Within these regions the gas is nearly self-gravitating (see Fig.~\ref{Pgps01}).
\item 
two low molecular SFE regions were selected with a $SFE_{\rm H_{2}}$ below 0.49 Gyr$^{-1}$, corresponding to  $1 \sigma$ below the average $SFE_{\rm H_{2}}$ of NGC~4501: (i) a U-shaped region with a low $SFE_{\rm H_{2}}$ in the southeast (green triangles and green contours in Fig.~(\ref{totgas++}) and (ii) a northwestern low $SFE_{\rm H_{2}}$ region (orange diamonds in Fig.~\ref{plot4501} and orange contours in Fig.~\ref{totgas++}). These two regions do not deviate from the SFR-$\Sigma_{\rm g}$ and the $R_{\rm mol}$-$P_{\rm tot}$ relations, but show a $\sim$ $0.1$~dex higher Toomre $Q$ (see lower panel of Fig.~\ref{plot4501_3}) than  the regions of high atomic gas surface density.
\item 
a high $R_{\rm mol}/P_{\rm tot}$ region in the southwestern quadrant (red crosses  in Fig.~\ref{plot4501} and red contours in Fig.~\ref{totgas++}). In this region $R_{\rm mol}$ deviates by more than $1 \sigma$ from the $R_{\rm mol}$-$P_{\rm tot}$ relation. The molecular $SFE_{\rm H_{2}}$ is slightly but not significantly lower than the mean in this region. The ISM pressure is dominated by the stellar gravitational potential as $\Sigma_{\rm g} \sim 20$ M$_{\odot}$pc$^{-2}$ (see Fig.~\ref{Pgps01})
\end{itemize}

In NGC~4568 two regions were selected with a molecular $SFE_{\rm H_{2}}$ that is at least $1 \sigma$ below the established correlation ($SFE_{\rm H_{2}}<0.61$~Gyr$^{-1}$ compared to the mean $SFE_{\rm H_{2}}$ of $0.74$~Gyr$^{-1}$):
\begin{itemize}
\item the first region is located in the northwestern ridge of the interaction between the two galaxies (yellow contours and diamonds in Fig.~\ref{totgas4567} and \ref{plot4568}). 
\item the second region is located in the south of NGC~4568 (purple triangles and contours in Fig.~\ref{totgas4567} and \ref{plot4568}). 
\end{itemize}
Both regions have H{\sc i} surface densities of $\sim 8.$ M$_{\odot} $pc$^{-2}$. They do not deviate  from the SFR-$\Sigma_{\rm g}$ or from the $R_{\rm mol}$-$P_{\rm tot}$ relation. We note that the second low molecular $SFE_{\rm H_{2}}$ region has a rather high Toomre $Q$ (lower panel of Fig.~\ref{plot4568_3}).

We conclude that continuous low $SFE_{\rm H_{2}}$ regions show a somewhat higher $Q$ compared to the mean, but they do not deviate from the $R_{\rm mol}/P_{\rm tot}$ relation.
On the other hand, regions where the gas is nearly self-gravitating have a lower Q, a higher molecular $SFE_{\rm H_{2}}$, and a lower $R_{\rm mol}/P_{\rm tot}$ with respect to the mean.

\section{ Analytical model \label{analy}}

\begin{figure*}
\centering

\includegraphics[width=18cm]{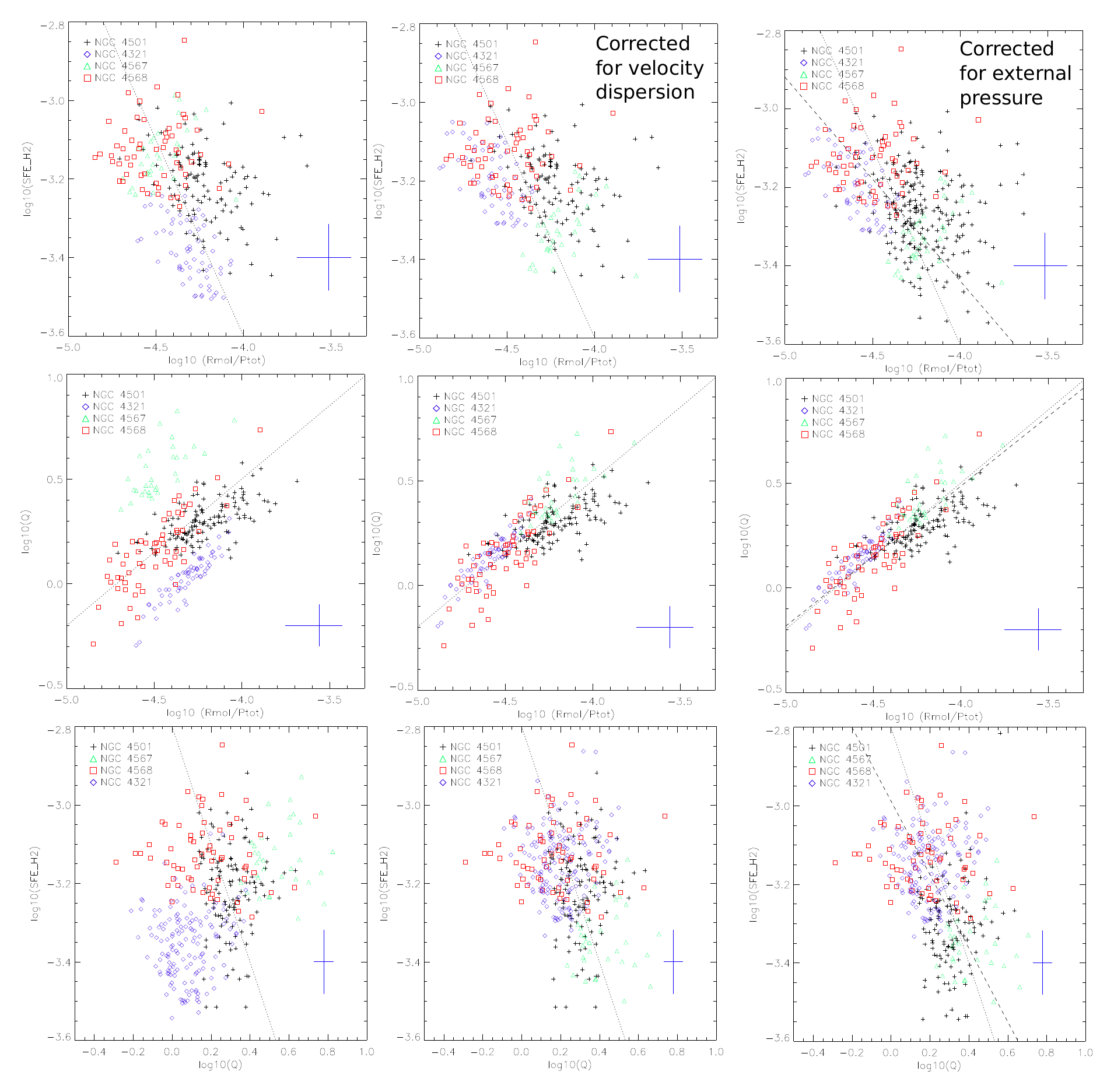}

   \caption{Left (from top to bottom) for NGC~4501, NGC~4567/68, and NGC~4321: $SFE_{\rm H_{2}}$-$R_{\rm mol}/P_{\rm tot}$, $Q$-$R_{\rm mol}/P_{\rm tot}$, and $SFE_{\rm H_{2}}$-$Q$. Middle: Relations corrected for the unknown velocity dispersion (see text). Right: With the $SFE_{\rm H_{2}}$ of NGC~4501 corrected for the external pressure (leading to a 0.1~dex lower $SFE_{\rm H_{2}}$). The dotted lines correspond to the correlations predicted by the analytical model: $\log(SFE_{\rm H_2})/\log(R_{\rm mol}/P_{\rm tot}=-1$, $\log(R_{\rm mol}/P_{\rm tot})/\log(Q)=1.4$, and $\log(SFE_{\rm H_2})/\log(Q)=-1.5$. The dashed lines correspond to correlations fitted to the data: $\log(SFE_{\rm H_2})/\log(R_{\rm mol}/P_{\rm tot}=-0.5$, $\log(R_{\rm mol}/P_{\rm tot})/\log(Q)=1.5$, and $\log(SFE_{\rm H_2})/\log(Q)=-0.9$.}

\label{4gal255}%
\end{figure*}

In the previous sections we  investigate correlations of physical parameters within the galaxies. In a further step we now compare the correlations between the galaxies.
The relations $R_{\rm mol}/P_{\rm tot}$-$Q$, $SFE_{\rm H_{2}}$-$Q$, and $SFE_{\rm H_{2}}$-$R_{\rm mol}/P_{\rm tot}$ for the four galaxies are presented in the left panels of Fig.~\ref{4gal255}.
%\begin{figure}
%\centering
%\includegraphics[width=7cm]{FFIG/4gal_SFE_Rmp.pdf}
%\includegraphics[width=7cm]{FFIG/4gal_Rmol_P_Q.pdf}
%\includegraphics[width=7cm]{FFIG/4gal_SFE_Q.pdf}
%   \caption{From top to bottom for NGC~4501, NGC~4567/68 and NGC~4321: $SFE_{\rm H_{2}}$-$R_{\rm mol}/P_{\rm tot}$, $Q$-$R_{\rm mol}/P_{\rm tot}$ and $SFE_{\rm H_{2}}$-$Q$.
%   The dotted lines correspond to the correlations predicted by the analytical model: $\log(SFE_{\rm H_2})/\log(R_{\rm mol}/P_{\rm tot}=-1$,
%   $\log(R_{\rm mol}/P_{\rm tot})/\log(Q)=1.4$, and $\log(SFE_{\rm H_2})/\log(Q)=-1.5$.}
%\label{4gal2}%
%\end{figure}

In NGC~4321 nearly half of the galaxy has values below $Q=1$ indicating that either the turbulent velocity dispersion is underestimated (see Eq.~\ref{Qequa}) or that the gas surface density, i.e., the CO-H$_{2}$ conversion factor, is overestimated. Since the metallicity of NGC~4321 is comparable to that of NGC~4501 (Skillman et al. 1996), we do not expect a variation of the CO-H$_{2}$ conversion factor between the galaxies. Thus, NGC~4321  most probably has a velocity dispersion in excess of $10$~km\,s$^{-1}$. 

A clear correlation between $R_{\rm mol}/P_{\rm tot}$ and $Q$ is found in all galaxies individually where $R_{\rm mol}/P_{\rm tot}$ increases with Q. Whereas NGC~4501 and NGC~4568 follow approximately the same relation, NGC~4321 and NGC~4567 show offsets of $\sim 0.3$~dex in opposite directions. We do not detect clear signs of correlation between
the SFE and $Q$ within the individual galaxies. The molecular SFE of NGC~4321 is about $0.2$~dex lower than that of NGC~4501, NGC~4567, and NGC~4568. 
The resulting $SFE_{\rm H_{2}}$-$R_{\rm mol}/P_{\rm tot}$ is dominated by the offset of NGC~4321.

We used the analytical model described in Vollmer \& Leroy (2011) to investigate these relations in detail. This analytical model considers the ISM (warm, cold, and molecular phases) as a single turbulent gas. This gas is assumed to be in vertical hydrostatic equilibrium (Eq.~\ref{eqq5}).
In this picture, the warm, cold, and molecular phases of the ISM are a single entity.
Locally, the phase of the gas may depend on the local pressure, metallicity, stellar radiation field, stellar winds, and
shocks. These factors are viewed as secondary by making a few simplifying assumptions. 

The equilibrium between the different
phases of the ISM and the equilibrium between turbulence and star formation depends on three local timescales: the turbulent
crossing time $t_{\rm turb}$, the molecule formation timescale ${\rm t}_{\rm mol}$, and the local free-fall timescale ${\rm t}_{\rm ff}$ of a gas cloud.

The free-fall ${\rm t}_{\rm ff}$ and the H$_2$ formation timescale ${\rm t}_{\rm mol}$ are
\begin{eqnarray}
{\rm t}_{\rm ff}&=&\sqrt{\frac{3\pi}{32 G \rho \phi_{\rm v}^{-1}}}, \label{anal1}\\
{\rm t}_{\rm mol}&=&\frac{\alpha}{\rho \phi_{\rm v}^{-1}},      
\end{eqnarray}
with $\rho$ the average gas density, $\phi_{\rm v}$ the volume filling factor, G the gravitational constant, and $Z \propto 1/\alpha$ the metallicity computed assuming a closed box model:
\begin{equation}
\alpha = 7.2 \times 10^7 \times (\log(\frac{\Sigma_{\star}+\Sigma_{\rm g}}{\Sigma_{\rm g}}))^{-1}, {\rm M_{\odot} yr^{-1} pc^{-3}}\ .
\end{equation}

The molecular fraction can be written as
\begin{eqnarray}
R_{\rm mol}&=&\frac{{\rm t}_{\rm ff}}{{\rm t}_{\rm mol}}\ {\rm and}\\
{\rm f}_{\rm mol}&=&\frac{\Sigma_{\rm H_{2}}}{\Sigma_{\rm g}}=\frac{1}{R_{\rm mol}^{-1}+1}\ .
\label{anal2}
\end{eqnarray}
This gives $\Sigma_{\rm H{_2}}={\rm f}_{\rm mol}\times \Sigma_{\rm g}$ and $\Sigma_{\rm HI}=(1-{\rm f}_{\rm mol})\times \Sigma_{\rm g}$.
In addition, we assumed a conversion factor $N_{\rm H_2}/I_{\rm CO} \propto \alpha \propto Z^{-1}$. 

The model assumes that the gas is turbulent, so that the turbulent velocity dispersion exceeds the sound speed throughout the disk. This turbulence is driven by
SNe which input their energy in turbulent eddies that have a characteristic length scale, $l_{\rm driv}$, and a characteristic velocity, $v_{\rm turb}$.
It is assumed that the energy input rate into the ISM due to SNe is cascaded to smaller scales without losses caused by turbulence. At scales smaller than the
size of the largest self-gravitating clouds the energy is dissipated via cloud contraction and star formation.
One can connect the energy input into the ISM by SNe directly to the SFR assuming a constant initial mass function independent of environment (Eq.~\ref{EF}). 
In addition, we assume energy flux conservation (Eq.~\ref{EF}), an SFR recipe (Eq.~\ref{SFRR}), and a prescription for the turbulent viscosity (Eq.~\ref{TV}):
\begin{eqnarray}
\xi\dot{\Sigma_{\star}}&=&\Sigma_{\rm g}\frac{v_{\rm turb}^3}{{\rm l}_{\rm driv}}, \label{EF}\\
\dot{\Sigma_{\star}}&=&\phi_{\rm v}\frac{\rho}{{\rm t_{\rm SF}^l}}{\rm l_{\rm driv}}, \label{SFRR}\\
\Sigma_{\rm g}&=&\rho H,\\
\nu&=&v_{\rm turb}{\rm l}_{\rm driv} \label{TV}.
\end{eqnarray}
Here $v_{\rm turb}$ is the gas turbulent velocity dispersion, l$_{\rm driv}$ the turbulent driving length scale, $\rho$ the disk mid-plane gas density, $H$ the thickness of the gas disk, $\xi=4.6\times 10^{-8}$ (pc$^2$ yr$^{-1}$) the constant relating SN energy input into star formation, $\nu$ the viscosity, and $\dot{\Sigma_{\star}}$ the star formation rate. 

The resulting properties of the galactic gas disk depend on the stellar surface density $\Sigma_*$, the angular velocity $\Omega$, the disk radius $R$, and three free parameters: (1) the mass accretion rate $\dot{\rm M}=2\pi\nu\Sigma_{\rm g}$ of the disk, (2) the unknown scaling factor $\delta$ relating l$_{\rm driv}$ to the size of gravitationally bound clouds, and (3) the Toomre $Q$ parameter of the disk. In addition, we allow for a varying X$_{\rm co}$ conversion factor. 

This set of equations was solved radially and the relations $R_{\rm mol}/P_{\rm tot}$-Q, $SFE_{\rm H_{2}}$-Q, and $SFE_{\rm H_{2}}$-$R_{\rm mol}/P_{\rm tot}$) were computed.
We assumed a constant turbulent velocity dispersion $v_{\rm turb}$ and thus $Q \propto \Omega/\Sigma_{\rm g}$. This assumption will be dropped later.
 
\begin{figure*}
   \centering
\includegraphics[width=15cm]{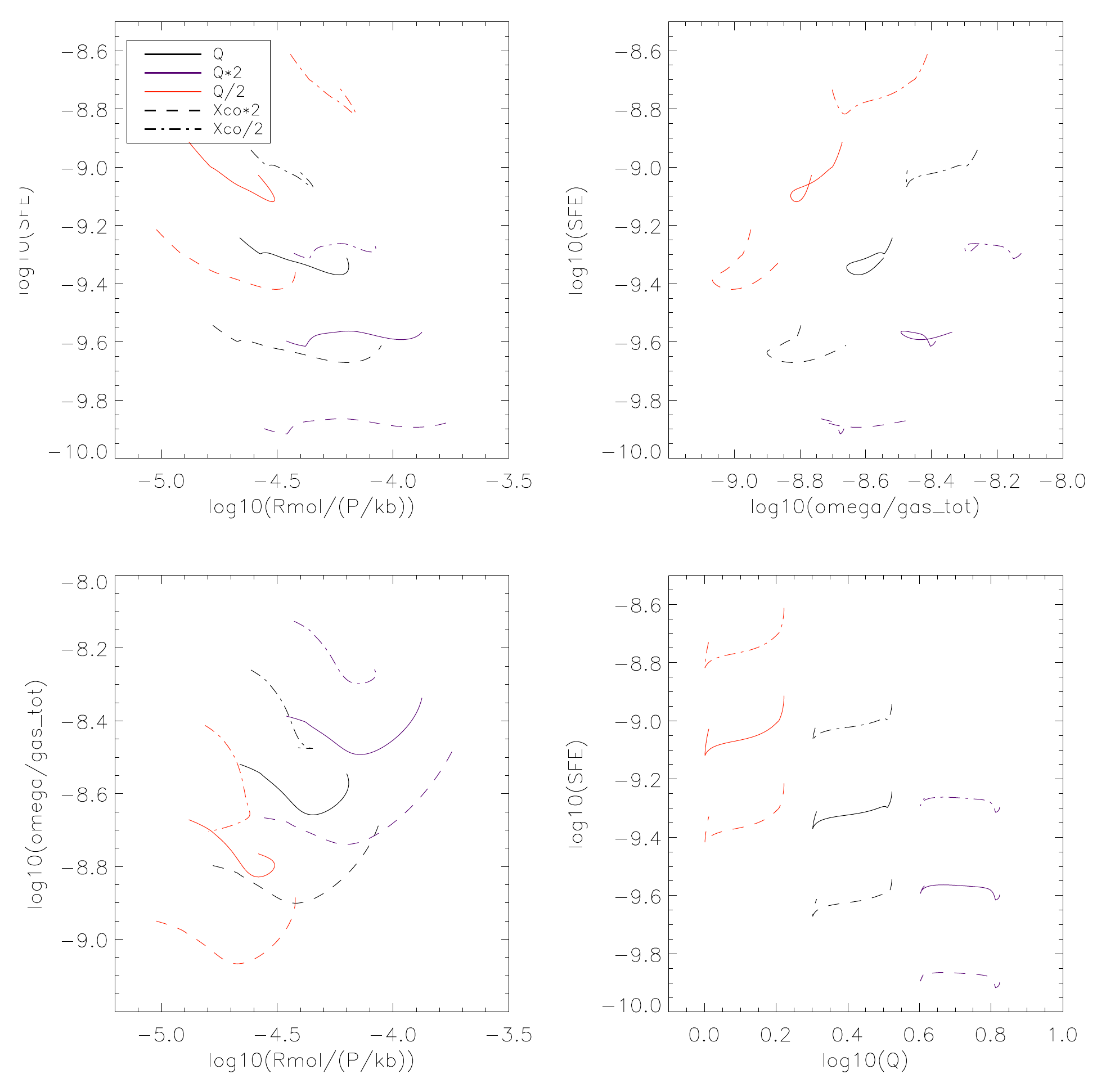}
   \caption{Analytical model. {\em Top left panel}: $SFE_{\rm H_{2}}$-$R_{\rm mol}/P_{\rm tot}$ for three different values of $Q$ (colors) and variation of X$_{\rm co}$ (dotted and dash-dotted lines). {\em Top right}: $SFE_{\rm H_{2}}$-$Q$. {\em Lower left}: $Q$-$R_{\rm mol}/P_{\rm tot}$. {\em Lower right}: $SFE_{\rm H_{2}}$-$Q$.}
\label{analmodel}%
\end{figure*}

Leroy et al. (2008) and Schruba et al. (2011) observed variations in the molecular $SFE_{\rm H_{2}}$ among galaxies in the HERACLES sample. Overall variations of the CO to H$_2$ conversion factor can significantly affect the measured molecular $SFE_{\rm H_{2}}$. To investigate the effect of  X$_{\rm CO}$ variations on the model results, we varied X$_{\rm CO}$ by a factor of $\pm$ 0.3 dex. We assume a radial profile of $Q$, which is constant over most of the galactic disk and increases toward the center and at the edge of the disk (see Fig. 9 of Leroy et al. 2008 and Fig. 2 of Vollmer \& Leroy 2011).
For the CO-H$_2$ conversion factor, we varied $Q$ by a factor of $\pm$ 0.3 dex. The model results are presented in Fig.~\ref{analmodel}.

In the framework of this model, we note that   $R_{\rm mol}/P_{\rm tot}$ is relatively insensitive to X$_{\rm co}$ variations. 
The scatter due to  X$_{\rm co}$ variations is perpendicular to the SFE-Q relations, whereas it is highly inclined with respect to the $R_{\rm mol}/P_{\rm tot}$-Q relation.
Thus, the scatter due to  X$_{\rm co}$ variations easily masks an underlying correlation between SFE and Q.
We found the following model slopes: $\log(SFE_{\rm H_2})/\log(Q)=-1.5$, $\log(R_{\rm mol}/P_{\rm tot})/\log(Q)=1.4$,  and  $\log(SFE_{\rm H_2})/\log(R_{\rm mol}/P_{\rm tot})=-1.0$.

The analytical model helps to interpret the observed relations.
The slopes of the individual observed $R_{\rm mol}/P_{\rm tot}$-$Q$ relations are in agreement within the error bars with the prediction of the analytical model (middle panels of 
Figs.~\ref{plot4501_3}, \ref{plot4567_3}, \ref{plot4568_3}, and \ref{plot4321_3}). Except for NGC~4567, the individual observed $SFE$-$R_{\rm mol}/P_{\rm tot}$ relations show
the trend predicted by the analytical model (upper panels of Figs.~\ref{plot4501_3}, \ref{plot4567_3}, \ref{plot4568_3}, and \ref{plot4321_3}). 
However, the predicted slope of the  $SFE$-$Q$ relation is not observed in our data, because it is masked by variations of the CO-H$_{2}$ conversion factor
(lower panels of Figs.~\ref{plot4501_3}, \ref{plot4567_3}, \ref{plot4568_3}, and \ref{plot4321_3}).
Thus, the slope of the observed $R_{\rm mol}/P_{\rm tot}$-$Q$ relation (Fig.~\ref{4gal255}) is comparable to that of the analytical model for all galaxies (Fig.~\ref{analmodel}):
$R_{\rm mol}/P_{\rm tot}$ increases with increasing Toomre $Q$.
However, there are global offsets between the galaxies in the $R_{\rm mol}/P_{\rm tot}$-Q relation (Fig.~\ref{4gal255}).

\begin{figure}
\centering
\includegraphics[width=9cm]{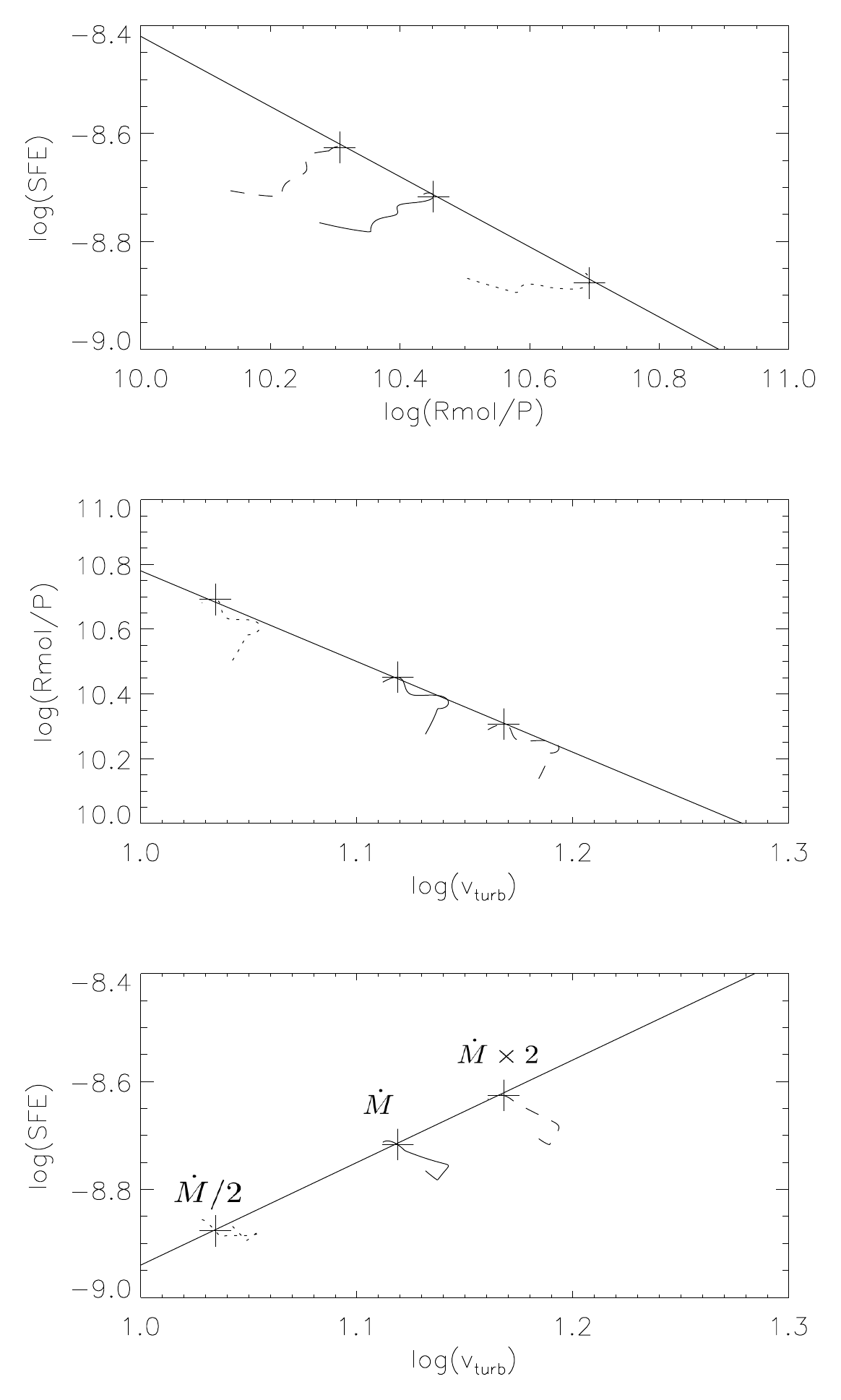}
   \caption{Analytical model. {\em Top panel}: molecular $SFE_{\rm H_{2}}$ as a function of $R_{\rm mol}/P_{\rm tot}$ for different mass accretion rates. The solid line corresponds to $y=-0.65x+1.92$. {\em Middle panel}: $R_{\rm mol}/P_{\rm tot}$ as a function of the turbulent velocity. The solid line corresponds to $y=-2.8x+13.58$. {\em Lower panel}: $SFE_{\rm H_{2}}$ as a function of the turbulent velocity. The solid line corresponds to  $y=1.9x+10.84$.}
\label{analmodel2}%
\end{figure}

These offsets can be explained by the variation of the second free parameter of the analytical model, the mass accretion rate within the gas disk $\dot{M}$.
With increasing $\dot{M}$ the turbulent velocity dispersion of the disk $v_{\rm turb}$ increases. To investigate the influence of the mass accretion rate on
the observed relations, we varied the mass accretion rate by $\pm 0.3$~dex (Fig.~\ref{analmodel2}). The associated velocity dispersions are $v_{\rm turb}=10.7$, $13.2$, and $14.8$~km\,s$^{-1}$.

A linear fit to the model results yields $\log(R_{\rm mol}/P_{\rm tot})=-2.8 \log(v_{\rm turb})+13.58$, $\log(SFE_{\rm H_{2}})=1.9 \log(v_{\rm turb})+10.84$, and
$\log(SFE_{\rm H_{2}})=-0.65 \log(R_{\rm mol}/P_{\rm tot})+1.92$. 
Thus, for increasing v$_{\rm turb}$ the $SFE_{\rm H_{2}}$ increases and $R_{\rm mol}/P_{\rm tot}$ decreases. 

To correct the observed relations for the unknown velocity dispersion,
we assumed a 0.1~dex higher velocity dispersion for NGC~4321 (so that the observed Toomre $Q$ exceeds unity) and a 0.1~dex lower v$_{\rm turb}$ for NGC~4567.
These assumptions are chosen to align all galaxies in the $R_{\rm mol}/P_{\rm tot}$-Q relation (middle panels of Fig.~\ref{4gal255}). 
By doing this, the SFE of NGC~4321 increases and that of NGC~4567 decreases and a correlation between the $SFE_{\rm H_{2}}$-Q and $SFE_{\rm H_{2}}$-$R_{\rm mol}/P_{\rm tot}$ appears.
As previously noted, the latter correlation is partially masked by variations of the CO-H$_{2}$ conversion factor.
%\begin{figure}
%\centering
%\includegraphics[width=7cm]{FFIG/4gal_SFE_Rmp_dex1.pdf}
%\includegraphics[width=7cm]{FFIG/4gal_Rmol_P_Q_dex1.pdf}
%\includegraphics[width=7cm]{FFIG/4gal_SFE_Q_dex1.pdf}
%   \caption{As Fig.~\ref{analmodel2} with the observed relations ``corrected'' for the unknown velocity dispersion (see text).
%   The dotted lines correspond to the correlations predicted by the analytical model: $\log(SFE_{\rm H_2})/\log(R_{\rm mol}/P_{\rm tot}=-1$,
%   $\log(R_{\rm mol}/P_{\rm tot})/\log(Q)=1.4$, and $\log(SFE_{\rm H_2})/\log(Q)=-1.5$.}
%\label{4gal2dex1}%
%\end{figure}

\begin{figure}
\centering
\includegraphics[width=9cm]{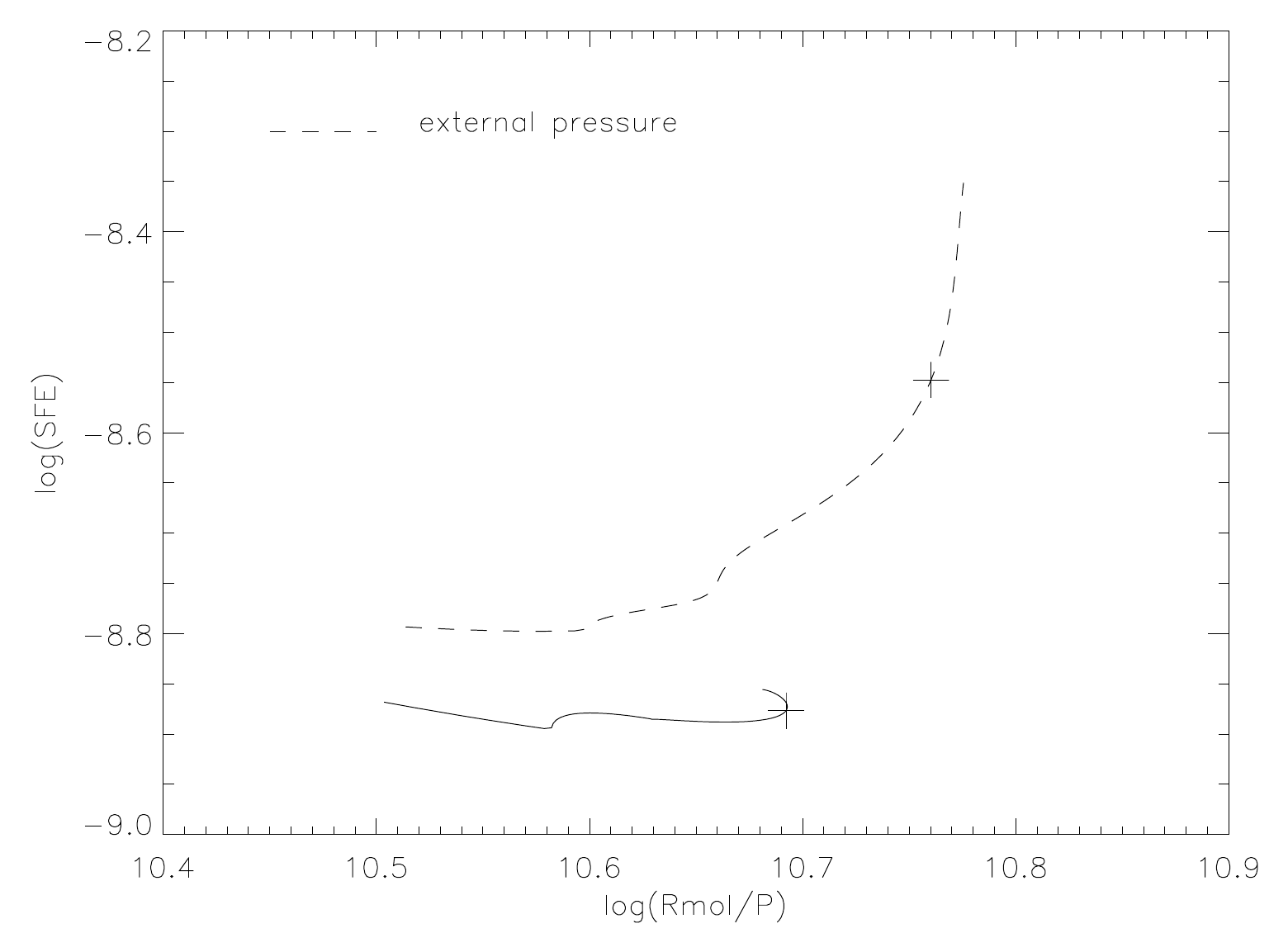}
   \caption{Analytical model. Molecular $SFE_{\rm H_{2}}$ as a function of $R_{\rm mol}/P_{\rm tot}$. The dashed line is for a galaxy with an external isotropic pressure applied to the ISM.}
\label{analmodel3}%
\end{figure}

In the upper panel of Fig.~\ref{4gal255} NGC~4501 seems to have a 0.1~dex higher molecular SFE than the other galaxies. This can be explain by adding an external pressure in the model (see Fig.~\ref{analmodel3}). 
By adding an external isotropic pressure $P_0=7 \times 10^{-12}$~erg\,cm$^{-3}$ the molecular SFE increases by 0.1~dex whereas $R_{\rm mol}/P_{\rm tot}$ stays almost unchanged. Thus, if we assume an addition external pressure for NGC~4501 the observed correlations are reasonably close to the analytical predictions (right panels of Fig.~\ref{4gal255}):
$\log(R_{\rm mol}/P_{\rm tot})/\log(Q)=1.5$, $\log(SFE_{\rm H_{2}})/\log(Q)=-0.9$, and $\log(SFE_{\rm H_{2}})/\log(R_{\rm mol}/P_{\rm tot})=-0.5$.

We thus conclude that our observations are consistent with the prediction of the analytical model if different global velocity dispersions are assumed for the galaxies. 
%\begin{figure}
%\centering
%\includegraphics[width=7cm]{FFIG/4gal_SFE_Rmp_dex2.pdf}
%\includegraphics[width=7cm]{FFIG/4gal_Rmol_P_Q_dex2.pdf}
%\includegraphics[width=7cm]{FFIG/4gal_SFE_Q_dex2.pdf}
%   \caption{As Fig.~\ref{4gal2dex1} with the $SFE_{\rm H_{2}}$ of NGC~4501 ``corrected'' for the external pressure (leading to a 0.1~dex lower $SFE_{\rm H_{2}}$).
%     The dotted lines correspond to the correlations predicted by the analytical model: $\log(SFE_{\rm H_2})/\log(R_{\rm mol}/P_{\rm tot}=-1$,
%     $\log(R_{\rm mol}/P_{\rm tot})/\log(Q)=1.4$, and $\log(SFE_{\rm H_2})/\log(Q)=-1.5$. 
%     The dashed lines correspond to correlations fitted to the data: $\log(SFE_{\rm H_2})/\log(R_{\rm mol}/P_{\rm tot}=-0.5$,
%     $\log(R_{\rm mol}/P_{\rm tot})/\log(Q)=1.5$, and $\log(SFE_{\rm H_2})/\log(Q)=-0.9$.
%}
%\label{4gal2dex2}%
%\end{figure}

Is an external isotropic pressure of $P_0=7 \times 10^{-12}$~erg\,cm$^{-3}$ realistic for NGC~4501?
The thermal pressure of the ICM can be estimated from the dynamical model. At the time of interest NGC~4501 is located at a distance of $\sim 420$~kpc from the cluster center. At this distance the ICM density is $n_{\rm ICM} \sim 3 \times 10^{-4}$~cm$^{-3}$. With an ICM temperature of $T_{\rm ICM} \sim 3 \times 10^{7}$~K, the thermal pressure of the ICM is $P_{\rm therm} \sim 3 \times 10^{-12}$ ~erg\,cm$^{-3}$. In addition, ram pressure exerts an isotropic pressure on the windward face of the galaxy $P_0 \sim P_{\rm ram} \sin(i)$, where $i=30^{\circ}$ is the angle between the disk plane and the ram pressure wind direction. The dynamical model yields a ram pressure at the time of interest $P_{\rm ram} = 10^{-11}$~erg\,cm$^{-3}$ and thus $P_0=5 \times 10^{-12}$~erg\,cm$^{-3}$. To remove the gas from the outer edge of the gas disk a ram pressure of $P_{\rm ram} \sim 2 \times 10^{-10}$~erg\,cm$^{-3}$ is needed. An isotropic pressure of $P_0=7 \times 10^{-12}$~erg\,cm$^{-3}$ is thus plausible for NGC~4501.

%%%%%%%%%%%%%%%%%%%%%%%%%%%%%%%%%%%%%%%%%%%%%%%%%%%%%%%%%%%%%%%%%%%%%%%%%%%%%%%

\section{ Dynamical model \label{model}}

In the following the relations discussed in the previous section are studied using a dynamical model for NGC~4501 and NGC~4567/68. 
We used the N-body code described in Vollmer et al. (2001), which consists of two components: a non-collisional component that simulates the stellar bulge/disk 
and the dark halo, and a collisional component that simulates the ISM. A scheme for star formation was implemented where stars were formed
during cloud collisions and then evolved as non-collisional particles (see Vollmer et al. 2012).

The non-collisional component consists of 81\,920 particles that simulate the galactic halo, bulge, and disk. The characteristics of the different galactic components 
are adapted to the observed properties. We adopted a model where the ISM is simulated as a collisional component, i.e., as discrete particles that possess a
mass and a radius and can have inelastic collisions (sticky particles). The 20\,000 particles of the collisional component represent
gas cloud complexes that evolve in the gravitational potential of the galaxy. 
During the disk evolution, the particles can have inelastic collisions, the outcome of which (coalescence, mass exchange, or fragmentation) is simplified following Wiegel (1994). 
This results in an effective gas viscosity in the disk.

As the galaxy moves through the ICM, its clouds are accelerated by ram pressure. Within the galaxy’s inertial system, its clouds are exposed to a wind coming from a direction 
opposite to that of the galaxy’s motion through the ICM. The temporal ram pressure profile has the form of a Lorentzian, which is realistic
for galaxies on highly eccentric orbits within the Virgo cluster (Vollmer et al. 2001). The effect of ram pressure on the clouds is simulated by an additional force on the 
clouds in the wind direction. Only clouds that are not protected by other clouds against the wind are affected. Since the gas cannot develop instabilities,
the influence of turbulence on the stripped gas is not included in the model. The mixing of the intracluster medium into the ISM is very crudely approximated by a finite 
penetration length of the intracluster medium into the ISM, i.e., up to this penetration length the clouds undergo an additional acceleration caused by ram pressure.

The 3D dynamical model has the advantage of giving direct access to the volume density and velocity dispersion, two quantities that are very difficult to observe.
The analytical model (Eq.~\ref{anal1} and Eq.~\ref{anal2}) was used to compute $\Sigma_{\rm H_{2}}$ and $\Sigma_{\rm HI}$ in the simulations.
In this dynamical model, there is no explicit energy dissipation and a large-scale compression is adiabatic.

In the simulations the SFR is traced by cloud-cloud collisions. When a collision between gas particles occurs, a new stellar particle is created and the information of the time of creation is stored. The FUV emission of a stellar particle is reconstructed using single stellar population models from STARBURST99 (Leitherer et al. 1999). Once the SFR derived from FUV flux and the molecular fraction are computed, the observable parameters ($\Sigma_{\rm HI}$, $\Sigma_{\rm H_{2}}$, $\Sigma_{\star}$, $R_{\rm mol}$, $P_{\rm tot}$, Toomre Q, SFR, $SFE_{\rm H_{2}}$) can be determined and model images and data cubes can be produced with the resolution of the observations.

\subsection{NGC~4501 dynamical simulation}

\begin{figure}
   \centering
\includegraphics[width=8cm]{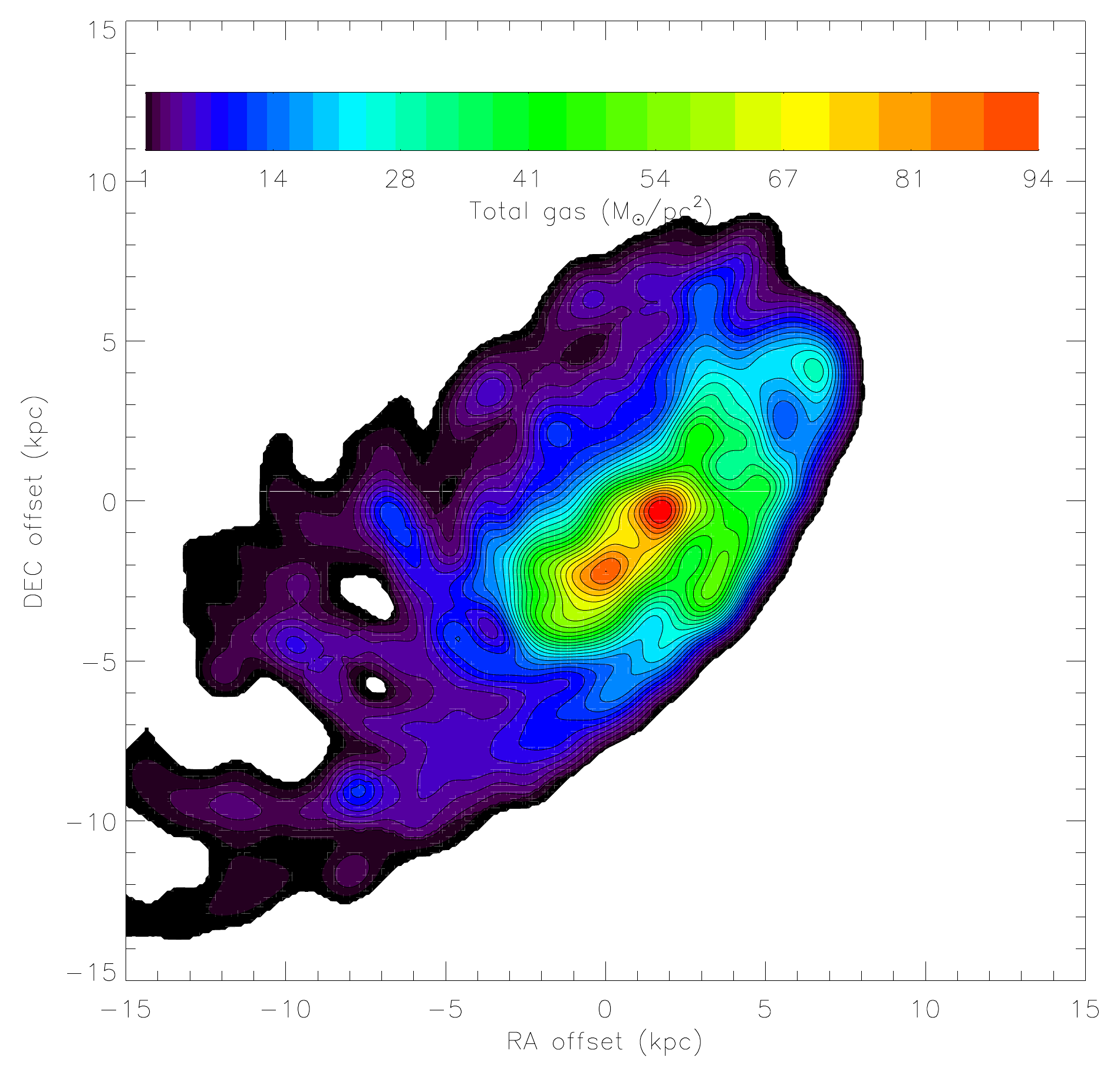}
\includegraphics[width=8cm]{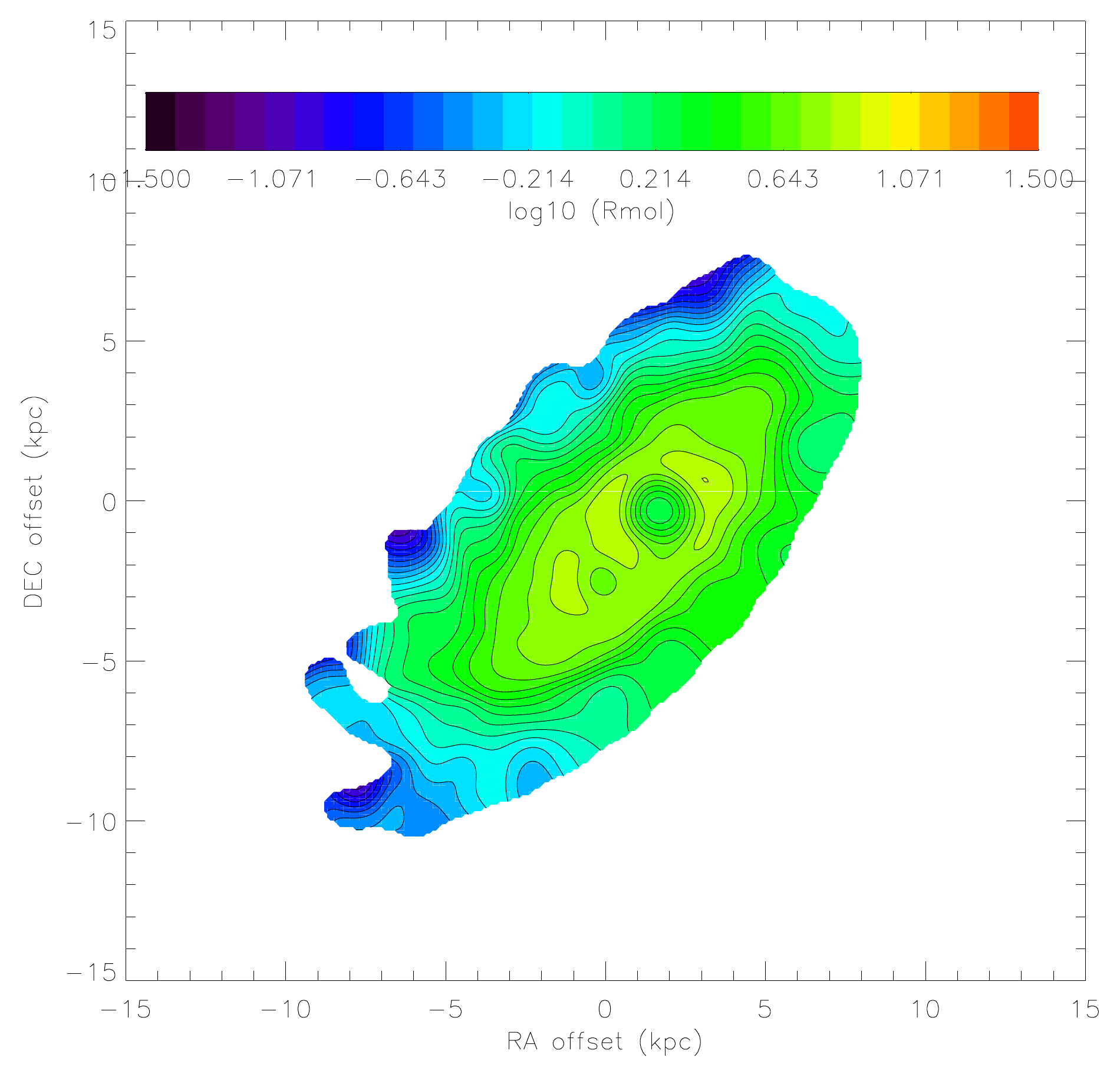}
   \caption{NGC~4501 simulation A. {\em Upper panel}: Total gas distribution. {\em Lower panel}: Molecular fraction.}
\label{simmap1}%
\end{figure}

\begin{figure}
   \centering
\includegraphics[width=8cm]{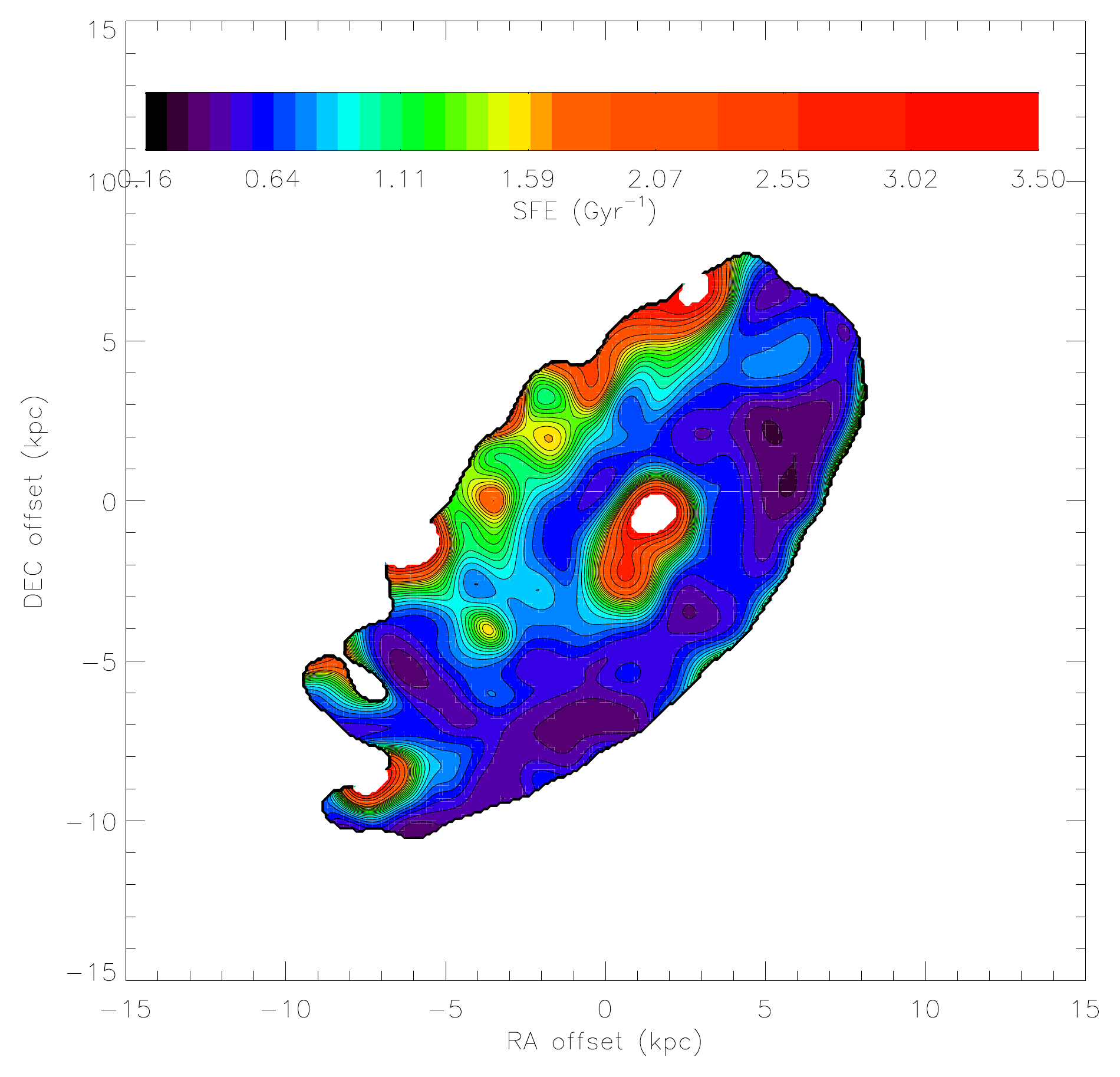}
\includegraphics[width=8cm]{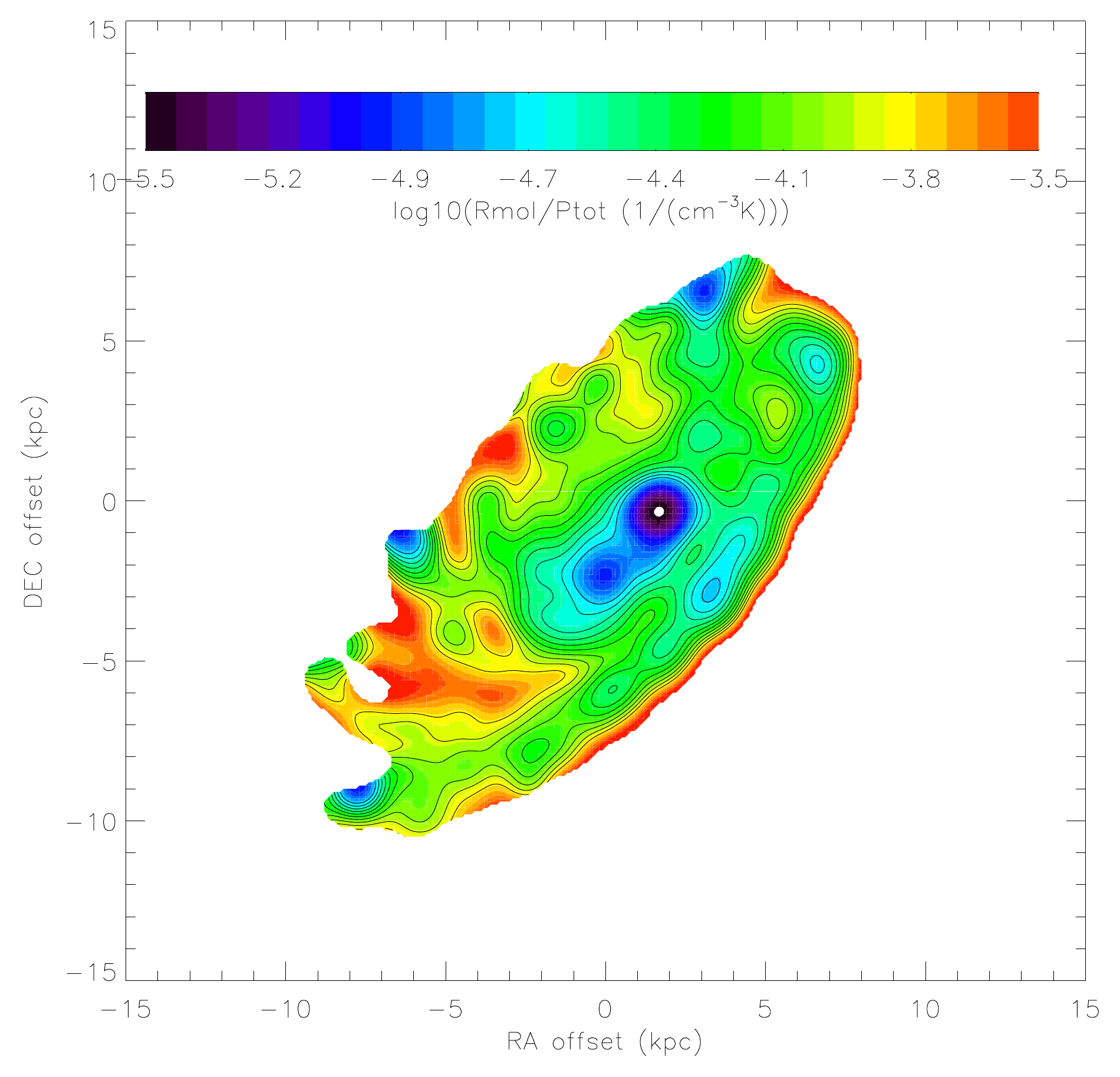}
   \caption{NGC~4501 simulation A. {\em Upper panel}: $SFE_{\rm H_{2}}$. {\em Lower panel}: $R_{\rm mol}/P_{\rm tot}$.}
\label{simmap2}%
\end{figure}

\begin{figure}
\centering
\includegraphics[width=8cm]{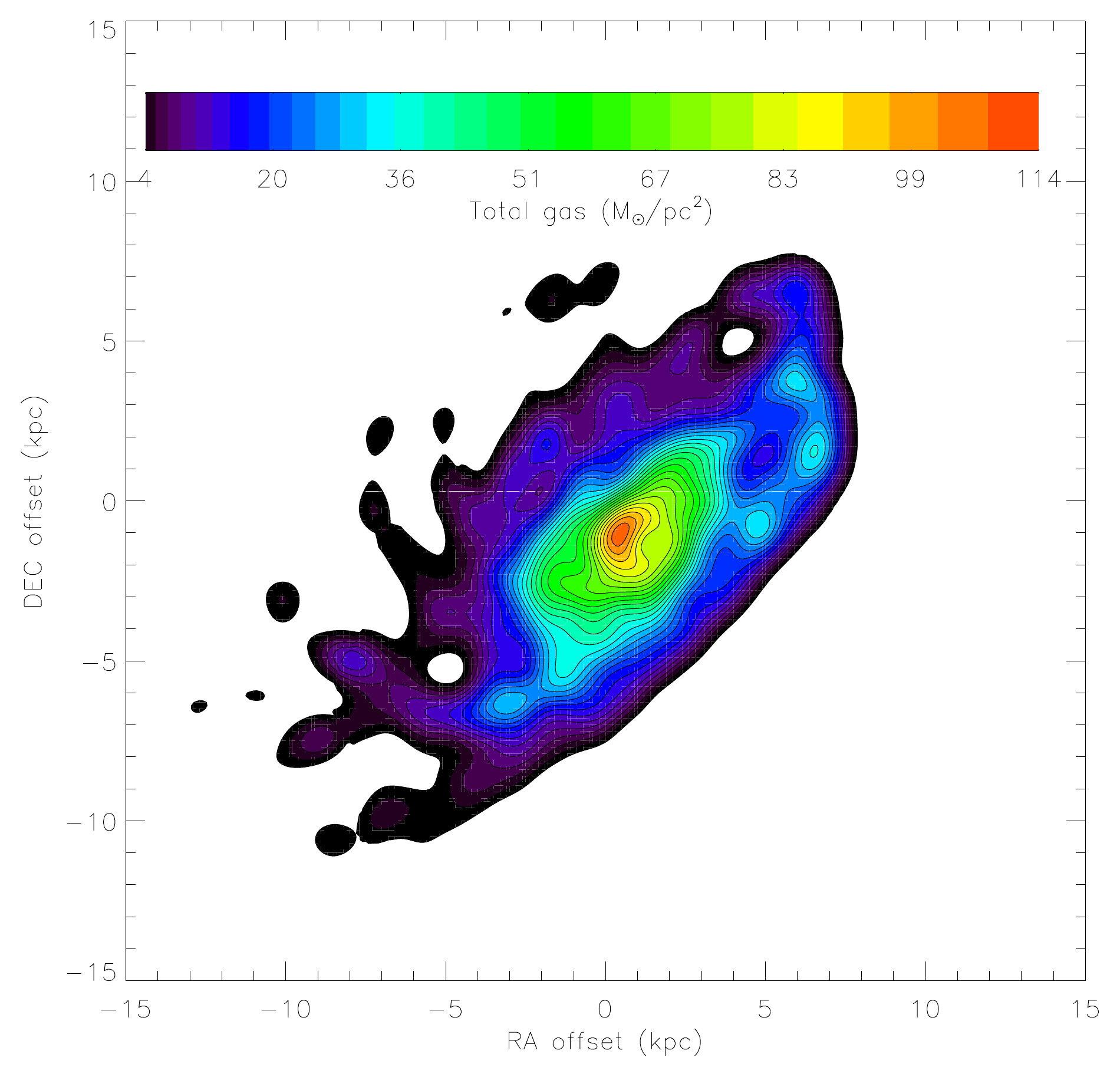}
\includegraphics[width=8cm]{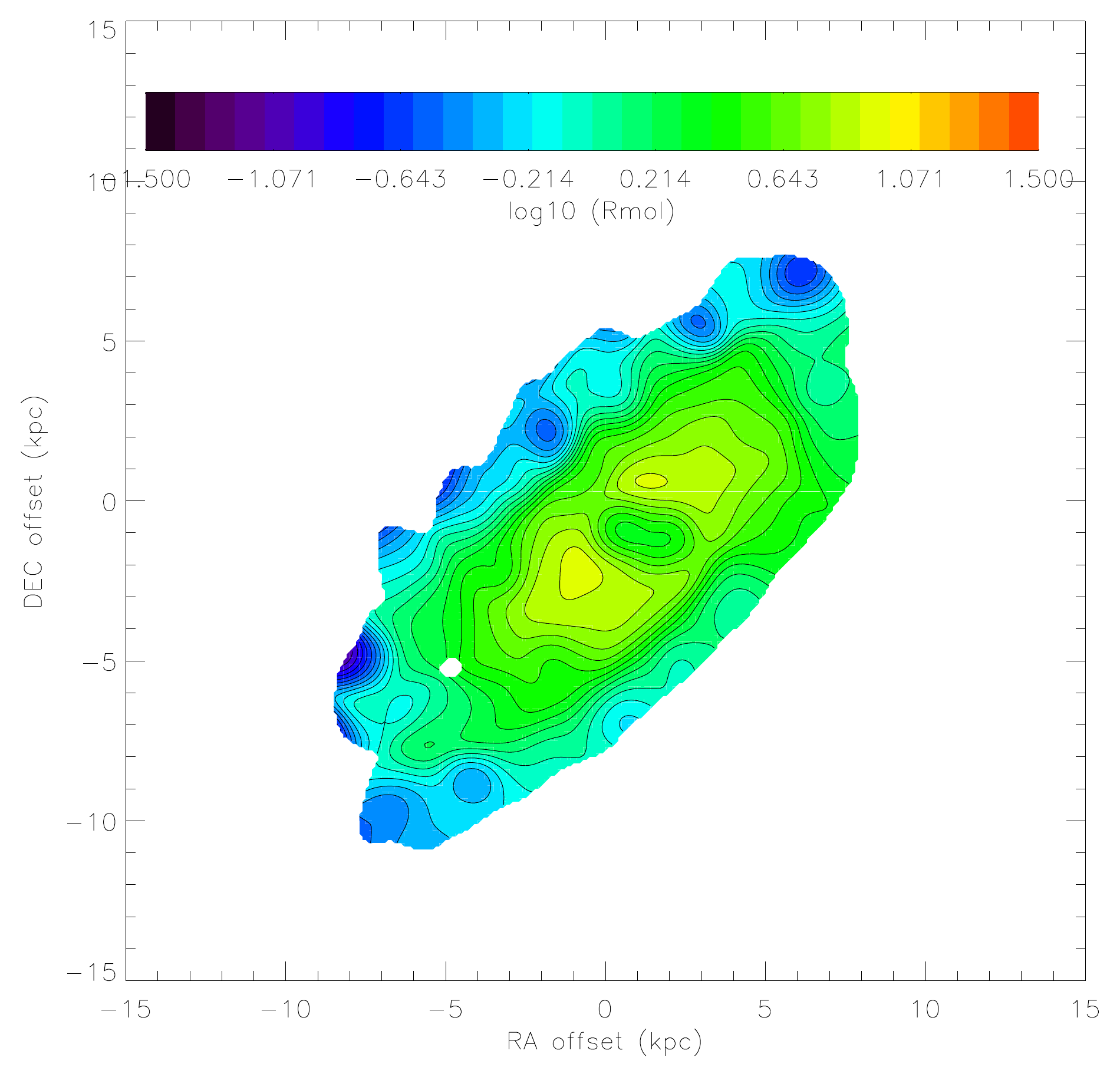}
\caption{NGC~4501 simulation B. {\em Upper panel}: Total gas distribution. {\em Lower panel}: Molecular fraction.}
\label{simmap1_vari}
\end{figure}

For  NGC~4501, we made two simulations with identical temporal ram pressure profiles and different times of peak ram pressure (simulations A and B). 
With the delayed time of peak ram pressure we want to investigate the influence of galactic structure, i.e., spiral arms, on the resulting disturbed gas distribution.
The resulting gas distribution of simulation A, molecular fraction, $R_{\rm mol}/P_{\rm tot}$, and molecular $SFE_{\rm H_{2}}$ maps are presented in Fig.~\ref{simmap1} and \ref{simmap2}, respectively. The gas distribution and molecular fraction of simulation B are presented in Fig. \ref{simmap1_vari}. 

As expected, the overall gas distribution of simulation B (Fig.~\ref{simmap1_vari}) is very close to that of simulation A  (Fig. \ref{simmap1}).
In simulation A a prominent gas spiral arm is found in the compressed region leading to a high surface density west of galaxy center; however, this is not the case for simulation B 
where only a small enhancement of the gas surface density is observed northwest of the galaxy center. We thus conclude that the detailed observed gas distribution 
depends on galactic structure, i.e., whether a spiral arm is located in the compressed region.

The simulated gas surface density of simulation A reproduces (i) the well-defined compressed western edge, (ii) the three low surface density spiral arms in the eastern part of the disk extending to the north, (iii) the regions of high gas surface density in the northern part of the disk, and (vi) the high molecular fraction in the western compressed side of the disk.
The $SFE_{\rm H_{2}}$ map is quantitatively reproduced, the higher $SFE_{\rm H_{2}}$ in the northwestern quadrant of the disk is present in the simulation as is the U-shaped low $SFE_{\rm H_{2}}$ region in the south. Moreover, the modeled $SFE_{\rm H_{2}}$ is lower in the north of the compressed region as has been observed. However, the model does not reproduce the observed higher $SFE_{\rm H_{2}}$ west of the center.

Within the western compression region, the reason for the low molecular star formation efficiency in the low surface density gas is the following. In a quiet model without interaction, 
the velocity dispersion of the gas particles is proportional to the square root of the gas density. Once the gas is compressed, it adapts its velocity dispersion to the new situation within a gas particle collision timescale.
In the compressed regions of low surface density the particle density and thus the cloud collision frequency is smaller than the compression timescale and the equilibrium 
$v_{\rm turb} \propto \sqrt{\rho}$ is not yet reached at the time of interest. The molecular star formation efficiency is thus lower. On the other hand, in the high surface density region the 
collision timescale is short and the equilibrium is reached rapidly, i.e., the  molecular star formation efficiency is unchanged.
We believe that, in reality, the molecular star formation efficiency  decreases once the compression becomes adiabatic, i.e., once the turbulent dissipation timescale $t_{\rm turb} \sim l_{\rm driv}/v_{\rm turb}$
is longer than the compression timescale $t_{\rm comp} = \rho/(\partial \rho/\partial t)$. Even if the physics in the model and in reality that lead to a decreasing molecular star formation efficiency are quite 
different, the mechanisms are similar because in a collisional model the dissipation timescale is proportional to the collision timescale. Therefore, our simple dynamical model is able to predict
regions of decreased molecular star formation efficiency due to adiabatic compression. In the future, we will test this prediction with simulations of NGC~4438 and the Taffy galaxies, where
adiabatic compression decreases $SFE_{\rm mol}$ by a factor of four to five (Braine et al. 2003, Vollmer et al. 2009, 2012).

The global model $R_{\rm mol}/P_{\rm tot}$ is higher than observed. Nevertheless, we reproduce the main observational characteristics: (i) a low $R_{\rm mol}/P_{\rm tot}$ in the northwestern quadrant and in the compressed region and (ii) a high $R_{\rm mol}/P_{\rm tot}$ in the northwestern quadrant.  

A pixel-by-pixel analysis was made for the NGC~4501 simulation A.
The fitting results and the Spearman rank correlation coefficients $\rho$ are presented in Table~\ref{tabsim}.
\begin{table*}
\caption{Results of the fitting analysis for the dynamical simulations of NGC~4501 and NGC~4567/68.}           % title of Table
\label{tabsim}    
\centering      
\begin{tabular}{ l c c c c c c c c r}       
\hline\hline    
Relations            &  \multicolumn{3}{c}{NGC~4501 simulation}  & \multicolumn{3}{c}{NGC~4567 simulation}  &  \multicolumn{3}{c}{NGC~4568 simulation}             \\  
 ...                                                           & $\rho$ & slope  &  offset      & $\rho$ & slope   &  offset          & $\rho$& slope&  offset \\
\hline        
SFR-$\Sigma_{{\rm H}_2}$                                       & +0.905 & +1.02  & -3.12        & +0.967  & +1.54  & -4.06            &+0.977 &+1.30 &-3.77 \\
SFR-$\Sigma_{\rm HI}$                                          & +0.747 & +2.65  & -3.53        & +0.865  & +1.98  & -4.05            &+0.686 &+3.00 &-4.97 \\
SFR-$\Sigma_{\rm g}$                                           & +0.935 & +1.44  & -3.69        & +0.934  & +1.52  & -4.32            &+0.971 &+1.78 &-4.70 \\
$R_{\rm mol}$-$P_{\rm tot}$                              & +0.870 & +0.67  & -2.69        & +0.769  & +0.27  & -1.09            &+0.911 &+0.67 &-2.85 \\
$R_{\rm mol}$-$\Sigma_{\star}$                                 & +0.954 & +0.89  & -1.94        & +0.839  & +0.48  & -0.88            &+0.951 &+0.76 &-1.58 \\
$R_{\rm mol}$-$\Sigma_{\rm g}$                                 & +0.797 & +1.06  & -0.60        & +0.754  & +0.40  & -0.24            &+0.873 &+1.20 &-1.28 \\
$P_{\rm g}/P_{\rm s}$-$R_{\rm mol}$/$P_{\rm tot}$         & -0.957 & -0.56  & -3.52        & -0.956  & -0.53  & -3.36            &+0.822 &-0.46 &-2.41 \\
$\Sigma_{\rm g}$-$R_{\rm mol}$/$P_{\rm tot}$             & -0.827 &  ...   & ...          & -0.977  & ...    &  ...             &-0.866 & ...  & ...  \\
$Q$-$R_{\rm mol}/P_{\rm tot}$                                  & +0.901 & +0.73  & -3.82        & +0.944  & +0.80  & +4.42            &+0.669 &+1.40 &+6.60\\
SFE$_{H_{2}}$-$R_{\rm mol}/P_{\rm tot}$                  & -0.153 &  ...   & ...          & -0.656  & ...    &  ...             &-0.375 & ...  & ... \\
SFE$_{H_{2}}$-$Q$                                                & +0.082 &  ...   & ...          & -0.474  & ...    &  ...             &-0.473 & ...  & ... \\
\hline                               %inserts single line
\end{tabular}
\end{table*}

We find slopes in agreement with observations for the following relations: SFR - $\Sigma_{{\rm H}_2}$, SFR - $\Sigma_{\rm g}$, $R_{\rm mol}$-$P_{\rm tot}$, $R_{\rm mol}$ - $\Sigma_{\star}$, $R_{\rm mol}$ - $\Sigma_{\rm g}$, and $R_{\rm mol}$/$P_{\rm tot}$ -Q. However, the slope of the $R_{\rm mol}$ - $P_{\rm tot}$ relation is lower in our simulation of NGC~4501 ($0.7$) than the observed slope of $1.0$. We found a correlation between $R_{\rm mol}/P_{\rm tot}$ - $Q$ (Spearman $\rho=0.9$) with a slope of $\sim 0.7$ in agreement with the observed slope $0.84 \pm 0.36$.

We thus conclude that the dynamical simulation of NGC~4501 is in good agreement with observations.

\subsection{NGC~4567/68 dynamical simulation}

For simulations of colliding galaxies the parameter space (impact parameter, relative velocity, disk inclinations, prograde/retrograde encounter) is huge and degenerated. Our aim was to simulate NGC~4568 in a first place. 

The size and rotation velocity of the model galaxies were adapted to the observed properties of NGC~4567 and NGC~4568. For the search of the orbit of NGC~4567 with respect to
NGC~4568 we made several dozens of test simulations without a collisional component. We aimed at reproducing the position of NGC~4567 with respect to
NGC~4568 together with a western compression ridge in NGC~4568. We have chosen a retrograde encounter with respect to NGC~4568. To produce a visible effect, the impact
parameter is small ($\Delta x = 6$~kpc). The relative velocity at closest approach is $\Delta v = 460$~km\,s$^{-1}$. NGC~4567 approaches NGC~4568 from the south/southwest. 
The time of interest is $10$~Myr after the closest approach. In this stage the stellar disks overlap and a compression region forms in the western half of the disk of NGC~4568 and in the eastern half of the disk of NGC~4567. To avoid the overlap of the gas disk, we truncated the gas disk of NGC~4567. We are aware  that this simulation does not perfectly reproduce the observed properties of the system. However, it can serve as a guideline to interpret observations.

\begin{figure*}
   \centering
\includegraphics[width=7.cm]{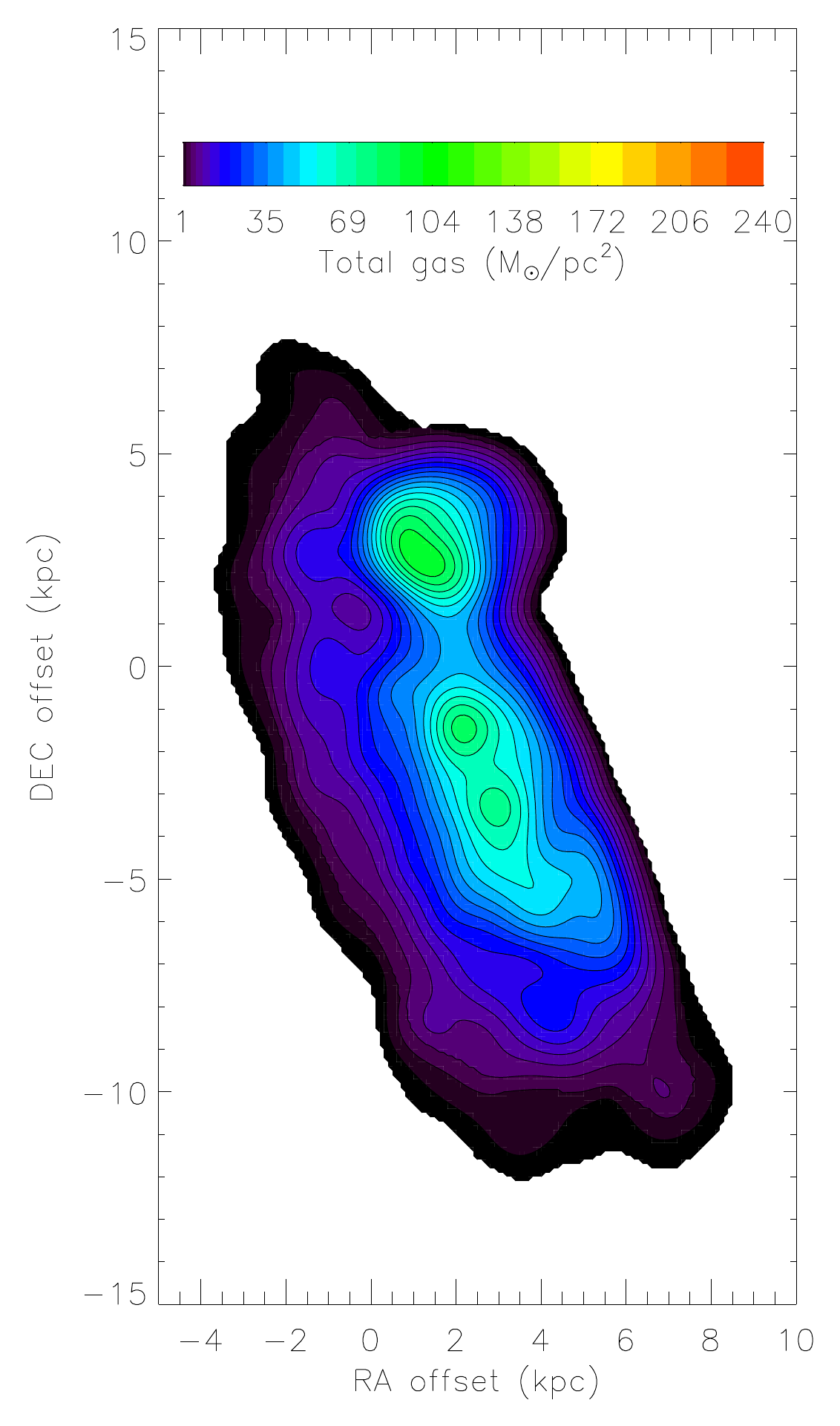}
\includegraphics[width=7.cm]{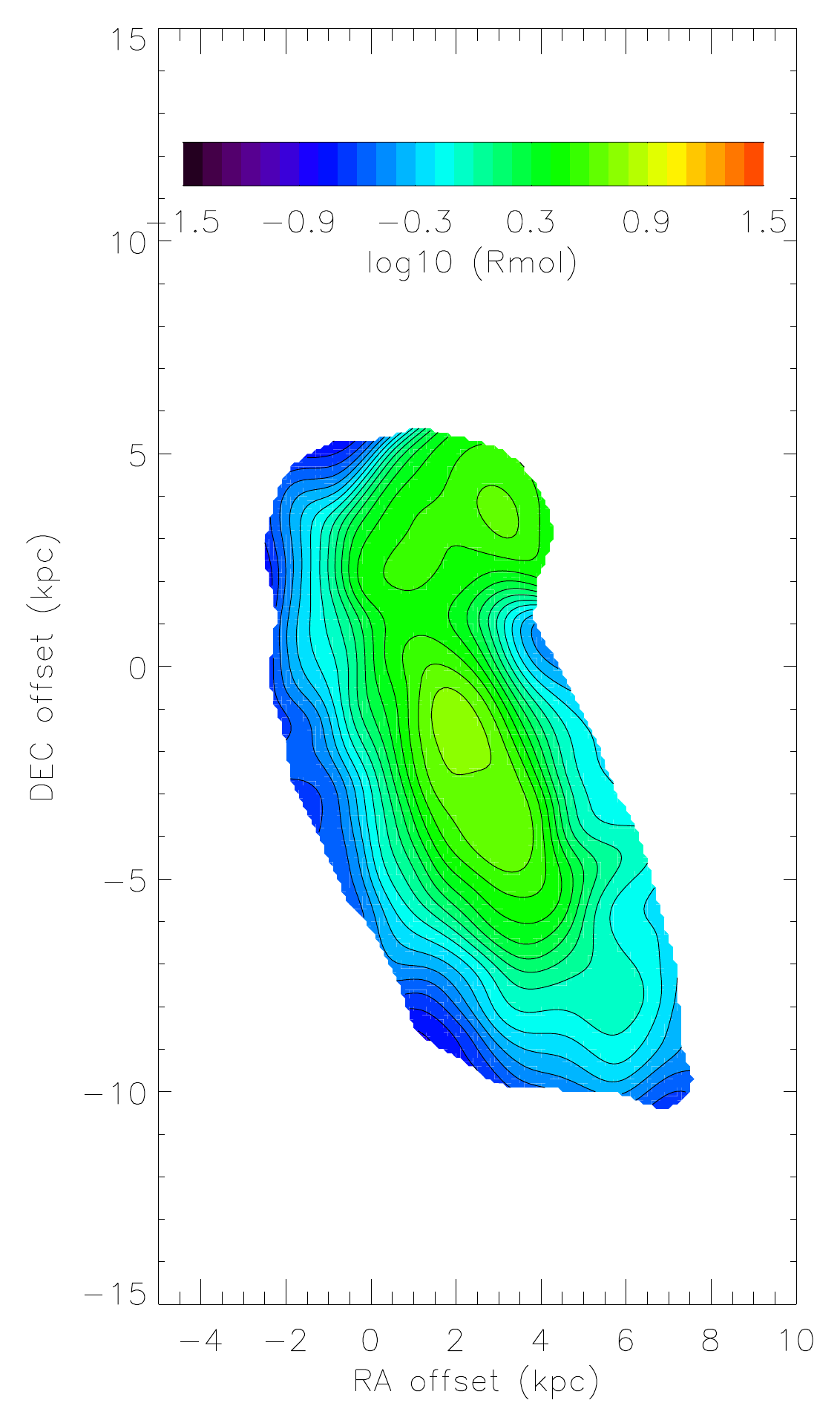}
\includegraphics[width=7.cm]{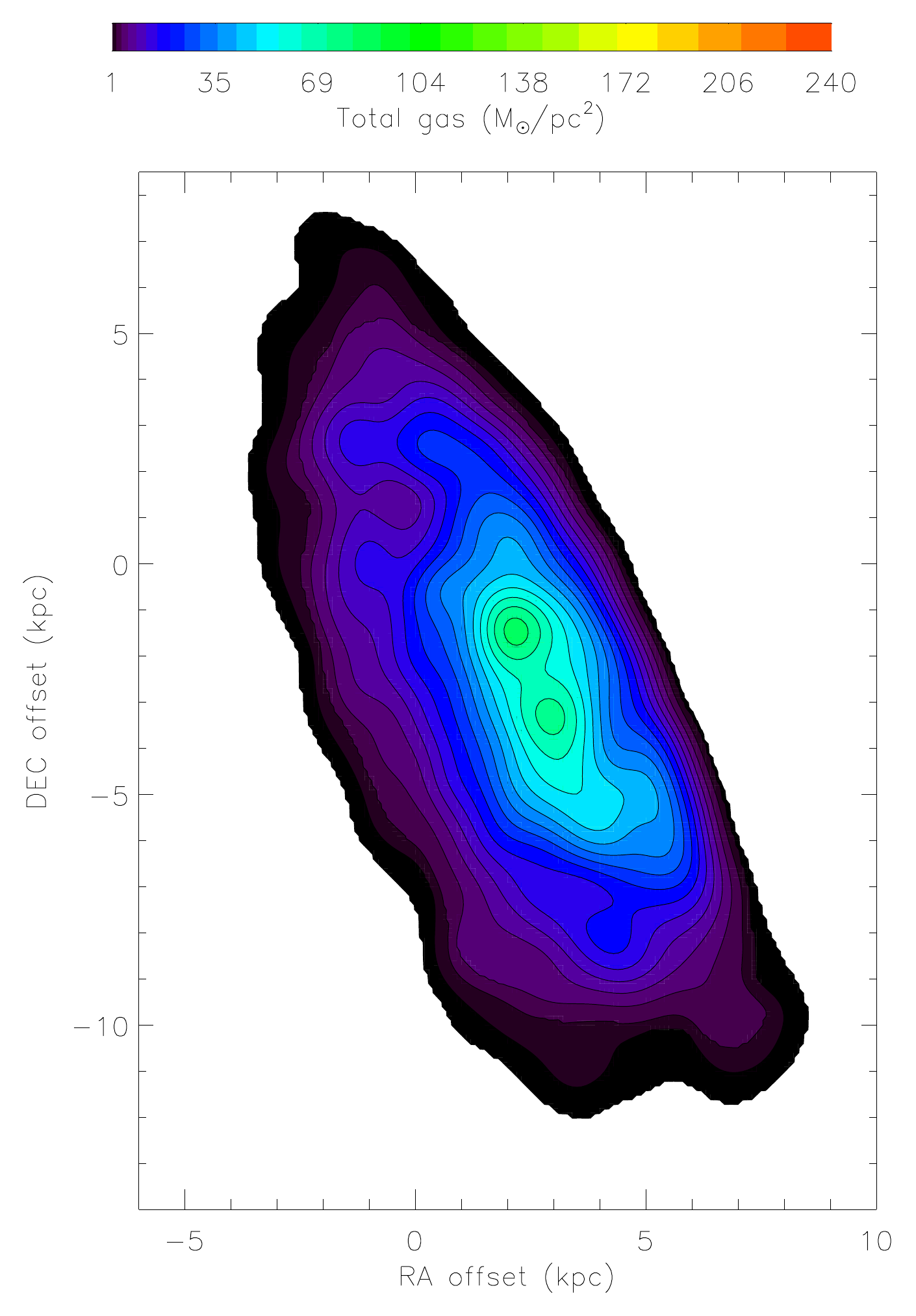}
\includegraphics[width=7.cm]{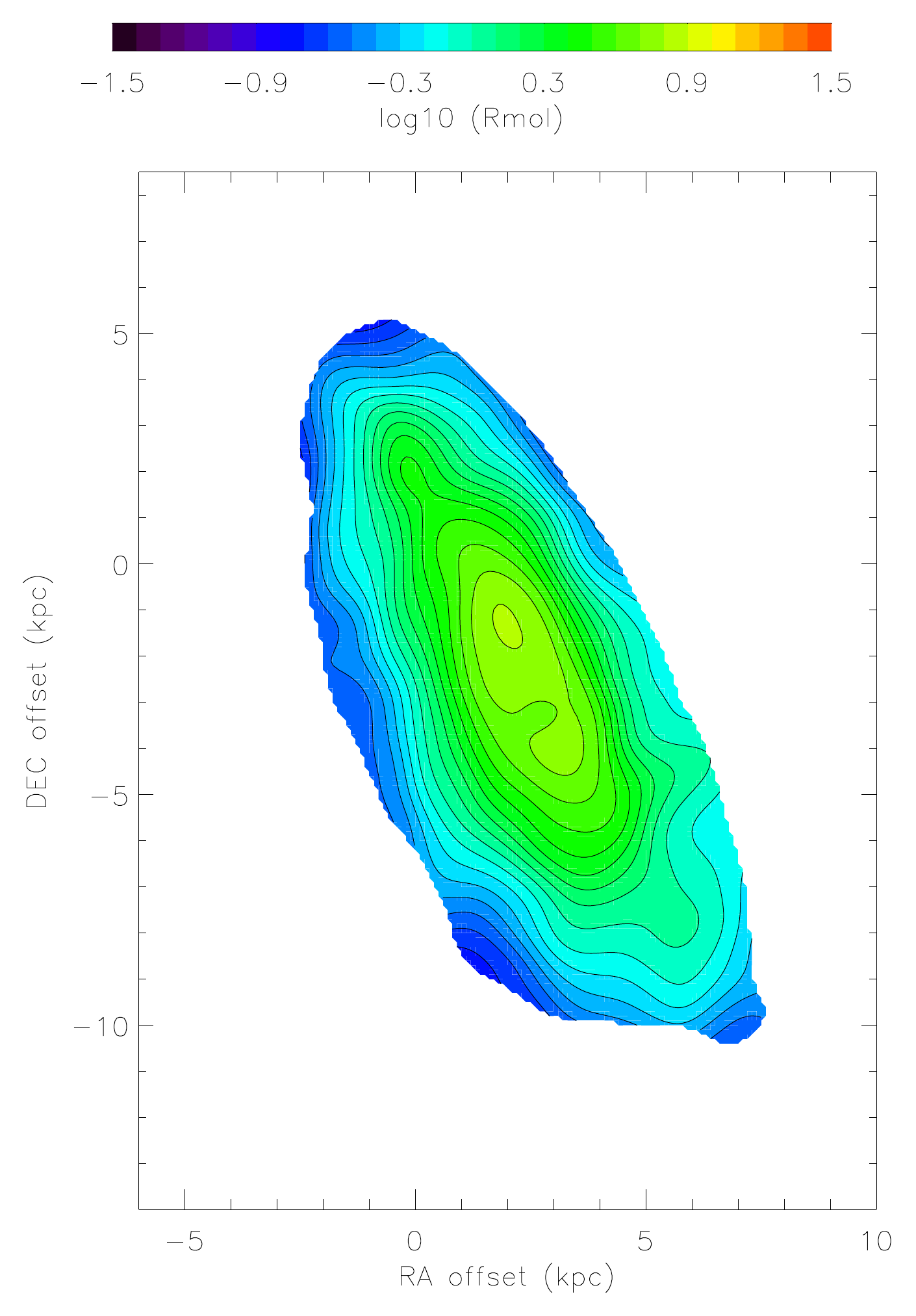}
\includegraphics[width=4.5cm]{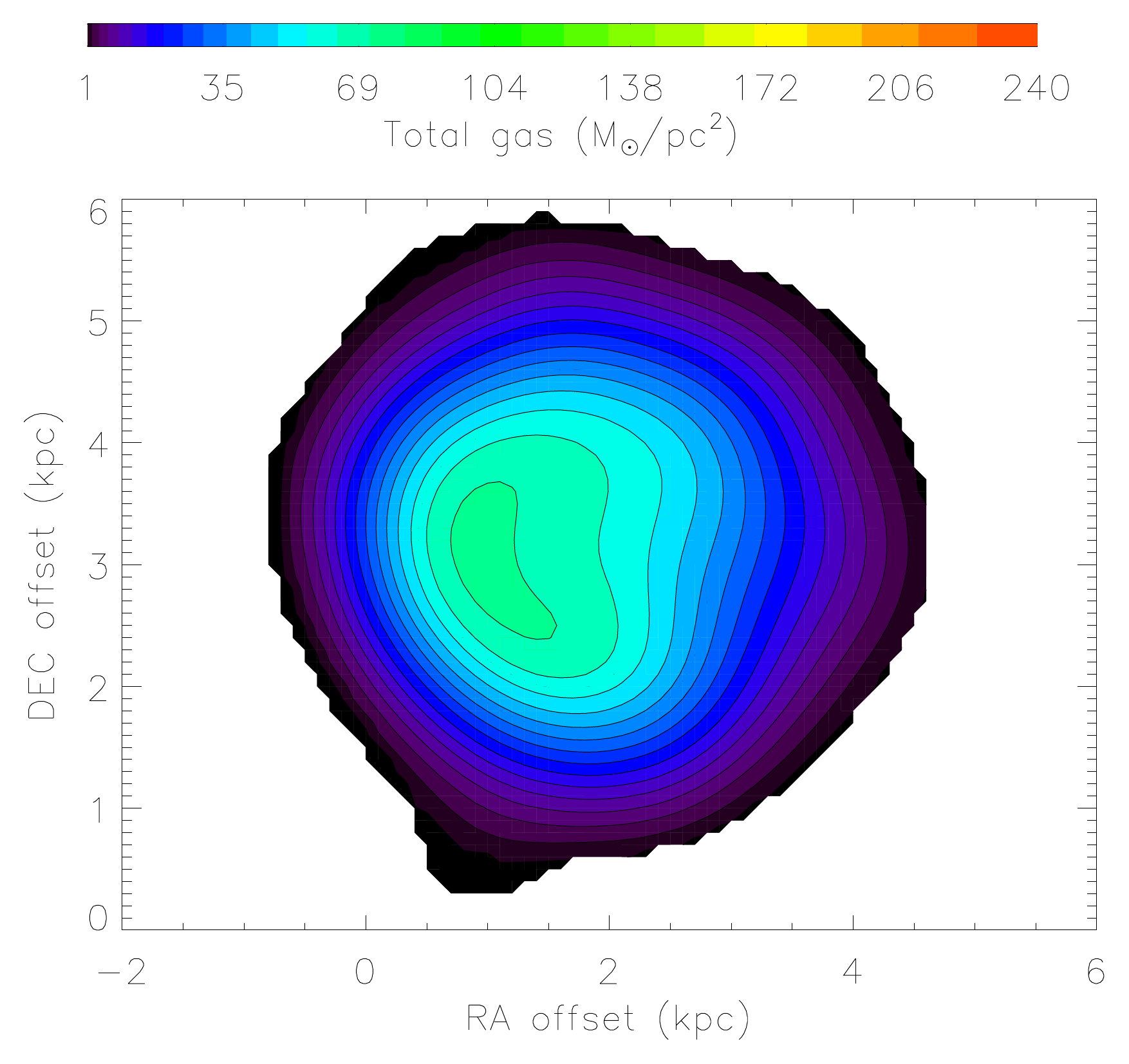}
\includegraphics[width=4.5cm]{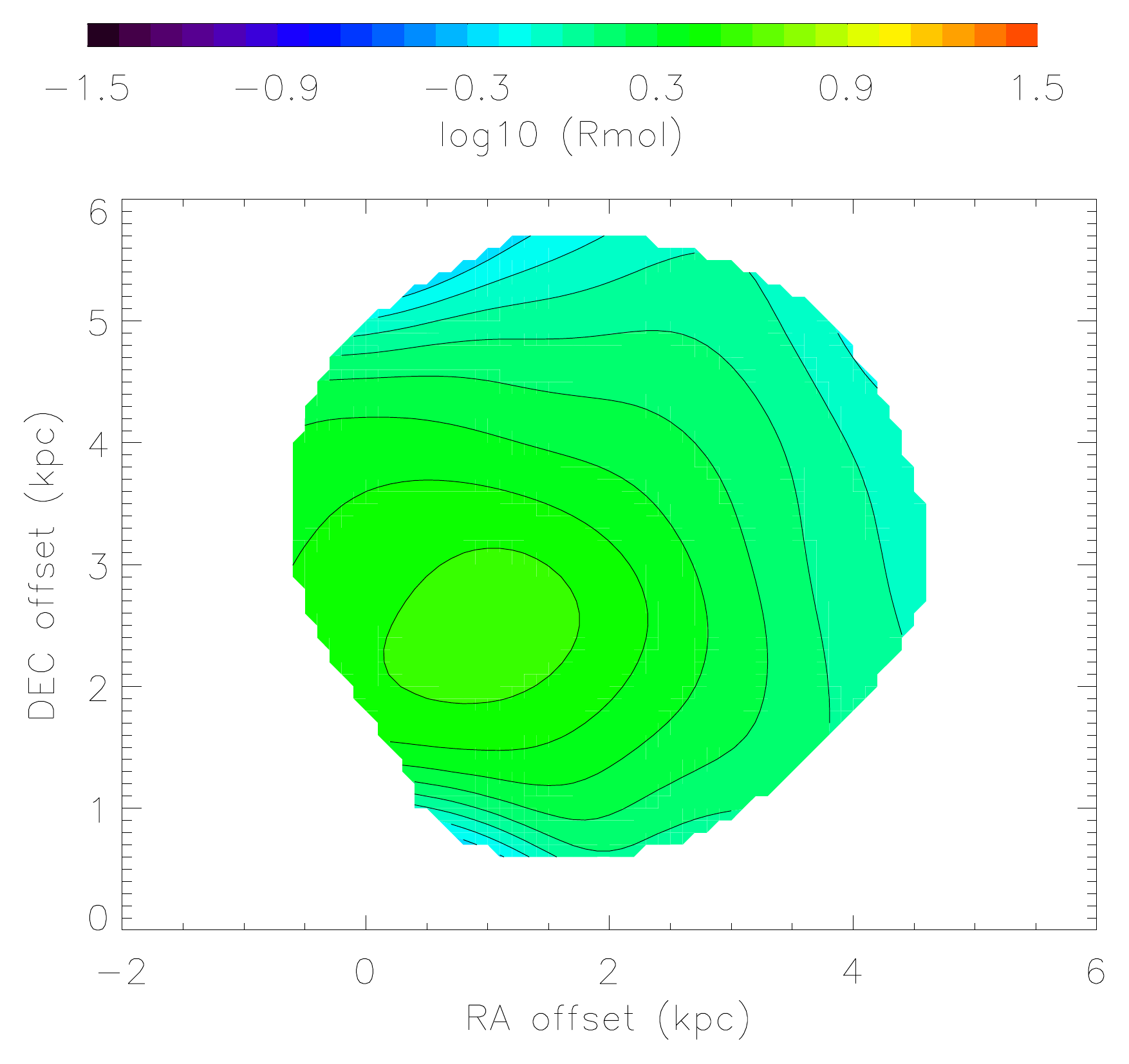}
   \caption{NGC~4567/68 simulation. {\em Upper left panel}: Total gas distribution of both galaxies. {\em Middle left panel}: Total gas distribution of NGC~4568m. {\em Lower left panel}: Total gas distribution of NGC~4567m. {\em Upper right panel}: Molecular fraction of both galaxies. {\em Middle right panel}: Molecular fraction of NGC~4568m. {\em Lower right panel}: Molecular fraction of NGC~4567m.}
\label{simmap3_a}%
\end{figure*}

\begin{figure*}
   \centering
\includegraphics[width=5.5cm]{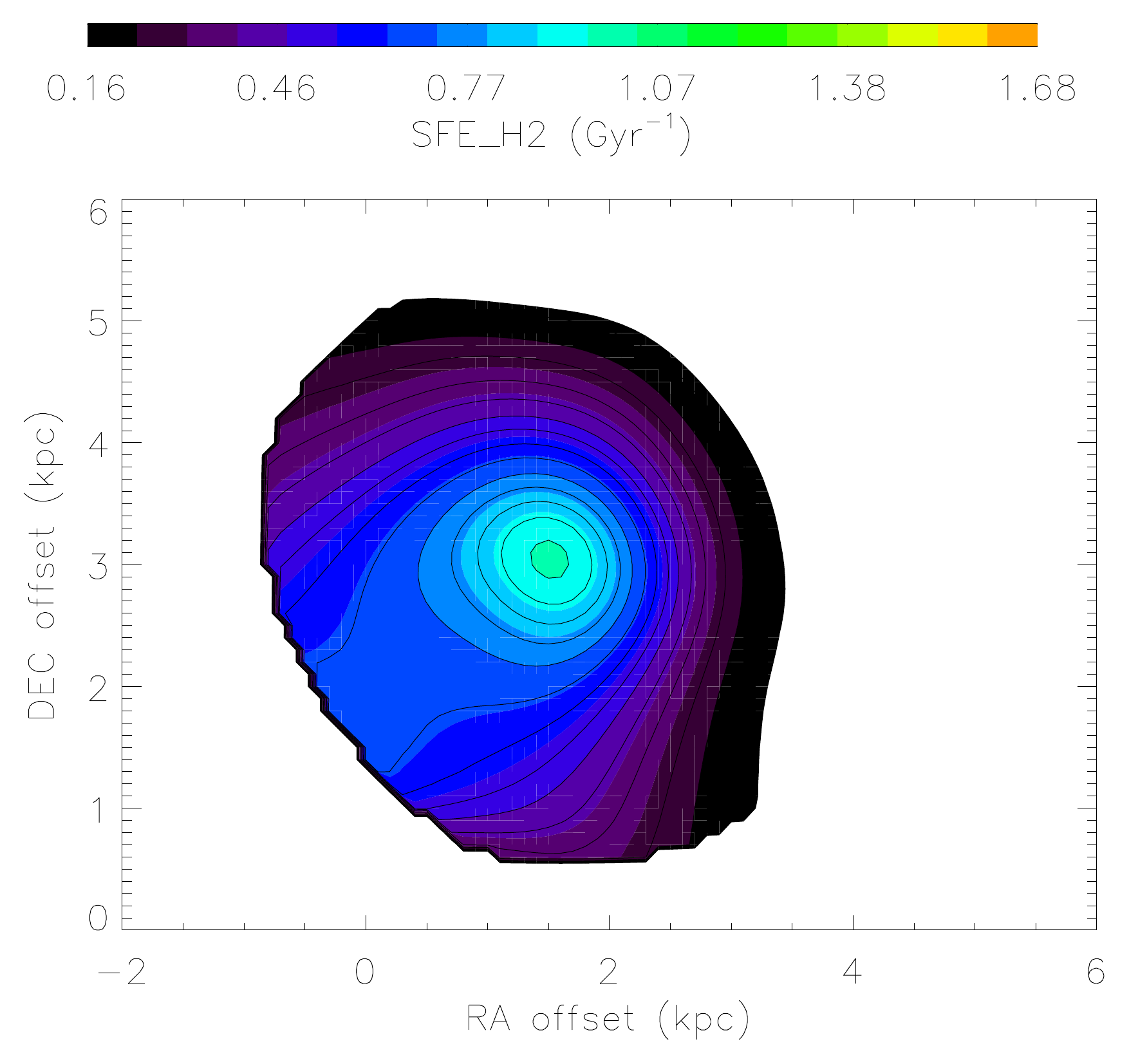}
\includegraphics[width=5.5cm]{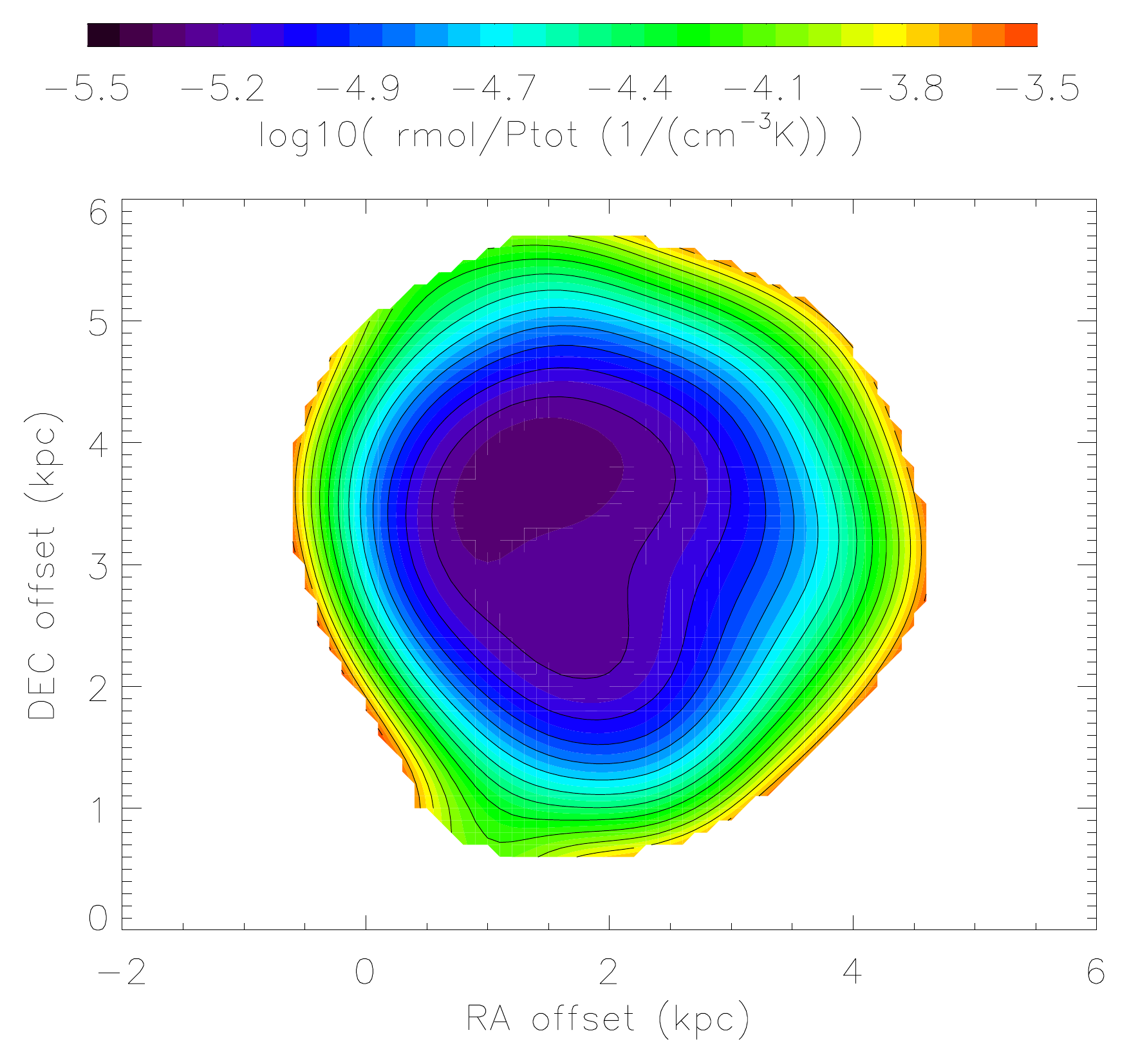}
\includegraphics[width=7.5cm]{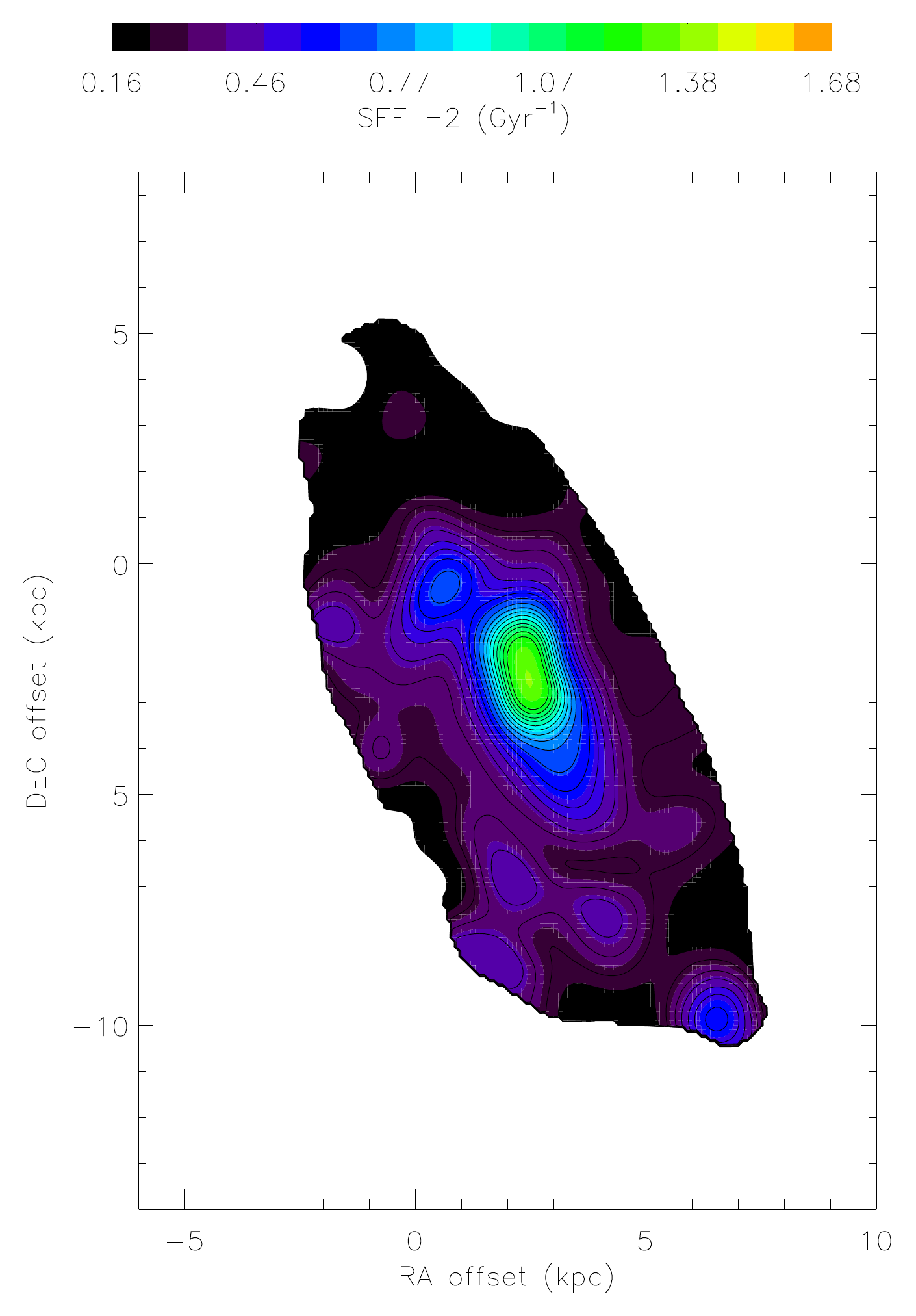}
\includegraphics[width=7.5cm]{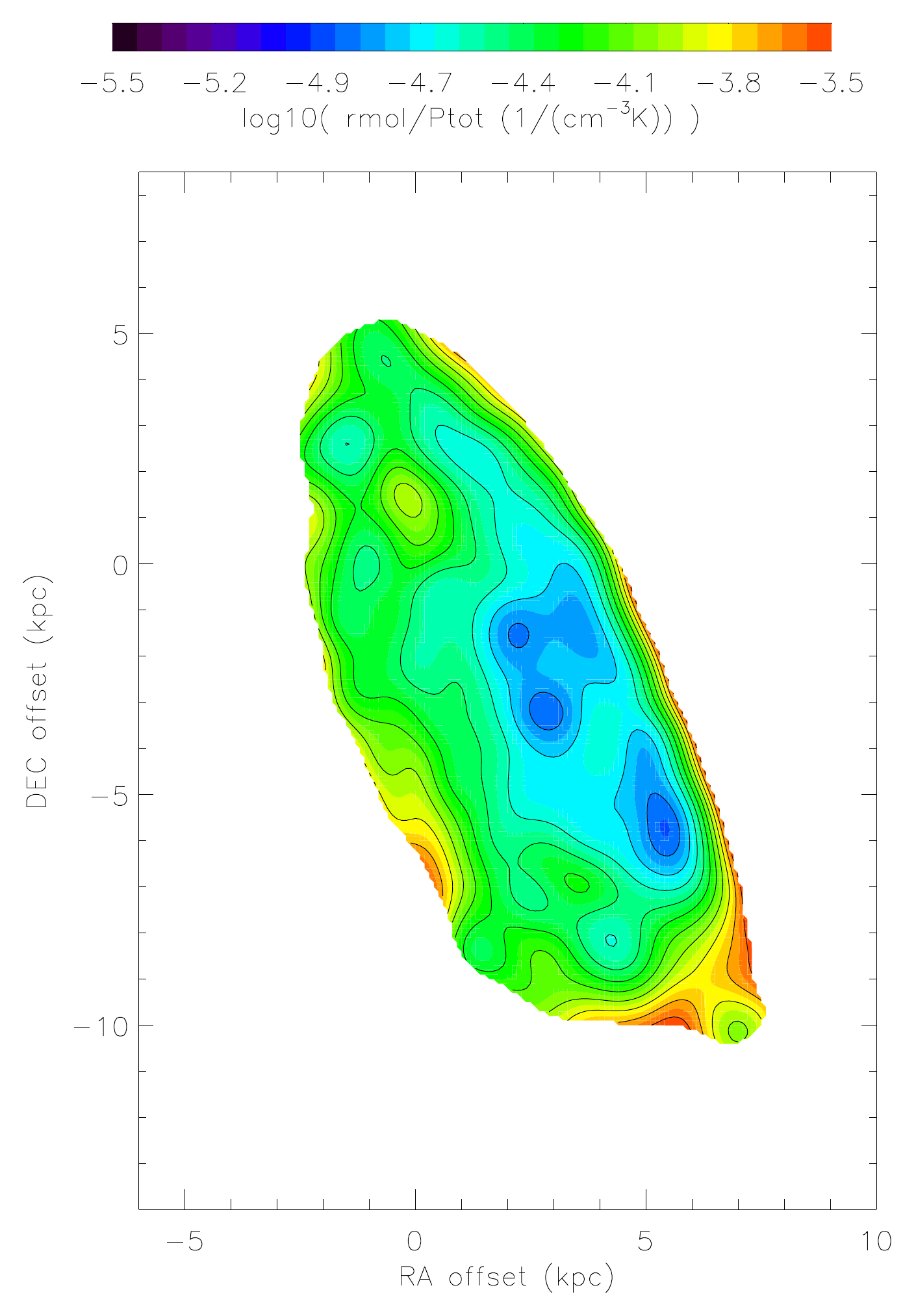}
   \caption{ NGC~4567/68 simulation. {\em Upper left panel}: $SFE_{\rm H_{2}}$ of NGC~4567m. {\em Lower left panel}: Molecular $SFE_{\rm H_{2}}$ of NGC~4568m. {\em Upper right panel}: $R_{\rm mol}$/P$_{\rm h}$ of NGC~4567m. {\em Lower right panel}: $R_{\rm mol}$/P$_{\rm h}$ of NGC~4568m.}
\label{simmap4_a}%
\end{figure*}

The 3D simulations have the advantage that the two galaxies can be separated properly avoiding superposition effects. The maps of the gas distribution, molecular fraction, $R_{\rm mol}/P_{\rm tot}$, and molecular $SFE_{\rm H_{2}}$ are presented in Figs.~\ref{simmap3_a} and \ref{simmap4_a}.

In the following we name the model galaxies NGC~4567m and NGC~4568m.
NGC~4568m shows a southern spiral arm curved to the west and an east-west asymmetry within the southern half of the disk; the southeastern part of the disk shows lower surface densities than the southwestern part. Thus, the southern part of the galactic disk of NGC~4568 (Fig.~\ref{totgas4567}) is  reproduced well by the model.
The northern spiral arm of NGC~4568 is also reproduced by the model, but located somewhat farther to the west. 
The simulated gas surface density in the center of NGC~4568m is lower than the observed value by a factor of $\sim$ 2.
The total gas distribution of NGC~4567m does not show the observed spiral arms.
The highest gas surface densities of NGC~4567m are not found in the center as in NGC~4567, but east of the center, where the gas is compressed by the tidal field.
The overall gas surface density of NGC~4567m is significantly higher than that of NGC~4567.

The simulated overall distribution of the molecular fraction of NGC~4568m (middle right panel of Fig.~\ref{simmap3_a}) is in rough agreement with observations (Fig.~\ref{rmol4567}): it is symmetric in the inner parts of the galactic disks. In addition, there is an excess of the molecular fraction in the southwestern part of the disk of NGC~4568m.
The region between the galaxies shows a higher molecular fraction in the simulation than in the observations.
The molecular fractions of NGC~4567m and NGC~4568m do not show the observed high values in the galaxy centers. This is due to a gas hole in the simulated galaxies that we introduced for computational reasons. 

The simulated overall distribution of the molecular star formation efficiency of NGC~4568m (lower left panel of Fig.~\ref{simmap4_a}) is also in rough agreement with observations (Fig.~\ref{SFEh24567}).
Only the model $SFE_{\rm H_{2}}$ of the galaxy centers is significantly higher than observed.
As observed, the $SFE_{\rm H_{2}}$ distribution of NGC~4568m shows regions of higher $SFE_{\rm H_{2}}$ on the eastern side of the galactic disk than on the western side. However, the absolute values of
these regions are significantly lower in the model than in the observed galaxy. The observed low $SFE_{\rm H_{2}}$ region in the southwestern edge of NGC~4568 is also present in the simulation.

The $R_{\rm mol}/P_{\rm tot}$ distribution of NGC~4568m (lower right panel of Fig.~\ref{simmap4_a}) shows (i) a lower $R_{\rm mol}/P_{\rm tot}$ in the center as has been observed and (ii) low $R_{\rm mol}/P_{\rm tot}$ in the western part of NGC~4568m
where the gas is compressed by the tidal field.
 The minimum $R_{\rm mol}/P_{\rm tot}$ is located in the south of the compression ridge in NGC~4568m, whereas it is located in the northwest of the galactic disk in NGC~4568. 
The local minimum of  $R_{\rm mol}/P_{\rm tot}$ west of the center of NGC~4568m is not present in the observations.
As has been observed, the region with the lowest  $R_{\rm mol}/P_{\rm tot}$ in the compression ridge has a normal molecular star formation efficiency. 
Owing to a strong tidal compression, the molecular fraction of NGC~4567m is much higher than that of NGC~4567. In addition, NGC~4567m lacks the observed spiral structure with a low molecular fraction in the spiral arms. 
In the light of the dynamical simulation we conclude that the relatively low molecular fraction of NGC~4567 and NGC~4568 compared to that of NGC~4321 is probably caused by tidal compression of the two galactic disks.     
The fact that the western side of the molecular fraction distribution of NGC~4568m does not reproduce the observed structure might again be due to galactic structure, as in NGC~4501. We will test this hypothesis within a subsequent project.

Thus, NGC~4567/68 is similar to NGC~4501 with respect to gas compression: the compression of the ISM increases the gas surface density and the molecular fraction, and makes 
$R_{\rm mol}/P_{\rm tot}$ decrease. 
Moreover, the $SFE_{\rm H_{2}}$ is normal, i.e., $\sim 2$~Gyr in high surface density compressed regions and somewhat lower ($\sim 0.2$~dex) than the mean in regions of low gas surface density. 

A pixel-by-pixel analysis was preformed in the same way as for the observations. The results of the fit and the Spearman rank correlation coefficient $\rho$ are presented in Table~\ref{tabsim}.

We found steeper slopes of the molecular KS relation for NGC~4567 ($1.5$ compared to an observed slope of $0.9 \pm 0.1$) and for NGC~4568 ($1.3$ compared to $1.0 \pm 0.1$). 
For the SFR-$\Sigma_{\rm g}$ relation the model slopes are also steeper for NGC~4567 ($1.5$ compared to an observed slope of $1.1 \pm 0.2$) and for NGC~4568 ($1.8$ compared to $1.3 \pm 0.1$).
The slope of the $R_{\rm mol}$-$P_{\rm tot}$ correlation of NGC~4568 is  reproduced well (slopes of $\sim 0.7$). 
For the $R_{\rm mol}$ - $\Sigma_{\star}$ and $R_{\rm mol}$ - $\Sigma_{\rm g}$ relations the slopes are on the lower side, but within the errors of the observed values.
As observed, we found a correlation between $Q$ - $R_{\rm mol}$/$P_{\rm tot}$ with a comparable slope for NGC~4567 ($0.8$ compared to the observed slope of $0.7$) and a much steeper slope for NGC~4568 ($1.4$ compared to $0.7$). 

We conclude that our observations are in reasonable agreement with observations.         
                  
\subsection{What can be learned from the dynamical models}

The standard slopes of the $SFR$-$\Sigma_{\rm H_2}$, $R_{\rm mol}$-$P_{\rm tot}$, and $R_{\rm mol}$-$\Sigma_*$ relations 
(e.g., Bigiel et al. 2008, Leroy et al. 2008, Blitz \& Rosolowsky 2006) are reproduced by the dynamical models for NGC~4501 and NGC~4568.   
This makes us confident that the model predictions are reliable.

Furthermore, the dynamical models correctly reproduce the enhancement of the molecular fraction in the compressed regions 
of NGC~4501 and NGC~4568. Since in the analytical model the molecular fraction mainly depends on the gas density, we conclude that 
the dynamical model correctly predicts the enhancement of the gas density due to large-scale tidal or ram pressure compression.

As observed, the ratio between the molecular fraction divided by the total ISM pressure increases with $Q$: $R_{\rm mol}/P_{\rm tot} \propto Q^{1.5}$
in the dynamical models of NGC~4501 and NGC~4568 (Fig.~\ref{4gal255}). 
The model $SFE_{\rm H_2}$-$R_{\rm mol}/P_{\rm tot}$ relation is also broadly consistent with our observations:
$SFE_{\rm H_2} \propto (R_{\rm mol}/P_{\rm tot})^{-0.5}$. However, the expected $SFE_{\rm H_2}$-$Q$ correlation is not present in the dynamical models.

 The dynamical model of NGC~4501 correctly reproduces the spatial distribution of all observed quantities; however, the model of NGC~4568
reproduces only the main characteristics of our observations without a good match between the modeled and observed spatial distribution.
Nevertheless, both simulations show the same trends.
A gravitationally induced ISM compression thus has the same consequences as the ram pressure compression that can also be observed in galactic spiral arms
(NGC~4321): (i) an increasing gas surface density, (ii) an increasing molecular fraction, and (iii) a decreasing $R_{\rm mol}/P_{\rm tot}$ in the compressed region 
due to the presence of nearly self-gravitating gas.
The response of star formation and thus $SFE_{\rm H_{2}}$ to compression is more complex. In regions of high gas surface density the molecular
star formation efficiency stays close to the mean, whereas in regions of low gas surface density, $SFE_{\rm H_2}$ decreases.
We suggest that partial adiabatic compression is responsible for this behavior: in the high surface density gas the turbulent dissipation timescale is shorter
than the compression timescale, the excess energy injected by compression can be dissipated, and the molecular star formation efficiency stays constant.
On the other hand, in a compressed low surface density gas the compression timescale is shorter than the turbulent dissipation timescale, the molecular clouds 
become overpressured, and the star formation activity decreases leading to a decrease in $SFE_{\rm H_2}$.
Since we only observe a mild decrease in $SFE_{\rm H_2}$ in the compressed low surface density gas, we conclude that in NGC~4501 and NGC~4568 the
turbulent dissipation and compression timescales are not very different.

\section{Summary and conclusions \label{conclu}}

New IRAM 30m HERA CO(2-1) observations of the  Virgo spiral galaxies NGC~4501 and NGC~4567/68 are presented. The average rms noise level is less then 8 mK in the 10.4 km\,s$^{-1}$ velocity channel at a resolution of $12''$. 

NGC~4501 shows signs of ongoing nearly edge-on ram pressure stripping. Molecular gas is detected extending over almost the entire optical disk (R$_{25}=4.35'$). Moreover, the molecular gas distribution shows a well-defined southwestern edge where ram pressure compresses the ISM and a more diffuse northeastern edge. NGC~4501 shows a higher molecular fraction ($R_{\rm mol}=\Sigma_{\rm H_{2}}/\Sigma_{\rm HI}$) in the compressed southwestern edge. In this particular region the compressed gas is at the limit of self-gravitation. The compression of the ISM does not change  the overall molecular or the total gas Schmidt-Kennicutt relation at a kpc resolution (SFR$\propto\Sigma_{\rm g}^{1.53\pm0.27}$ and SFR$\propto\Sigma_{\rm H_{2}}^{0.91\pm0.08}$). However, continuous regions of low molecular star formation efficiency are present in the disk: the molecular star formation efficiency ($SFE_{\rm H_{2}}$) map is 0.2 dex lower in a U-shaped southeastern region and in the western part of the disk. 

For the gravitationally interacting Virgo spiral galaxy NGC~4568, we find that the southern molecular gas distribution is more extended than the northern distribution. Moreover, a peculiar spiral arm is detected in the south curving toward its companion, NGC~4567. Within the latter galaxy the molecular gas distribution shows a symmetric two-arm spiral structure. The molecular fraction increases in the region between the two galaxies. However, this could be due to projection effects.  As in NGC~4501, the overall Schmidt-Kennicutt relations at $\sim1.5$ kpc resolution are not affected by the compression of the ISM, but continuous regions of low/high molecular star formation efficiency are present in the disk:
the $SFE_{\rm H_{2}}$ is significantly higher than the mean in the eastern part of NGC~4568 and somewhat lower in the northwestern interacting part and in the southwestern edge of NGC~4568.

We note that when computing the total ISM pressure $P_{\rm tot}$ the self-gravitating gas term is not negligible in gas-rich galaxies and has been taken into account. The molecular fraction of the galaxies is correlated with the total ISM pressure
$R_{\rm mol} \propto P_{\rm tot}^{0.7-1.0}$. The offset of the molecular fraction-ISM  pressure relation is sensitive to the gas content of the galaxies. In agreement with Leroy et al. (2008), we found that the molecular fraction is better predicted by the stellar surface density than by the total gas surface density. We introduced a new parameter $R_{\rm mol}/P_{\rm tot}$, which is related to the gas self-gravitation and is poorly sensitive to variations of the CO-H$_2$ conversion factor. For all galaxies in our sample $R_{\rm mol}/P_{\rm tot}$ is correlated with the Toomre $Q$ parameter $R_{\rm mol}/P_{\rm tot} \propto Q^{1.4-1.6}$. We found that galactic regions which are at the limit of self-gravitation (low Toomre $Q$ and low $R_{\rm mol}/P_{\rm tot}$) tend to have higher $SFE_{\rm H_{2}}$.

The analytical model described in Vollmer \& Leroy (2011) was used to investigate the dependence of the molecular star formation efficiency on self-gravitation. The model correctly reproduces the correlations between $R_{\rm mol}/P_{\rm tot}$, $SFE_{\rm H_{2}}$, and $Q$ if different global turbulent velocity dispersions are assumed for the different galaxies. In addition,  an isotropic external pressure might increase the molecular star formation efficiency slightly. We found that variations in the conversion factor can mask most of the correlation between $SFE_{\rm H_{2}}$ and the Toomre $Q$ parameter.     

Dynamical simulations were used to compare the effects of ram pressure and tidal ISM compression. These models give direct access to the volume density.

The gas distribution of NGC~4501 is  reproduced well by the model, as is the increase in the molecular fraction in the compressed part of the disk. We found that the gas distribution at the time of interest is sensitive to galactic structure, i.e., the position of the spiral arms with respect to the gas compression.
The gas surface density is highest if a spiral arm is present in the compressed region. The simulated $R_{\rm mol}/P_{\rm tot}$ map  reproduces the observed map fairly well, with lower values in the compressed regions where the gas becomes almost self-gravitating. The model successfully reproduces the lower $SFE_{\rm H_{2}}$ in the northwest and in the southwestern part of the compressed region. In these regions the collisional timescale is longer than the compression timescale. Thus, the gas cannot adapt its turbulent velocity fast enough to the new conditions in the compressed region. The effect of adiabatic compression is similar to this mechanism in our collisional model. We therefore suggest that adiabatic compression quenches star formation in low surface density parts of the compressed region by making the gas clouds overpressured.  

For the gravitationally interacting system NGC~4567/68, the simulation correctly reproduces the gas distribution of NGC~4568. The model shows a small increase in the molecular fraction in the western compressed side of the disk as NGC~4501. The gravitationally induced compression increases the gas density and lowers $R_{\rm mol}/P_{\rm tot}$. Moreover, the lower $SFE_{\rm H_{2}}$ in the southwestern part of the disk is reproduced in NGC~4568. In the light of the dynamical model, we confirm that NGC~4567/68 is a young interacting pair with a compression region west of the center of NGC~4568.

We conclude that, a gravitationally induced ISM compression has the same consequences as ram pressure compression, which can also be observed in galactic spiral arms: (i) an increasing gas surface density, (ii) an increasing molecular fraction, and (iii) a decreasing $R_{\rm mol}/P_{\rm tot}$ in the compressed region due to the presence of nearly self-gravitating gas. The response of star formation and thus $SFE_{\rm H_{2}}$ to compression is more complex. While in violent ISM-ISM collisions (Taffy galaxies, NGC~4438) the interaction decreases star formation by a factor  of $4$-$5$ (Braine et al. 2003, Vollmer et al. 2009, 2012 ), we only detect an $SFE_{\rm H_{2}}$ variation of $\sim 50$\,\%  in the compressed regions of NGC~4501 and NGC~4568. We suggest that partial adiabatic compression is responsible for the quenched molecular star formation efficiency in all systems.
 The compression timescale is much smaller than the turbulent dissipation timescale in violent ISM-ISM collisions leading to a strong quenching of the
star formation activity, whereas the two timescales are not very different in NGC~4501 and NGC~4568. 

%________________________________________________________________

\begin{acknowledgements}
Based on IRAM observations. IRAM is supported by INSU/CNRS (France), MPG (Germany), and IGN (Spain).
\end{acknowledgements}

%____________________________________________________________________________________________________________%
  
\clearpage
\appendix

\section{CO(2-1) observations: rms \label{Ap1}}

\begin{figure} 
  \centering
\includegraphics[width=9cm]{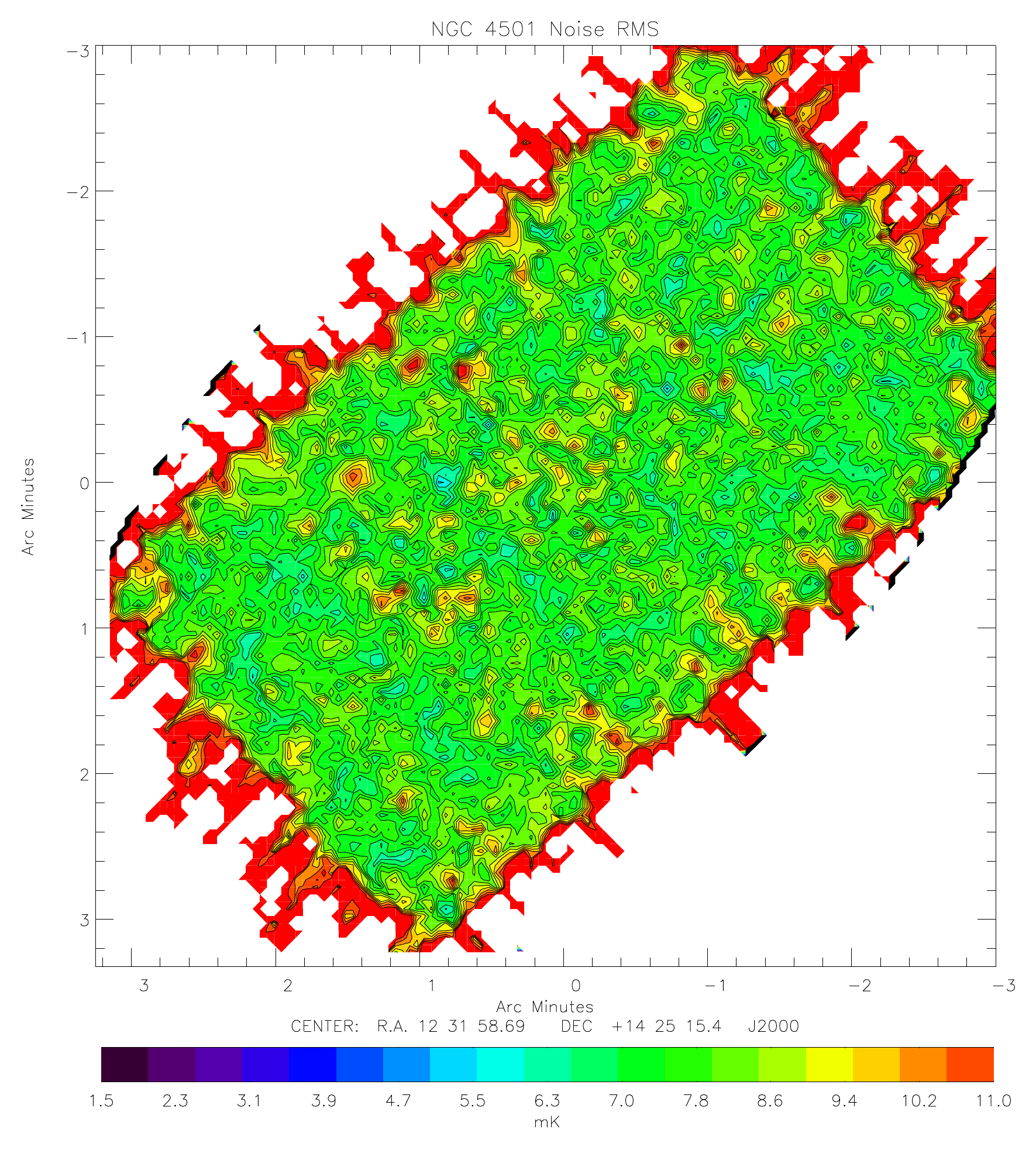}
\includegraphics[width=9cm]{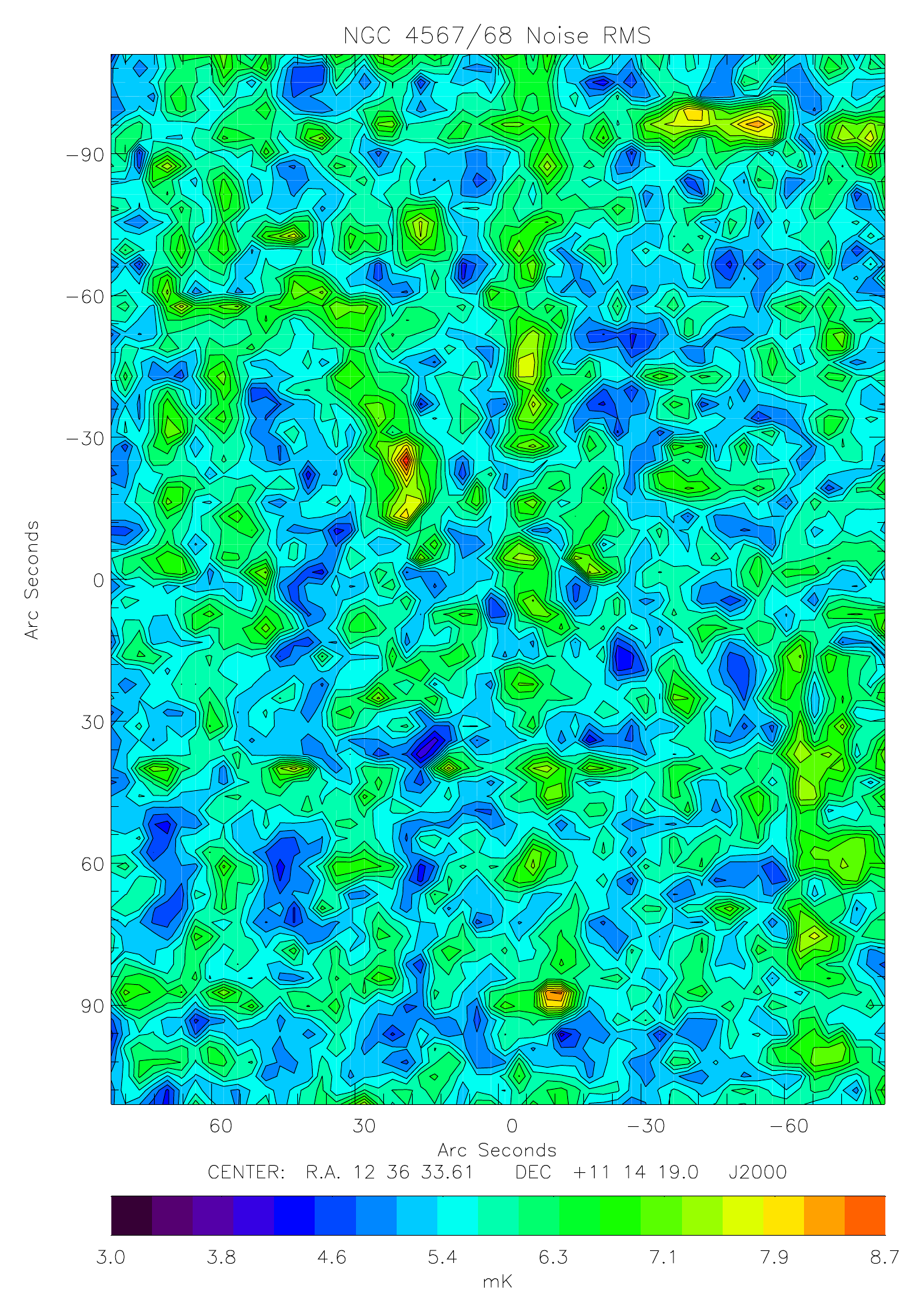}
   \caption{CO(2-1) rms noise maps of NGC~4501 (upper panel) and NGC~4567/68 (lower panel). The color bar is in units of mK. The spectral velocity resolution is 
$10.4$~km\,s$^{-1}$. }
\label{RMS4501}%
\end{figure}

We computed the rms of the CO(2-1) spectra for each position. The CO(2-1) emission line signal in each spectrum was excluded from the rms calculation. The rms of the column density was defined as  $\Delta{\Sigma_{{\rm H}_{2}}} = rms \times \Delta{ v } \times \sqrt{ {N}_{ {\rm CO(2-1)}}}$, where $rms$ is the noise level of the CO spectra, $\Delta{v}$ the spectral velocity resolution ($10.4$~km\,s$^{-1}$), and $N_{\rm CO(2-1)}$ the number of channels containing a CO(2-1) line signal. The CO(2-1) noise maps of NGC~4501 and NGC~4567/68 (Fig.~\ref{RMS4501}) do not show any systematic behavior.  

\section{SFR comparison}

\subsection{Total IR \label{TTIIRR}}

\begin{figure*} 
 \centering
\includegraphics[width=12cm]{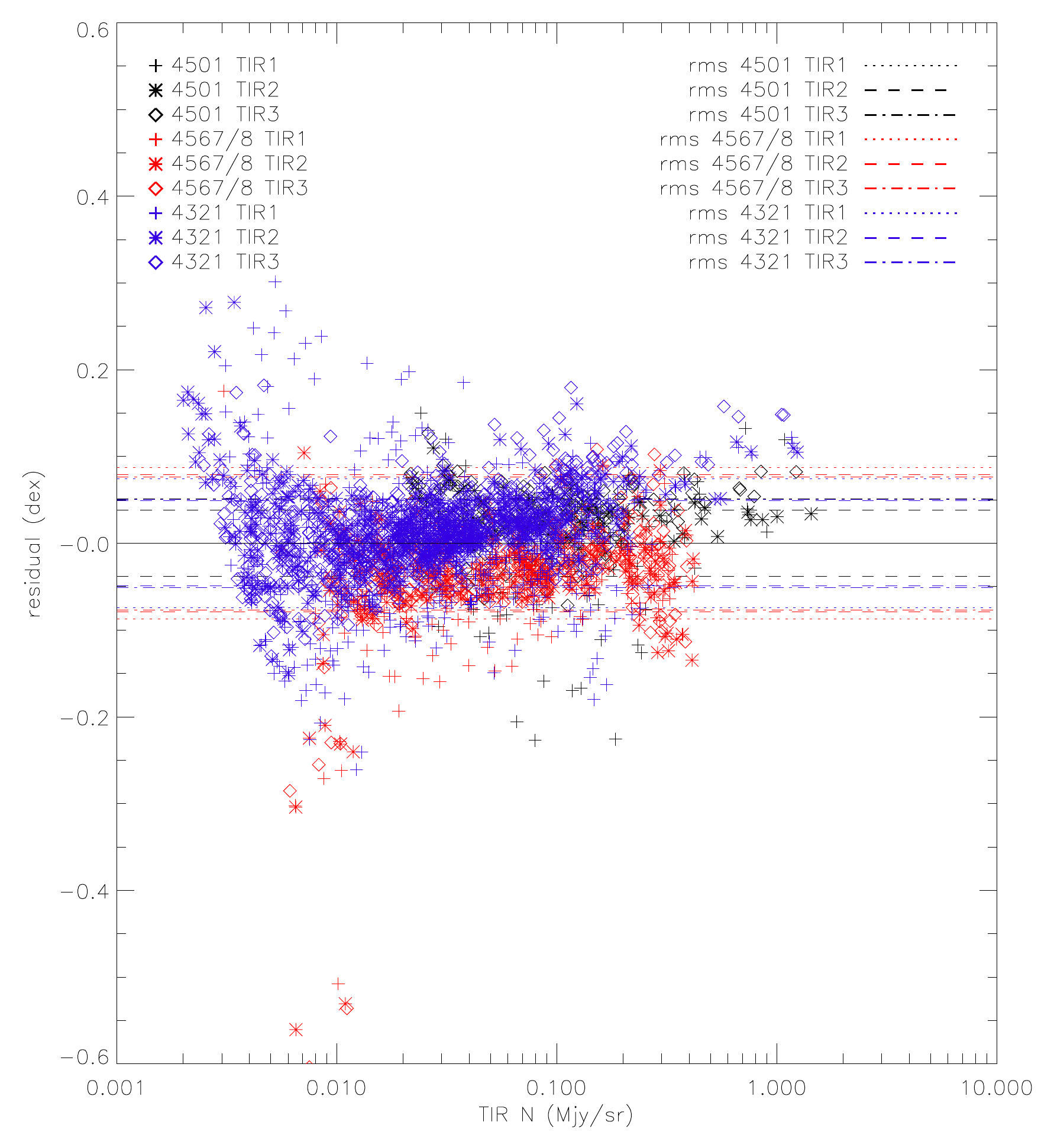}
   \caption{Total infrared (TIR) correlations with $I_{\nu}(24 \mu{\rm m})$ for NGC~4567/68,  NGC~4501, and NGC 4321. The pixel size is $18"$. Residuals of the relations $I_{\nu}(24 \mu{\rm m})\times3.2\times10^{-3}/8.1\times10^{-2}$ (see Eq.~\ref{eqqfuv}) as a function of $TIR1\times0.46$, $TIR2\times0.46$, and $ TIR3\times0.46$ (see Eq.~\ref{SFRTIR}). The dashed and dotted lines are at $\pm 1 \times rms$ in the same color as the corresponding galaxy.}
\label{TIRc}%
\end{figure*}

The FUV heated dust emission can be traced using the total infrared emission (TIR) (see Hao et al. 2011 and Galametz et al. 2013).  
We tried three different recipes to compute the TIR emission:
\begin{itemize}
\item TIR1 based on the 3.6, 8, 24, and 70~$\mu$m from Spitzer (Eqs.~\ref{PAH}, \ref{eq24}, and \ref{eqTIR1}).
\item TIR2 based on the 3.6, 24~$\mu$m from Spitzer and 100, 160~$\mu$m from {\em Herschel} PACS (Eq.~\ref{eqTIR2}).
\item TIR3 based on the 3.6, 24, and 70~$\mu$m from Spitzer and 100, 160~$\mu$m from {\em Herschel} PACS (Eq.~\ref{eqTIR3}).
\end{itemize}
\begin{eqnarray}
 I_{\nu}(PAH 8 \mu{\rm m}) & = & I_{\nu}(8 \mu{\rm m})-0.232 \, I_{\nu}(3.6 \mu{\rm m}) \label{PAH}\\
 I_{\nu}(24 \mu{\rm m}) & = & I_{\nu}(24 \mu{\rm m})-0.032 \, I_{\nu}(3.6 \mu{\rm m}) \label{eq24}\\
 I({\rm TIR1}) & = & 0.95\, \nu I_{\nu}({\rm PAH} 8 \mu{\rm m})+1.15 \times \nonumber \\ 
               &   & \nu I_{\nu}(24 \mu{\rm m})+  2.3 \, \nu  I_{\nu}(70 \mu{\rm m}) \label{eqTIR1} \\
I({\rm TIR2}) & = & 2.708 \, \nu I_{\nu}(24 \mu{\rm m}) + 0.734 \times \nonumber \\ 
              &   &  \nu I_{\nu}(100 \mu{\rm m})+0.739 \, \nu I_{\nu}(160 \mu{\rm m}) \label{eqTIR2} \\
 I({\rm TIR3}) & = & 2.064 \, \nu I_{\nu}(24 \mu{\rm m})+ 0.539 \times \nonumber \\ 
               &   &   \nu I_{\nu}(70 \mu{\rm m})+0.277 \times  \nonumber \\ 
               &   &   \nu I_{\nu}(100 \mu{\rm m})+ 0.938 \, \nu I_{\nu}(160 \mu{\rm m})  \label{eqTIR3}\\
\dot{\Sigma_{\star}}(TIR) & = & 8.1 \, 10^{-2} I( {\rm FUV})+0.46 \, I({\rm TIR})  \label{SFRTIR}
\end{eqnarray}
The TIR1, TIR2, and TIR3 recipes of NGC~4501, NGC~4567/68, and NGC 4321 were compared in a pixel-by-pixel analysis with the 24~$\mu$m emission (Fig.~\ref{TIRc}). 
The three TIR recipes tightly correlate with the 24~$\mu$m emission ($I_{\nu}(24 \mu{\rm m})$). The rms of these correlations are smaller than 0.09~dex. 
%As a conclusion the impact of using only the 24 microns to correct the $H\alpha$ or the FUV flux from dust absorption is negligible.

\subsection{SFR from H$\alpha$}
\label{HA}
\begin{figure*}
\centering
\includegraphics[width=12cm]{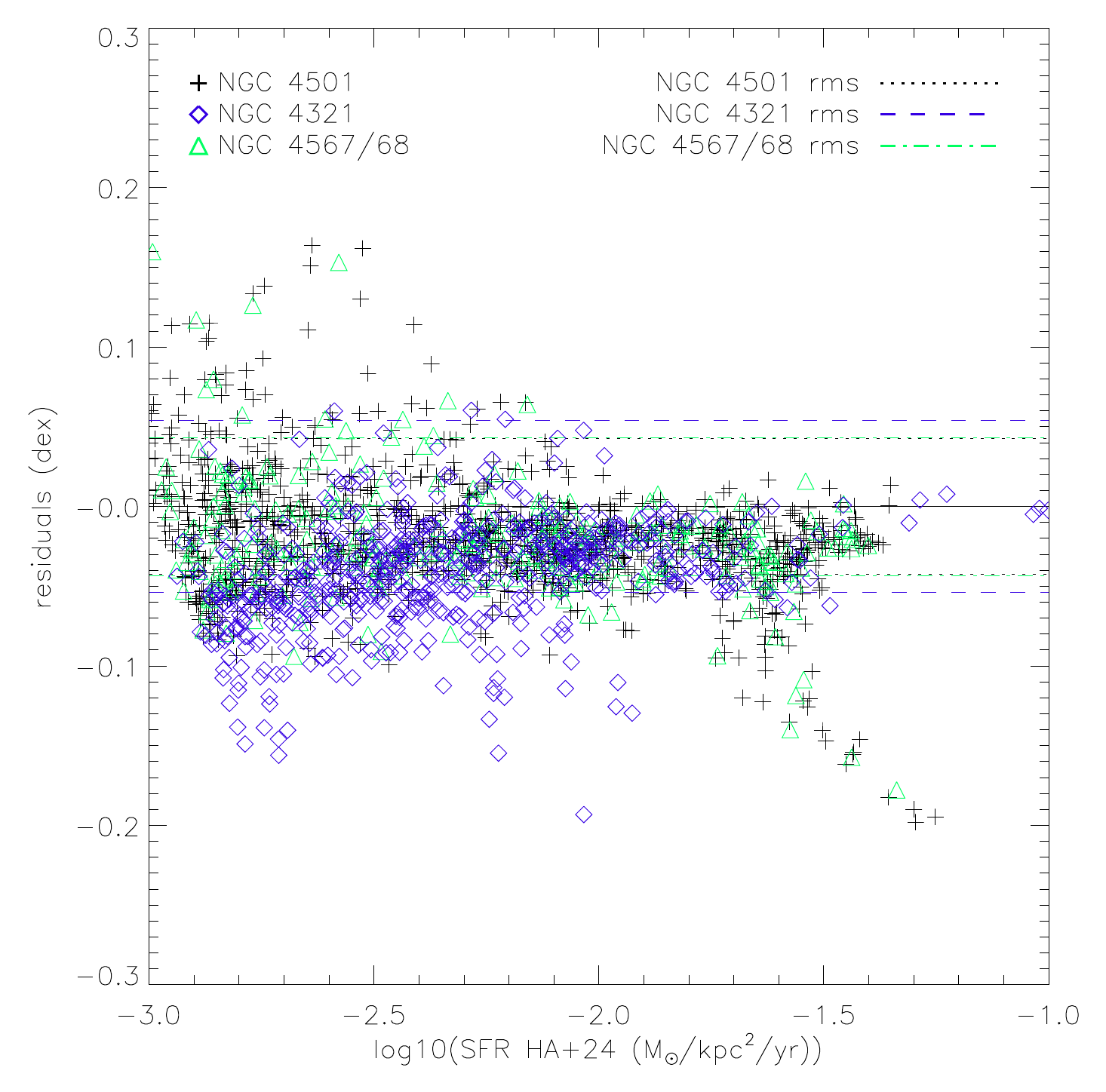} %compareSFR_tot.pdf
   \caption{Comparison of star formation rates computed with FUV+$24\mu m$ and H$_{\alpha}$+$24\mu m$. Residuals of the relations in dex for NGC~4501, NGC~4567/68, and NGC 4321. The continuous line corresponds to a linear relation between the two SFR indicators, while the dashed and dotted lines correspond to $\pm 1 \sigma$ for each galaxy.}
\label{SFRcompare}%
\end{figure*}

We calculated the star formation rate based on ${\rm H}_{\alpha}$ emission: $\dot{\Sigma_{\star}}(\rm H\alpha+24 \mu {\rm m})$ (Eq.~\ref{eqqHA}),
\begin{equation}
\dot{\Sigma_{\star}}(\rm H\alpha+24 \mu {\rm m})=(634\,f_{\rm H\alpha}+2.5 \, 10^{-3}\, {I}_{24\mu{\rm m}}) \cos(i)\ ,    \label{eqqHA}
\end{equation}
where $f_{\rm H\alpha}$ is the H$\alpha$ flux density (ergs s$^{-1}$ cm$^{-2}$ {\AA}$^{-1}$) corrected for N{\sc ii} line contamination (${\rm NII}/{\rm H}\alpha$ ratio of 0.54; see Kennicutt et al. 2008). 
The H$\alpha$ data for NGC~4501, NGC~4567/68, and NGC 4321 are part of the GOLD-mine database (Gavazzi et al. 2003).
$\dot{\Sigma_{\star}}(\rm H\alpha+24 \mu {\rm m})$ and  $\dot{\Sigma_{\star}}(\rm FUV+24 \mu {\rm m})$ were compared pixel-by-pixel  (see Fig.~\ref{SFRcompare}). A significant deviation is found in the center of NGC~4501 where the $\dot{\Sigma_{\star}}$ derived from H$\alpha$ exceeds that derived from the FUV. The rms of the linear relations are $0.056$~dex, $0.023$~dex, and $0.054$~dex for NGC~4501, NGC~4567/68, and NGC 4321, respectively.

\section{Individual plots I}
\label{annex44}

\begin{figure*}
   \centering
\includegraphics[width=15cm]{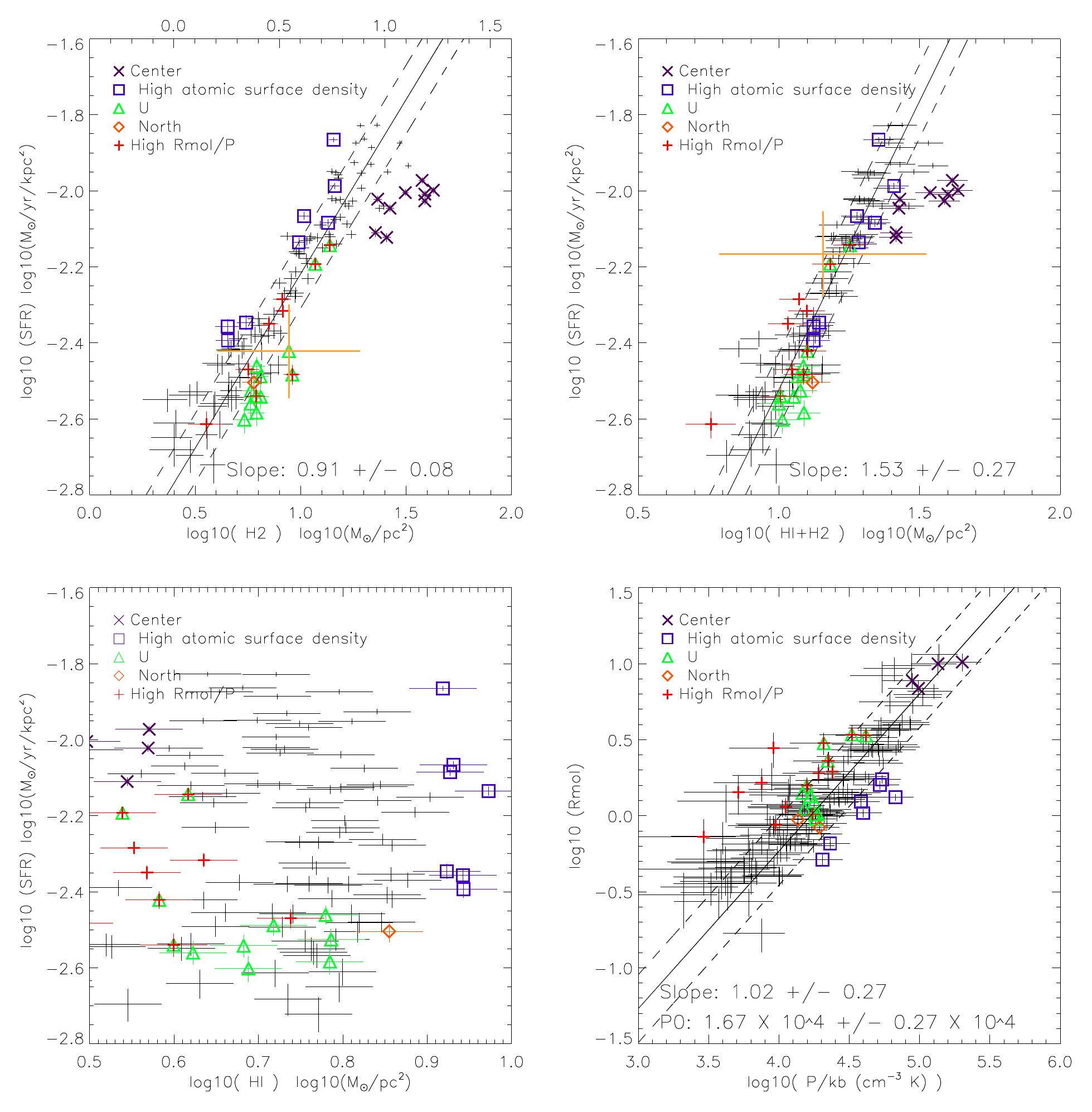}
   \caption{NGC~4501. {\em Upper left panel}: SFR as a function of $\Sigma_{H_2}$. {\em Upper right panel}: SFR as a function of total gas surface density. {\em Lower left panel}: SFR as a function of atomic gas surface density. {\em Lower right}: Molecular fraction versus ISM pressure $P_{\rm tot}$.
The orange error bars take into account systematic uncertainties in the calibration of the $SFR$ ($0.1$~dex) and the $N_{\rm H_2}/I_{\rm CO}$ conversion factor ($0.3$~dex).}
\label{plot4501}%
\end{figure*}
\begin{figure*}
   \centering
\includegraphics[width=15cm]{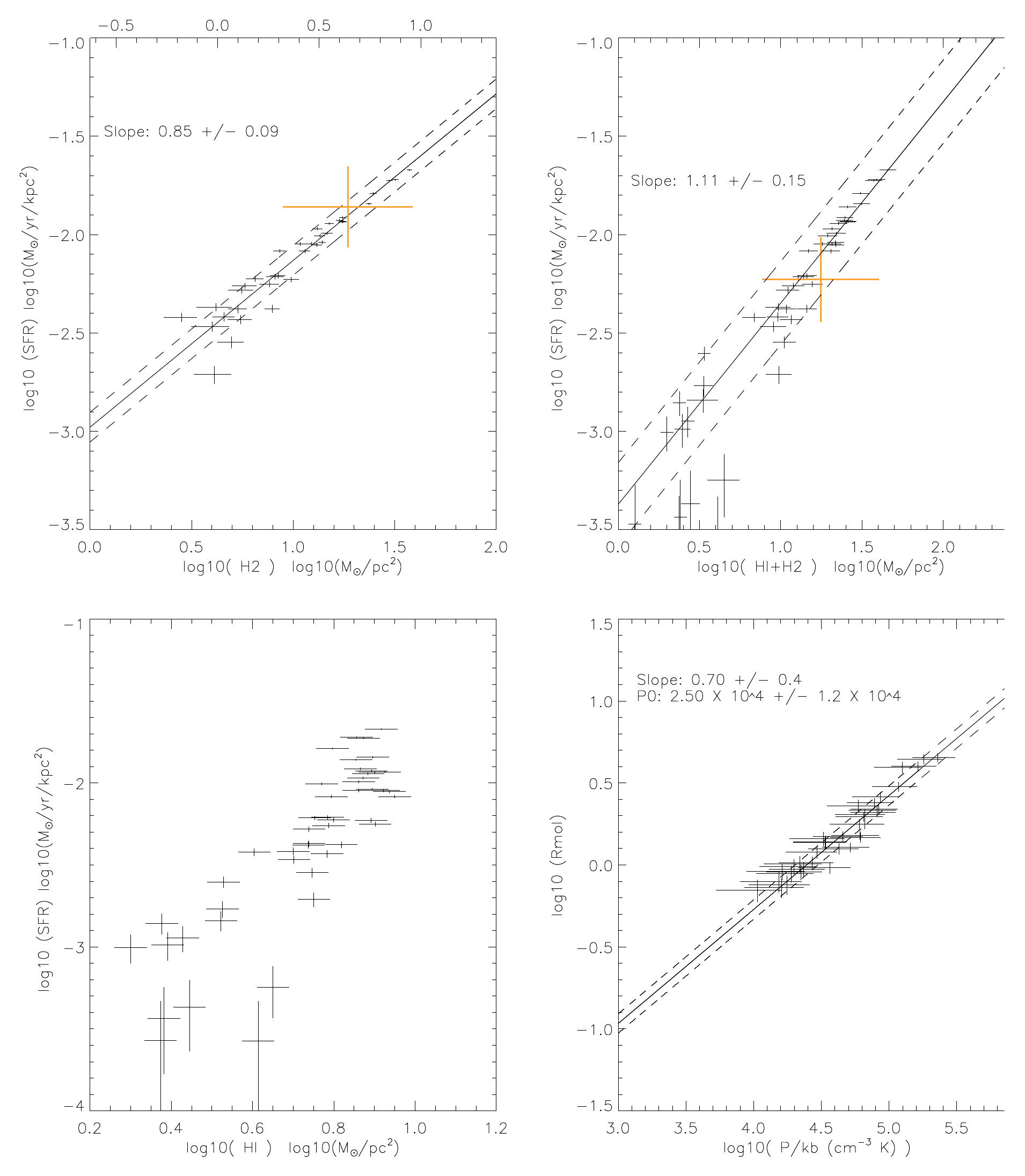}
   \caption{NGC~4567. {\em Upper left panel}: SFR as a function of $\Sigma_{H_2}$. {\em Upper right panel}: SFR as a function of total gas surface density.  {\em Lower left panel}: SFR as a function of atomic gas surface density. {\em Lower right}: Molecular fraction versus the ISM pressure $P_{\rm tot}$.
The orange error bars take into account systematic uncertainties in the calibration of the $SFR$ ($0.1$~dex) and the $N_{\rm H_2}/I_{\rm CO}$ conversion factor ($0.3$~dex).}
\label{plot4567}%
\end{figure*}
\begin{figure*}
   \centering
\includegraphics[width=15cm]{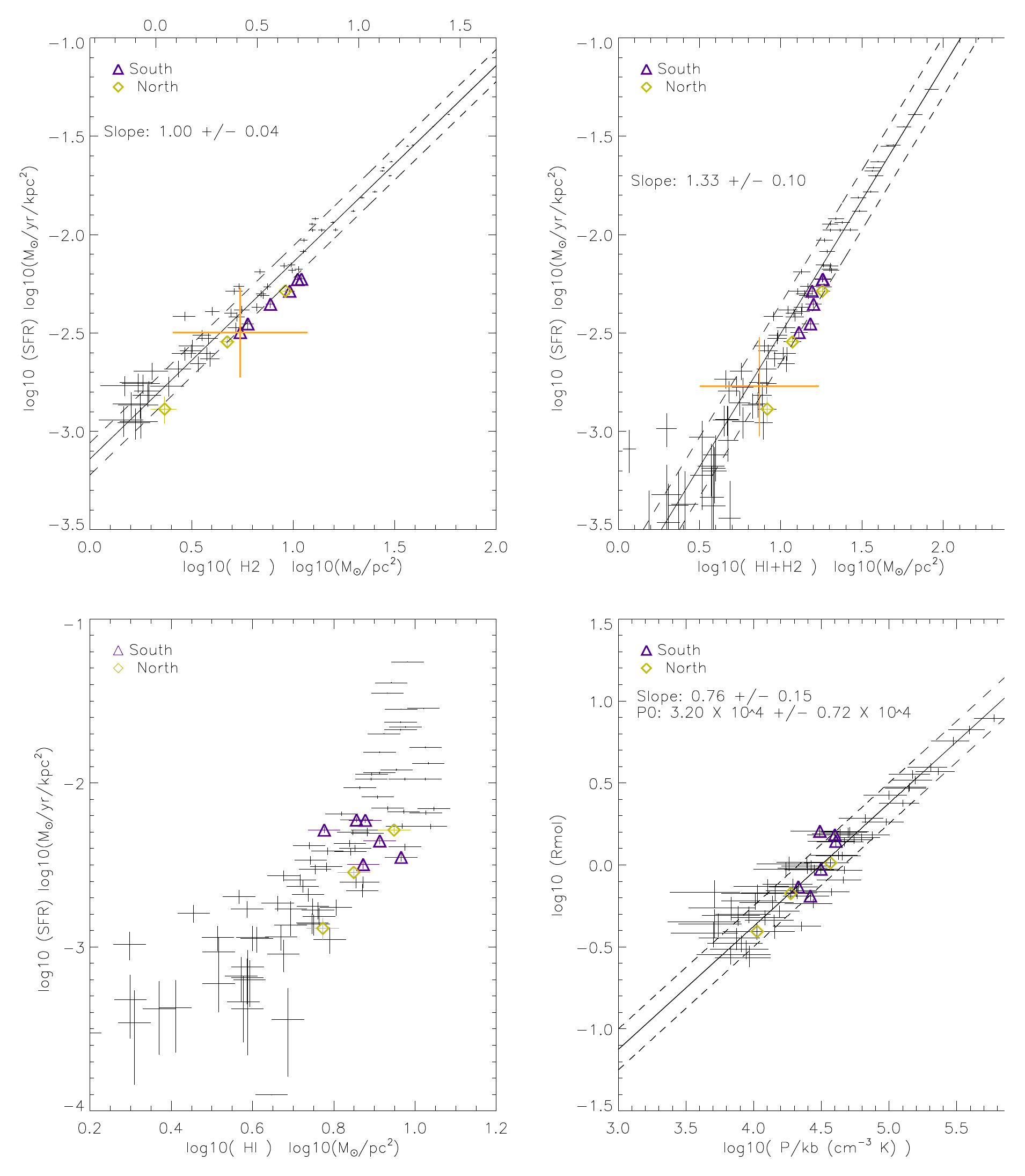}
   \caption{NGC~4568. {\em Upper left panel}: SFR as a function of $\Sigma_{H_2}$. {\em Upper right panel}: SFR as a function of total gas surface density.  {\em Lower left panel}: SFR as a function of atomic gas surface density. {\em Lower right}: Molecular fraction versus the ISM pressure $P_{\rm tot}$.
The orange error bars take into account systematic uncertainties in the calibration of the $SFR$ ($0.1$~dex) and the $N_{\rm H_2}/I_{\rm CO}$ conversion factor ($0.3$~dex).}
\label{plot4568}%
\end{figure*}
\begin{figure*}
   \centering
\includegraphics[width=15cm]{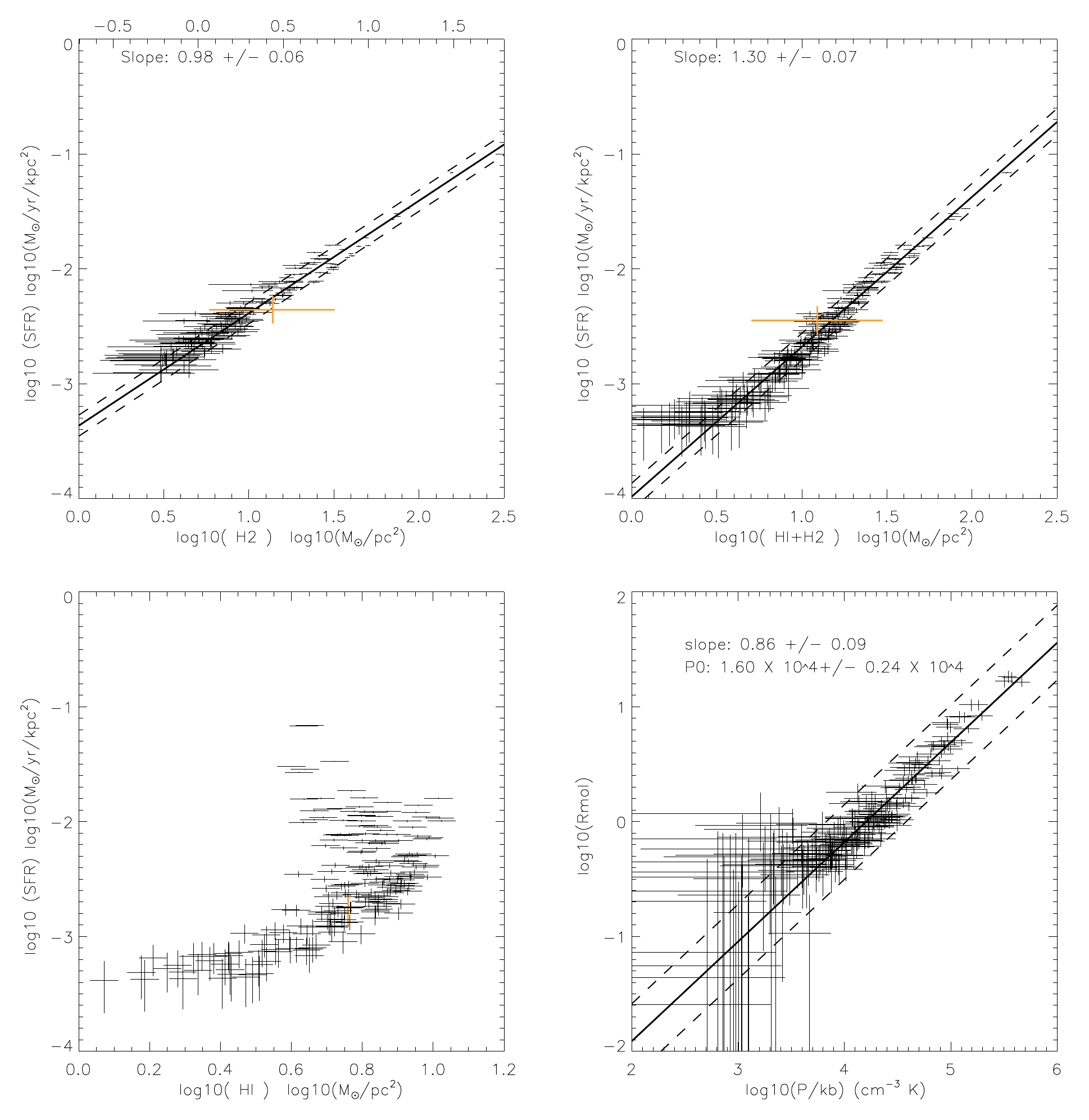}
   \caption{NGC~4321. {\em Upper left panel}: SFR as a function of $\Sigma_{H_2}$. {\em Upper right panel}: SFR as a function of total gas surface density.  {\em Lower left panel}: SFR as a function of atomic gas surface density. {\em Lower right}: Molecular fraction versus the ISM pressure $P_{\rm tot}$.
The orange error bars take into account systematic uncertainties in the calibration of the $SFR$ ($0.1$~dex) and the $N_{\rm H_2}/I_{\rm CO}$ conversion factor ($0.3$~dex).}
\label{plot4321}%
\end{figure*}

\clearpage
\section{Individual plots II}
\label{annexe55}

\begin{figure*}
   \centering
\includegraphics[width=7cm]{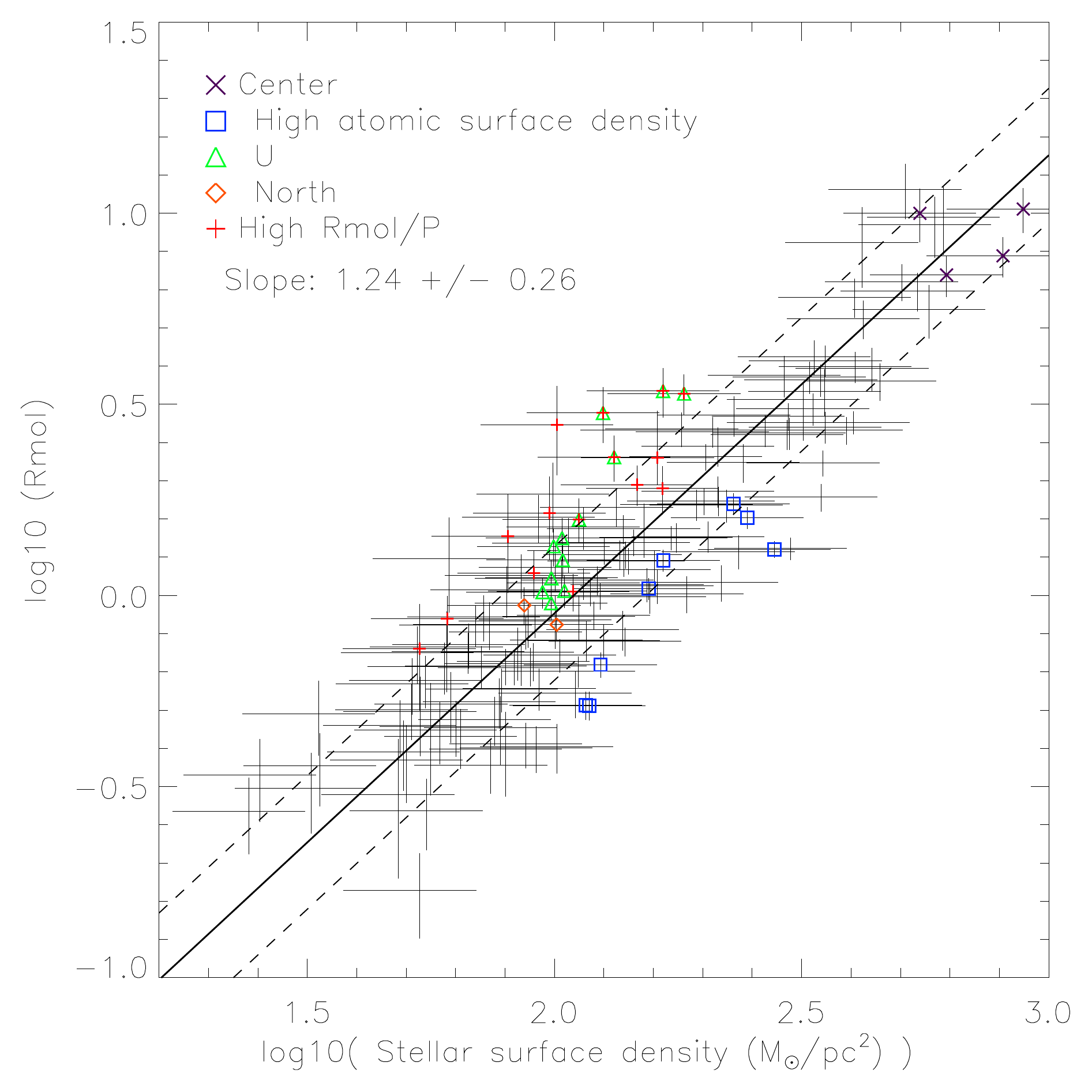}
\includegraphics[width=7cm]{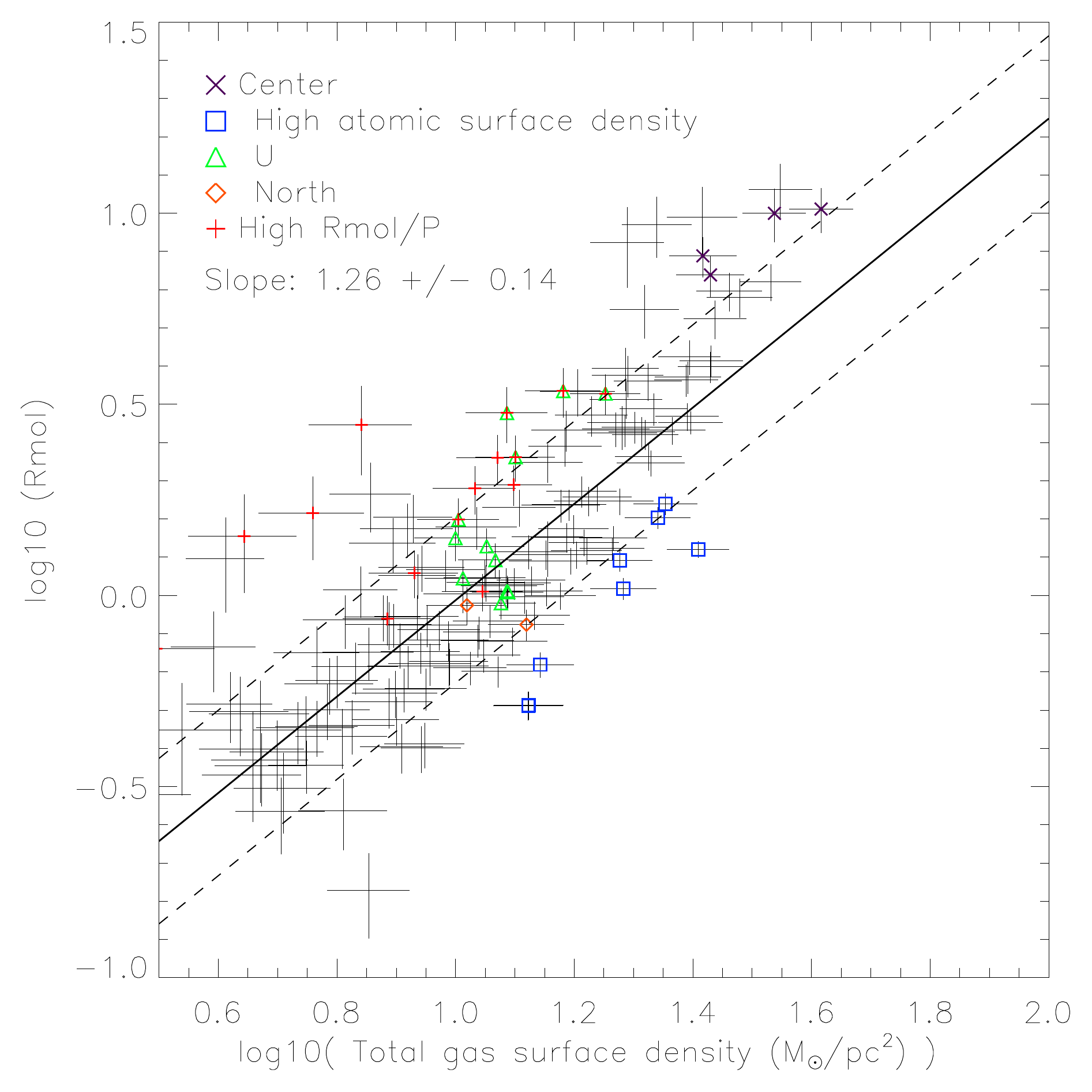}
\includegraphics[width=7cm]{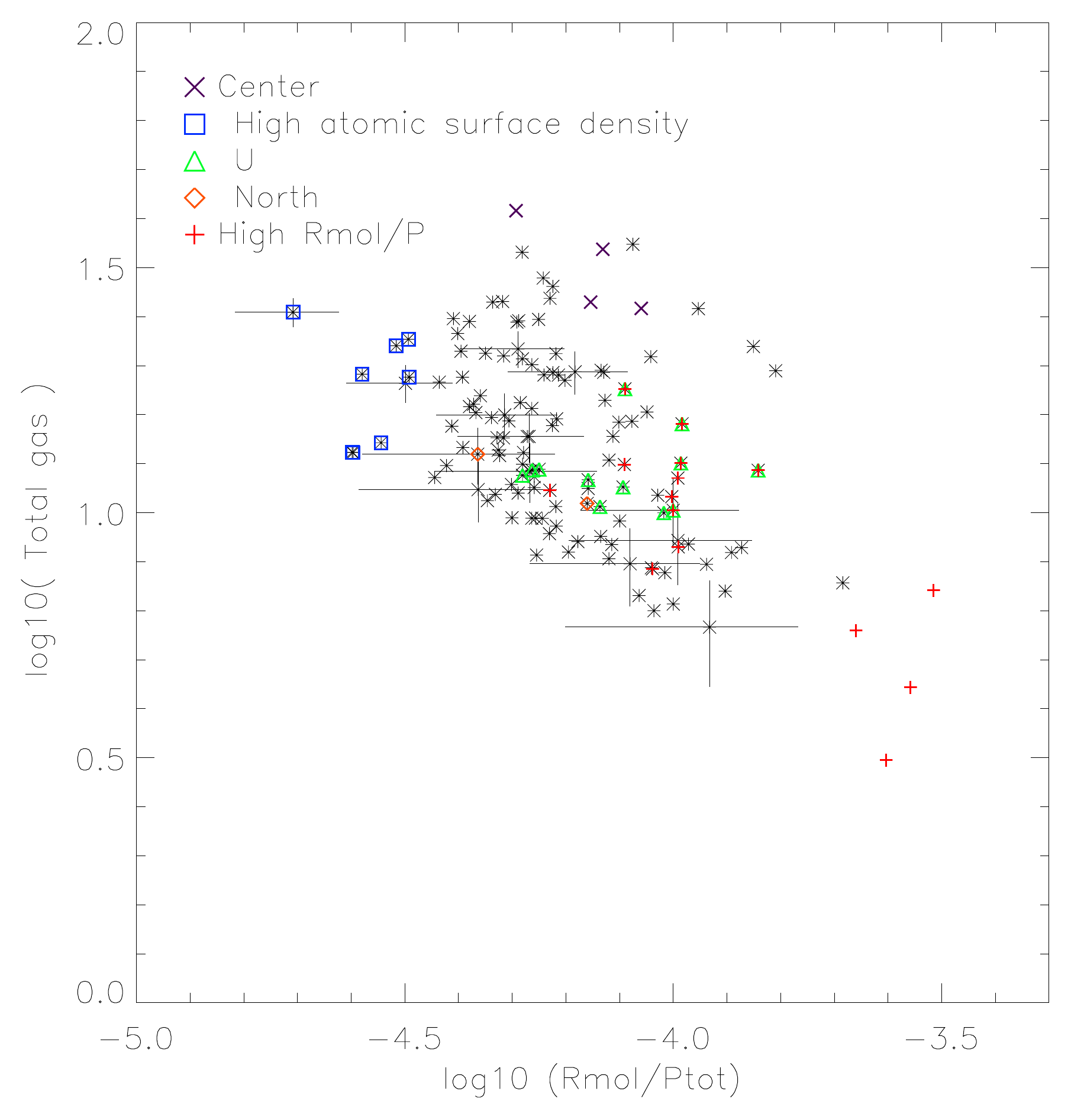}
\includegraphics[width=7cm]{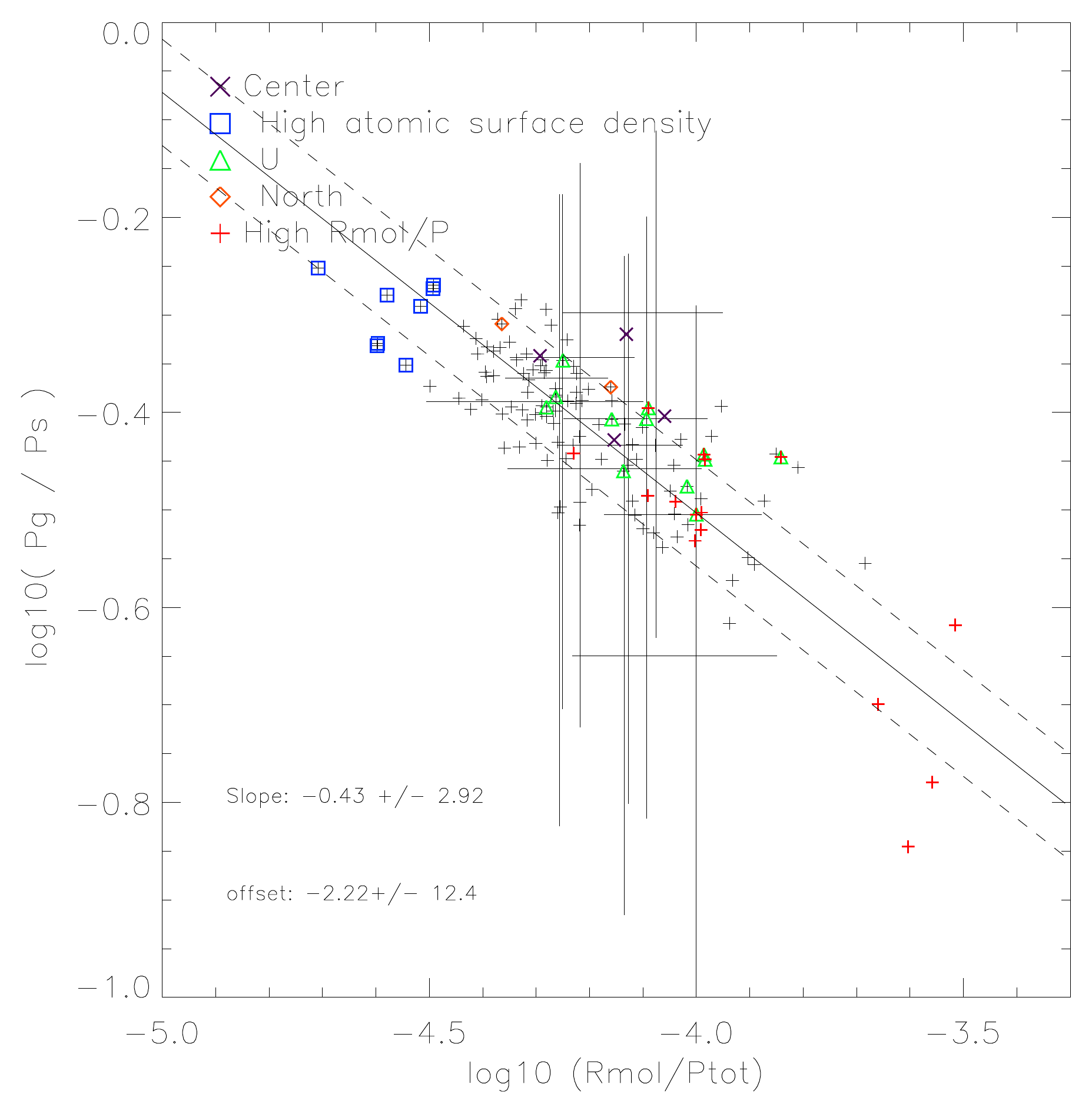}
   \caption{NGC~4501. {\em Upper left panel}: Molecular fraction as a function of stellar surface density. {\em Upper right panel}: Molecular fraction as a function of the total gas surface density.  {\em Lower left panel}: Total gas as a function of molecular fraction divided by ISM pressure. {\em Lower right}: Pressure due to the gravitational potential of the gas divided by the pressure due to the potential of the stellar disk ($P_{\rm g}/P_{\rm s}$) as a function of the molecular fraction divided by ISM pressure. For clarity we only show a few errors bars.}
\label{plot4501_2}%
\end{figure*}
\begin{figure*}
   \centering
\includegraphics[width=7cm]{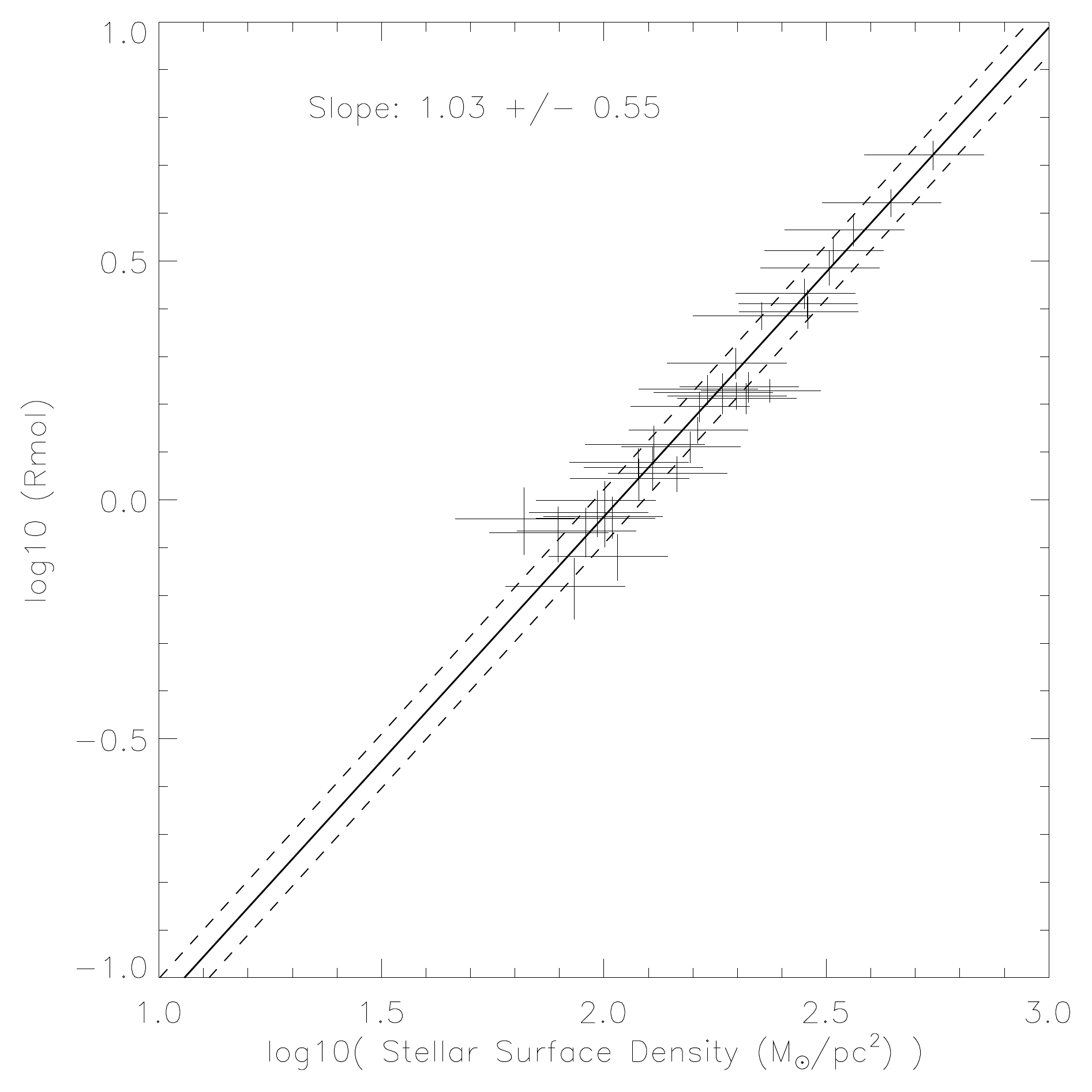}
\includegraphics[width=7cm]{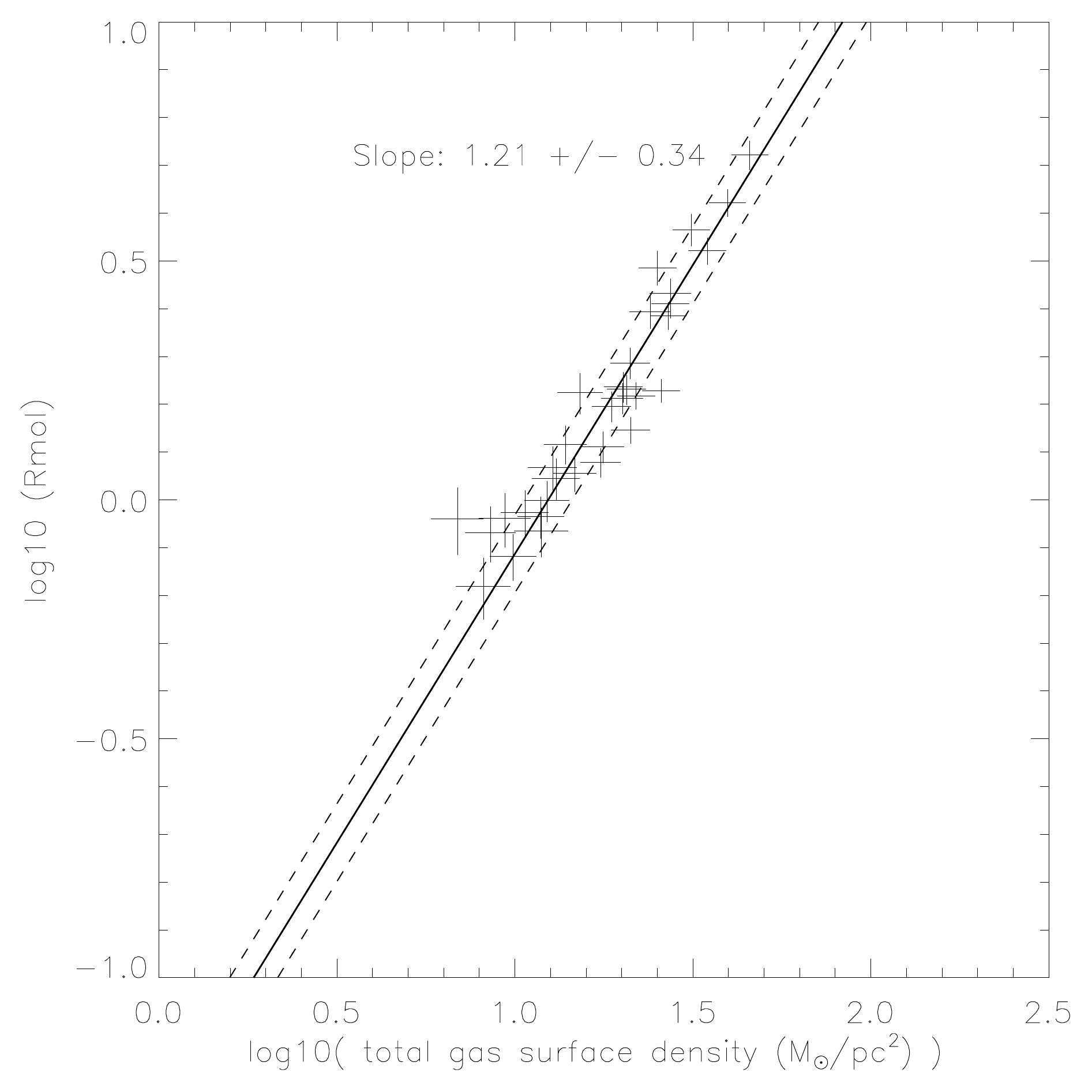}
\includegraphics[width=7cm]{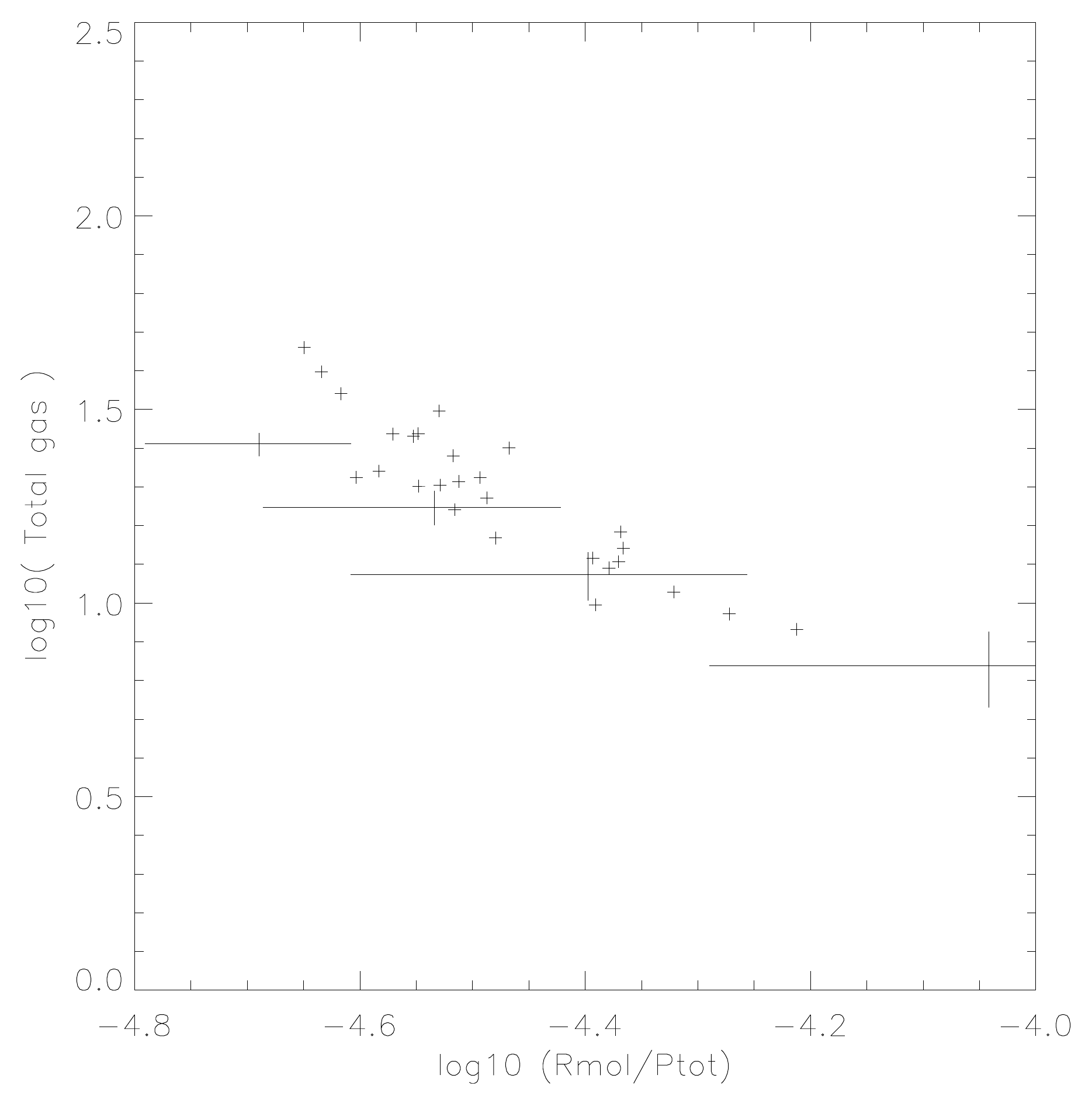}
\includegraphics[width=7cm]{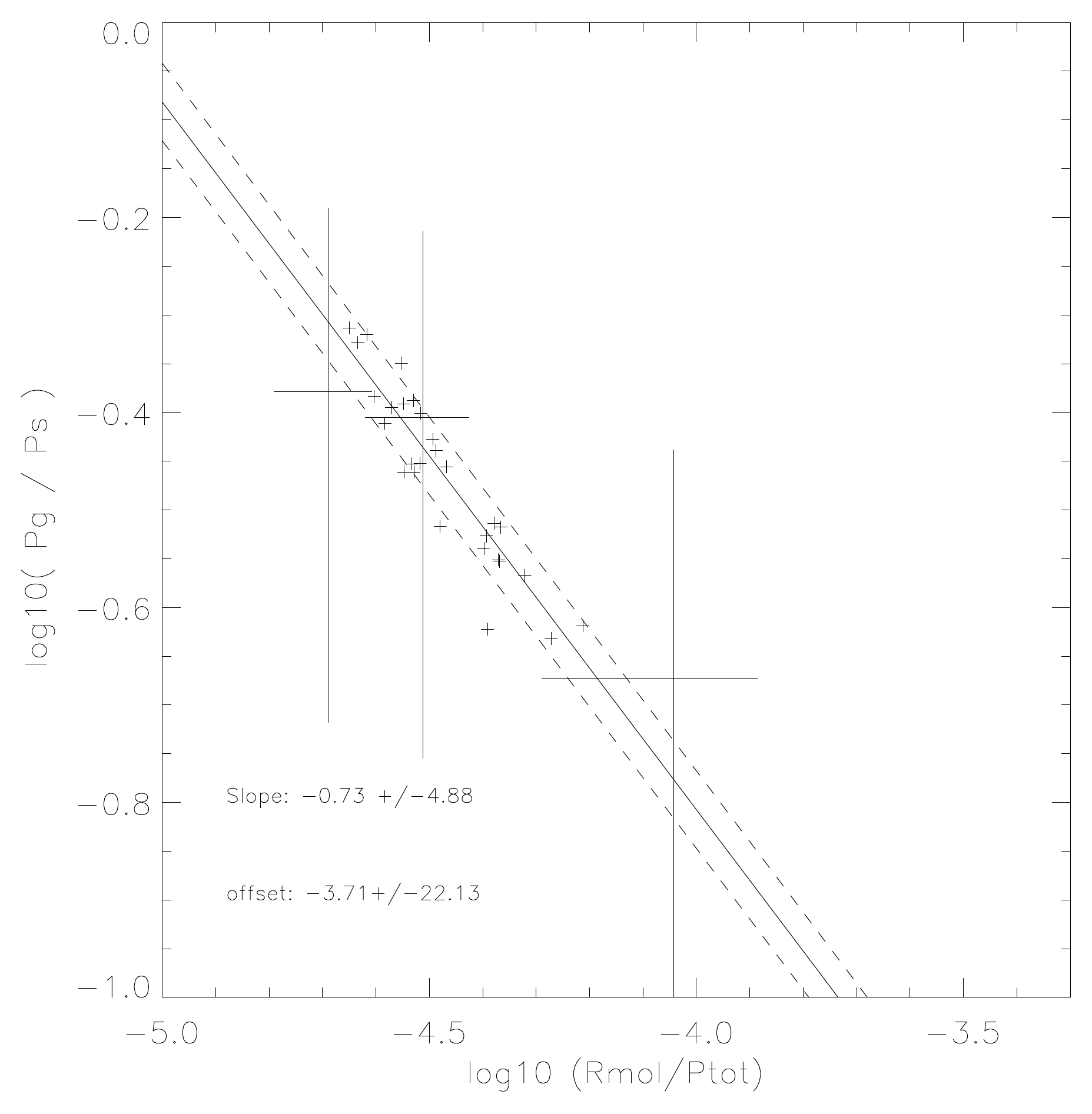}
   \caption{NGC~4567. Same panels as in Fig.~\ref{plot4501_2}.}
\label{plot4567_2}%
\end{figure*}
\begin{figure*}
   \centering
\includegraphics[width=7cm]{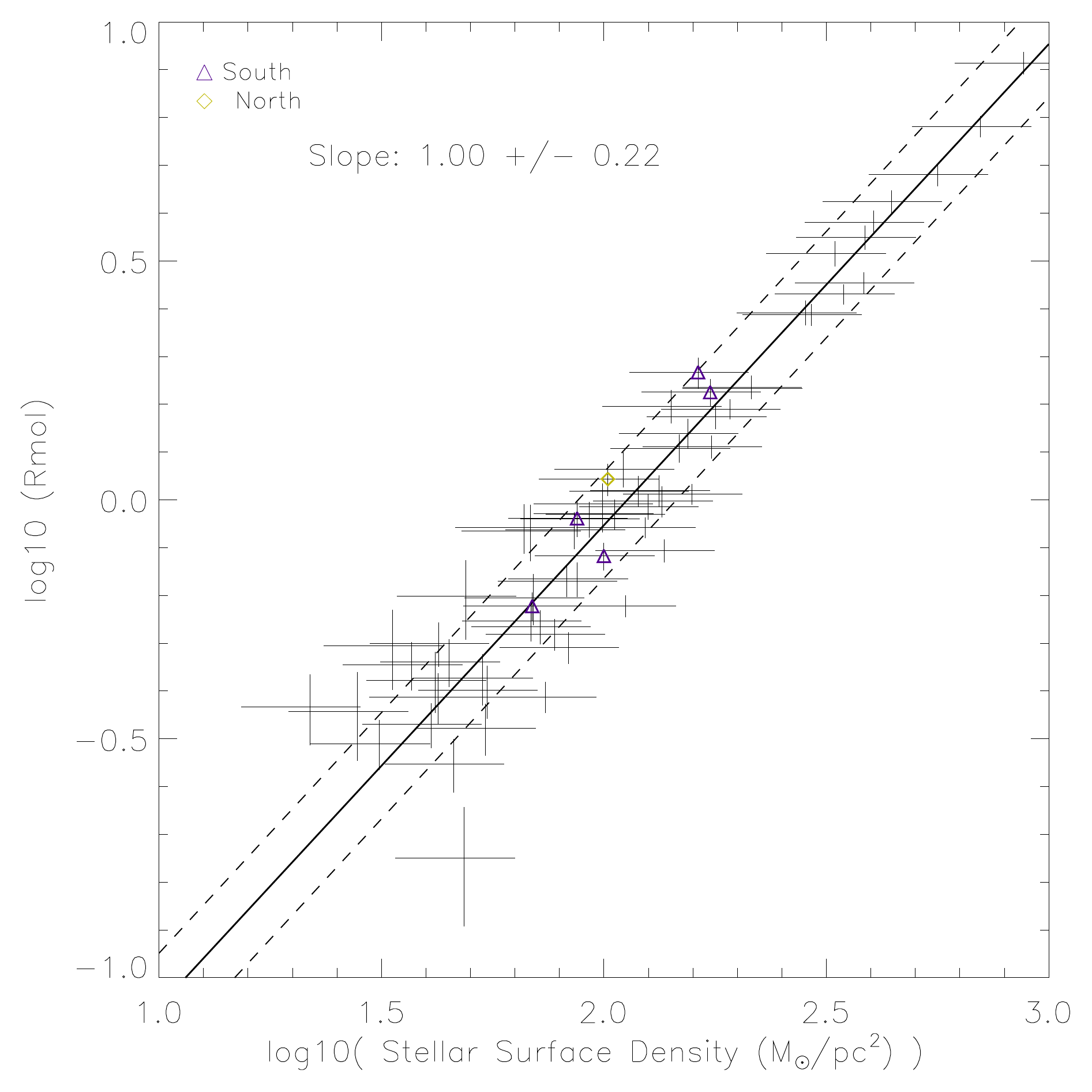}
\includegraphics[width=7cm]{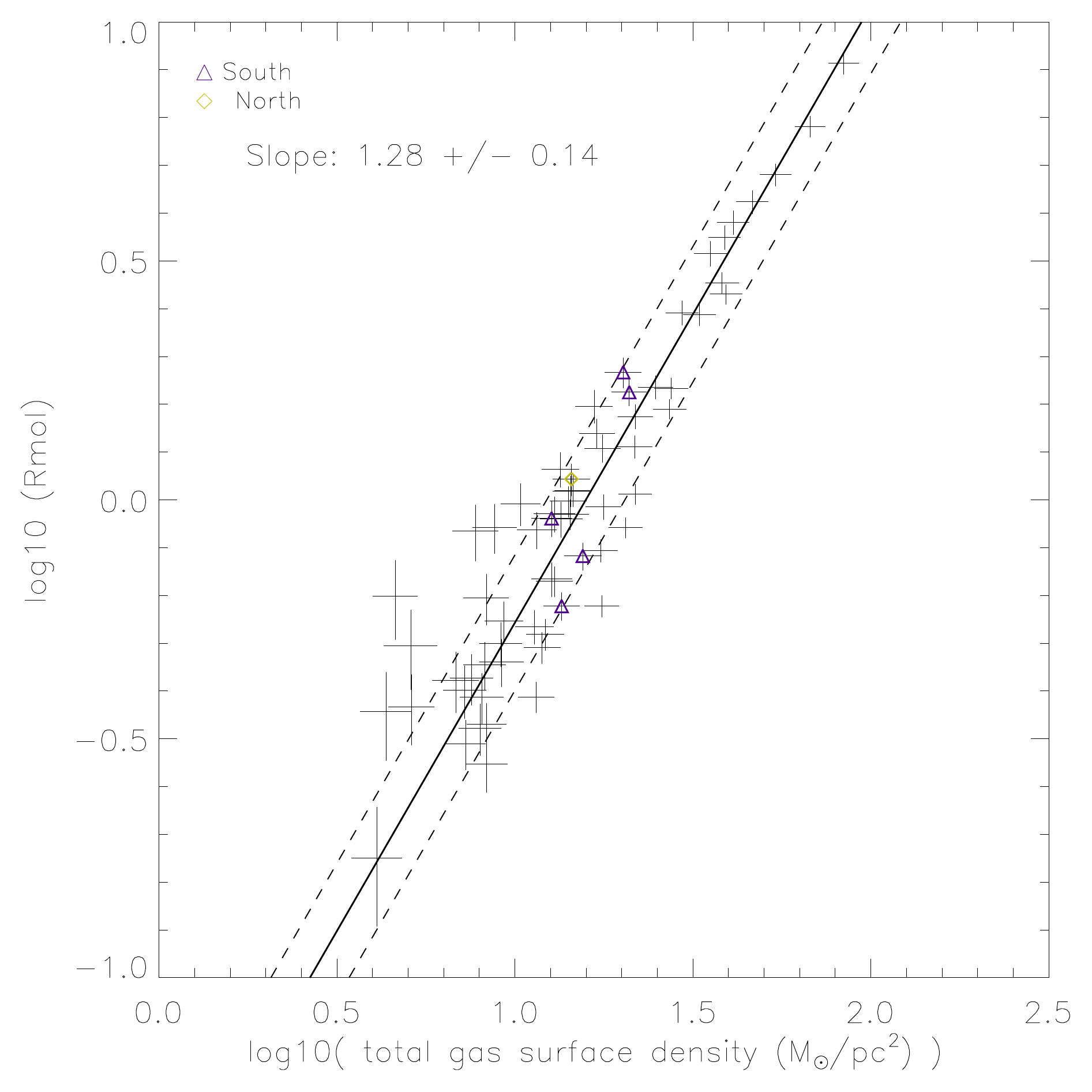}
\includegraphics[width=7cm]{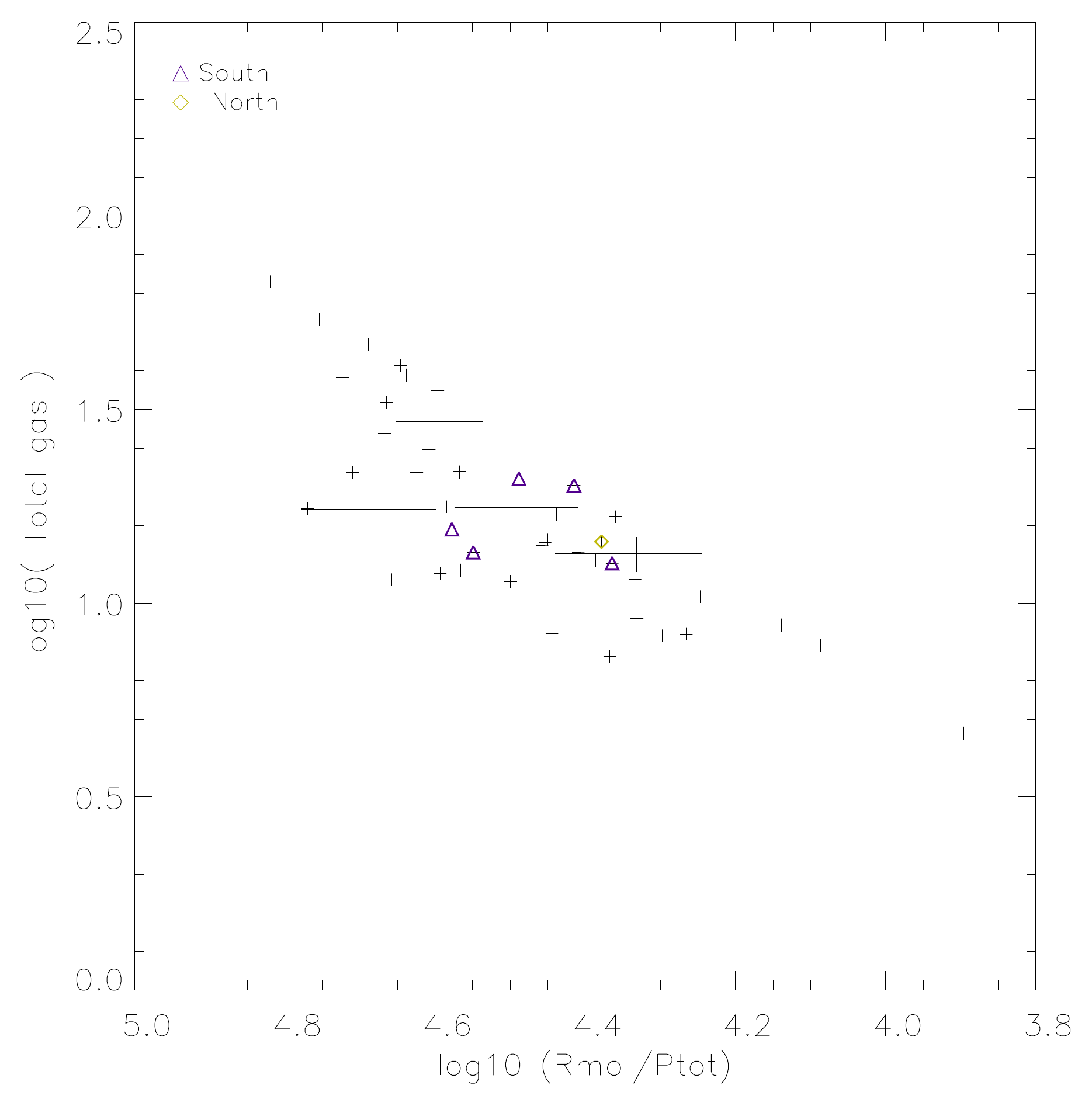}
\includegraphics[width=7cm]{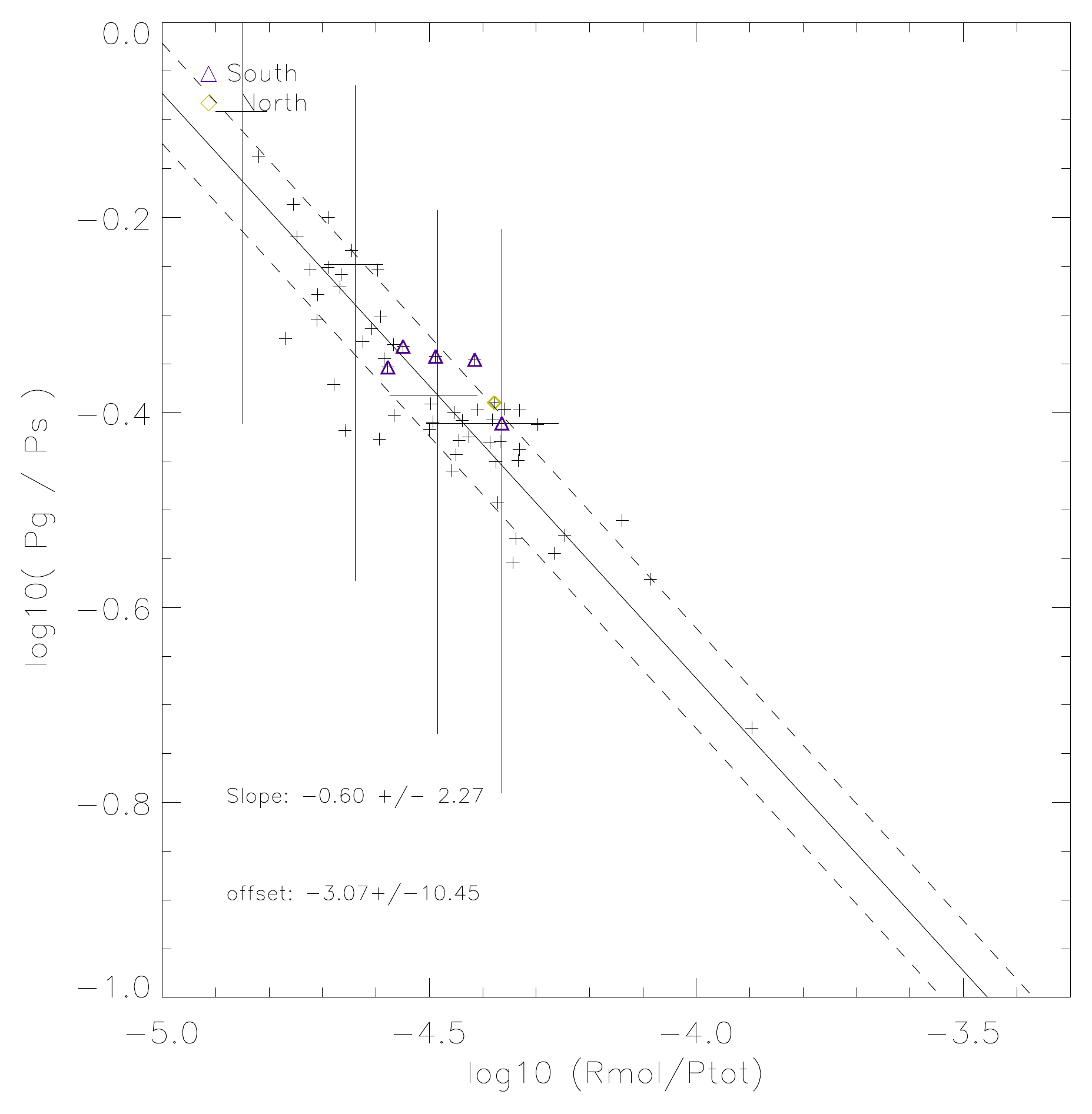}
   \caption{NGC~4568. Same panels as in Fig.~\ref{plot4501_2}.}
\label{plot4568_2}%
\end{figure*}
\begin{figure*}
   \centering
\includegraphics[width=7cm]{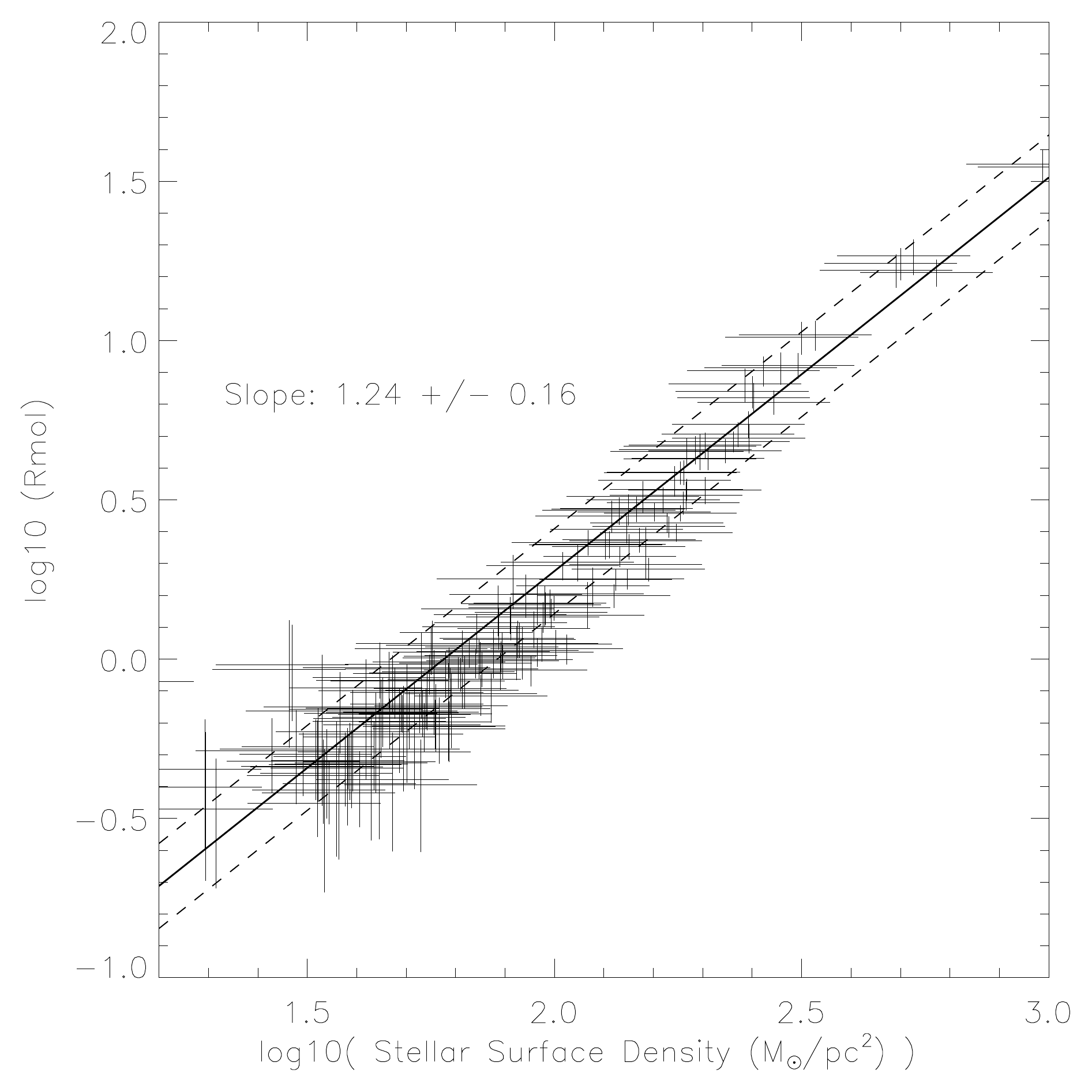}
\includegraphics[width=7cm]{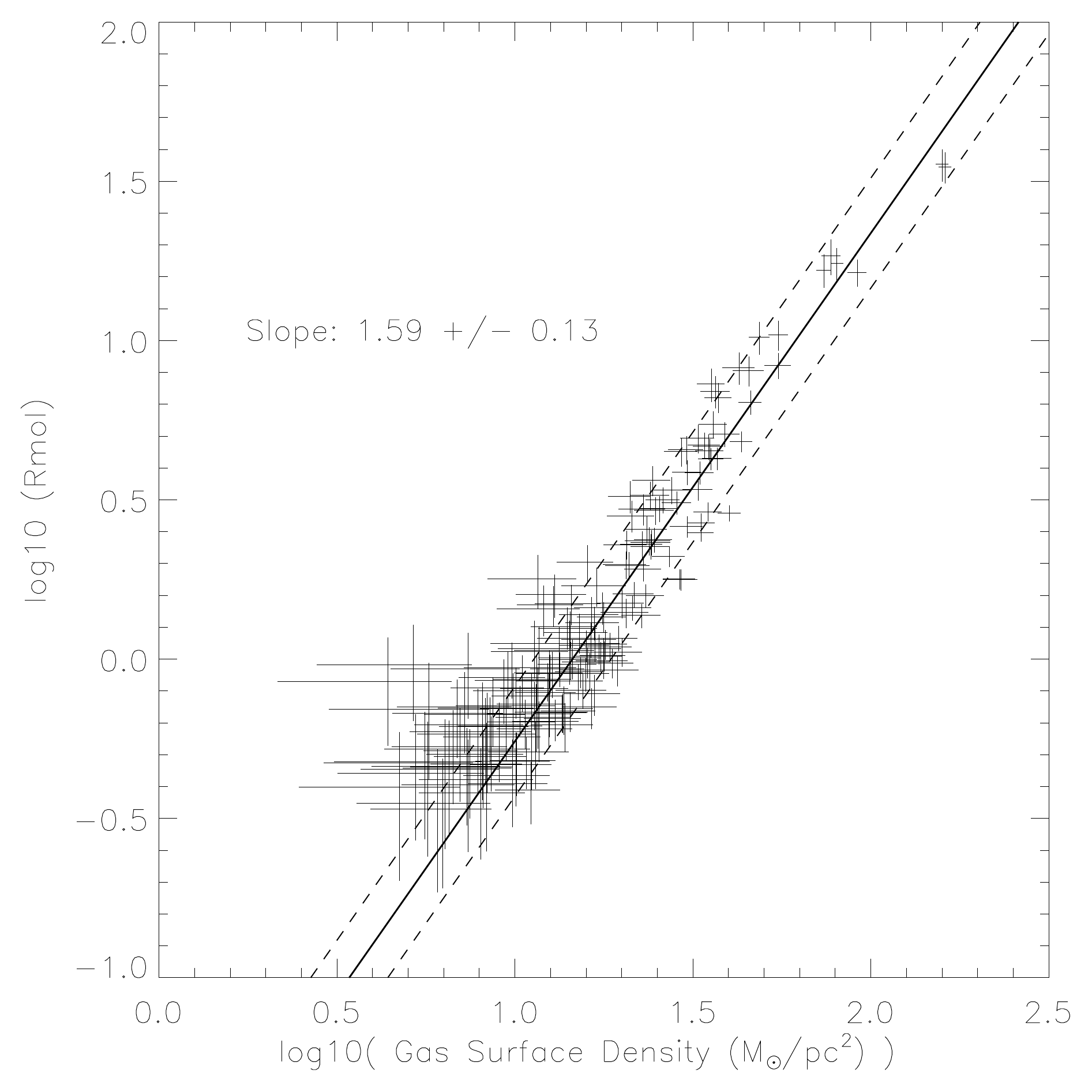}
\includegraphics[width=7cm]{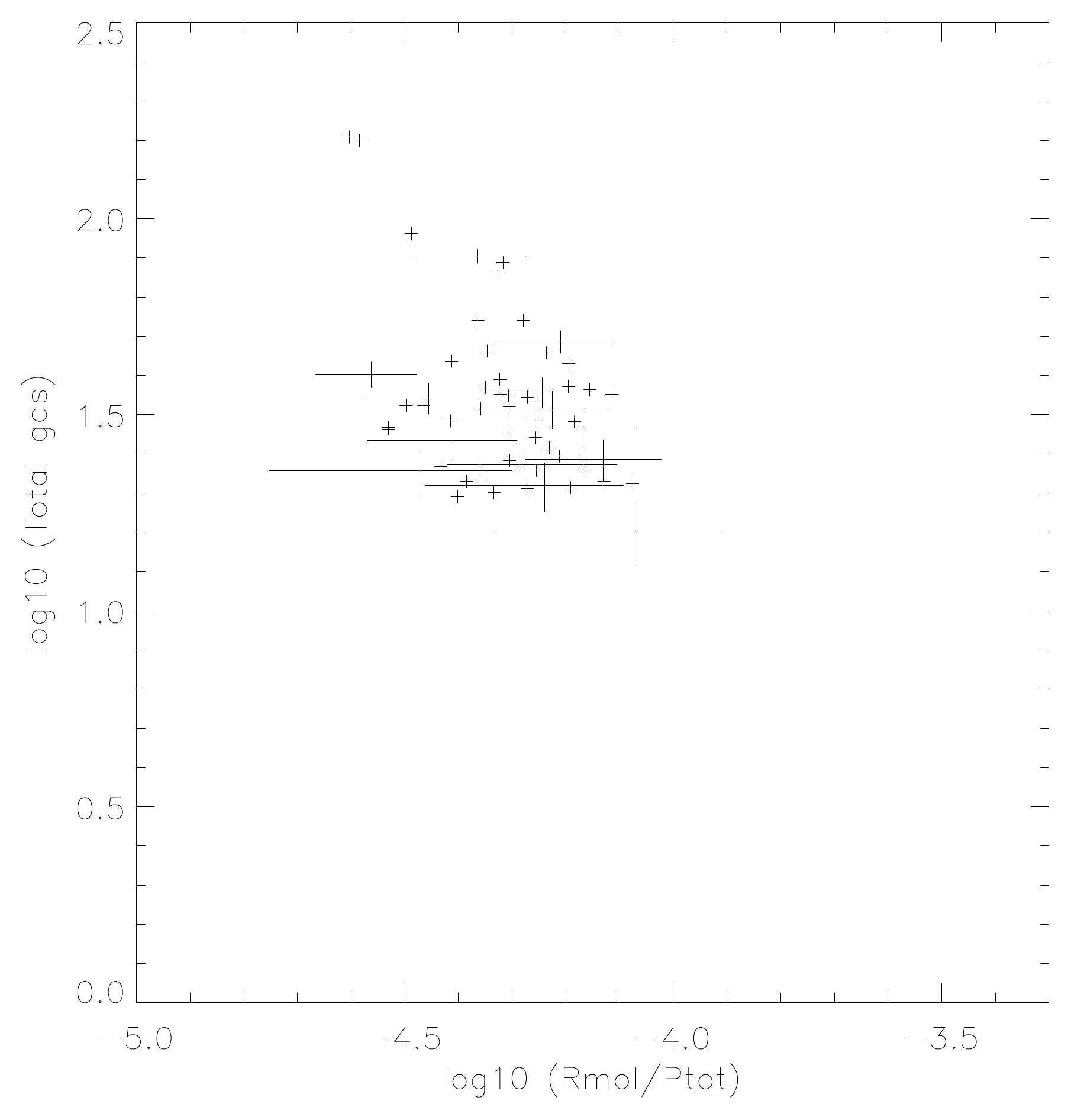}
\includegraphics[width=7cm]{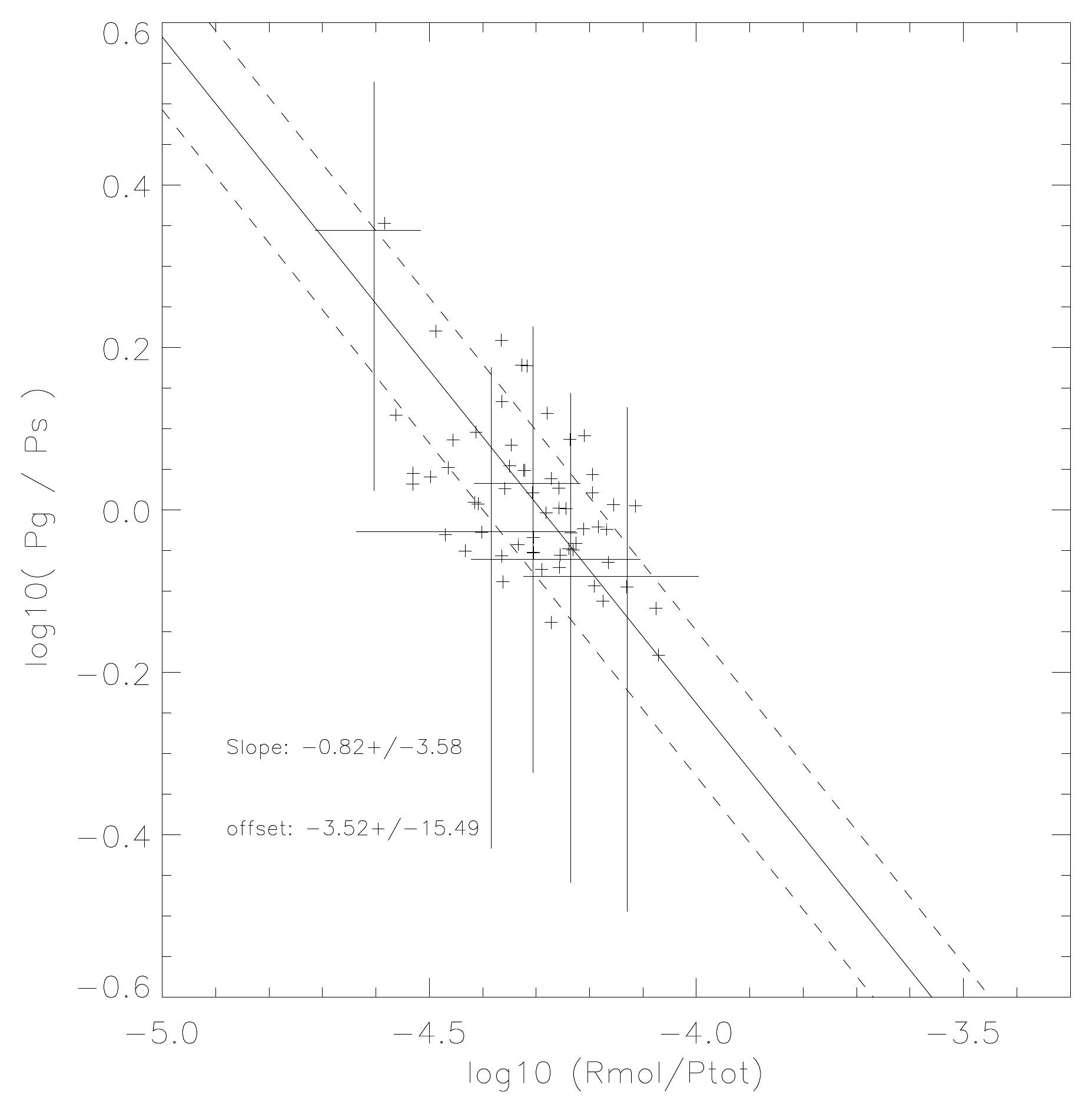}
   \caption{NGC~4321. Same panels as in Fig.~\ref{plot4501_2}}
\label{plot4321_2}%
\end{figure*}

\clearpage

\section{Individual plots III}
\begin{figure}
\includegraphics[width=7cm]{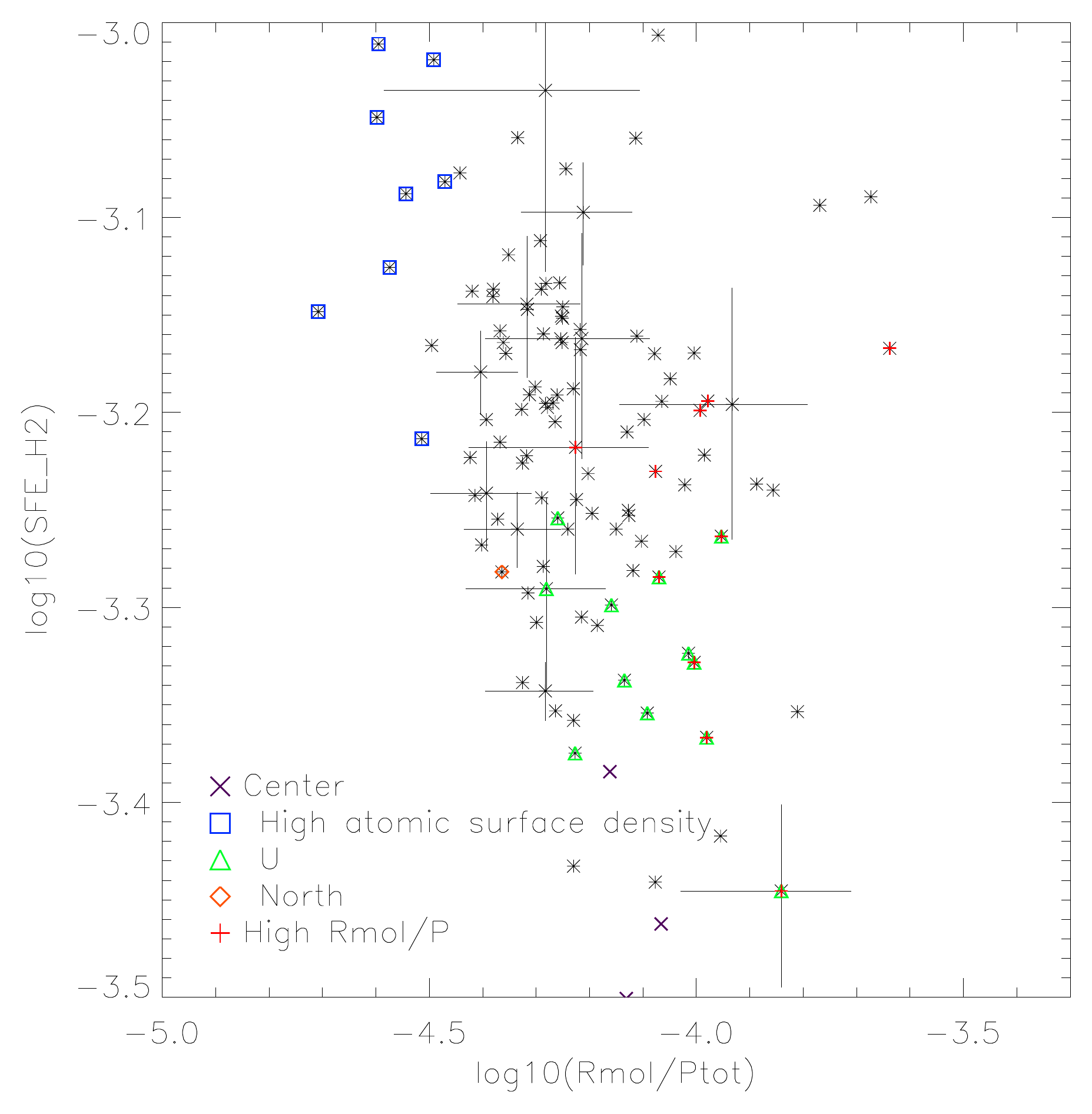}
\includegraphics[width=7cm]{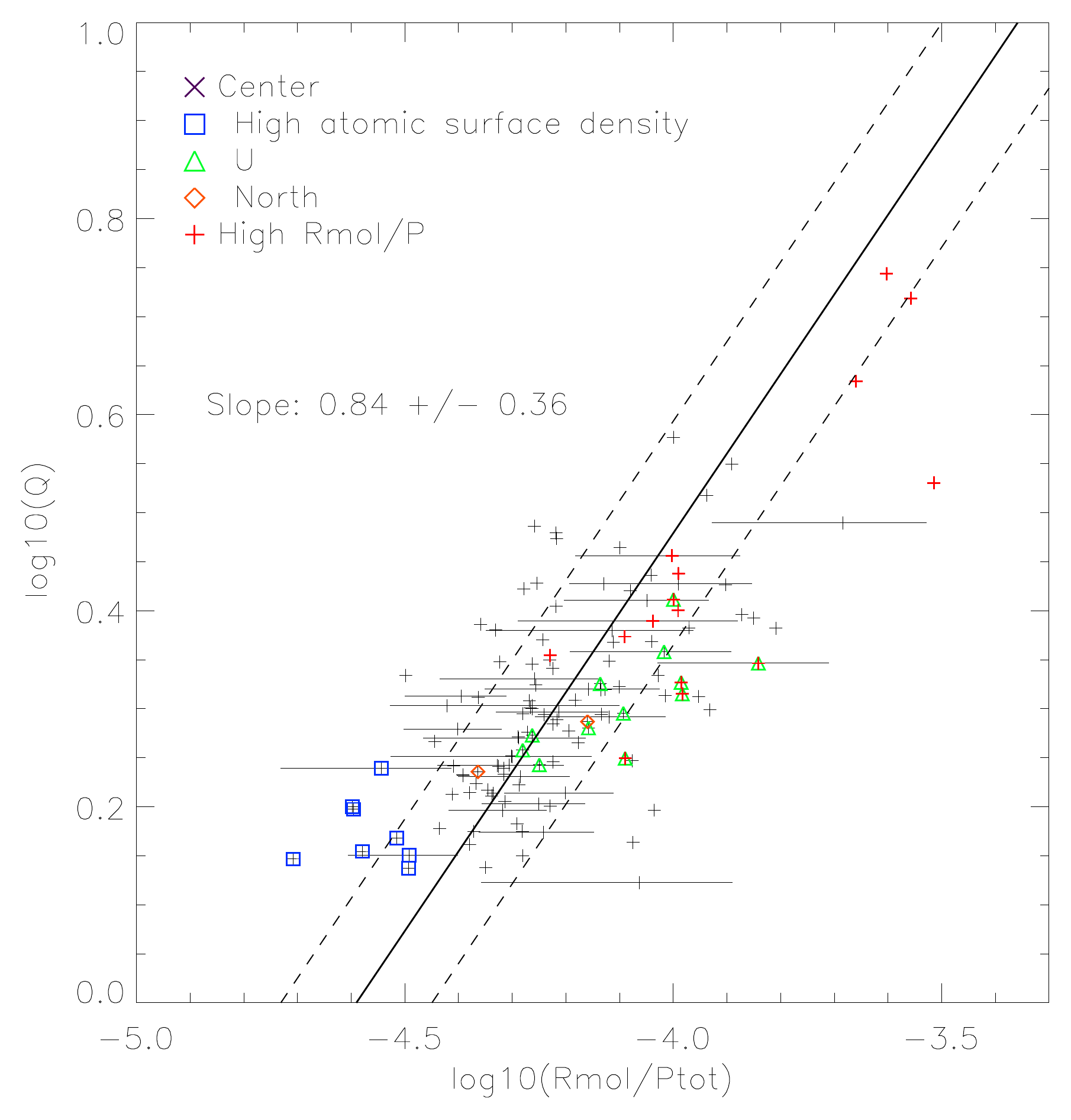}
\includegraphics[width=7cm]{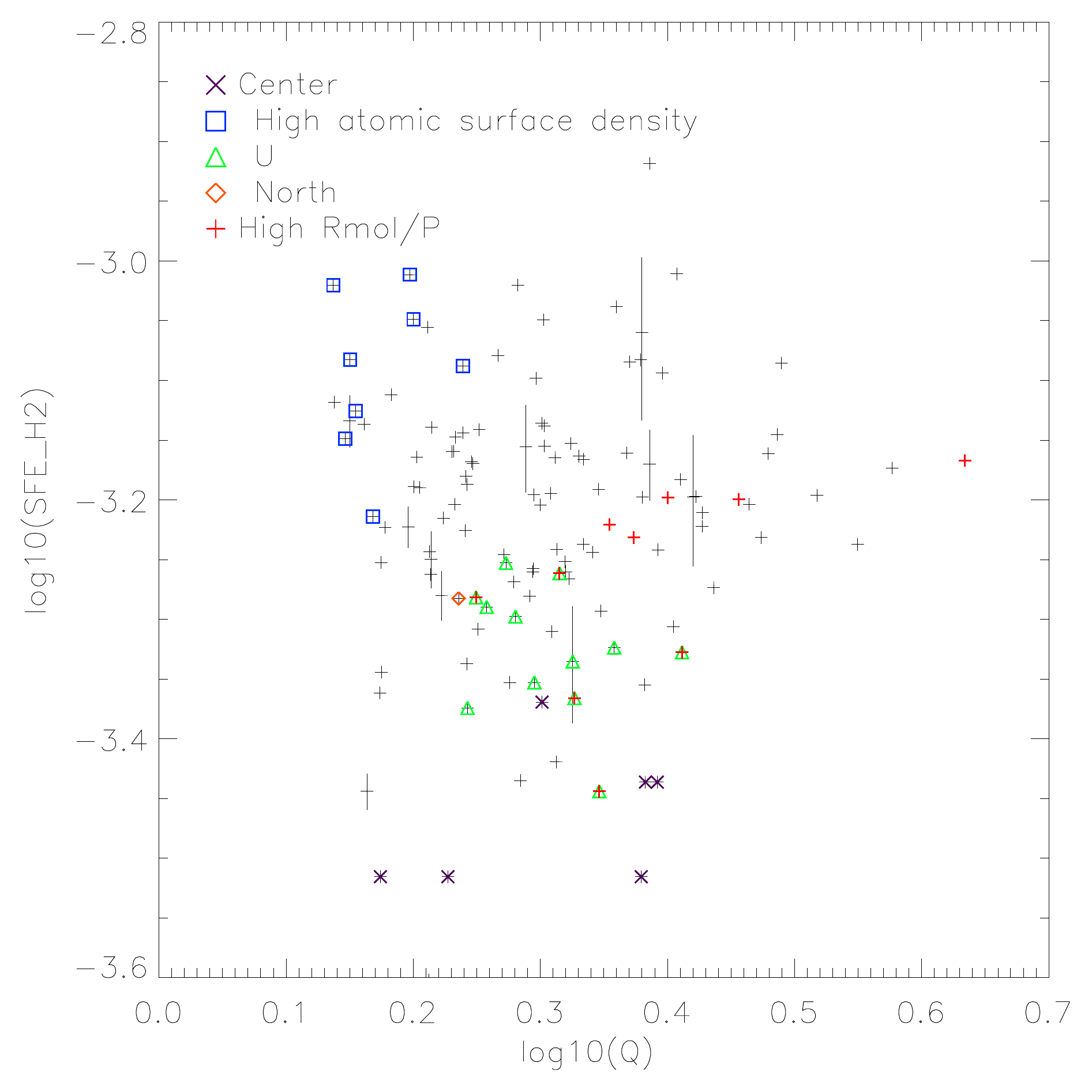}
\caption{NGC~4501. {\em Upper panel}: $SFE_{\rm H_{2}}$ as a function of the molecular fraction over ISM pressure. {\em Middle panel}: Toomre $Q$ as a function of $R_{\rm mol}/P_{\rm tot}$. {\em Lower panel}: $SFE_{\rm H_{2}}$ as a function of Toomre Q.}
\label{plot4501_3}%
\end{figure}
\begin{figure}
\includegraphics[width=7cm]{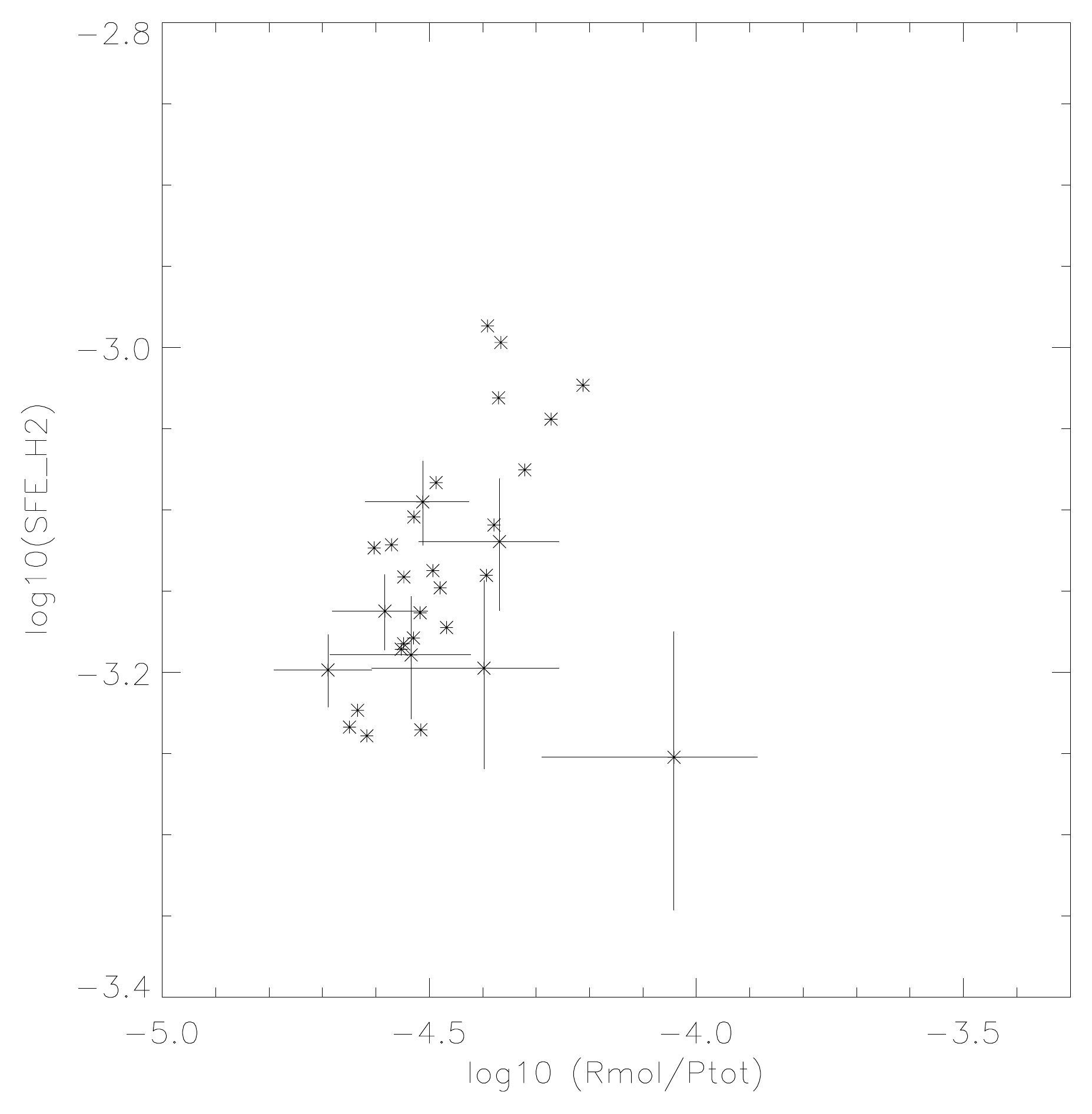}
\includegraphics[width=7cm]{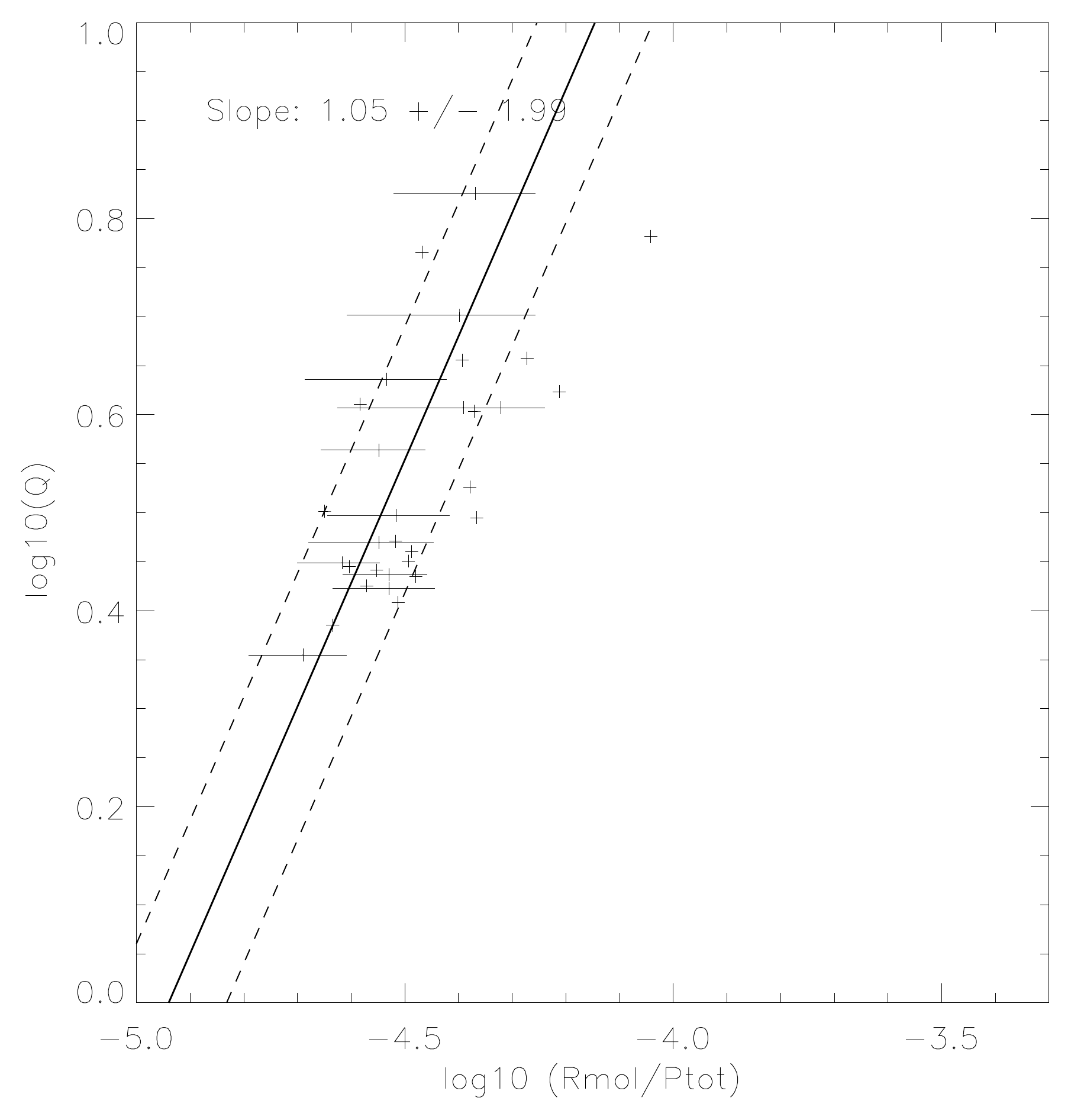}
\includegraphics[width=7cm]{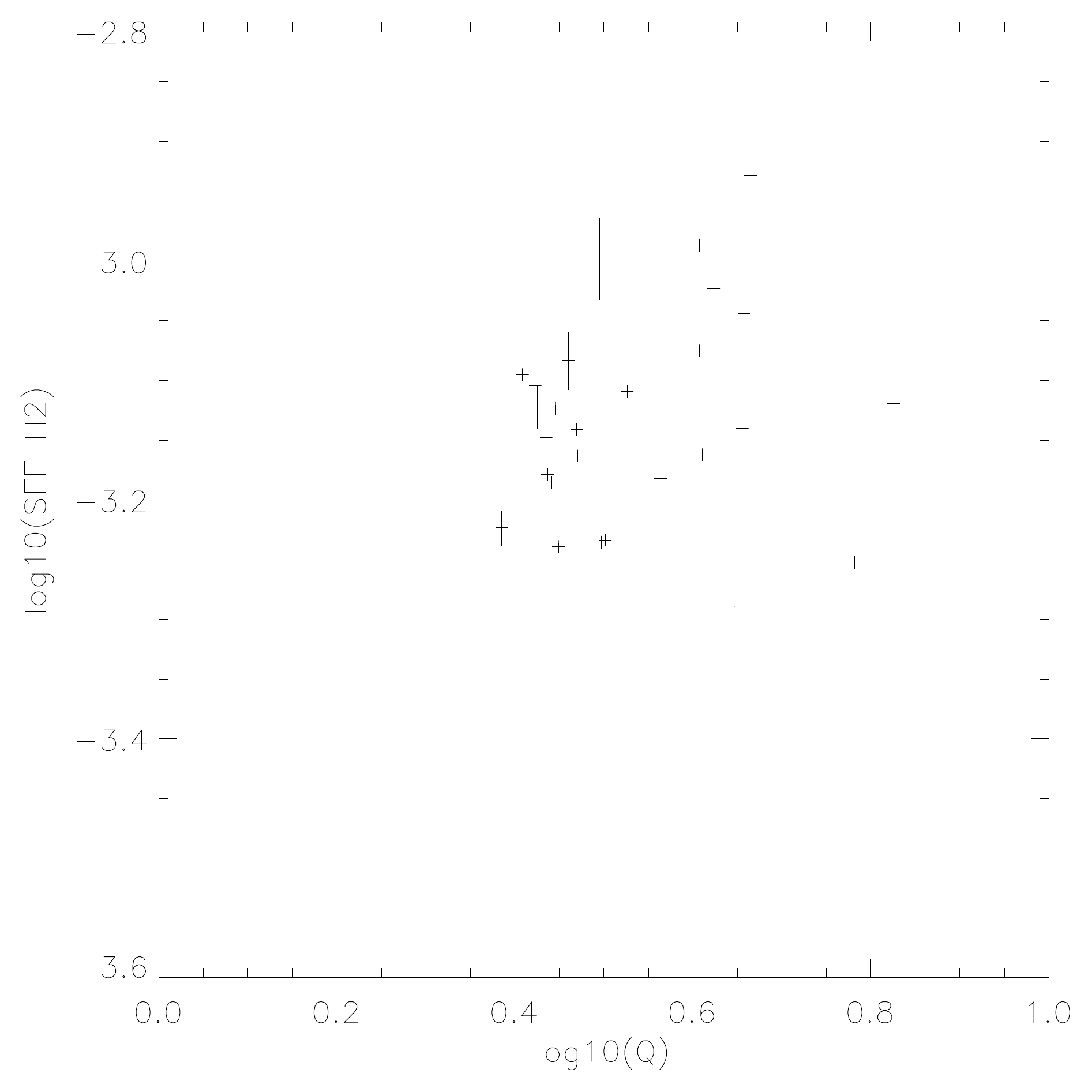}
   \caption{NGC~4567. Same panels as in Fig.~\ref{plot4501_3} }
\label{plot4567_3}%
\end{figure}
\begin{figure}
\includegraphics[width=7cm]{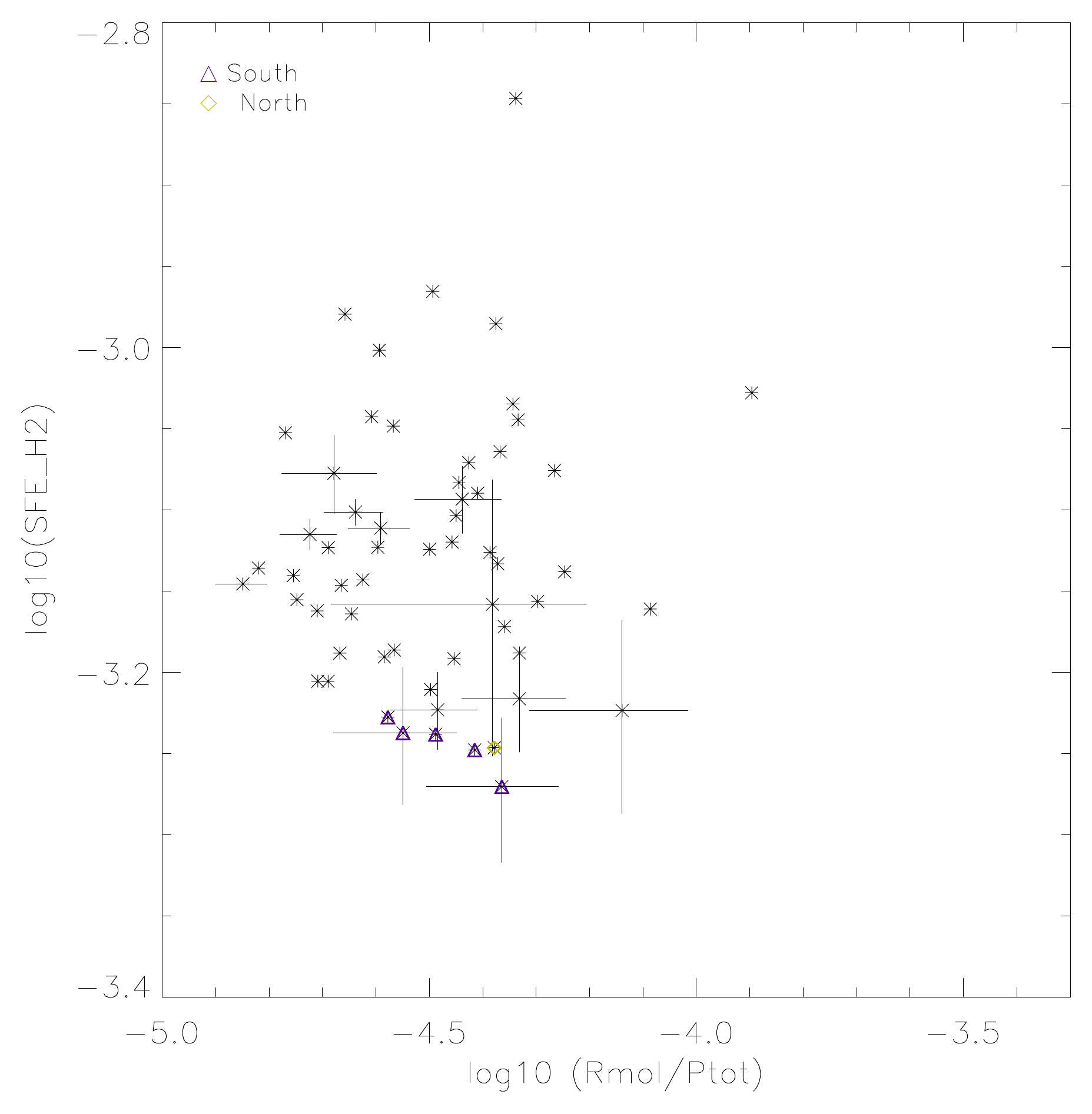}
\includegraphics[width=7cm]{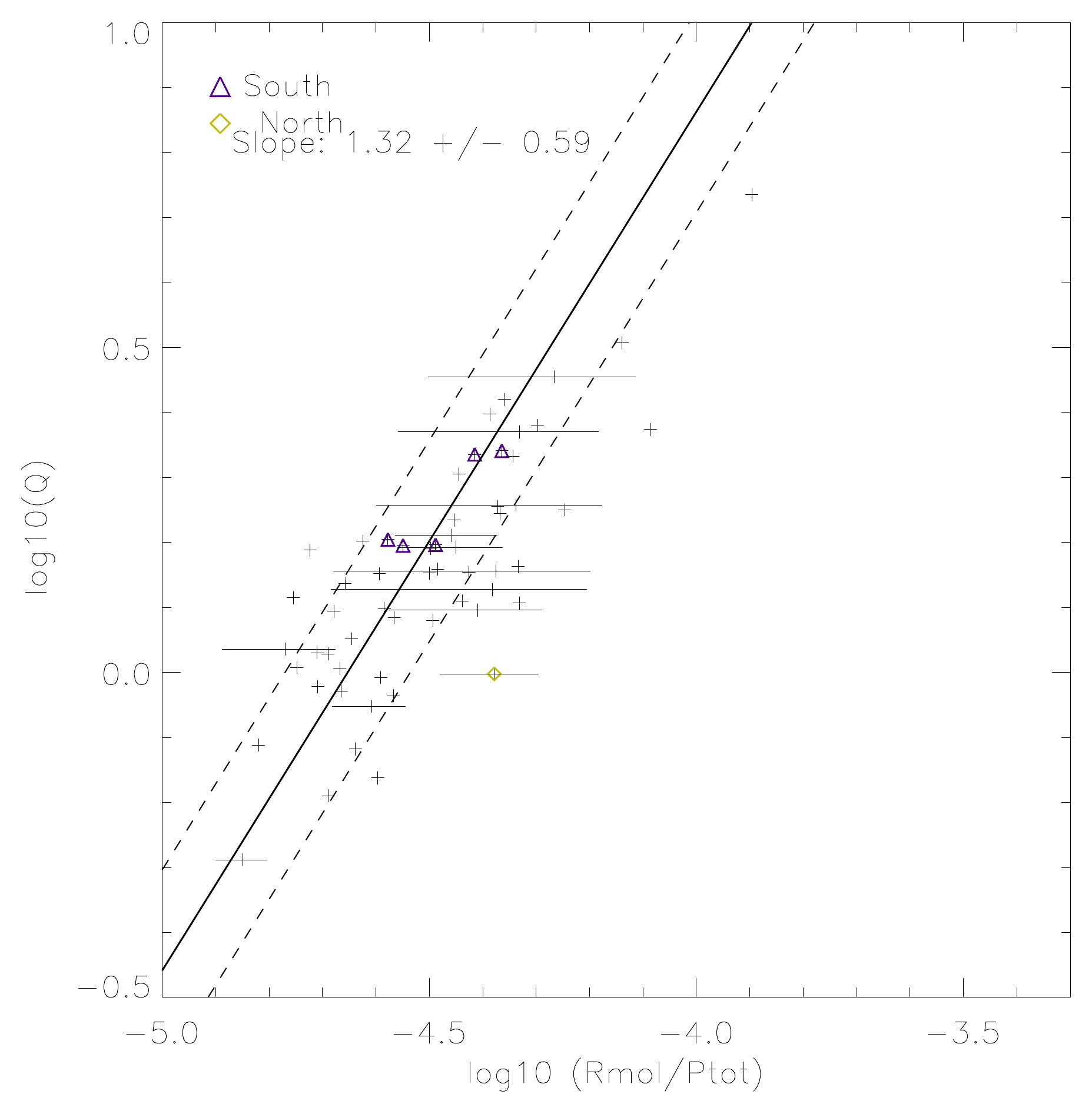}
\includegraphics[width=7cm]{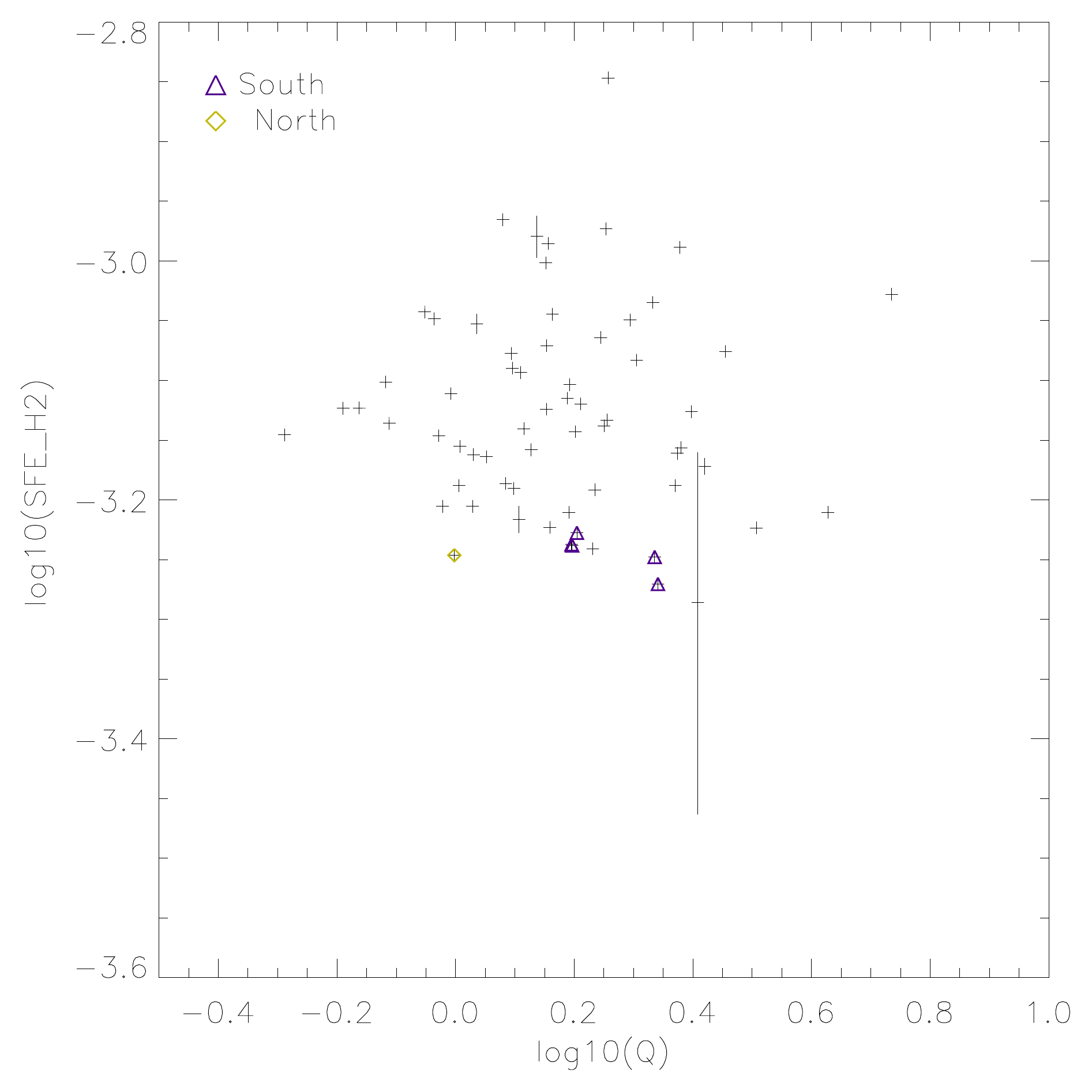}
   \caption{NGC~4568. Same panels as in Fig.~\ref{plot4501_3}}
\label{plot4568_3}%
\end{figure}
\begin{figure}
\includegraphics[width=7cm]{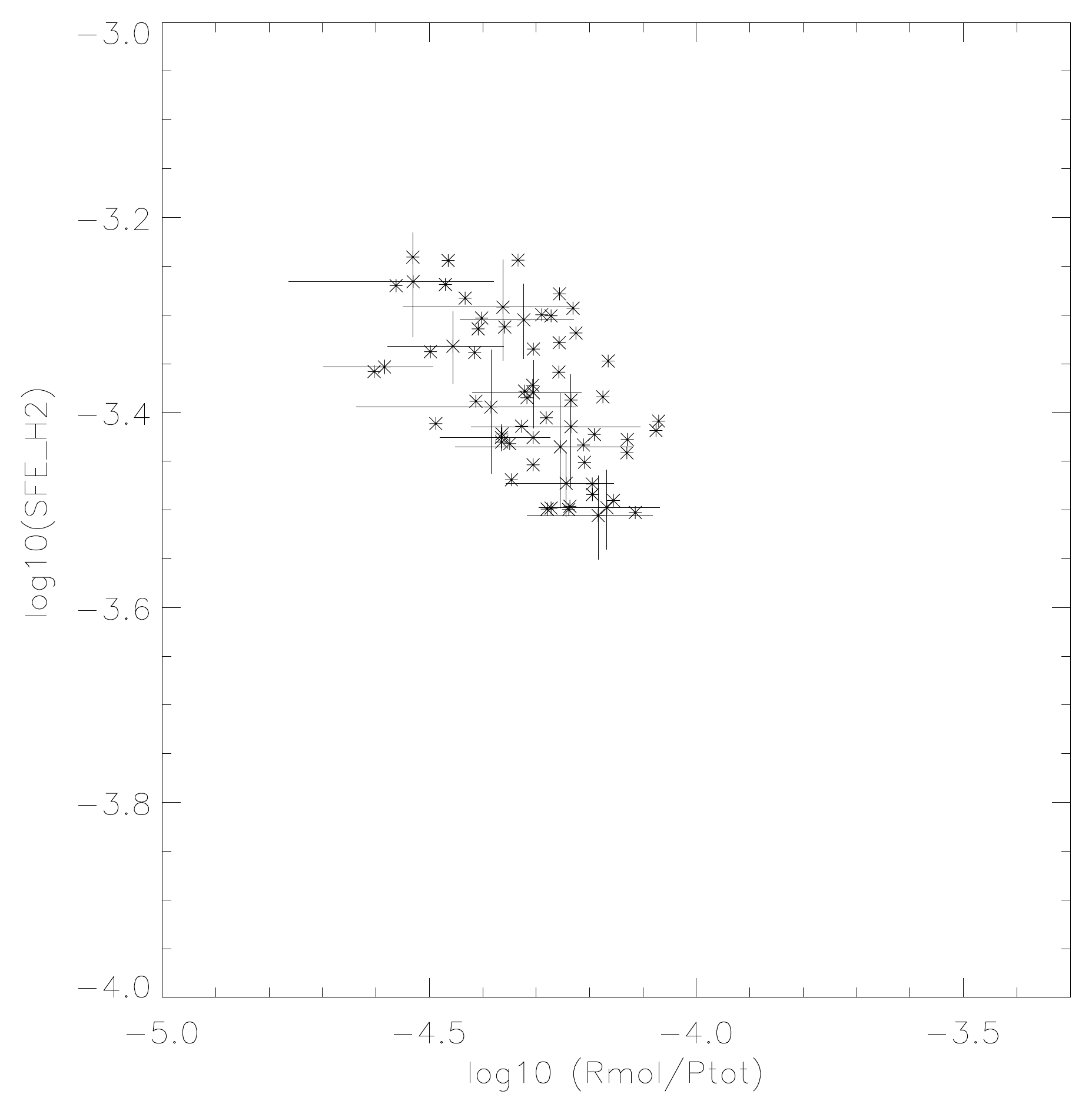}
\includegraphics[width=7cm]{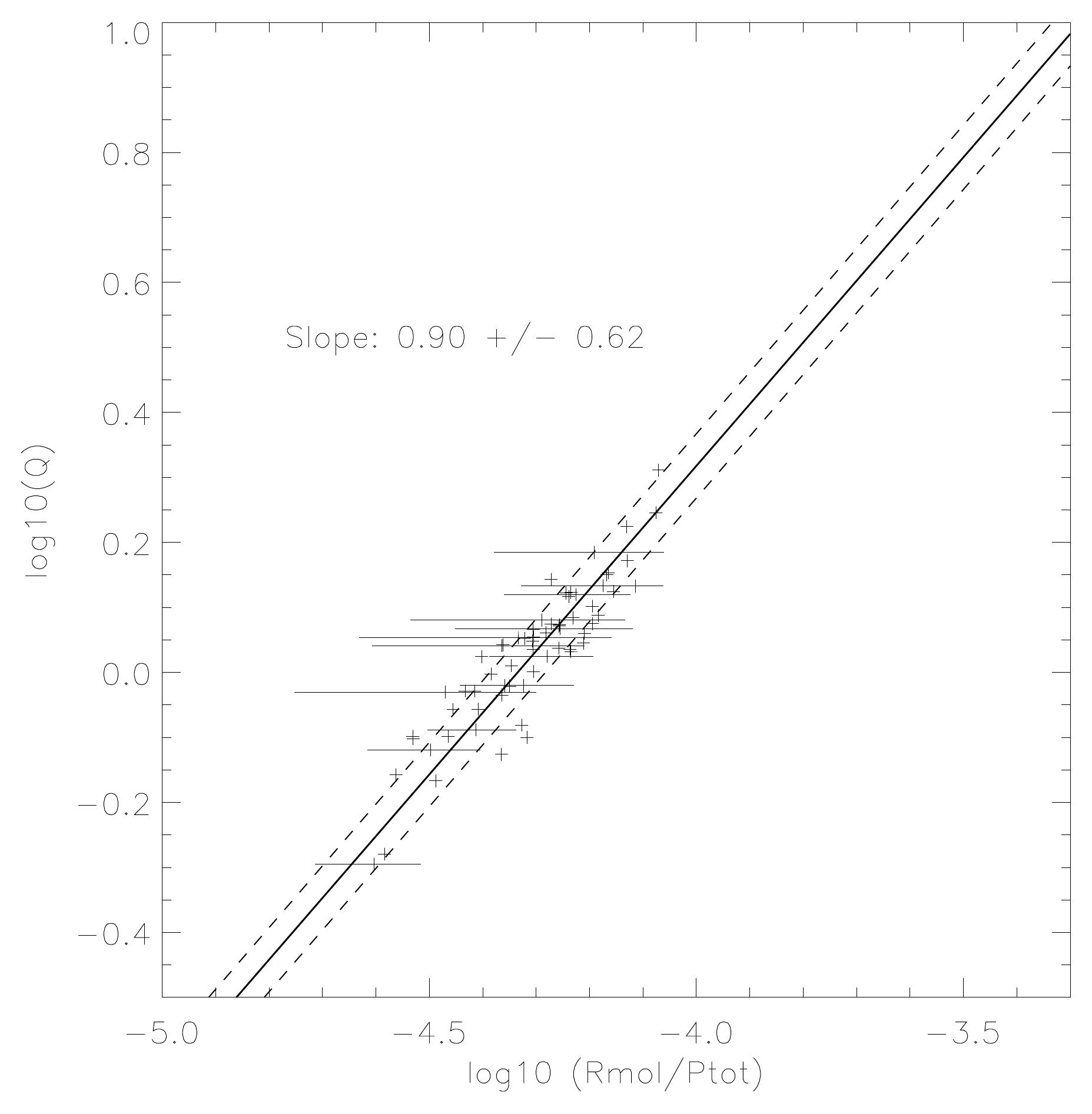}
\includegraphics[width=7cm]{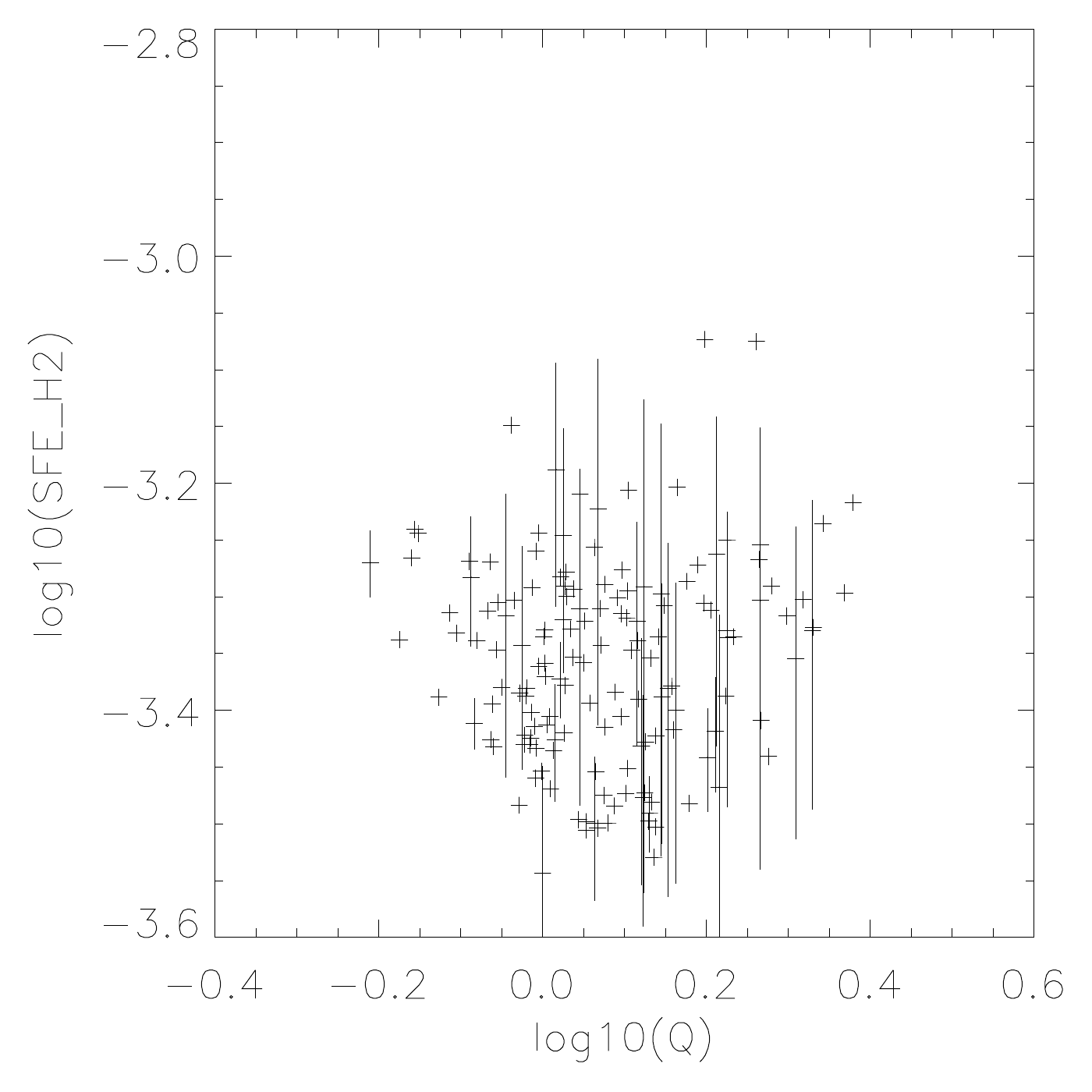}
   \caption{NGC~4321. Same panels as in Fig.~\ref{plot4501_3}}
\label{plot4321_3}%
\end{figure}

\end{document}